\documentclass[12pt]{cernrep}
\usepackage{subfigure}
\usepackage{floatflt}
\usepackage{multirow}
\usepackage{amssymb}
\usepackage[english]{babel}
\usepackage{axodraw}
\usepackage{epsfig}
\usepackage{here}
\usepackage{booktabs}
\usepackage{cite}
\usepackage{here}
\usepackage{graphicx}
\usepackage{color}
\begin{document}
\ifx\href\undefined\else\hypersetup{linktocpage=true}\fi
\setcounter{tocdepth}{0}
\pagestyle{myheadings}
\thispagestyle{empty}

\title{{\small \hfill{\mdseries SLAC-PUB-10365}}\\ \vskip 0.3cm 
LES HOUCHES ``PHYSICS AT TEV COLLIDERS 2003''
BEYOND THE STANDARD MODEL WORKING GROUP: SUMMARY REPORT}
\author{B.C.~Allanach$^{*1}$ (editor),      
A.~Aranda$^{2}$,
H.~Baer$^3$,            
C.~Bal\'azs$^4$,        
M.~Battaglia$^{*5}$,
G.~B\'elanger$^1$,
F.~Boudjema$^1$, 
K.~Desch$^{42}$,
J. L. D\'{\i}az-Cruz$^{7}$,
M. Dittmar$^{8}$,
A.~Djouadi$^9$, 
G.~Dewhirst$^{10}$,
D.~Dominici$^{11}$,
M.~Escalier$^{12}$,
L.~Fan$\grave{o}^{13}$,
S.~Ferrag$^{12,24}$,
S. Gascon-Shotkin$^{7}$,
S.~Gennai$^{14}$,
R.~Godbole$^{15}$,
J.~Guasch$^{16}$,
M.~Guchait$^{17}$,
J.~Gunion$^{18}$,
S.~Heinemeyer$^{11,19}$,      
J.~Hewett$^{20}$,
J.~Kalinowski$^{21}$,
K.~Kawagoe$^{22}$,
W.~Kilian$^{6}$,          
J-L.~Kneur$^9$,       
S.~Kraml$^{11}$,
R.~Lafaye$^{23}$,
B.~Laforge$^{12}$,
C.G.~Lester$^{25}$,
Y.~Mambrini$^{26}$,
K.~Mazumdar$^{17,11}$,
F.~Moortgat$^{27}$,     
G.~Moortgat-Pick$^{28}$,
S.~Moretti$^{29}$,      
M.~M\"uhlleitner$^{16}$, 
A.-S.~Nicollerat$^{8}$,
A.~Nititenko$^{11}$,
M.~Nojiri$^{*30}$,
T.~Plehn$^{11}$,
G.~Polesello$^{31}$,
W.~Porod$^{32}$,        
D.~Prieur$^{23}$,
A.~Pukhov$^{33}$,       
O. Ravat$^{34}$,
P.~Richardson$^{28}$,
T.G.~Rizzo$^{*20}$,
A.~de Roeck$^{*11}$,
S.~Schumann$^{35}$,     
P.~Skands$^{36}$,
P.~Slavich$^{37}$,      
M.~Spira$^{16}$, 
M.~Spiropoulu$^{*38}$,
K.~Sridhar$^{17}$,
D.R.~Tovey$^{*39}$,
G.~Weiglein$^{28}$,
J.D.~Wells$^{40}$,
D.~Zerwas$^{41}$
}

\institute{\vspace{1cm}
$^*$ convenor of {\em Beyond the Standard Model} working group\\
$^1$ LAPTH, Annecy-le-Vieux, France\\
$^2$ Facultad de Ciencias, Universidad de Colima, Colima, Mexico\\
$^3$ Dept. of Physics, Floride State University, Tallahassee, USA\\
$^4$ HEP Division, Argonne National Laboratory, Argonne, USA\\
$^5$ University of California, Berkeley, USA\\
$^6$ Theory Group, DESY, Hamburg, Germany\\
$^{7}$ Instituto de F\'{\i}sica, BUAP, Puebla, Mexico\\
$^{8}$ IPP, ETH Zurich, Zurich, Switzerland\\
$^9$ LPMT, Universit\'{e} Montpellier II, Montpellier, France\\
$^{10}$ Imperial College, London, UK\\
$^{11}$ CERN, Geneva, Switzerland\\
$^{12}$ Universit\'{e} Paris VI \& VII, Paris, France\\
$^{13}$ INFN, Perugia, Italy\\
$^{14}$ INFN, Pisa, Italy\\
$^{15}$ CTS, Indian Institute of Science, Bangalore, India\\
$^{16}$ Paul Scherrer Institut, Villigen PSI, Switzerland\\
$^{17}$ Tata Institute of Fundamental Research, Mumbai, India\\
$^{18}$ Physics Dept., University of California, Davis, USA\\
$^{19}$ LMU Munich, Munich, Germany\\
$^{20}$ SLAC, Stanford, USA\\
$^{21}$ Institute of Theoretical Physics, Warsaw University, Warsaw, Poland\\
$^{22}$ YITP, Kyoto University, Japan\\
$^{23}$ LAPP, Annecy-le-Vieux, France\\
$^{24}$ Dept. of Physics, University of Oslo, Oslo, Norway\\
$^{25}$ Cavendish Laboratory, University of Cambridge, Cambridge, UK\\
$^{26}$ LPT, Universit\'{e} Paris-Sud, Orsay, France\\
$^{27}$ Universtity of Antwerp, Antwerp, Belgium\\
$^{28}$ IPPP, University of Durham, Durham, UK \\
$^{29}$ School of Physics and Astronomy, University of Southampton,
  Southampton, UK\\
$^{30}$ Department of Physics, Kobe University, Kyoto, Japan\\
$^{31}$ INFN, Pavia, Italy\\
$^{32}$ Institut fur Theoretische Physik, University of Zurich, Zurich,
  Switzerland\\
$^{33}$ Moscow State University, Russia\\
$^{34}$ Institut de Physique Nucl\'eaire de Lyon, Villeurbanne, France\\
$^{35}$ Institut fur Theoretische Physik, TU Dresden, Dresden, Germany\\
$^{36}$ Theoretical Physics, Lund University, Lund, Sweden\\
$^{37}$ Max Planck Institut fur Physik, Munchen, Germany\\
$^{38}$ Fermilab, Batavia, USA\\
$^{39}$ Department of Physics and Astronomy, University of Sheffield,
  Sheffield, UK\\
$^{40}$ MCTP, University of Michigan, Ann Arbor, USA\\
$^{41}$ LAL, Orsay, France\\
$^{42}$ Institut f\"ur Experimentalphysik, Universit\"at Hamburg,
  Hamburg, Germany\\
\vspace{1cm}
}
\maketitle
\begin{abstract}
The work contained herein constitutes a 
report of the ``Beyond the Standard Model'' working group for the Workshop 
``Physics at TeV Colliders", Les Houches, France, 26 May--6 June, 2003.
The research presented is original, and was performed specifically 
for the workshop. 
Tools for calculations in the minimal supersymmetric standard model are
presented, including a comparison 
of the dark matter relic density predicted by public codes.
Reconstruction of supersymmetric particle masses at the LHC and a future
linear collider facility is examined. Less orthodox supersymmetric signals
such as non-pointing photons and R-parity violating signals are studied. 
Features of extra dimensional models are examined next, including measurement
strategies for radions and Higgs', as well as the virtual effects of Kaluza
Klein modes of gluons. An LHC search strategy for a heavy top found in many
little Higgs model is presented and
finally, there is an update on LHC $Z'$ studies. 
\end{abstract}

\vfill

\section*{Acknowledgements}

We would like to heartily thank the funding bodies, organisers, staff and other
participants of the Les Houches workshop for providing a stimulating and
lively environment in which to work. 

\newpage

\tableofcontents

\newpage

\part{Introduction}
{\it B.C.~Allanach}

\maketitle \vspace{1cm}

The workshop took place at the \'{e}cole de physique in the lee of Mont
Blanc, and lasted for two weeks. Computer systems were installed by the
helpful LAPP staff for use by participants throughout the workshop. 
The first two days consisted of plenary talks intended to stimulate ideas and
the proposition of projects. A couple of further plenary seminars on 
hot topics occured sporadically later in the workshop.
The Beyond the Standard Model working group (convened by M Battaglia, M Nojiri,
T Rizzo, A de Roeck, D Tovey, M Spiropoulu and the author)
held a meeting on the second evening
to set individual projects for small groups of interested parties. 
This report contains a summary of the fruits of participants' labour during
and after the workshop on those projects. The projects were phenomenological
studies of both supersymmetric and non-supersymmetruc models and.
The report first discusses the studies of
supersymmetric models and then those of extra dimensions. We close with an
update on $Z'$ studies.

At the time of the workshop the Tevatron and DESY collider runs were
proceeding, and the start-up of the Large Hadron Collider (LHC)
was eagerly awaited. Many hopes are
concentrating upon the production and detection of supersymmetric (SUSY)
particles at these colliders. Although an immense amount of literature has
been accumulated on SUSY phenomenology, there is still a large amount of work
to do 
because of its complexity and the abundance of models of
SUSY breaking. The models predict different cascade
chains leading to radically differing signals in experiments. 
In order to facilitate the phenomenological study of SUSY models, we need
to both calculate the sparticle spectrum and also we need to simulate events. 
These two calculations are typically 
performed by seperate calculational tools. For each calculation, there exist
several competing tools performing the same task with different
approximations or assumptions. A common interface between the different tools
has clear advantages, and much time was 
spent at (and after) the workshop arguing, debating and negotiating the
various conventions. The write-up of this ``SUSY Les Houches Accord''
constitutes part~\ref{accord12} of this report. A form-based web tool is
presented in part~\ref{susycom} in which one can determine the spectra from
mSUGRA models via the different 
public sparticle spectrum generators. The difference between the predictions
of the different generators gives an idea of the theoretical uncertainties
involved in the calculation. 
Many
recent works have used the WMAP determination of the cold dark matter density
$\Omega_{CDM}$ to vastly restrict the minimal supersymmetric standard model
(MSSM) parameter space. 
An initial study on the uncertainty induced from sparticle masses upon
the prediction of the dark matter density is then entered in part~\ref{dm}. 
Supposing supersymmetric particles are measured in colliders, a $\chi^2$ fit
to the observables (such as the masses) will restrict the SUSY breaking
parameter space, and discriminate models of SUSY breaking. A tool which
enables one to perform this fitting efficiently is presented in part~\ref{sfitter}.
A new code to determine the branching ratios and decays of SUSY
particles is presented in part~\ref{sdecay}.

The report then turns to issues surrounding the measurement of sparticle mass
and mixing parameters. The measurement of the lightest chargino mass in a
universal minimal supergravity (mSUGRA) model is presented in
part~\ref{chargino}. 
It is pointed out in part~\ref{leshouches_fin} that
{\em without}
assumptions about SUSY breaking,
information coming 
from a future linear collider facility could be extremely useful when
analysing neutralino and chargino signals at the LHC.
 A bold new reconstruction technique for SUSY
processes at the LHC is proposed in part~\ref{new6}. Normally, it is not
possible 
to reconstruct the neutralino momenta involved in R-parity conserving events
since they remain undetected and the overall energy of the hard collision is
not known. In the new technique, particularly long SUSY cascade chains are
identified and lead to an over-constrained system when pairs of events are
considered. The idea is pushed further to ensembles of more than 20 events in
part~\ref{chris}. If the idea stands up to further scrutiny, 
the new method would be the one yielding the most information about sparticle
masses (provided the relevant decay chain(s) is(are) present).

We next consider non-standard supersymmetric signatures, such as (part~\ref{prieur})
non-pointing 
photons at the CERN LHC, which are often predicted in gauge mediated SUSY
breaking models.  R-parity violation provides an oppurtunity to understand
neutrino masses without the need for adding gauge singlets, and also to
correlate neutrino oscillation observables with SUSY collider signatures. 
Part~\ref{LH_BRPV} shows that the scenario predicts a relation between branching ratios
of lightest SUSY particle (LSP) decay modes given the atmospheric neutrino
mixing angle, providing a 
useful test. In part~\ref{BMS_les}, resonant slepton production at the LHC is examined
in scenarios with ultra-light gravitinos. 

Models and signals incorporating extra dimensions are then considered.
Many extra dimension models
predict an additional higgs-like scalar (the radion), which
stabilizes the branes. 
In part~\ref{hewett}, the mixing between radion and Higgs' in a two-Higgs
doublet model is investigated. 
The discovery potential for two decay modes of the
radion is determined in part~\ref{paper}. Models with flat dimensions often predict
Higgs decays into invisible graviscalars. These decays would provide a signal
for the extra dimensions via an invisible Higgs width, and they are
investigated in part~\ref{inv_proc_3}. Also, certain models have the lightest Higgs as a
mixture of brane and bulk scalars. This unfortunately would suppress Tevatron
Run II or 500 GeV linear collider Higgs signals, but would enhance production
at the LHC or CLIC. 
Such issues are examined in part~\ref{XDH}. Part~\ref{proceeding} examines the sensitivity of
the LHC for gluonic Kaluza-Klein states by their effects on dijet production. 
We next present an LHC search study of the heavy supplementary top quark 
present in many little Higgs models. 
An update on $Z'$ studies at the LHC
is presented in the final
part~\ref{zprime}. The analysis focuses on the combination of several
measurements in order to distinguish models. A modification of the leptonic
$A_{FB}$ measurement proves to be very useful in this respect.

\setcounter{figure}{0}
\setcounter{table}{0}
\setcounter{section}{0}
\setcounter{equation}{0}
\clearpage

\newcommand{\DRbar}{\ensuremath{\overline{\mathrm{DR}}}}
\newcommand{\MSbar}{\ensuremath{\overline{\mathrm{MS}}}}
\newcommand{\ttt}[1]{\texttt{#1}}
\newcommand{\mrm}[1]{\mathrm{#1}}
\newcommand{\ti}[1]{\tilde{#1}}
\newcommand{\half}{\ensuremath{\textstyle \frac12}}
\newcommand{\st}{\ensuremath{^{\mathrm{st}}}}
\newcommand{\nd}{\ensuremath{^{\mathrm{nd}}}}
\newcommand{\rd}{\ensuremath{^{\mathrm{rd}}}}
\newcommand{\neut}{\ensuremath{\ti{\chi}^0}}
\newcommand{\grav}{\ensuremath{\ti{\G}}}
\newcommand{\charg}{\ensuremath{\ti{\chi}^+}}
\newcommand{\abs}[1]{\ensuremath{\left|#1\right|}}
\newcommand{\brat}[1]{\ensuremath{\langle\ \! #1\!\ |}}
\newcommand{\tket}[1]{\ensuremath{|\ \! #1\ \!\rangle}}
\renewcommand{\a}{\mathrm{a}}
\renewcommand{\b}{\mathrm{b}}
\renewcommand{\c}{\mathrm{c}}
\renewcommand{\d}{\mathrm{d}}
\newcommand{\e}{\mathrm{e}}
\newcommand{\f}{\mathrm{f}}
\newcommand{\g}{\mathrm{g}}
\newcommand{\hrm}{\mathrm{h}}
\newcommand{\lrm}{\mathrm{l}}
\newcommand{\n}{\mathrm{n}}
\newcommand{\p}{\mathrm{p}}
\newcommand{\q}{\mathrm{q}}
\newcommand{\s}{\mathrm{s}}
\renewcommand{\t}{\mathrm{t}}
\renewcommand{\u}{\mathrm{u}}
\newcommand{\A}{\mathrm{A}}
\renewcommand{\B}{\mathrm{B}}
\newcommand{\D}{\mathrm{D}}
\newcommand{\F}{\mathrm{F}}
\renewcommand{\G}{\mathrm{G}}
\renewcommand{\H}{\mathrm{H}}
\newcommand{\J}{\mathrm{J}}
\newcommand{\K}{\mathrm{K}}
\renewcommand{\L}{\mathrm{L}}
\newcommand{\Q}{\mathrm{Q}}
\newcommand{\R}{\mathrm{R}}
\newcommand{\T}{\mathrm{T}}
\newcommand{\W}{{W}}
\newcommand{\Z}{{Z}}
\newcommand{\sqt}{\ensuremath{\ti{\t}}}
\newcommand{\GeV}{\mathrm{GeV}}
\newcommand{\twovec}[2]{\ensuremath{
\left(\begin{array}{c}#1\\#2\end{array}\right)}
}
\newcommand{\snumentry}[2]{
\begin{minipage}[t]{1.2cm}\flushright\ttt{#1}\end{minipage}
\hspace{2mm}:
\begin{minipage}[t]{11cm}\noindent
#2\end{minipage}\\[1mm]
}
\newcommand{\numentry}[2]{
\begin{minipage}[t]{1.3cm}\flushright\ttt{#1}\end{minipage}
\hspace{5mm}: 
\begin{minipage}[t]{13cm}\noindent
#2\end{minipage}\\[2mm]
}

\newcommand{\arrdes}[1]{
\begin{center}
\framebox{\parbox{0.94\textwidth}{#1}}
\end{center}}
\newcommand{\arrdec}[2]{
\framebox{\parbox{#1\textwidth}{#2}}
}
\newcommand{\mgut}{\ensuremath{M_{\mathrm{input}}}}
\newcommand{\mmess}{\ensuremath{M_{\mathrm{mess}}}}
\newcommand{\glu}{\ensuremath{{\ti{\g}}}}
\newcommand{\sqd}{\ensuremath{{\ti{\d}}}}
\newcommand{\sqdL}{\ensuremath{{\ti{\d}_L}}}
\newcommand{\sqdR}{\ensuremath{{\ti{\d}_R}}}
\newcommand{\squ}{\ensuremath{{\ti{\u}}}}
\newcommand{\squL}{\ensuremath{{\ti{\u}_L}}}
\newcommand{\squR}{\ensuremath{{\ti{\u}_R}}}
\newcommand{\sqs}{\ensuremath{{\ti{\s}}}}
\newcommand{\sqsL}{\ensuremath{{\ti{\s}_L}}}
\newcommand{\sqsR}{\ensuremath{{\ti{\s}_R}}}
\newcommand{\sqc}{\ensuremath{{\ti{\c}}}}
\newcommand{\sqcL}{\ensuremath{{\ti{\c}_L}}}
\newcommand{\sqcR}{\ensuremath{{\ti{\c}_R}}}
\newcommand{\sqb}{\ensuremath{{\ti{\b}}}}
\newcommand{\sqbL}{\ensuremath{{\ti{\b}_1}}}
\newcommand{\sqbH}{\ensuremath{{\ti{\b}_2}}}
\newcommand{\sqtL}{\ensuremath{{\ti{\t}_1}}}
\newcommand{\sqtH}{\ensuremath{{\ti{\t}_2}}}
\newcommand{\se}{\ensuremath{{\ti{\e}}}}
\newcommand{\seL}{\ensuremath{{\ti{\e}_L}}}
\newcommand{\seR}{\ensuremath{{\ti{\e}_R}}}
\newcommand{\snu}{\ensuremath{{\ti{\nu}}}}
\newcommand{\sneL}{\ensuremath{{\ti{\nu}_{\e L}}}}
\newcommand{\sneR}{\ensuremath{{\ti{\nu}_{\e R}}}}
\newcommand{\smu}{\ensuremath{{\ti{\mu}}}}
\newcommand{\smL}{\ensuremath{{\ti{\mu}_L}}}
\newcommand{\smR}{\ensuremath{{\ti{\mu}_R}}}
\newcommand{\snm}{\ensuremath{{\ti{\nu}_\mu}}}
\newcommand{\snmL}{\ensuremath{{\ti{\nu}_{\mu L}}}}
\newcommand{\snmR}{\ensuremath{{\ti{\nu}_{\mu R}}}}
\newcommand{\stau}{\ensuremath{{\ti{\tau}}}}
\newcommand{\stL}{\ensuremath{{\ti{\tau}_1}}}
\newcommand{\stH}{\ensuremath{{\ti{\tau}_2}}}
\newcommand{\snt}[1]{\ensuremath{{\ti{\nu}_{\tau #1}}}}
\newcommand{\sntL}{\ensuremath{{\ti{\nu}_{\tau_1}}}}
\newcommand{\sntH}{\ensuremath{{\ti{\nu}_{\tau_2}}}}

\part{The SUSY Les Houches Accord Project}
{\it P.~Skands,
B.C.~Allanach,      
H.~Baer,            
C.~Bal\'azs,        
G.~B\'elanger,
F.~Boudjema, 
A.~Djouadi, 
R.~Godbole, 
J.~Guasch,
S.~Heinemeyer,      
W.~Kilian,          
J-L.~Kneur,       
S.~Kraml,
F.~Moortgat,     
S.~Moretti,      
M.~M\"uhlleitner, 
W.~Porod,        
A.~Pukhov,       
P.~Richardson,
S.~Schumann,     
P.~Slavich,      
M.~Spira, 
G.~Weiglein%
\label{accord12}
}
\maketitle
\begin{abstract}
An accord specifying a unique set of conventions for supersymmetric
extensions of the Standard Model together with 
generic file structures for 
(1) supersymmetric model specifications and input parameters, 
(2) electroweak scale supersymmetric mass and coupling spectra, and 
(3) decay tables is defined, to provide a universal interface between
spectrum calculation programs, decay packages, and high energy physics
event generators. 
\end{abstract}

\section{INTRODUCTION}

An increasing number of advanced programs for the calculation of the
supersymmetric (SUSY) mass and coupling spectrum are appearing
\cite{Allanach:2001kg,Baer:1993ae,Djouadi:2002ze,Heinemeyer:1998yj,Porod:2003um} 
in step with the more and more refined approaches which are taken in the
literature. Furthermore, these programs are often interfaced to
specialized decay packages, 
\cite{Beenakker:1996ed,Djouadi:1998yw,Muhlleitner:2003vg}, relic density
calculations 
\cite{Belanger:2001fz,Gondolo:2002tz}, and (parton--level) event generators
\cite{Baer:1999sp,Corcella:2000bw,Gleisberg:2003xi,Pukhov:1999gg,Sjostrand:2000wi,Sjostrand:2003wg,Tanaka:1997qn,Kilian:2001qz}, 
in themselves fields with a proliferation of philosophies and,
consequentially, programs.
  
At 
present, a small number of specialized interfaces exist between various
codes. Such tailor-made interfaces are not easily generalized
and are time-consuming to construct and test
for each specific implementation. A universal interface would 
 clearly be an advantage here.  
However, since the codes involved are not all written in the same
programming language, the question naturally arises how to make such an
interface work across languages. At this point, we deem an inter--language
runtime linking solution too fragile to be set loose among the particle
physics community. Instead, we advocate a less elegant but more robust
solution, exchanging information between FORTRAN and C(++) codes
via three ASCII files, one for model input, one for model input plus 
spectrum output, and one for model input plus spectrum output plus decay
information. The detailed structure of these files is described in 
 \cite{Skands:2003cj}. 
Briefly stated, the purpose of this Accord is thus the
following: 
\begin{enumerate}
\item To present 
a set of generic definitions for an input/output file structure which 
provides a universal framework for interfacing SUSY spectrum calculation
programs.
\item To present a generic file structure for the transfer of decay
  information between decay calculation packages and event generators.
\end{enumerate}
Note that different codes may have different implementations of 
how SUSY Les Houches Accord (SLHA) input/output is \emph{technically} 
achieved. The details of how to `switch on' SLHA input/output with a
particular program should be
described in the manual of that program and are not covered here.

\section{CONVENTIONS \label{sec:conventions}}

One aspect of supersymmetric calculations that has often given rise to
confusion and consequent inconsistencies in the past is the multitude of ways
in which the parameters can be, and are being, defined. Hoping to minimize
both the extent and impact of such confusion, we have chosen to adopt one
specific set of self-consistent conventions for the parameters appearing in
this Accord. These conventions are described in the following subsections. As
yet, we only consider R--parity and CP conserving scenarios, with the particle
spectrum of the MSSM.

\subsection{STANDARD MODEL PARAMETERS\label{sec:smconv}}

In general, the SUSY spectrum calculations impose low--scale boundary
conditions on the renormalization group equation (RGE) flows to ensure that
the theory gives correct predictions for low--energy observables. Thus,
experimental measurements of masses and coupling constants at the electroweak
scale enter as inputs to the spectrum calculators.

In this Accord, we choose a specific set of low--scale input parameters,
letting the electroweak sector be fixed by 
\begin{enumerate}  
\item The conventional electromagnetic coupling at the $\Z$ pole,
$\alpha_\mrm{em}(m_\Z)$:
\begin{equation}
\alpha_\mrm{em}(m_\Z) = \frac{\alpha}{1 -
   \Delta\alpha_\mrm{lep}(m_\Z) -\Delta\alpha_\mrm{had}^{(5)}(m_\Z) -
   \Delta\alpha_\mrm{top}(m_\Z)},
\end{equation}
where $\alpha$ is the fine structure constant,
$\Delta\alpha_\mrm{lep}(m_\Z)$ and $\Delta\alpha_\mrm{top}(m_\Z)$ represent
the quantum corrections coming from leptons and top quarks, respectively (see
\cite{Steinhauser:1998rq,Bardin:1999gt}), and
$\Delta\alpha_\mrm{had}^{(5)}(m_\Z)$ is the contribution  
from the five light quark flavours (see e.g.~\cite{Hagiwara:2002fs}). 
\item The Fermi constant determined from muon decay, $G_F$.
\item The $\Z$ boson pole mass, $m_\Z$.
\end{enumerate} 
All other electroweak parameters, such as $m_W$ and $\sin^2\theta_W$, should
be derived from these inputs if needed. 

The strong interaction strength is fixed by $\alpha_s(m_\Z)^{\MSbar}$
(five--flavour), and the third 
generation Yukawa couplings are obtained from the top and tau pole masses,
and from $m_b(m_b)^{\MSbar}$, see \cite{Hagiwara:2002fs}. 
The reason we take $m_b(m_b)^{\MSbar}$ rather than a pole mass definition is
that the latter suffers 
from infra-red sensitivity problems, hence the former is the quantity which
can be most accurately related to experimental measurements. If required,
relations between running and pole quark masses may be found in
\cite{Melnikov:2000qh,Baer:2002ek}. 
 
It is also important to note that all the parameters mentioned here
should be the `ordinary' ones obtained from SM fits,
i.e.~with no SUSY corrections included.  The spectrum calculators themselves
are then assumed to convert these parameters into ones appropriate to an MSSM
framework. 

Finally, while we assume $\MSbar$ running quantities with the SM as the
underlying theory as input, all running parameters in the \emph{output} of
the spectrum calculations are defined in the modified dimensional reduction
($\DRbar$) scheme \cite{Siegel:1979wq,Capper:1980ns,Jack:1994rk}, with
different spectrum calculators possibly using different prescriptions for the
underlying effective field content. More on this in section
\ref{sec:regularization}.

\subsection{SUPERSYMMETRIC PARAMETERS \label{sec:susypar}}

The chiral superfields of the MSSM have the following $SU(3)_C\otimes
SU(2)_L\otimes U(1)_Y$ quantum numbers
\begin{eqnarray}
L:&(1,2,-\half),\quad {\bar E}:&(1,1,1),\qquad\, \textstyle
Q:\,(3,2,\frac16),\quad
{\bar U}:\,(\bar{3},1,-\frac{2}{3}),\nonumber\\ {\bar D}:&(\bar{3},1,\frac13),\quad
H_1:&(1,2,-\half),\quad  H_2:\,(1,2,\half)~.
\label{fields}
\end{eqnarray}
Then, the superpotential (omitting RPV terms) is written as
\begin{eqnarray}
W&=& \epsilon_{ab} \left[ 
  (Y_E)_{ij} H_1^a    L_i^b    {\bar E}_j 
+ (Y_D)_{ij} H_1^a    Q_i^{b}  {\bar D}_{j} 
+ (Y_U)_{ij} H_2^b    Q_i^{a}  {\bar U}_{j}  
- \mu H_1^a H_2^b \right]~.
\label{eq:superpot}
\end{eqnarray}

We denote $SU(2)_L$ fundamental representation
indices by $a,b=1,2$ and generation indices by $i,j=1,2,3$. Colour indices
are everywhere suppressed.
$\epsilon_{ab}$ is the antisymmetric tensor, with
$\epsilon_{12}=\epsilon^{12}=1$. Lastly, we will use ${t,b,\tau}$ to
denote the $i=j=3$ entries of mass or coupling matrices (top, bottom and tau).

The Higgs vacuum expectation values (VEVs) are $\langle H_i^0 \rangle =
v_i/\sqrt{2}$, and $\tan\beta=v_2/v_1$. We also use the notation 
$v=\sqrt{v_1^2+v_2^2}$. Different choices of 
renormalization scheme and scale are possible for defining $\tan\beta$. 
For the input to the spectrum calculators, we adopt by default the commonly
encountered definition
\begin{equation}
\tan\beta(m_\Z)^{\DRbar},
\end{equation}
i.e.~the $\tan\beta$ appearing in the input is defined as a \DRbar\ running
parameter given at the scale $m_\Z$. However, an option is included to allow 
$\tan\beta$ to be input at a different scale, $\tan\beta(\mgut\ne
m_\Z)^{\DRbar}$. Lastly, the spectrum calculator may be
instructed to write out one or several values of $\tan\beta(Q)^{\DRbar}$ at
various scales $Q_i$, see \cite{Skands:2003cj}.

Finally, the MSSM \DRbar\ gauge couplings are: $g'$
(hypercharge gauge coupling in Standard Model normalization), $g$ ($SU(2)_L$
gauge coupling) and $g_3$ (QCD gauge coupling).

\subsection{SUSY BREAKING PARAMETERS\label{sec:susybreak}}

We now tabulate the notation of the soft SUSY breaking parameters.
The trilinear scalar interaction potential is
\begin{equation}
V_3 = \epsilon_{ab} \sum_{ij}
\left[
(T_E)_{ij} H_1^a \tilde{L}_{i_L}^{b} \tilde{e}_{j_R}^* +
(T_D)_{ij} H_1^a               \tilde{Q}_{i_L}^{b}  \tilde{d}_{j_R}^* +
(T_U)_{ij}  H_2^b \tilde{Q}_{i_L}^{a} \tilde{u}_{j_R}^*
\right]
+ \mrm{h.c.}~,
\label{eq:trilinear} 
\end{equation}
where fields with a tilde are the scalar components of the superfield
with the identical capital letter. 
In the literature the T matrices are often decomposed as 
\begin{equation}
\frac{T_{ij}}{Y_{ij}} = A_{ij}~~~~~; (\mathrm{no~sum~over~}i,j)~,
\end{equation}
where $Y$ are the Yukawa matrices and A the soft supersymmetry breaking
trilinear couplings. 
 
The scalar bilinear SUSY breaking terms are contained in the potential
\begin{eqnarray}
V_2 &=& m_{H_1}^2 {{H^*_1}_a} {H_1^a} + m_{H_2}^2 {{H^*_2}_a} {H_2^a} +
{\tilde{Q}^*}_{i_La} (m_{\tilde Q}^2)_{ij} \tilde{Q}_{j_L}^{a} +
{\tilde{L}^*}_{i_La} (m_{\tilde L}^2)_{ij} \tilde{L}_{j_L}^{a}  
+ \nonumber \\ &&
\tilde{u}_{i_R} (m_{\tilde u}^2)_{ij} {\tilde{u}^*}_{j_R} +
\tilde{d}_{i_R} (m_{\tilde d}^2)_{ij} {\tilde{d}^*}_{j_R} +
\tilde{e}_{i_R} (m_{\tilde e}^2)_{ij} {\tilde{e}^*}_{j_R} -
(m_3^2 \epsilon_{ab} H_1^a H_2^b + \mrm{h.c.})~.
\end{eqnarray}
Instead of $m_3^2$ itself, we use the more convenient
parameter $m_A$, defined by:
\begin{equation}
m_A^2 = \frac{m_3^2}{\sin\beta\cos\beta},
\end{equation}
which is identical to the pseudoscalar Higgs mass at tree level in our
conventions.

Writing the bino as ${\tilde b}$, the
unbroken $SU(2)_L$ gauginos as ${\tilde w}^{A=1,2,3}$, and the gluinos as
${\tilde g}^{X=1\ldots8}$, the gaugino mass terms are contained in the
Lagrangian 
\begin{equation}
{\mathcal L}_G = \frac{1}{2} \left( M_1 {\tilde b}{\tilde b} + M_2 {\tilde
    w}^A{\tilde w}^A 
+ M_3 {\tilde g}^X {\tilde g}^X \right) + \mrm{h.c.}~. \label{eq:LG}
\end{equation}

\subsection{MIXING MATRICES\label{sec:mixing}}

In the following, we describe in detail our conventions for neutralino,
chargino, sfermion, and Higgs mixing. Essentially all SUSY spectrum
calculators on the market today work with
mass matrices which include higher--order corrections. 
Consequentially, a formal depencence on the renormalization scheme and scale,
and on the external momenta appearing in the corrections, 
enters the definition of the corresponding mixing matrices. Since, at the
moment, no consensus exists on the most convenient definition to use here, the
mixing matrices should be thought of as `best choice' solutions, at the
discretion of each spectrum 
calculator. For example, one program may output on--shell
parameters at vanishing external momenta in these blocks while
another may be using $\DRbar$ definitions at certain `characteristic'
scales. For details on specific prescriptions,
the manual of the particular spectrum calculator should be consulted.

Nonetheless, for obtaining loop--improved tree--level results,
these parameters can normally be used as is. They can
also be used for consistent cross section and decay width calculations
at higher orders, but then the renormalization prescription
employed by the spectrum calculator must match or be consistently matched to
that of the intended higher order calculation.

Finally, different spectrum calculators may disagree on the overall sign of
one or more rows in a mixing matrix, owing to different 
diagonalization algorithms. Such differences do
not lead to inconsistencies, only the relative sign between entries on the
same row is physically significant, for processes with interfering amplitudes.

\subsubsection{NEUTRALINO MIXING \label{conv:nmix}} 

The Lagrangian contains the (symmetric) neutralino mass matrix as
\begin{equation}
\mathcal{L}^{\mrm{mass}}_{\neut} =
-\frac12{\tilde\psi^0}{}^T{\cal M}_{\tilde\psi^0}\tilde\psi^0 +
\mathrm{h.c.}~,
\end{equation}
in the basis of 2--component spinors $\tilde\psi^0 =$
$(-i\tilde b,$ $-i\tilde w^3,$
$\tilde h_1,$ $\tilde h_2)^T$. We define the unitary 4 by 4
neutralino mixing matrix $N$, such that:
\begin{equation}
-\frac12{\tilde\psi^0}{}^T{\cal M}_{\tilde\psi^0}\tilde\psi^0
= -\frac12\underbrace{{\tilde\psi^0}{}^TN^T}_{{\neut}{}^T} \underbrace{N^*{\cal
    M}_{\tilde\psi^0}N^\dagger}_{\mathrm{diag}(m_{\neut})}
\underbrace{N\tilde\psi^0}_{\neut}~,  \label{eq:neutmass}
\end{equation}
where the (2--component) neutralinos $\neut_i$ are defined such that their
absolute masses
increase with increasing $i$. Generically, the resulting mixing matrix $N$
may yield complex entries in the mass matrix,
$\mathrm{diag}(m_{\neut})_i=m_{\neut_i}e^{i\varphi_i}$. If so,
we absorb the phase into the definition of the corresponding eigenvector,
$\neut_i \to \neut_i e^{i\varphi_i/2}$, making
the mass matrix strictly real:
\begin{equation}
\mathrm{diag}(m_{\neut})\equiv\left[N^*
  {\cal M}_{\tilde\psi^0} N^\dagger\right]_{ij}=m_{\neut_i}\delta_{ij}.
\end{equation}
Note, however, that a
special case occurs when CP violation is absent and one or more of the
$m_{\neut_i}$ turn out to be negative. In this case, we allow for maintaining 
a strictly real mixing matrix $N$, instead writing the \emph{signed} mass
eigenvalues in the output. 
Thus, a negative $m_{\neut_i}$ in the output implies that the
physical field is obtained by the rotation $\neut_i \to \neut_i e^{i\pi_i/2}$.

\subsubsection{CHARGINO MIXING \label{conv:chmix}} 

We make the identification ${\tilde w}^\pm = ({\tilde w^1} \mp i{\tilde w^2}
) / \sqrt{2}$ for the charged winos and ${\tilde h_1^-}, {\tilde h_2^+}$ for
the charged higgsinos.  The Lagrangian contains the chargino mass matrix as
\begin{equation}
\mathcal{L}^{\mrm{mass}}_{\charg} = -\frac12 {\tilde\psi^-}{}^T{\cal M}_{\tilde\psi^+}\tilde\psi^+ +
\mrm{h.c.}~,
\end{equation}
in the basis of 2--component spinors $\tilde\psi^+ = (-i\tilde
w^+,\ \tilde h_2^+)^T,\ \tilde\psi^-= (-i\tilde w^-,\ \tilde h_1^-)^T$.  We
define the unitary 2 by 2 chargino mixing matrices, $U$ and
$V$, such that:
\begin{equation}
-\frac12{\tilde\psi^-}{}^T{\cal M}_{\tilde\psi^+}\tilde\psi^+
= -\frac12
  \underbrace{{\tilde\psi^-}{}^TU^T}_{{\ti{\chi}^-}{}^T}
  \underbrace{U^*{\cal M}_{\tilde\psi^+}V^\dagger}_{\mathrm{diag}(m_{\charg})}
  \underbrace{V\tilde\psi^+}_{\charg}~,  \label{eq:chargmass}
\end{equation}
where the (2--component) charginos $\charg_i$ are defined such that their
absolute masses increase with increasing $i$ and such that the mass matrix,
$m_{\charg_i}$, is strictly real:
\begin{equation}
\mathrm{diag}(m_{\charg})\equiv\left[U {\cal M}_{\tilde\psi^+}
  V^T\right]_{ij} = m_{\charg_i}\delta_{ij}~.
\end{equation}

\subsubsection{SFERMION MIXING \label{conv:sfmix}}

At present, we restrict our attention to left--right mixing in the third
generation sfermion sector only.  The convention we use is, for the
interaction eigenstates, that $\ti{f}_L$ and $\ti{f}_R$ refer to the $SU(2)_L$
doublet and singlet superpartners of the fermion $f\in\{t,b,\tau\}$,
respectively, and, for the mass eigenstates, that $\ti{f}_1$ and $\ti{f}_2$
refer to the lighter and heavier mass eigenstates, respectively. With this
choice of basis, the spectrum output should contain the elements of the
following matrix: 
\begin{equation}
\twovec{{\ti{f}_1}}{{\ti{f}_2}} = \left[
\begin{array}{cc}
F_{11} & F_{12} \\
F_{21} & F_{22}
\end{array}\right]\twovec{{\ti{f}_L}}{{\ti{f}_R}}~,
\end{equation}
whose determinant should be $\pm 1$. We here deliberately avoid notation
involving mixing angles, to prevent misunderstandings which could arise due
to the different conventions for these angles used in the literature.  The
mixing matrix elements themselves are unambiguous, apart from the overall
signs of rows in the matrices, see above.  

\subsubsection{HIGGS MIXING \label{conv:hmix}}

The conventions for $\mu$, $v_1$, $v_2$, $v$, $\tan\beta$, and $m_A^2$ were
defined above in sections
\ref{sec:susypar} and \ref{sec:susybreak}. The angle
$\alpha$ we define by the rotation matrix:
\begin{equation}
\twovec{{{H}^0}}{{{h}^0}} = \left[
\begin{array}{cc}
\cos\alpha & \sin\alpha \\
-\sin\alpha & \cos\alpha
\end{array}\right]\twovec{{{H}^0_1}}{{{H}^0_2}} ~,
\end{equation}
where ${H^0_1}$ and ${H^0_2}$ are the CP--even neutral Higgs scalar
interaction eigenstates, and $h^0$ and $H^0$ the corresponding mass
eigenstates (including any higher order corrections present in the spectrum
calculation), with $m_{h^0}<m_{H^0}$ by definition. 

\subsection{RUNNING COUPLINGS \label{sec:regularization}}

In contrast to the effective definitions adopted above for the mixing
matrices, we define the gauge couplings, the Yukawa couplings, and
the soft breaking Lagrangian terms which appear in the output 
as \DRbar~running parameters, computed at a user--specifiable scale $Q$ (or
grid of scales $Q_i$, see below).

That the \DRbar\ scheme is adopted for the output of running parameters is
simply due to the fact that this scheme substantially simplifies many SUSY
calculations (and hence all spectrum calculators use it).
However, it does have drawbacks which for some applications are serious.  For
example, the \DRbar\ scheme violates mass factorization as used in QCD
calculations~\cite{Beenakker:1988bq}. For consistent calculation beyond
tree--level of processes relying on this factorization, e.g.~cross sections
at hadron colliders, the \MSbar\ scheme is the only reasonable choice. At the
present level of calculational precision, this is fortunately not an
obstacle, since at one loop, a set of parameters calculated in either of the
two schemes can be consistently translated into the other
\cite{Martin:1993yx}, see also \cite{Skands:2003cj} for explicit
prescriptions.  

Note, however, 
that different spectrum calculators use different choices for the
underlying particle content of the effective theory. The programs
\textsc{Softsusy} (v.~1.8), \textsc{SPheno} (v.~2.1), and \textsc{Suspect}
(v.~2.2) use the full MSSM spectrum at all scales, whereas in \textsc{Isajet}
(v.~7.69) a more involved prescription is followed, with different particles
integrated out of the effective theory at different scales. Whatever the
case, these couplings should \emph{not} be used `as is' in calculations
performed in another renormalization scheme or where a different effective
field content is assumed.

Unfortunately, ensuring consistency of the field content assumed in the
effective theory must still be done on a per program basis, though
information on the prescription used by a particular spectrum calculator may
conveniently be given as comments, when running parameters are
provided.

Technically, we treat running parameters in the output in the following
manner: since programs outside the spectrum calculation will not normally be
able to run parameters with the full spectrum included, or at least less
precisely than the spectrum calculators themselves, an option is included to
allow the spectrum calculator to write out values for each running parameter
at a user--defined number of logarithmically spaced scales, i.e.\ to give
output on running parameters at a grid of scales, $Q_i$, where the lowest
point in the grid will normally be $m_\Z$ and the highest point is
user--specifiable. A complementary possibility is to let the spectrum
calculator give output for the running couplings at one or more scales equal
to specific sparticle masses in the spectrum.

\section{DEFINITIONS OF THE INTERFACES}

The following general structure for the SLHA files is proposed:
\begin{itemize}
\item All quantities with dimensions of energy (mass) are implicitly
  understood to be in GeV (GeV$/c^2$).
\item Particles are identified by their PDG particle codes. See
  \cite{Skands:2003cj} for lists of these, relevant to the MSSM. 
\item The first character of every line is reserved for control and
  comment statements. Data lines should have the first character empty.
\item In general, formatted output should be used for write-out, to avoid
  ``messy-looking'' files, while a free format should be used on read-in, to
  avoid misalignment etc.~leading to program crashes. 
\item A ``\ttt{\#}'' 
mark anywhere means that the rest of the line is intended as a comment
to be ignored by the reading program. 
\item All input and output is divided into sections in the form of named
  ``blocks''. A ``\ttt{BLOCK xxxx}'' (with the ``\ttt{B}'' being the first
character on the line) marks the beginning of entries belonging to 
the block named ``\ttt{xxxx}''. E.g.\
``\ttt{BLOCK MASS}'' marks that all following lines until the next
``\ttt{BLOCK}'' (or ``\ttt{DECAY}'') statement contain mass values, to be read 
in a specific format, intrinsic to the \ttt{MASS} block. The order of blocks
is arbitrary, except that input blocks should always come before output
blocks. 
\item
Reading programs should skip over blocks that are not recognized, issuing a
warning rather than crashing. Thereby, stability is increased and 
private blocks can be constructed, for instance \ttt{BLOCK MYCODE} could
contain some parameters that only the program \textsc{MyCode} (or a special
hack of 
it) needs, but which are not recognized universally.
\item A line with a blank first character is a data statement, to be
  interpreted according to what data the current
  block contains. Comments and/or descriptions added after the data values, 
e.g.\ ``\ttt{\ ... \# comment}'', should always be
added, to increase readability of the file for human readers. 
\end{itemize}
Finally, program authors are advised to check that any parameter relations 
they assume in their code (implicit or explicit) are obeyed by the parameters
in the files. For instance, tree--level relations should not be used with
loop--corrected parameters.

For the technical specifications of the blocks contained in the 
SUSY Les Houches Accord files the full writeup
\cite{Skands:2003cj} should be consulted. 

\section{OUTLOOK}

The present Accord \cite{Skands:2003cj} specifies a unique set of conventions
together with ASCII file formats for 
model input and spectrum output for most commonly investigated supersymmetric
models, as well as a decay table file format for use with decay
packages. 

With respect to the model parameter input file, mSUGRA, mGMSB, and mAMSB
scenarios can be handled, with some options for non-universality. However,
this should not discourage users desiring to investigate alternative models;
the definitions for the spectrum output file are at present capable of
handling any CP and R--parity conserving supersymmetric model, with the
particle spectrum of the MSSM. Specifically, this includes the so-called SPS
points \cite{Allanach:2002nj}.
 
Also, these definitions are not intended to be static solutions. Great
efforts have gone into ensuring that the Accord may accomodate essentially
any new model or new twist on an old one with minor modifications required
and full backwards compatibility. Planned issues for future extensions of the
Accord are, for instance, to 
include options for R--parity violation and CP violation,
and possibly to include definitions for an NMSSM. Topics which are at
present only implemented in a few codes, if at all, will be taken up as the
need arises. Handling RPV and CPV should require very minor modifications to
the existing structure, while the NMSSM, for which there is at present not
even general agreement on a unique definition, will require some additional
work. \\[0.4cm] 
\section*{ACKNOWLEDGEMENTS}

The authors are grateful to the organizers of the Physics at TeV
Colliders workshop (Les Houches, 2003) and to the organizers of the
Workshop on Monte Carlo tools for the LHC (MC4LHC, CERN, 2003). The
discussions and agreements reached at those two workshops
constitute the backbone of this writeup.

This work has been supported in part by CERN, by the 
``Collider Physics'' European Network under contract 
HPRN-CT-2000-00149, and 
by the Swiss Bundesamt f\"ur Bildung  und Wissenschaft. W.P.~is supported by
the Erwin Schr\"odinger 
fellowship 
No.~J2272 of the `Fonds zur F\"orderung der wissenschaftlichen
Forschung' of Austria and partly by the Swiss 'Nationalfonds'.

\setcounter{figure}{0}
\setcounter{table}{0}
\setcounter{section}{0}
\setcounter{equation}{0}
\clearpage

\part{Web Tool For The Comparison Of Susy Spectrum Computations \label{susycom}}
{\it B.~C.~Allanach, S.~Kraml}
\maketitle

\begin{abstract}
We present and describe an internet resource which allows the user to compare
different calculations of MSSM spectra. After providing (currently mSUGRA)
SUSY breaking input parameters, the spectra predicted by the publicly
available programs {\tt ISASUGRA}, {\tt
  SOFTSUSY}, {\tt SPHENO} and {\tt SUSPECT} are output by the resource. 
The variance and range of results is also produced. 
\end{abstract}

\section{INTRODUCTION}

Several publicly-available computer programs exist that calculate the MSSM
spectrum consistent with current data on particle masses and gauge couplings,
and a theoretical boundary condition on SUSY breaking.
Given the experimental accuracies that are expected for SUSY analyses 
at both the LHC and a future $e^+e^-$ Linear Collider, theoretical 
uncertainties in spectrum computations are important to consider 
in the total uncertainty of any fit to a SUSY breaking pattern. 

As was pointed out in Ref.~\cite{Allanach:2003jw}, 
important sources of such uncertainties are 
the treatement of thresholds in the renormalization group (RG) running,  
and SUSY loop corrections to the top and bottom Yukawa couplings. 
There has in fact been much progress recently in improving 
the spectrum calculations in commonly used 
public codes around `tricky' corners of the SUSY parameter 
space, such as large $\tan\beta$ or large $m_0$.
However, depending on the specific parameter point chosen,  
the differences in the results of various state-of-the-art codes 
may still be of the same order as or even larger than the expected 
experimental accuracies. Differences in earlier program versions 
tend to be significantly larger. 
 
\section{ONLINE SPECTRUM COMPARISON}

A pragmatic approach, which was also used in Ref.~\cite{Allanach:2003jw},
is to estimate the to-date uncertainty as the spread in the results of 
the most advanced public codes. 
As mentioned above, this `computational uncertainty' varies over the SUSY 
parameter space and should therefore be evaluated for each particular 
benchmark point. There also exist several private RG codes, 
which their authors might like to compare to the available public ones 
in an easy way.  
Moreover, it can be useful to check the results of older program versions 
against newer ones. 

For these reasons we have set up a web application  
which allows to compare the results of 
Isajet~\cite{Baer:1999sp}, 
Softsusy~\cite{Allanach:2001kg}, 
Spheno~\cite{Porod:2003um}, and 
Suspect~\cite{Djouadi:2002ze} online. 
The location is 
\begin{center}
   {\tt http://cern.ch/kraml/comparison/} 
\end{center}
Here the user can input a mSUGRA parameter point\,\footnote{At the moment 
of writing, only the mSUGRA model is supported. Other models may be 
added at a later stage.} 
and choose the program versions to compare. 
On clicking the {\tt submit} button he then gets a list of sparticle 
masses from the four codes together with the mean, the range
and the variance of the results. Note that for the Standard Model 
input the default values of the various codes are used. 

\begin{figure}[t]
\begin{center}
\includegraphics[width=12cm]{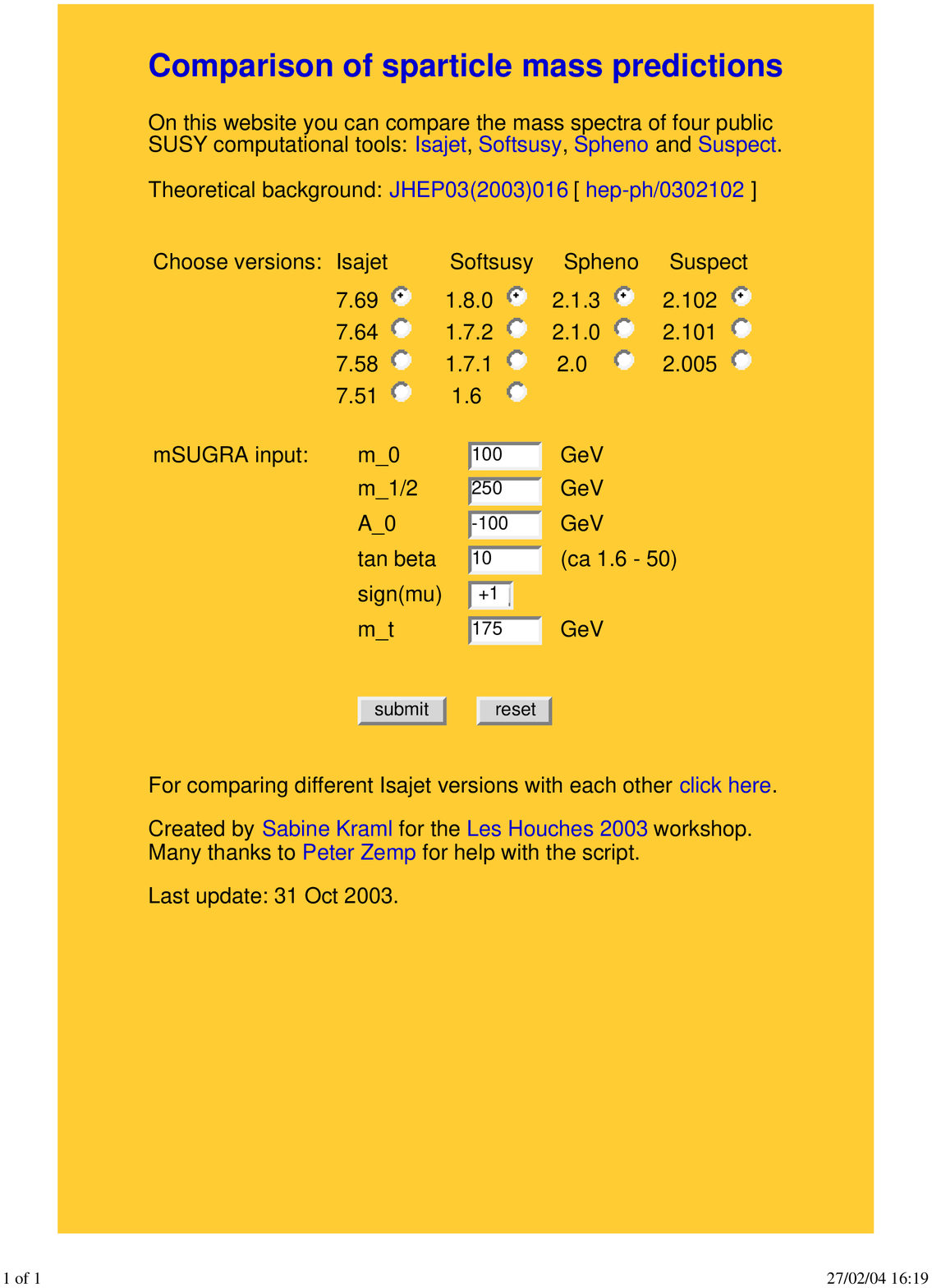} 
\caption{Screenshot of the online spectrum comparison webpage.
\label{fig:screenshot}}
\end{center}
\end{figure}

Figure~\ref{fig:screenshot} shows a screenshot of the webpage. 
The application was set up for the Les Houches workshop in June 2003. 
By 31 Oct 2003, it was used by over 30 different users 
about twice a day on average. 

\section{RESULTS FOR SPS1A AND SPS2}

In order to give a concrete example, we list in Table~\ref{tab:SPS1aMasses} 
some sparticle masses as obtained by today's most recent program versions 
for the SPS1a benchmark point~\cite{Allanach:2002nj}
($m_0=100$~GeV, $m_{1/2}=250$~GeV, $A_0=-100$~GeV, $\tan\beta=10$, 
$\mu>0$, $m_t=175$~GeV). 
As can be seen, the relative differences amout to about 1--2\% at SPS1a. 

The agreement is less good for neutralino and chargino masses at SPS2 
($m_0=1450$~GeV, $m_{1/2}=300$~GeV, $A_0=0$~GeV, $\tan\beta=10$, 
$\mu>0$, $m_t=175$~GeV), as shown in Table~\ref{tab:SPS2Masses}. 
The differences amout to 3\,--\,7\% due to the notoriously 
difficult calulation of the $\mu$ parameter for large $m_0$. 
Here note that a variation of the input $m_t$ by 1~GeV has a similar 
effect on the $\tilde\chi^0$ and $\tilde\chi^\pm$ masses. 
The reason is that large cancellations make $\mu$ extremely 
sensitive to the precise value of the top Yukawa coupling. 
Table~\ref{tab:SPS2Masses} shows, however, an order-of-magnitude 
improvement compared to older program versions, where huge discrepancies 
have been encountered at large $m_0$. 

We note that the effect of going from 2 to 3-loop 
renormalisation group evolution \cite{Jack:2003sx} is comparable in size to
the differencies  we find between the latest 2-loop RGE codes.

\begin{table}
\caption{Sparticle masses as obtained by Isajet\,7.69, 
   Softsusy\,1.8, Spheno\,2.1.3 and Suspect\,2.1.0.2 for SPS1a. 
   The uncertainty $\Delta$ is calculated as $\Delta(x)=0.5[\mbox{max}(x)-\mbox{min}(x)]$. 
   All values are in GeV. 
\label{tab:SPS1aMasses}}
\begin{center}
\begin{tabular}{l|c|c|c|c|c|c|c|c|c|c|c}
  & $\tilde\chi^0_1$ & $\tilde\chi^0_2$ 
  & $\tilde e_R$ & $\tilde e_L$ & $\tilde\tau_1$ & $\tilde\tau_2$ 
  & $\tilde u_R$ & $\tilde u_L$ & $\tilde t_1$ & $\tilde b_1$ & $\tilde g$\\ 
\hline
  Isajet   & 95.5& 181.7& 143.1& 204.7& 134.5& 207.7& 548.3& 564.7& 401.3& 514.8& 611.7\\
  Softsusy & 96.3& 179.3& 143.3& 200.7& 133.9& 204.8& 546.5& 563.0& 399.5& 513.7& 608.8\\
  Spheno   & 97.7& 183.1& 143.9& 206.6& 134.5& 210.4& 547.8& 564.9& 398.8& 516.3& 594.3\\
  Suspect  & 96.5& 183.0& 144.9& 204.4& 135.5& 208.2& 552.6& 572.5& 412.9& 522.0& 617.3\\
\hline
  \quad $\Delta$ & 1.1& 1.9& 0.9& 3.0& 0.8& 2.8& 3.0& 4.7& 7.0& 4.1& 11.5\\
\end{tabular}
\end{center}
\end{table}

\begin{table}
\caption{Neutralino masses as obtained by Isajet\,7.69, 
   Softsusy\,1.8, Spheno\,2.1.3 and Suspect\,2.1.0.2 for SPS2 
   ($m_{\tilde\chi^\pm_1}\simeq m_{\tilde\chi^0_2}$, 
    $m_{\tilde\chi^\pm_2}\simeq m_{\tilde\chi^0_4}$). 
   The uncertainty $\Delta$ is calculated as $\Delta(x)=0.5[\mbox{max}(x)-\mbox{min}(x)]$. 
   All values are in GeV. 
\label{tab:SPS2Masses}}
\begin{center}
\begin{tabular}{l|c|c|c|c}
  & $\tilde\chi^0_1$ & $\tilde\chi^0_2$ 
  & $\tilde\chi^0_3$ & $\tilde\chi^0_4$ \\ 
\hline
  Isajet   & 120.1& 235.1& 431.3& 448.0 \\
  Softsusy & 118.4& 233.0& 490.1& 509.8 \\
  Spheno   & 124.5& 237.2& 456.8& 472.4 \\
  Suspect  & 123.5& 247.6& 495.9& 509.8 \\
\hline
  \quad $\Delta$ & 3.1& 7.3& 32.3& 30.9 \\
\end{tabular}
\end{center}
\end{table}

\section*{ACKNOWLEDGEMENTS}
We would like to thank F.~Boudjema and D.~Zerwas 
for a useful exchange, and CERN for hosting the site.

\setcounter{figure}{0}
\setcounter{table}{0}
\setcounter{section}{0}
\setcounter{equation}{0}
\clearpage

\def\bsgamma{b\ra s\gamma}
\def\bsmu{B_s\ra \mu^+\mu^-}
\def\feynhiggs{{\tt FeynHiggsfast}~}
\def\micro{{\tt micrOMEGAs1.2}~}
\def\isajet{{\tt Isajet7.69}~}
\def\suspect{{\tt Suspect2.2}~}
\def\softsusy{{\tt SOFTSUSY1.8.3}~}
\def\spheno{{\tt SPHENO2.20}~}
\def\mneuto{m_{\tilde{\chi}_1^0}}
\def\stauo{\tilde{\tau}_1}
\def\staut{\tilde{\tau}_2}
\def\staul{\tilde{\tau}_L} 
\def\staur{\tilde{\tau}_R}
\def\mstauo{m_{\stauo}}

\def\noi{\noindent}
\def\nn{\noindent}

\def\sinb{\sin\beta}
\def\cosb{\cos\beta}
\def\sinbb{\sin (2\beta)}
\def\cosbb{\cos (2 \beta)}
\def\tgb{\tan \beta}
\def\tgbt{$\tan \beta\;\;$}
\def\tgbsq{\tan^2 \beta}
\def\sel{\tilde{e}_L}
\def\ser{\tilde{e}_R}
\def\msel{m_{\sel}}
\def\mser{m_{\ser}}
\def\mslr{m_{\tilde{l}_R}}
\def\m0{M_0}
\def\mhf{M_{1/2}}
\def\omegah{$\Omega h^2$}
\def\amu{\delta a_\mu}

 \part{Uncertainties in Relic Density Calculations in mSUGRA\label{dm}}
{\it B. Allanach, G.~B\'elanger, F.~Boudjema,  A. Pukhov, W.
Porod}
\begin{abstract}
We compare the relic density of neutralino dark matter within 
the minimal supergravity model (mSUGRA)
using  four different public codes for  supersymetric spectra
evaluation.
\end{abstract}

\section{INTRODUCTION}

One of the most stringent constraints on supersymmetric models with 
R-parity
conservation 
arises from the upper limit on the relic density of dark matter.
This is particularly true with the recent precise measurements of the cosmological
parameters realised by WMAP. 
It is therefore crucial to quantify the theoretical uncertainties that
enter the calculation of the relic density of the lightest supersymmetric particle (LSP) and to see how they reflect
 on the
allowed parameter space.
We do not attempt to answer  this question fully here. We will only
consider  one aspect:  the uncertainty introduced by the calculation of the weak scale
SUSY parameters  using renormalization group equations  (RGE) within the 
context of the mSUGRA model. 
As a measure of the theoretical uncertainty on 
the  mSUGRA parameters, we use
 the four public state-of-the-art RGE codes: \isajet\cite{Baer:2003mg},
 \softsusy\cite{Allanach:2001kg}, \spheno\cite{Porod:2003um} and 
\suspect\cite{Djouadi:2002ze}, link them to \micro
\cite{Belanger:2001fz} and compare estimates for the relic density.
At this point no attempt is made to estimate the uncertainties that
could arise directly in the calculation of the relic density itself.

\section{RGE CODES AND  RELIC DENSITY CALCULATION}

A detailed study of theoretical uncertainties on the supersymmetric
 spectra as
obtained by RGE codes was presented in \cite{Allanach:2003jw}.
It was shown that differences in masses less than a few percent are usually 
found, although
some corners of parameter space are still difficult to tackle and can display
much larger differences. 
The discrepancies can be traced back to   
the level of approximation used in the weak-scale boundary conditions.
The large $\tan\beta$ region and the 
focus point region  (large $\m0$) are 
still subject to large 
theoretical errors. 
Both of these regions are 
precisely 
where one can find 
cosmologically
interesting 
values for the relic density,
$\Omega h^2< .128$.  
In the focus point region, the LSP is mainly a Higgsino and annihilates
efficiently into gauge bosons. At large $\tan\beta$, even rather heavy
neutralinos can annihilate into $b\overline{b}$ pairs via s-channel
exchange of a heavy Higgs.
The coannihilation region where the Next-to-Lightest supersymmetric particle
(NLSP)
is nearly degenerate in mass with the LSP, is another  cosmologically relevant region. 
Although it  is a priori not difficult to handle by the RGE
codes, the value of the relic density  depends sensitively on the mass 
difference between the NLSP and
the LSP and even shifts of ${\cal O}(1)$ GeV can cause large
shifts in the relic density.
The other  cosmologically viable  mSUGRA region, the bulk region, 
shows a much smaller induced sensitivity upon the MSSM mass spectrum.

The  link between \micro and the RGE codes  is done within the spirit of the  SUSY Les Houches Accord \cite{Skands:2003cj} : common
input values are chosen 
and pole masses, mixing matrices, the  $\mu$ parameter  and the
trilinear couplings are calculated by the RGE codes.
All parameters are  read  by \micro~1.2. The annihilation cross-sections
 are
then evaluated at tree-level. Important  radiative corrections to the
Higgs widths  and in particular the $\Delta m_b$ correction
 are taken into account. 

\section{RESULTS}

For  the numerical results as default values we have fixed
 $m_t=175$ GeV,
$\alpha_s(M_Z)^{\overline{MS}}=.1172$ and $m_b(m_b)^{\overline{MS}}=4.16$ GeV.
 This corresponds to $m_b(M_Z)^{\overline{DR}}=2.83$ GeV.
We concentrate on the three regions where the relic density 
is within the WMAP range and where potentially large discrepancies can be
observed: the focus point region, the large $\tan\beta$ region
and the coannihilation region. 

\subsection{Coannihilation\\
$\m0=150$ GeV, $A_0=0$, $\tan\beta=10$, $\mu>0$}

\begin{figure}
\begin{center}
\includegraphics[width=14.5cm]{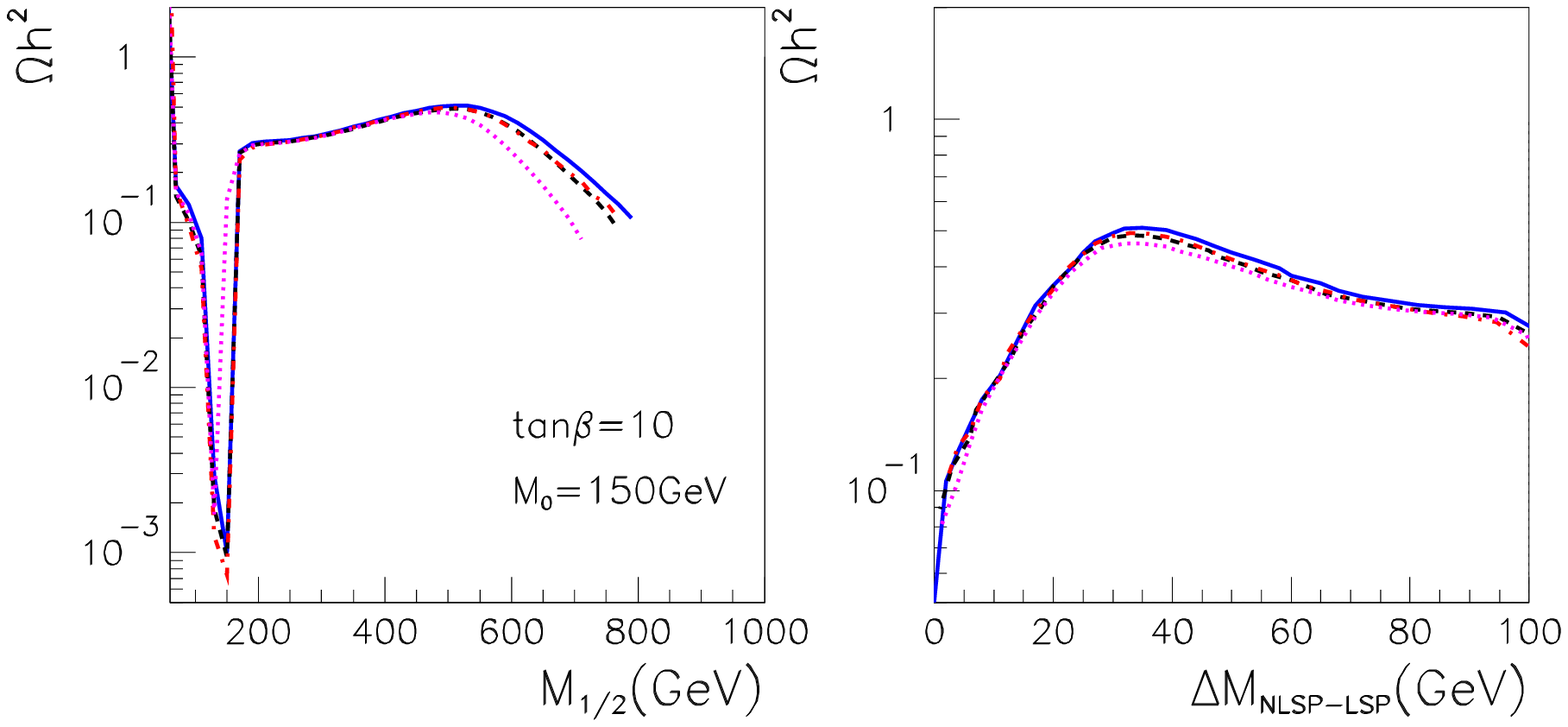} 
\vspace{-1cm}
\caption{\label{figcoan} a) $\Omega h^2$ for $\m0=150$ GeV, $\tan\beta=10$, $A_0=0$, $\mu>0$
for \softsusy (full), \isajet (dashed), \suspect (dash-dotted),
and \spheno (dotted). At large $\mhf$, \isajet and \suspect give nearly 
identical results. b) $\Omega h^2$ vs $\mstauo-\mneuto$ for the same set of
parameters as a).}
\end{center}
\end{figure}

The small $\mhf$ region corresponds to the so-called  bulk region where the bino-LSP
annihilates into lepton pairs via s-channel $Z$ or Higgs exchange or t-channel
slepton exchange. Here one finds very good agreement between the values of
\omegah  using the different RGE codes (see Fig.~\ref{figcoan}a) since 
the predicted values for slepton and neutralino masses are in good agreement
(within a few GeV).
The exact position of the  $Z$
pole (corresponding to the big dip in \omegah) is slightly shifted for \spheno
but the range of values of $\mhf$ for which $\Omega h^2<.128$ are basically
identical. Note that the $Z$ pole region is ruled out by the LEP constraints on
neutralinos within the context of mSUGRA models.

As one moves up in $\mhf$, one reaches the so-called coannihilation 
region where
the $\tilde{\tau}$ is the NLSP and is nearly degenerate with the neutralino,
as in Fig.~\ref{figcoan}b. 
Coannihilation with the $\tilde{\tau}$, 
 and to a lesser extent the selectron and smuon,
brings the relic density in the desired range.
For a given value of $\mhf$, differences between the codes can reach a factor 
2, the largest differences are found between \spheno and
{\tt SOFTSUSY1.8.3}. However very good agreement is found between all codes
when the relic density is plotted as a function of the mass difference between
the LSP and the NLSP (here the $\tilde{\tau}$). All codes obtain  values of
\omegah compatible with WMAP for mass differences $\mstauo-\mneuto\approx
4$~GeV (at the extreme left of Fig.~\ref{figcoan}b),
even though the corresponding value of the neutralino mass can differ. 
The value of $\mhf$ for which the relic density becomes compatible with WMAP
  varies from 670 GeV ({\tt SPHENO2.20}) to 790 GeV ({\tt SOFTSUSY1.8.3}), a
  12\%   difference  on $\mhf$.
 
\subsection{Focus point\\
$\mhf=300$ GeV, $A_0=0$, $\tan\beta=10$, $\mu>0$}

In addition to the small $\m0$ (bulk/coannihilation region) where annihilation into
leptons is important, the cosmologically relevant region is found at values of 
$\m0$ well above $1$TeV. As one approaches the region where electroweak 
symmetry breaking is forbidden, the  $\mu$ parameter approaches zero. 
This means  that the LSP is mainly Higgsino. This LSP can then annihilate
 very efficiently into gauge bosons (WW/ZZ) and to a lesser extent into $Zh$. 
 The parameter $\mu$ is however very sensitive~\cite{Allanach:2000ii}
to the top
Yukawa coupling, $h_t$ (which is also reflected in a sensitivity to the value
 of the top quark mass) and huge differences between  codes were 
 observed\cite{Allanach:2003jw}.
The impact on
the relic density and on the exclusion region is likewise very significant.
\begin{figure}
\begin{center}
\includegraphics[width=14.5cm]{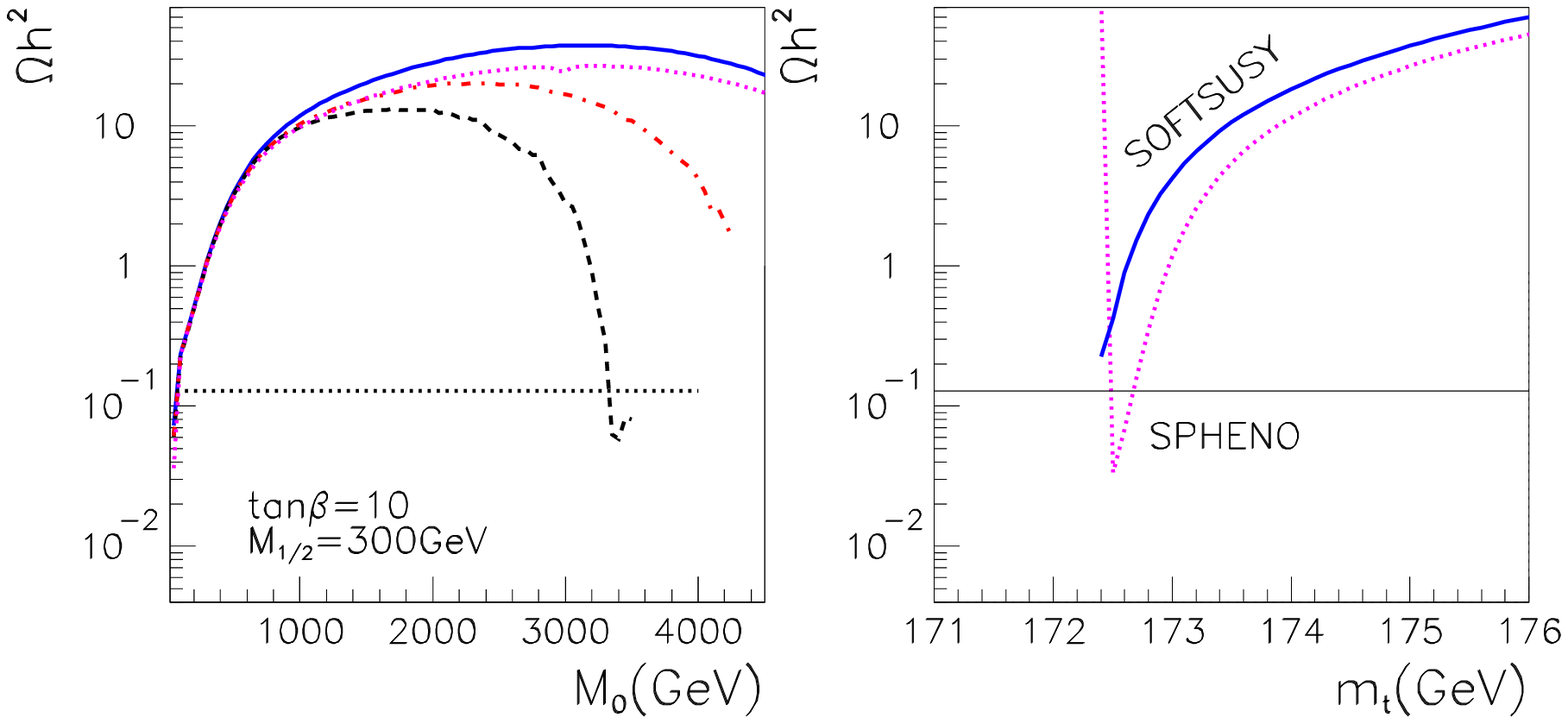} 
\vspace{-1cm}
\caption{\label{largem0}a) $\Omega h^2$ as a function of $\m0$ for $\mhf=300$ GeV,
$\tan\beta=10$, $A_0=0$ and $\mu>0$ and $m_t=175$. Same labels as in Fig. 1.
b) Dependence of the relic density  on $m_t$ for \softsusy (full) and
\spheno (dash) .}
\end{center}
\end{figure}

As can be seen in Fig.~\ref{largem0}, all codes agree very well for $\m0<1$TeV
 but  as one gets to large values of $\m0$,
  more than one order of magnitude differences in \omegah~ can be
found. For $m_t=175$ GeV, only Isajet finds a large drop in the $\mu$ parameter
as one moves to $\m0\approx 3000$ GeV, this is when  $\Omega h^2$ drops 
below the upper limit from WMAP. The other codes do 
not find this  
 drop in $\mu$ and do not obtain a cosmologically interesting region
 for $\m0<4000$ GeV.
These large  differences between codes  however
 are just a reflection of the sensitivity to the top Yukawa, 
$h_t(M_{SUSY})$ which is proportional to $m_t$. We show in Fig. 2b, the variation of $\Omega h^2$ with
$m_t$ using \softsusy and \spheno for $\m0=3000$ GeV.
The value $\Omega h^2=.128$ found in 
\isajet for $m_t=175$ GeV can
be reproduced in \softsusy ({\tt SPHENO}) by changing the  input to
 $m_t=172.2(172.5)$ GeV.

\subsection{Large $\tan\beta$\\
$m_{1/2}=1500$ GeV, $A_0=0$,$\tan\beta=52$ $\mu>0$}

At large $\tan\beta$ the new feature is the annihilation of neutralinos into $b
\overline{b}$ via heavy Higgs exchange. With the current version of the RGE codes,
this is observed only for very large values of $\tan\beta$. The crucial
parameter here is $M_A/2\mneuto$ which must be close to unity to provide
sufficient annihilation of neutralinos. 
Large differences in the value of $M_A$ between the different RGE codes
occur because of the sensitivity of the RGE to the
 bottom Yukawa as well as from taking into account higher loop
effects.

As Fig. 3a shows,
all 4 programs predict a large drop in the relic density when the
neutralino mass gets close to $M_A/2$ although this drop occurs at much lower
values of $\mhf$ for {\tt SPHENO}, $\mhf\approx 1250$ GeV than for
\isajet , $\mhf\approx 1750$ GeV. However, here again the results are very sensitive
to the input parameters, in this case the value of the b-quark mass.
For $\mhf=1300$ GeV, we find an order of magnitude shift in $\Omega h^2$ 
for $m_b(m_b)=4-4.4$ GeV with the program {\tt SOFTSUSY1.8.3}.
By a slight shift of the b-quark mass we can find perfect agreement between
\spheno and {\tt SOFTSUSY1.8.3}, as shown in Fig. 3b.

\begin{figure}
\begin{center}
\includegraphics[width=14.5cm]{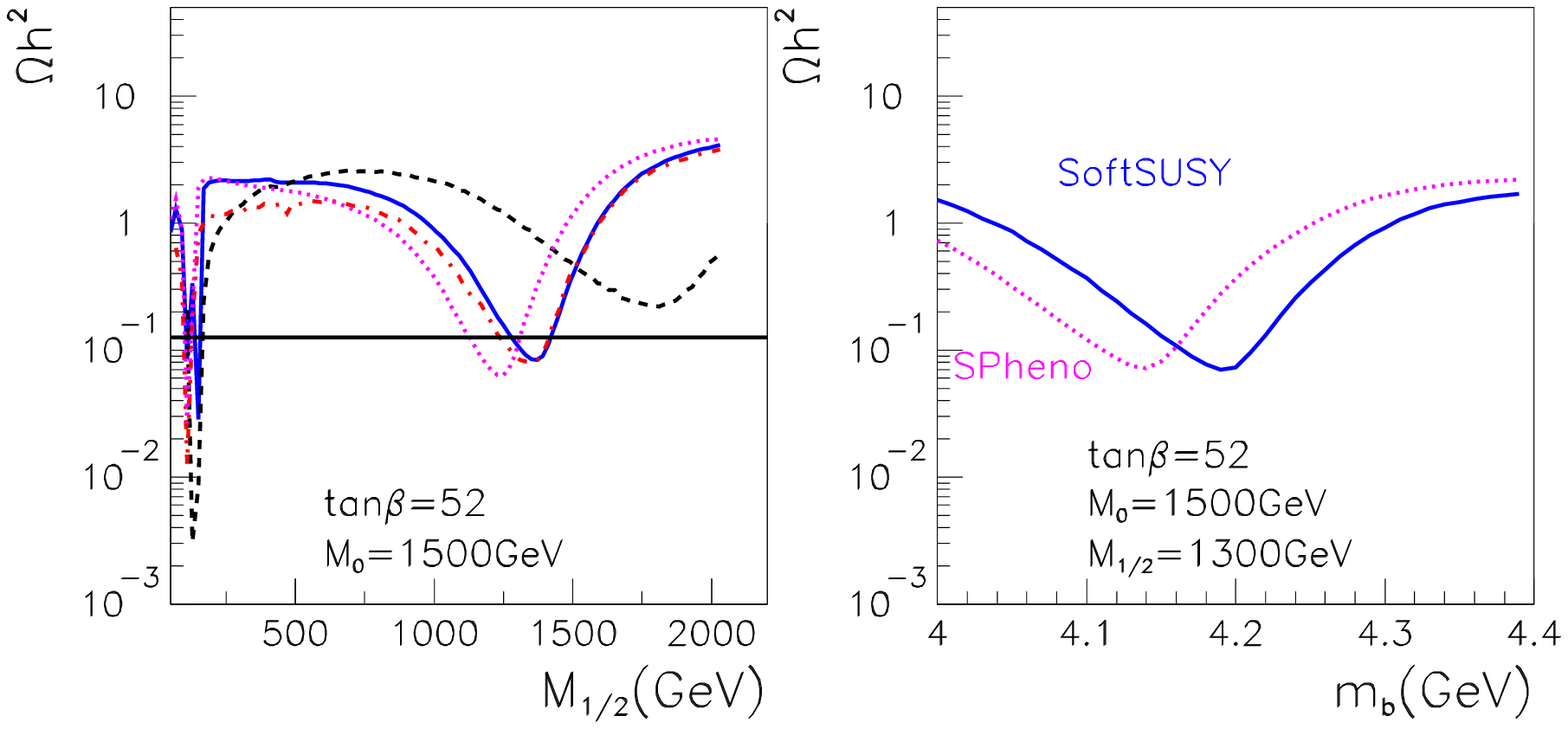} 
\vspace{-1cm}
\caption{\label{largetb} a) $\Omega h^2$ as a function of $\mhf$ for
$m_t=175$, 
$\tan\beta=52$, $A_0=0$ and $\mu>0$. Same labels as in Fig. 1.
b) Dependence of the relic density  on $m_b(m_b)$ for \softsusy (full) and
\spheno (dash).}
\end{center}
\end{figure}

\section{CONCLUSION}

While the predictions for the relic density of neutralinos are rather
stable in most of the mSUGRA space, it is in the most physically interesting region that
large discrepancies can be  observed, in particular the focus point/large
$\tan\beta$ and coannihilation regions. 
 It is however  reassuring to find that with 
the newer versions of the codes, the discrepancies in the sparticle spectra
tend to be reduced. 
More details on the theoretical uncertainties in the evaluation of the
relic density  arising from the standard model 
parameters, $\alpha_s,m_b,m_t$, used as input in a RGE code 
can be found in \cite{Allanach:2004}.

\section*{ACKNOWLEDGEMENTS}

This work was partially supported by  CNRS/PiCS-397, {\it Calculs automatiques de diagrammes de Feynman}.
We thank Jean-Loic Kneur for providing an improved version of \suspect.

\setcounter{figure}{0}
\setcounter{table}{0}
\setcounter{section}{0}
\setcounter{equation}{0}
\clearpage

\part{SFITTER: A Tool To Determine Supersymmetric Parameters \label{sfitter}}

{\it R. Lafaye, T. Plehn, D. Zerwas}

\maketitle

\begin{abstract}
SFITTER is a new tool to determine supersymmetric model parameters
from collider measurements. It allows to perform a grid search for
the minimal $\chi^2$ and/or a fit of a given model. Currently, the
model parameters in the general MSSM or in a gravity mediated SUSY
breaking model can be tested using a given set of mass, branching ratio
and cross section measurements.
\end{abstract}

\section{Introduction}

The most important task for the LHC as well as for any future Linear
Collider is to study in detail the mechanism which leads to
electroweak symmetry breaking. While the Standard Model describes all
available high energy physics experiments, it still has to be regarded
as an effective theory, valid at the weak scale. New physics are
expected to appear at the TeV energy scale. The minimal supersymmetric
extension of the Standard Model (MSSM) can provide a description of
physics up to the unification scale.

If supersymmetry or any other high-scale extension of the Standard
Model is discovered, it will be crucial to determine its fundamental
high-scale parameters from weak-scale
measurements~\cite{Blair:2000gy,Blair:2002pg}. The LHC and a future
Linear Collider will provide a wealth of measurements~\cite{Desch:2003vw},
which due to their complexity require a proper treatment to unravel
the corresponding high-scale physics. Even in the general weak-scale
MSSM without any unification or SUSY breaking assumptions the
measurements of masses and couplings are not likely to be independend
measurements; moreover, linking supersymmetric particle masses to
weak-scale SUSY parameters involves non-trivial mixing to mass
eigenstates in essentially every sector of the theory. On top of that,
for example in gravity mediated SUSY breaking scenarios (mSUGRA/cMSSM)
a given weak-scale SUSY parameter will always be sensitive to several
high-scale parameters which contribute through renormalization group
running. Therefore, a fit of the model parameters using all
experimental information available will lead to the best sensitivity
and make the most efficient use of the information available.

If the starting point of the
fit is not known and many parameters are involved, the allowed
parameter space might not be sampled completely in the fit
approach. To avoid boundaries imposed by non-physical parameter
points, which can confine the fit to a `wrong' parameter region,
combining the fit with an initial evaluation of a multi-dimensional
grid is the optimal approach. In the general
MSSM the weak-scale parameters can vastly outnumber the collider
measurements, so that a complete parameter fit is not possible and one
has to limit oneselve to a subset of parameters. 
In SFITTER both
grid and fit are realised and can be combined, including a general
correlation matrix and the option to exclude parameters of the model 
from the fit/grid by fixing them to a value.

\section{SFITTER --- Program Structure}

Currently, SFITTER uses the predictions for the supersymmetric masses
provided by SUSPECT~\cite{Djouadi:2002ze}, but the conventions of the
SUSY Les Houches accord~\cite{Skands:2003cj} could be helpful, if
provided as a common block/C-structure, to ease interfacing other
programs.  The branching ratios and $e^+$e$^-$ production cross
sections are provided by MSMlib~\cite{msmlib}, which has been used
extensively at LEP and cross checked with
Ref.~\cite{Barger:2001nu}. The next-to-leading order hadron collider
cross sections are computed using
PROSPINO~\cite{Beenakker:1997ch,Beenakker:1998ut,Beenakker:1999xh}. The
fitting program uses the MINUIT package~\cite{James:1975dr}. The
determination of $\chi^2$ includes a general correlation matrix
between measurements. For unphysical points in supersymmetric
parameter space, $\chi^2$ is set to 10$^{30}$.

\subsection{Initialization and Steering}

The program SFITTER is driven by two files: the first one sets up the
measurements and the corresponding errors. For each measurement one
specifies if it is to be used in the grid~(G) or in the MINUIT fit~(M)
or in both.
\begin{small} \begin{verbatim}
//set all errors to 0.5% of their central value
DATA_ERR = 0.005
//randomize the measurements around their nominal value 
RANDOMIZE = 1     
//Higgs mass and error to be used in the Fit only
m_h      = 112.6 +/- 0.1 [-/M]  
//Neutralino1 mass to be used in Grid and Fit
m_chi0_1 = 180.2 +/- 5.1 [G/M]  
//Correlation between two chargino mass measurements
CORR(m_chi+_1,m_chi+_2) = 0.03
\end{verbatim} \end{small}
The second file initializes everything related to the weak-scale
or high-scale MSSM model parameters. First the model (mSUGRA,
pMSSM etc) is specified, then the starting values of all MSSM parameters,
boundaries, stepsize and the number of points in the grid are specified. 
Moreover, the user
defines if a certain MSSM model parameter is included in the grid
and in the fit:
\begin{small} \begin{verbatim}
MODEL=MSUGRA    // use MSUGRA
// use the GRID (or not)
GRID=1
// M0 used in grid and fit, grid of 10+1 steps between 0 and 1000.
M0=500. [M/G] STEP=200. LOW=0. HIGH=1000. GRID=10 
// A0 used only in fit
A0=0. [M/-] STEP=200. LOW=-1000. HIGH=1000.    
\end{verbatim} \end{small}

\subsection{mSUGRA/cMSSM Parameter Determination}

Assuming that SUSY breaking is mediated by gravitational interactions
(mSUGRA/cMSSM) we fit four universal high-scale parameters to a toy
set of collider measurements: the universal scalar and gaugino masses,
$m_0$, $m_{1/2}$, the trilinear coupling $A_0$ and the ratio of the
Higgs vacuum expectation values, $\tan\beta$. The sign of the Higgsino
mass parameter $\mu$ is a discrete parameter and therefore fixed. The
assumed data set is the set of all supersymmetric particle masses for
the SUSY parameter point SPS1a~\cite{Allanach:2002nj,Ghodbane:2002kg},
as computed by SUSPECT. The errors on the toy mass measurements are
uniformly set to 0.5\%. The starting points for the mSUGRA parameters
are fixed to the mean of the lower and upper limit in the fit, {\sl
i.e.} they are not necessarily even close to the true SPS1a values.
The result of the fit is shown in Tab.~\ref{tab:msugra}. With SFITTER
the true parameter values were reconstructed well within the quoted
errors, in spite of starting values relatively far away from the true
ones. The measurement of $m_0$ and $m_{1/2}$ is very precise, while
the sensitivity of the masses on $\tan\beta$ and $A_0$ is
significantly weaker.

The correlations between the different high-scale SUSY parameters are
also given in Tab.~\ref{tab:msugra}. One can understand the
correlation matrix step by step~\cite{Drees:1995hj}: first, the
universal gaugino mass $m_{1/2}$ can be extracted very precisely from
the physical gaugino masses. The determination of the universal scalar
mass $m_0$ is dominated by the weak-scale scalar particle spectrum,
but in particular the squark masses are also strongly dependent on the
universal gaugino mass, because of mixing effects in the
renormalization group running. Hence, a strong correlation between the
$m_0$ and $m_{1/2}$ occurs. The universal trilinear coupling $A_0$ can
be measured through the third generation weak-scale mass parameters
$A_{b,t,\tau}$. However, the $A_{b,t,\tau}$ which appear for example
in the off-diagonal elements of the scalar mass matrices, also depend
on $m_0$ and $m_{1/2}$, so that $A_0$ is strongly correlated with
$m_0$ and $m_{1/2}$.

In the SPS1a scenario, the pseudoscalar Higgs is heavy and the Higgs
masses do not show a strong dependence on $\tan\beta$. Because of the
large mass difference between gauginos and Higgsinos they essentially
decouple, and the neutralino/chargino sector will not yield a good
determination of $\tan\beta$. The stop mixing is governed by $A_t$,
and not by $\mu/\tan\beta$, while the sbottom mixing is small
altogether.  Only the stau mixing is large and driven by $\mu
\tan\beta$ in the off-diagonal element of the stau mass matrix.  The
stau mass parameters are dominated by $m_0$, in particular the smaller
right handed stau mass.  Therefore, one expects $\tan\beta$ to be
strongly correlated with $m_0$ and less with $m_{1/2}$.  The result
from SFITTER as shown in Tab.~\ref{tab:msugra} is in agreement with
this prediction.  Thus, the results obtained with SFITTER can be
understood from the particular features of the SPS1a spectrum.

\begin{table}[t]
\begin{center}
\parbox[b]{7cm}{\begin{tabular}{|c|rrr|}
\hline
             & True & FitStart &  FitResult \\
\hline
$m_0$        & 100  & 500 & 100.01$\pm$0.58 \\
$m_{1/2}$    & 250  & 500 & 249.99$\pm$0.31 \\
$\tan\beta$  &  10  &  50 &  10.03$\pm$0.37 \\
$A_0$        & -100 &   0 & -100.1$\pm$5.26 \\
\hline
\end{tabular}}
\hspace*{1.5cm}
\parbox[b]{7cm}{\begin{tabular}{|c|rrrr|}
\hline
             & $m_0$ & $m_{1/2}$ &  $\tan\beta$ & $A_0$ \\
\hline
$m_0$        &     1 & -0.47 &  0.41 &  0.26 \\
$m_{1/2}$    &       &     1 & -0.07 & -0.30 \\
$\tan\beta$  &       &       &     1 &  0.35 \\
$A_0$        &       &       &       &     1 \\
\hline
\end{tabular}}
\end{center}
\caption{Left: summary of mSUGRA fit in SPS1a: true values, starting
values, fit values. As in SPS1a we fix $\mu>0$. All mass values are
given in GeV. Right: the (symmetric) correlation matrix of all SUSY
parameters in the mSUGRA fit.}
\label{tab:msugra}
\end{table}

\subsection{MSSM Parameter Determination}

In total 24 parameters describe the unconstrained weak-scale
MSSM. They are listed in Tab.~\ref{tab:mssm}: $\tan\beta$ just like in
mSUGRA, plus three soft SUSY breaking gaugino masses $M_i$, the
Higgsino mass parameter $\mu$, the pseudoscalar Higgs mass $m_A$, the
soft SUSY breaking masses for the right sfermions, $M_{\tilde{f}_R}$,
the corresponding masses for the left doublet sfermions,
$M_{\tilde{f}_L}$ and finally the trilinear couplings of the third
generation sfermions $A_{t,b,\tau}$.

In any MSSM spectrum, in first approximation, the parameters $M_1$,
$M_2$, $\mu$ and $\tan\beta$ determine the neutralino and chargino
masses and couplings. We exploit this feature to illustrate the option to
use a grid before the start the complete MINUIT fit. For testing
purposes, the error on all mass measurements is again set 0.5\%. The
starting values of the parameters are set to their nominal values, this 
study is thus less general than the one of mSUGRA. 
Then we minimize $\chi^2$ on a grid. For this grid
minimization the six chargino and neutralino masses are used as
measurements to determine the four SUSY parameters $M_1$, $M_2$, $\mu$
and $\tan\beta$ only. The step size of the grid is 10 for $\tan\beta$ and
100~GeV for the three mass parameters.  After the minimization, these
four parameters obtained from the minimum $\chi^2$ on the grid are
fixed and all remaining parameters are fitted. Only in a final run all
SUSY parameters are released and fitted, to give the final results
quoted in Tab.~\ref{tab:mssm}.

In Tab.~\ref{tab:mssm} the intermediate (after the grid evaluation)
results, the final results and the nominal values are shown. The final
fit values indeed converges to the correct central values within its
error. The central values of the fit are in good agreement with
generated values, except for the trilinear coupling $A_b$. As already
mentioned in the discussion of the mSUGRA fit, the mixing between the
two sbottom mass states is very small, so the assumed precision
of the 0.5\% is insufficient to determine the parameter from the mass
measurements alone. As $A_t$ enters in the calculation of the lightest
Higgs, additional sensitivity for this parameter comes from the mass
measurement of the lightest Higgs boson. 
The use of branching ratios and cross section
measurements should significantly increase the precision in future 
studies, especially for $A_{\tau}$ and $A_{b}$.

\begin{table}
\begin{center}
\begin{tabular}{|l|rrr||l|rrr|}
\hline
   & AfterGrid  & AfterFit   & SPS1a & & AfterGrid  & AfterFit   & SPS1a \\
\hline
$\tan\beta$          &         10 &      10.62$\pm$2.5  &       10 &
 $M_{\tilde{u}_R}$   &     528.03 &     528.06$\pm$2.8 &    532.1 \\
$M_1$                &        100 &     102.05$\pm$0.61 &    102.2 &
 $M_{\tilde{d}_R}$   &     525.12 &     525.14$\pm$2.8 &    529.3 \\
$M_2$                &        200 &     191.65$\pm$1.4  &    191.8 &
 $M_{\tilde{c}_R}$   &     528.03 &     528.06$\pm$2.8 &    532.1 \\
$M_3$                &     579.37 &     579.33$\pm$4.8  &    589.4 &
 $M_{\tilde{s}_R}$   &     525.12 &     525.15$\pm$2.8 &    529.3 \\
$\mu$                &        300 &     344.04$\pm$1.2  &    344.3 &
 $M_{\tilde{t}_R}$   &     417.36 &     415.44$\pm$5.7 &    420.2 \\
$m_A$                &     399.38 &     399.14$\pm$1.2  &    399.1 &
 $M_{\tilde{b}_R}$   &     524.59 &     523.99$\pm$2.9 &    525.6 \\
$M_{\tilde{e}_R}$    &     138.24 &     138.23$\pm$0.76 &    138.2 &
 $M_{\tilde{q}1_L}$  &     549.58 &     549.61$\pm$2.1 &    553.7 \\
$M_{\tilde{\mu}_R}$  &     138.24 &     138.23$\pm$0.76 &    138.2 &
 $M_{\tilde{q}2_L}$  &     549.58 &     549.61$\pm$2.1 &    553.7 \\
$M_{\tilde{\tau}_R}$ &     135.58 &     135.51$\pm$2.1  &    135.5 &
 $M_{\tilde{q}3_L}$  &     493.59 &     494.38$\pm$2.7 &    501.3 \\
$M_{\tilde{e}_L}$    &     198.74 &     198.75$\pm$0.68 &    198.7 &
 $A_{\tilde{\tau}}$  &    -724.25 &    -286.78$\pm$549 &   -253.5 \\
$M_{\tilde{\mu}_L}$  &     198.74 &     198.75$\pm$0.68 &    198.7 &
 $A_{\tilde{t}}$     &    -502.19 &    -495.19$\pm$15 &   -504.9 \\
$M_{\tilde{\tau}_L}$ &     197.79 &     197.81$\pm$0.89 &    197.8 &
 $A_{\tilde{b}}$     &     975.12 &     999.78$\pm$49 &   -799.4 \\
\hline
\end{tabular}
\end{center}
\caption{Result for the general MSSM parameter determination in
SPS1a. Shown are the nominal parameter values, the result after the
grid and the final result. The deviation in the squark sector of 1\%
is an artefact of differences between MSSM and mSUGRA part of the
renomalization group code~\cite{Djouadi:2002ze}. All masses are given
in GeV.}
\label{tab:mssm}
\end{table}

\section{Conclusions}

SFITTER is a new program to determine suspersymmetric parameters from
experimental measurements. The parameters can be extracted either
using a fit, a multi-dimensional grid minimisation, or a combination
of the two. Correlations between measurements can be specified and 
are taken into account in the calculation of the $\chi^2$.
SUSPECT, MSMlib and PROSPINO are used to calculate the
predictions for the masses, branching ratios and production cross
sections. A more realistic set of the measurements for example
assuming the SPS1a mass spectrum for the LHC and and a future Linear
Collider will be studied as a next step. The impact of correlations
between measurements on the estimated errors of MSSM parameters will
be studied in detail. In the future public version of the program we
will include different generators for the calculation of masses and
branching ratios.

\section*{Acknowledgements}

The authors would like to thank the organizers of the Les Houches
workshop and the convenors of the BSM working-group for the constructive
atmosphere in which SFITTER was born. 
TP would in particular like to thank Michael Spira for allowing him to 
participate in the Higgs and the BSM sessions at the Les Houches workshop.

\setcounter{figure}{0}
\setcounter{table}{0}
\setcounter{section}{0}
\setcounter{equation}{0}
\clearpage

\newcommand{\beq}{\begin{eqnarray}}
\newcommand{\eeq}{\end{eqnarray}}

\part{SDECAY: a Code for the Decays of the Supersymmetric Particles \label{sdecay}}
{\it A. Djouadi, Y. Mambrini and M. M\"uhlleitner}
%
%
%
\begin{abstract}
We present the Fortran code {\tt SDECAY}, a program which calculates the 
decay widths and branching ratios of all supersymmetric particles in the 
Minimal Supersymmetric Standard Model, including higher order effects. 
The usual two-body decays of sfermions and gauginos as well as the 
three-body decay modes of charginos, neutralinos and gluinos are included. 
Furthermore, the three-body and even the four-body decays of top squarks 
are calculated. The important loop-induced decays, the QCD corrections to 
the two-body widths involving strongly interacting particles and the 
dominant electroweak effects to all processes are evaluated as well.
\end{abstract}
%
%
\section{Introduction}
The search for new particles predicted by supersymmetric (SUSY) theories is a 
major goal of present and future colliders. In the Minimal Supersymmetric 
Standard Model (MSSM) \cite{Djouadi:1998di} there are still over 20 free 
parameters even in a phenomenologically viable model. It is therefore a very 
complicated task to deal with all the properties of the SUSY particles once 
they are found. Since their properties will be determined with an accuracy of 
a few per cent at the LHC and a precision at the per cent level or below at 
future $e^+ e^-$ linear colliders, the mass spectra, the various couplings, 
the decay branching ratios and the production cross sections have to be 
calculated with a rather high precision, also including higher order effects. 
The Fortran code {\tt SDECAY}\footnote{The code can be obtained at the url: http://people.web.psi.ch/muehlleitner/SDECAY} \cite{Muhlleitner:2003vg} which is 
presented here calculates the decays of SUSY particles in the MSSM, including 
the most important higher order effects. The Renormalization Group Equation 
(RGE) program {\tt SuSpect} \cite{Djouadi:2002ze} is used for the calculation 
of the mass spectrum and the soft SUSY-breaking parameters. [Of course, 
{\tt SDECAY} can be easily linked to any other RGE code.] Due to the limited 
space we refer for details of the notation, the description of the algorithm 
that is used in the code and the various higher order effects that have been 
included to the user's manual of {\tt SuSpect}. The program {\tt SDECAY} then 
evaluates the various couplings of the SUSY particles and MSSM Higgs bosons 
and calculates the decay widths and the branching ratios of all the two-body 
decay modes, including the QCD corrections to the processes involving 
coloured particles and the dominant electroweak effects to all processes. The 
loop-induced two-body decay channels as well as the possibly important higher 
order decays are included, such as the three-body decays of charginos, 
neutralinos, gluinos and top squarks and the four-body decays of the lighter 
top squark. In addition, the top quark SUSY decay widths and branching ratios 
are implemented. The program will be presented in the following.

\section{The decays of the supersymmetric particles}

\subsection{The tree level two-body decays}
The Fortran code {\tt SDECAY} includes the two-body decays 
of sfermions into a fermion and a gaugino,
as well as into a lighter sfermion of the same isodoublet and a gauge boson 
$V\!=\!W,Z$ or a Higgs boson $\Phi= h,H,A,H^\pm$
\beq
\tilde{f}_i & \to & \chi_j f^{(')} 
\eeq
\beq
\tilde{f}_i & \to & V \tilde{f}_j^{(')} \\
\tilde{f}_i & \to & \Phi \tilde{f}_j^{(')} 
\eeq  
For squarks heavier than the gluino the decay into a gluino-quark final state 
is also possible
\beq
\tilde{q}_i \to q \tilde{g} 
\eeq 
The heavier neutralino and chargino decays into the lighter chargino and 
neutralino states and gauge or Higgs bosons as well as the decays into 
fermion-sfermion pairs have been implemented
\beq
\chi_i & \to & \chi_j V \label{chi2bod1} \\ 
\chi_i & \to & \chi_j \Phi \label{chi2bod2} \\
\chi_i & \to & f \tilde{f}_j^{(')} \label{chi2bod3}
\eeq
For the gluinos the only relevant decay into a squark-quark pair is calculated 
\beq
\tilde{g} \to q \tilde{q}_i
\eeq 
In the case of a GMSB model the decays of the next-to-lightest SUSY particle 
(NLSP), which can be either the lightest neutralino $\chi_1^0$ or the 
lightest sfermion, in general the $\tilde{\tau}_1$, into a Gravitino 
$\tilde{G}$ and a photon, $Z$ or neutral Higgs 
boson (for $\chi_1^0$) and a $\tau$ (for $\tilde{\tau}_1$) are implemented
\beq
\chi_1^0 & \to & \gamma\tilde{G}, Z \tilde{G}, \Phi \tilde{G} \\
\tilde{\tau}_1 & \to & \tau \tilde{G}
\eeq 
The masses entering the phase space in the calculation of the widths are
the pole masses, but when they enter the various couplings they are - for the 
third-generation fermions - the running $\overline{{\rm DR}}$ masses at the 
scale of Electroweak Symmetry Breaking (EWSB). This is also the case for all 
soft SUSY-breaking parameters and the third generation sfermion mixing angles 
involved in the couplings. In addition, we have left the option for the QCD 
coupling constant and the bottom, top Yukawa couplings to be evaluated at the 
scale of the decaying superparticle or any other scale. In this case, only the 
standard QCD corrections are included in the running \cite{Djouadi:1996gt}.

\subsection{The QCD corrected two-body decays}
The one-loop QCD corrections to the following two-body decays involving 
(s)quarks and gluinos have been implemented using the formulae of Refs. 
\cite{Kraml:1996kz,Djouadi:1997wt}, 
\cite{Arhrib:1998nf,Bartl:1998xp,Bartl:1998pb,Bartl:1998xk} and 
\cite{Beenakker:1996dw,Beenakker:1997de}, respectively, 
\beq
\tilde{q}_i & \to & \chi_j q^{(')} \\
\tilde{q}_i & \to & \Phi \tilde{q}_j^{(')} \\
\tilde{q}_i & \to & q \tilde{g} \quad {\rm and} \quad 
\tilde{g} \to \tilde{q}_i q 
\eeq
All the corrections have been included in the $\overline{{\rm DR}}$ scheme.
The bulk of the electroweak radiative corrections due to the running of the 
gauge and third-generation fermion Yukawa couplings has been taken into 
account by evaluating these parameters at the EWSB scale.

\subsection{Loop-induced decays}
In case the two-body decays of the next-to-lightest neutralino are 
kinematically not allowed the loop-induced decay into the lightest 
supersymmetric particle (LSP) $\chi_1^0$ and a photon is calculated 
\cite{Haber:1989px,Ambrosanio:1996az,Ambrosanio:1997gz,Baer:2002kv}
\beq
\chi_2^0 \to \chi_1^0 \gamma
\eeq
For completeness, the loop-induced decay of a gluino into a gluon and the LSP 
has also been considered \cite{Ma:1988ns,Barbieri:1988ed,Baer:1990sc}
\beq
\tilde{g} \to g \chi_1^0
\eeq
If the tree-level stop two-body decays are kinematically closed the 
loop-induced decay into a charm and $\chi_1^0$ \cite{Hikasa:1987db} is 
calculated 
\beq
\tilde{t}_i \to c \chi_1^0 \label{stoploop}
\eeq

\subsection{Multibody decay modes}
If the two-body decays of the gauginos Eqs. (\ref{chi2bod1}-\ref{chi2bod3}) 
are kinematically forbidden the three-body decays into a lighter gaugino and 
a fermion pair and a gluino and two quarks are calculated 
\beq
\chi_i & \to & \chi_j f \bar{f}^{(')} \label{3bodb} \\
\chi_i & \to & \tilde{g} q \bar{q}^{(')}
\eeq
Analogously, the gluino three-body decays into a gaugino and two quarks are
considered when the gluino two-body modes are closed
\beq
\tilde{g} \to \chi_i q \bar{q}^{(')} \label{3bode}
\eeq
For the calculation of the processes Eqs.~(\ref{3bodb}-\ref{3bode}) we 
have used the formulae given in 
\cite{Bartl:1994bu,Bartl:1999iw,Baer:1997yi,Baer:1998sz,Djouadi:2001fa}.
Furthermore, the possibly important gluino decay into stop, bottom and a $W$ 
boson as well as the decay into stop, bottom and a charged Higgs boson have 
been implemented \cite{Porod:2002wz,Datta:2001qs}
\beq
\tilde{g} & \to & \tilde{t}_1 \bar{b} W^- \\
\tilde{g} & \to & \tilde{t}_1 \bar{b} H^-
\eeq
In case the stop two-body decays are not accessible, there are several 
three-body decay modes 
\cite{Porod:1997at,Porod:1998yp,Datta:1998dc,Djouadi:2001dx,Djouadi:2000aq,
Djouadi:2000bx} that can dominate over the loop-induced decay 
Eq.~(\ref{stoploop}) in rather large areas of the MSSM: the decays into a 
bottom, lightest neutralino and a $W$ or charged Higgs boson, the decay modes 
into bottom, lepton and slepton, the decays into the lightest sbottom and a 
fermion pair as well as for the heavy stop the possibility of decaying into 
the lighter stop and a fermion pair
\beq
\tilde{t}_i & \to & b W^+ \chi_1^0 \quad , \quad\quad\;\; b H^+ \chi_1^0 \\
\tilde{t}_i & \to & b l^+ \tilde{\nu}_l \quad {\rm and/or} \quad
b \tilde{l}^+ \nu_l \\
\tilde{t}_i & \to & \tilde{b}_1 f \bar{f}' \\
\tilde{t}_2 & \to & \tilde{t}_1 f \bar{f} 
\eeq
\noindent
{\tt SDECAY} evaluates the three-body decays if the two-body decays are closed,
taking into account all possible contributions of virtual particles, the 
radiatively corrected Yukawa couplings of third-generation fermions, the 
mixing pattern for their sfermion partners and the masses of the sparticles 
and gauge/Higgs bosons involved in the processes. Even the masses of the final 
state fermions have been included. The total decay widths of the exchanged 
particles have not been included in the propagators of the virtual particles.

If the stop three-body decay channels are kinematically forbidden the 
$\tilde{t}_1$ four-body decay mode into a bottom, the LSP and two massless 
fermions can become competitive with the loop induced decay into a charm and a 
neutralino, cf.~Eq.~(\ref{stoploop}), so that this channel \cite{Boehm:1999tr} 
has also been included in the program, 
\beq
\tilde{t}_1 \to b \chi_1^0 f \bar{f}'
\eeq

\subsection{Top quark decays}
For the top quark the following decays in the MSSM are calculated by 
{\tt SDECAY}
\beq
t & \to & b W^+ \\
t & \to & b H^+ \quad {\rm and} \quad \tilde{t}_1 \chi_1^0
\eeq

\section{How to use {\tt SDECAY}}
Apart from the files of the program {\tt SuSpect}, {\it i.e.} 
{\tt suspect2.in}, {\tt suspect2.f}, {\tt subh\_hdec.f}, {\tt feynhiggs.f}, 
{\tt hmsusy.f}, the program {\tt SDECAY} consists of three files: \newline
\phantom{h} 
1) \underline{The input file {\tt sdecay.in}} where one can 
choose the accuracy of the algorithm and the various options whether QCD 
corrections and multibody or loop decays are included or not, which scales and 
how many loops are used for the running couplings and if top and 
GMSB decays are calculated or not. \newline
\phantom{h} 
2) \underline{The main routine {\tt sdecay.f}} where the couplings 
of the SUSY and Higgs particles are evaluated and the decay branching ratios
and total widths are calculated. \newline
\phantom{h} 
3) \underline{The output file {\tt sdecay.out}} which gives the results
for the branching ratios and total widths, as well as the masses of the SUSY 
and Higgs particles, the mixing matrices and the gauge and third-generation 
Yukawa couplings at the EWSB or a chosen scale. The output is given in two 
possible formats, either in a simple and transparent form or according to the 
SUSY Les Houches Accord \cite{Skands:2003cj} which uses the PDG notation for 
the particles.  
\newline
All these files together with a makefile to compile the files can be found on 
the web page dedicated to {\tt SDECAY} at the address:
\vspace{-0.3cm}
\begin{center}
{\tt http://people.web.psi.ch/muehlleitner/SDECAY}
\end{center}

\section{Conclusions}
We have presented the Fortran code {\tt SDECAY}, which calculates the decay 
widths and branching ratios of all the two-body decays of the SUSY particles 
in the framework of the MSSM, including the QCD corrections to the decays 
involving strongly interacting particles, the three-body decays of the 
gauginos, gluinos and stops, as well as the four-body decays of the lightest 
top squark. Furthermore, the loop-induced decays of the gluino, the lightest 
neutralino and the lightest top squark, the decays of the next-to-lightest 
SUSY particle in GMSB models and the standard and SUSY decay modes of the top 
quark have been implemented. The dominant electroweak corrections due to the 
running of the gauge and fermion Yukawa couplings have been incorporated. The 
program which uses the RGE code {\tt SuSpect} can be easily linked to any 
other spectrum calculator. It is user-friendly, flexible for the choice of 
options and approximations and quite fast. The program is under rapid 
development and will be updated regularly.


\setcounter{figure}{0}
\setcounter{table}{0}
\setcounter{section}{0}
\setcounter{equation}{0}
\clearpage

\def\chioi{\tilde{\chi}^0_1}
\def\chioii{\tilde{\chi}^0_2}
\def\chipm{\tilde{\chi}^{\pm}_1}
\def\staui{\tilde{\tau}_1}
\def\slr{\tilde{l}_R}
\def\sql{\tilde{q}_L}
\def\sqr{\tilde{q}_R}
\def\ETM{E_T^{miss}}

\def\mg{m_{\tilde{g}}^2}
\def\mq{m_{\tilde{q}_L}^2}
\def\mT{m_{\tilde{\chi}_2^0}^2}
\def\ml{m_{\tilde{l}_R}^2}
\def\mlfour{m_{\tilde{l}_R}^4}
\def\mO{m_{\tilde{\chi}_1^0}^2}
\def\threshold{{\rm thres}}
\def\edge{{\rm edge}}
\def\max{{\rm max}}
\def\min{{\rm min}}
\def\MT2{M_{T2}}
\def\qn{q_{\rm n}}
\def\qf{q_{\rm f}}
\def\ln{\ell_{\rm n}}
\def\lf{\ell_{\rm f}}
\def\m0{${0}$}
\def\mhf{${1/2}$}
\def\tanb{tan${\beta}$}
\def\tg{{\tilde g}}
\def\tq{{\tilde q}}
\def\tu{{\tilde u}}
\def\td{{\tilde d}}
\def\ttop{{\tilde t}}
\def\tb{{\tilde b}}
\def\tchi{{\tilde\chi}}
\def\tl{{\tilde\ell}}
\def\sel{{\tilde e_L}}
\def\smul{{\tilde \mu_L}}
\def\sne{{\tilde\nu_e}}
\def\snm{{\tilde\nu_\mu}}
\def\lsp{{\tilde\chi_1^0}}

\part{Measuring The Mass Of The Lightest Chargino At The CERN LHC \label{chargino}}
{\it M.M. Nojiri, G. Polesello and D.R. Tovey}
\maketitle
\begin{abstract}
Results are presented of a feasibility study of techniques for
measuring the mass of the lightest chargino at the CERN LHC. These
results suggest that for one particular mSUGRA model a statistically
significant chargino signal can be identified and the chargino mass
reconstructed with a precision $\sim$ 11\% for $\sim$ 100 fb$^{-1}$ of
data.
\end{abstract}

\section{INTRODUCTION}

Much work has been carried out recently on measurement of the masses
of SUSY particles at the LHC
\cite{atlasTDR,Hinchliffe:1997iu,Hinchliffe:1999zc,Bachacou:1999zb,
Allanach:2000kt,Abdullin:1998pm}. These measurements can often be
considered to be `model-independent' in the sense that they require
only that a particular SUSY decay chain exists with an observable
branching ratio. A good starting point is often provided by the
observation of an opposite-sign same-flavour (OS-SF) dilepton invariant
mass spectrum end-point whose position measures a combination of the
masses of the $\chioii$, the $\chioi$ and possibly also the
$\tilde{l}^{\pm}$. Observation of end-points and thresholds in
invariant mass combinations of some or all of these leptons with
additional jets then provides additional mass constraints sufficient
to allow the individual sparticle masses to be reconstructed
unambiguously. A question remains however regarding how the mass of a
SUSY particle can be measured if it does not participate in a decay
chain producing an OS-SF dilepton signature. This problem has been
addressed for some sparticles (e.g. for the $\sqr$ \cite{LHCLC}) however
significant exceptions remain. Notable among these is the case of the
lightest chargino $\chipm$, which does not usually participate in
decay chains producing OS-SF dileptons due to its similarity in mass to
the $\chioii$.

In this paper we attempt to measure the mass of the $\chipm$ by
identifying the usual OS-SF dilepton invariant mass end-point arising
from the decay via $\chioii$ of the {\em other} initially produced
SUSY particle (i.e. not the one which decays to produce the
$\chipm$). We then solve the mass constraints for that decay chain to
reconstruct the momentum of the $\chioi$ appearing at the end of the
chain, and use this to constrain the momentum (via $\ETM$) of the
$\chioi$ appearing at the end of the decay chain involving the
$\chipm$. We finally use mass constraints provided by additional jets
generated by this chain to solve for the $\chipm$ mass. The technique
requires that both the decay chain 
$$
\sql \rightarrow \chioii q \rightarrow \slr lq \rightarrow \chioi llq
$$ 
and the decay chain 
$$
\sql \rightarrow \chipm q \rightarrow qW^{\pm} \chioi \rightarrow qq'q''
\chioi
$$ 
are open with significant branching ratios, and that the
masses of the $\chioi$, $\chioii$, $\slr$ and $\sql$ are known. No
other model-dependent assumptions are required however.

\section{SUSY MODEL AND EVENT GENERATION}

The SUSY model point chosen was that used recently by ATLAS for full
simulation studies of SUSY mass reconstruction \cite{fullsim}. This is a
minimal Supergravity (mSUGRA) model with parameters $m_0$ = 100 GeV,
$m_{1/2}$ = 300 GeV, $A_0$ = -300 GeV, $\tan(\beta)$ = 6 and
$\mu>0$. The mass of the lightest chargino is 218 GeV, while those of
the $\sql$, the $\slr$, the $\chioii$ and the $\chioi$ are $\sim$ 630
GeV, 155 GeV, 218 GeV and 118 GeV respectively. One of the
characteristics of this model is that the branching ratio of $\chipm
\rightarrow W^{\pm} \chioi$ is relatively large ($\sim$ 28
\%). Chargino mass reconstruction involving the decay $\chipm
\rightarrow \staui \nu_{\tau}$ (BR $\sim$ 68 \%) is likely to be very
difficult due to the additional degress of freedom provided by the
missing neutrino. Consequently the $W^{\pm}$ decay mode must be used.

The electroweak SUSY parameters were calculated using the ISASUGRA
7.51 RGE code \cite{Baer:1999sp}. SUSY events equivalent to an
integrated luminosity of 100 fb$^{-1}$ were then generated using Herwig
6.4 \cite{Corcella:2000bw,Moretti:2002eu} interfaced to the ATLAS fast
detector simulation ATLFAST 2.21 \cite{atlfast}. With the standard
SUSY selection cuts described below Standard Model backgrounds are
expected to be negligible. An event pre-selection requiring at least
two ATLFAST-identified isolated leptons was applied in order to reduce
the total volume of data.

\section{CHARGINO MASS RECONSTRUCTION}

Events were required to satisfy `standard' SUSY selection criteria
requiring a high multiplicity of high $p_T$ jets, large $\ETM$ and
multiple leptons:

\begin{itemize}

\item at least 4 jets (default ATLFAST definition \cite{atlfast}) with
$p_T$ $>$ 10 GeV, two of which must have $p_T$ $>$ 100 GeV,

\item $\left(\sum_{i=1}^4 p_{T(jet)}^i + \ETM\right)$ $>$ 400 GeV,

\item $\ETM$ $>$ max$\left(100\mathrm{GeV},0.2\left(\sum_{i=1}^4
p_{T(jet)}^i + \ETM\right)\right)$,

\item exactly 2 opposite sign same flavour isolated electrons or muons
with $p_T$ $>$ 10 GeV,

\item no b-jets or $\tau$-jets.

\end{itemize}

Events were further required to contain dileptons with an invariant
mass less than the expected $l^{\pm}l^{\mp}$ end-point position (100.2
GeV) and at least one dilepton + hard jet combination (one for each
combination of the dilepton pair with each of the two hardest jets)
with an invariant mass less than the expected $l^{\pm}l^{\mp}q$
end-point position (501.0 GeV). The smaller dilepton + hard jet
combination then defined which jet (assumed to be from the decay $\sql
\rightarrow \chioii q$) would be used together with the dileptons to
reconstruct the $\chioii$ production and decay chain.

The momentum of the $\chioi$ at the end of the $\chioii$ decay chain
was calculated by solving analytically the kinematic equations
relating the momenta of the decay products (including the $\chioi$) to
the masses of the SUSY particles, which were assumed to be known from
conventional end-point measurements
\cite{atlasTDR,Hinchliffe:1997iu,Hinchliffe:1999zc,Bachacou:1999zb,
Allanach:2000kt,Abdullin:1998pm}. This process is described in more
detail in Ref.~\cite{mmn} and results in two solutions for the
$\chioi$ momentum for each of the two possible mappings of the
reconstructed leptons to the sparticle decay products. In the present
analysis just one such mapping was assumed with no attempt being made
to select the correct assignment. Two possible solutions for the
$\chioi$ momentum were therefore obtained for each event.

\begin{figure}[thb]
\begin{center}
\epsfig{file=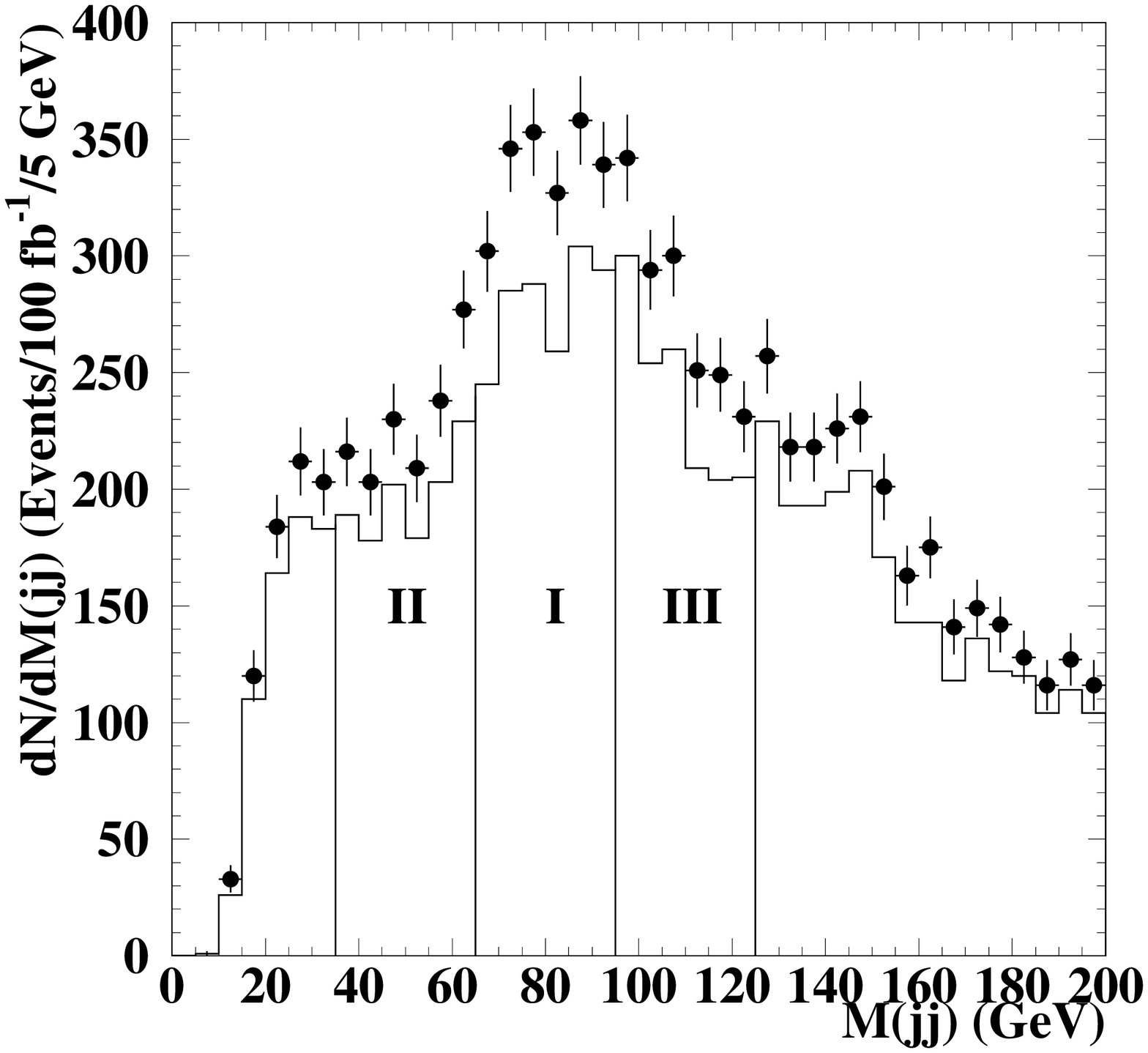,height=3in}
\caption{\label{fig1} {Reconstructed dijet invariant mass
distributions for all events (data points) and events not containing
the decay chain $\chipm \rightarrow W^{\pm} \chioi \rightarrow q'q''
\chioi$ selected using Monte Carlo truth. The signal band is labelled
`I' in the figure, while the two sideband are labelled `II' and `III'
respectively.}}
\end{center}
\end{figure}

The nest step in the reconstruction was to find the jet pair resulting
from a hadronic $W^{\pm}$ decay following production via $\chipm
\rightarrow W^{\pm} \chioi$. The potentially large combinatorial
background was reduced by rejecting jet combinations involving either
of the two hardest jets (since these were assumed to arise from $\sql$
decay) and by requiring that the harder(smaller) of the two jets
possessed $p_T$ greater than 40(20) GeV (i.e. selecting asymmetric jet
pairs consistent with a significant boost in the lab frame). A further
cut was applied on the invariant mass of the combination of the jet
pair with the hard jet giving the larger dilepton + jet mass (assumed
therefore to be the jet from the $\sql \rightarrow \chipm q$ decay
preocess). This invariant mass was conservatively required to be less
than that of the $\sql$.

\begin{figure}[thb]
\begin{center}
\unitlength=1in
\begin{picture}(5,5)
\put(0,2.5){\epsfig{file=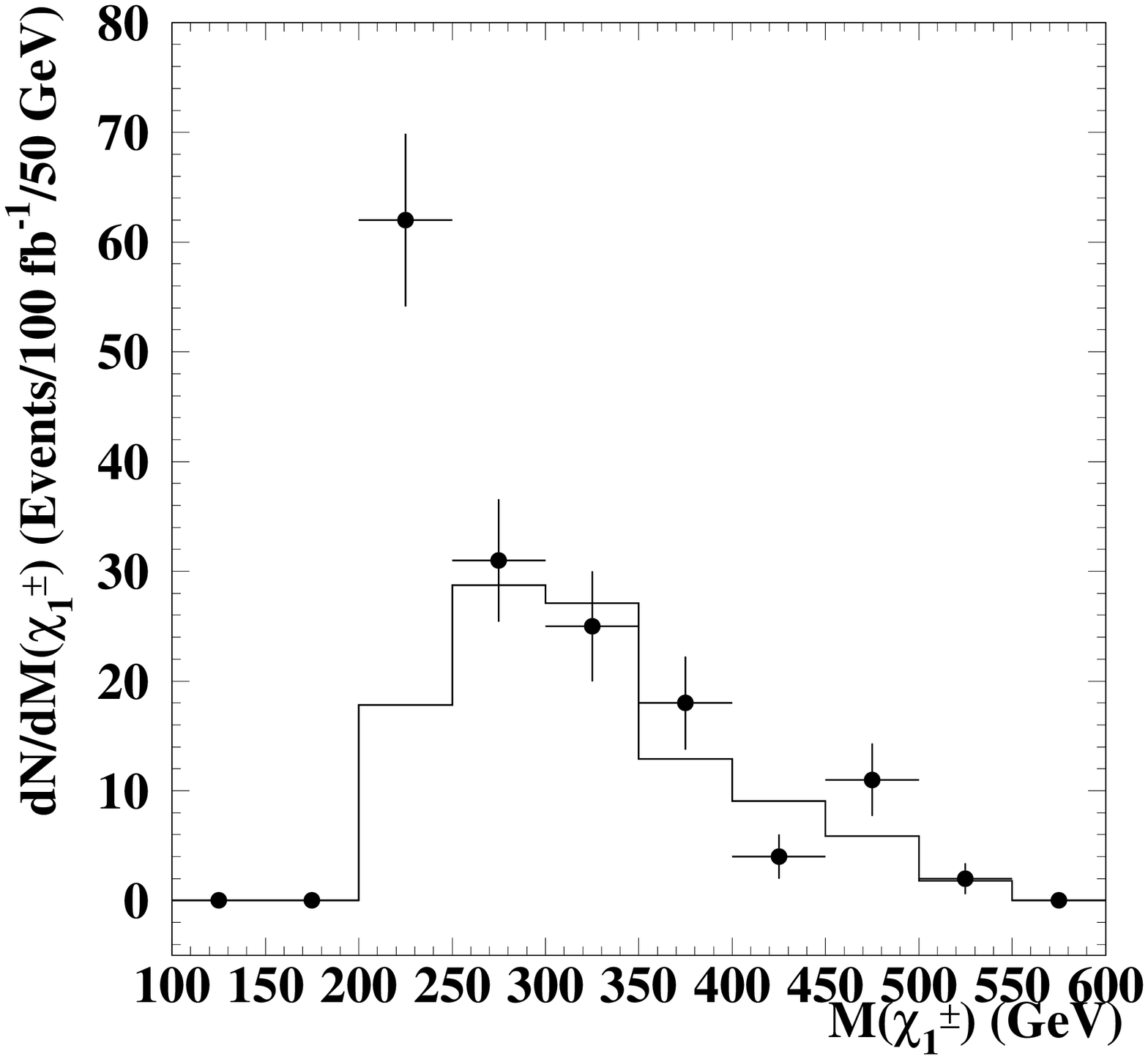,height=2.5in}}
\put(3,2.5){\epsfig{file=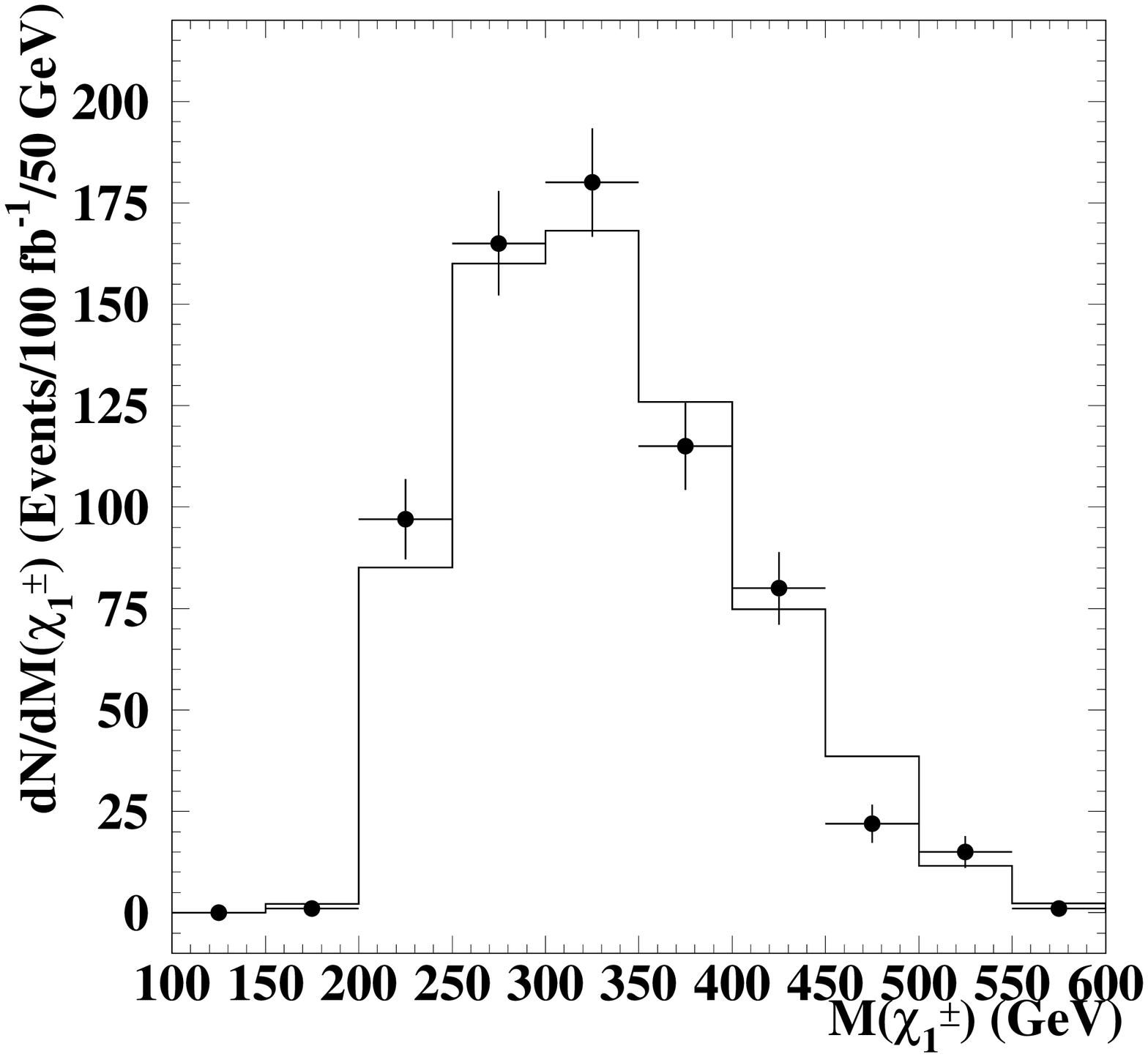,height=2.5in}}
\put(1.5,0){\epsfig{file=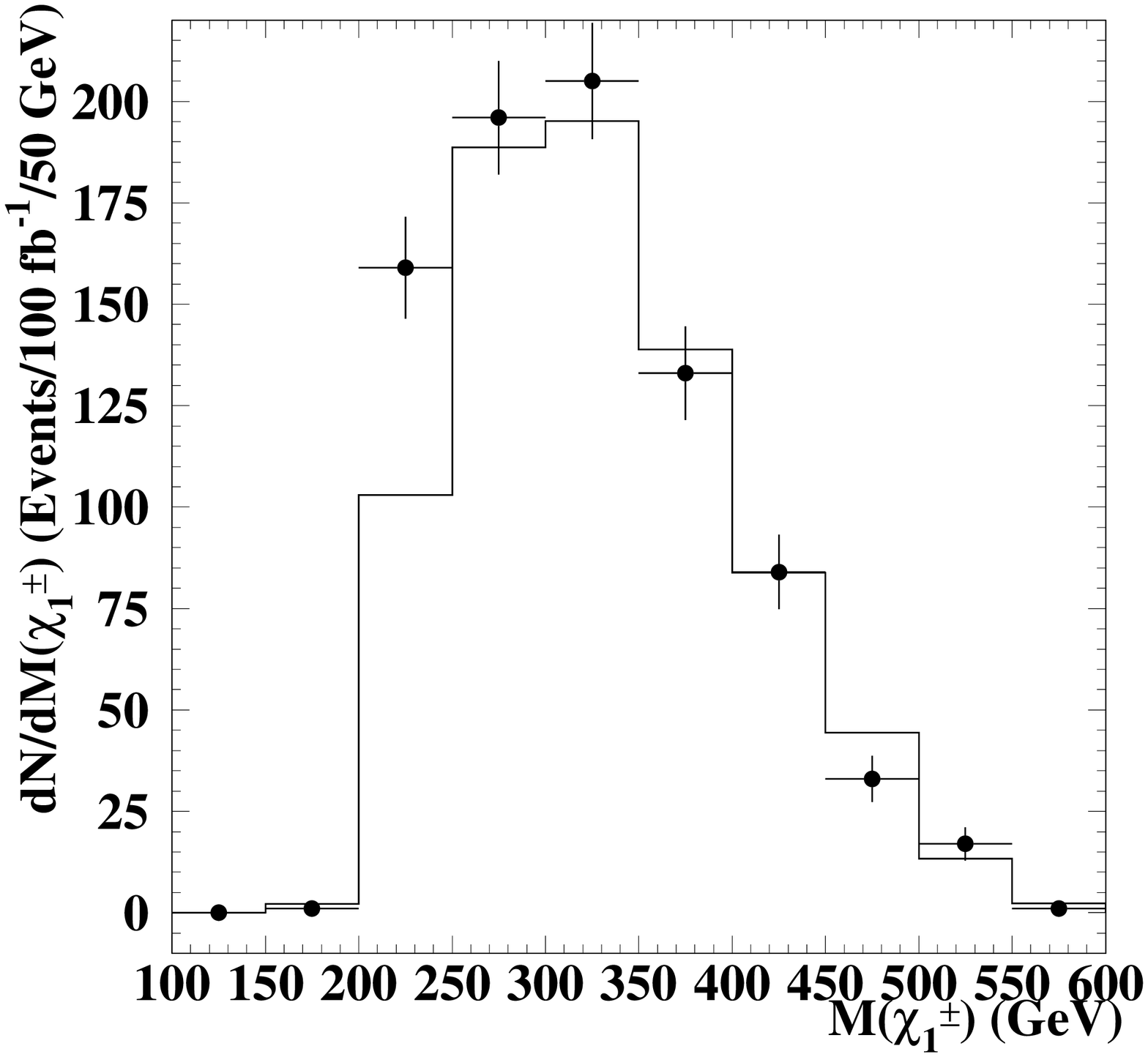,height=2.5in}}
\end{picture}
\caption{\label{fig2} {Reconstructed $\chipm$ mass distributions
showing signal distributions with $|m_{jj}-m_W|$ $<$ 15 GeV (data
points) and sideband distributions with 15 GeV $<$ $|m_{jj}-m_W|$ $<$
45 GeV (histograms). The left hand figure was obtained by selecting
events containing the decay chain $\chipm \rightarrow W^{\pm} \chioi
\rightarrow q'q'' \chioi$ using Monte Carlo truth. The central figure
was obtained by selecting background events not containing this decay
chain. The right hand figure was obtained by using all data.}}
\end{center}
\end{figure}

For each event any jet pairs satisfying the above criteria and
possessing $|m_{jj}-m_W|$ $<$ 15 GeV (Fig.~\ref{fig1}), were
considered to form $W$ candidates. For each event the candidate with
$m_{jj}$ nearest $m_W$ was then selected and used together with the
momentum of the hard jet identified previously and the two assumed $x$
and $y$ components of the $\chioi$ momentum (calculated from the two
solutions for the momentum of the $\chioi$ from the $\chioii$ decay
and $\ETM$) to calculate the chargino mass. Each of the two solutions
for the $\chioi$ momentum gives two possible solutions for
$m_{\chipm}$, the smaller of which is usually physical. Consequently
two possible values for $m_{\chipm}$ were obtained from each event
(plotted in Fig.~\ref{fig2}).

Following this procedure significant backgrounds remain from
combinatorics in SUSY signal events (due to their high average
multiplicity), and from SUSY background events (i.e. events in which
the decay process $\sql \rightarrow \chipm q \rightarrow qW^{\pm}
\chioi \rightarrow qq'q'' \chioi$ is not present). These backgrounds
(or at least those not involving a real $W^{\pm}$ decay) were removed
statistically using a sideband subtraction technique similar to that
described in Ref.~\cite{Hisano:2003qu}. All jet pairs satisfying all
the above selection criteria except the $|m_{jj}-m_W|$ requirement
were recorded if they satisfied the alternative requirement that 15
GeV $<$ $|m_{jj}-m_W|$ $<$ 45 GeV. This requirement then defined two
side-bands located on either side of the main signal band
($|m_{jj}-m_W|$ $<$ 15 GeV) of equal width 30 GeV. The momentum of
each jet pair was then rescaled such that the difference between its
rescaled mass and $m_W$ was the same as the difference between its
original mass and the centre of its sideband (50 or 110 GeV
respectively). Each jet pair was then given a weight of 1.3 (lower
sideband) or 1.0 (upper sideband) to account for the variation of the
background $m_{jj}$ distribution with $m_{jj}$
(Fig.~\ref{fig1}). Values for the chargino mass were then calculated
for each jet pair and used to create a sideband mass distribution
(Fig.~\ref{fig2}). Finally the sideband mass distribution was
subtracted from the signal mass distribution with a relative
normalisation factor of 0.7 to account for the differing efficiencies
for selecting sideband events and background events in the signal
region.

\section{RESULTS}

\begin{figure}[t]
\begin{center}
\epsfig{file=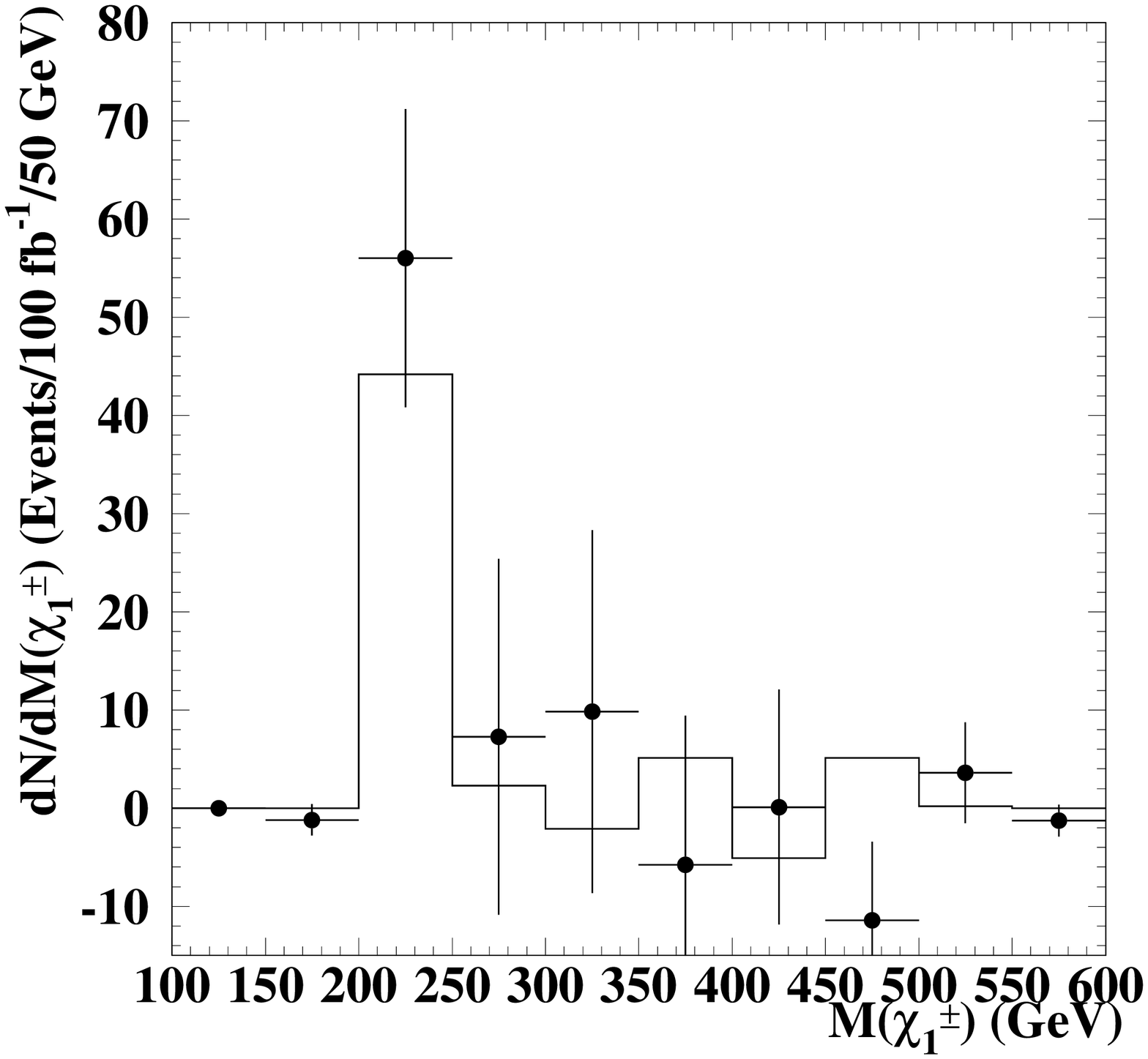,height=3in}
\caption{\label{fig3} {Reconstructed $\chipm$ mass distribution
for $\chipm \rightarrow W^{\pm} \chioi \rightarrow q'q'' \chioi$
signal events (histogram) and all events (points with errors).}}
\end{center}
\end{figure}

The sideband subtracted chargino mass distributions obtained from this
process are shown in Fig.~\ref{fig3}, both with and without a
selection requirement for $\chipm \rightarrow W^{\pm} \chioi
\rightarrow q'q'' \chioi$ obtained from Monte Carlo truth. In both
cases no events are observed at masses below the kinematic limit of
198 GeV ($=m_W+m_{\chioi}$) due to the origin of the mass values as
solutions to the kinematic mass relations. In the case where Monte
Carlo truth was used as input a clear peak is seen in the 200 GeV -
250 GeV bin, corresponding well to the actual mass of 218 GeV. At
higher mass values the sideband subtraction process has worked well
and the distribution is consistent with zero. In the case where no
Monte Carlo truth signal event selection has been performed (points
with errors) a clear peak is again seen in the vicinity of the
chargino mass, with few events at higher values. For 100 fb$^{-1}$ the
statistical significance of the peak is around 3 $\sigma$ indicating
that more integrated luminosity (or an improved event selection) would
be required to claim a 5 $\sigma$ discovery. Nevertheless it seems
reasonable to claim that if this data were generated by an LHC
experiment such as ATLAS, and that the observed signal were indeed not
a statistical fluctuation, then the mass of the lightest chargino
could be measured to a statistical precision $\sim$ $\pm$ 25 GeV
($\sim$ 11 \%). More work is needed to determine the likely systematic
error in this quantity arising from effects such as the statistical
and systematic uncertainty in the input sparticle masses used when
calculating the $\chioi$ momentum and $\chipm$ mass.

More work is needed to identify the optimum set of selection criteria
required to identify hadronic $W^{\pm}$ decays in this sample, with
the efficiency of the tau veto (required to remove $\chipm$ decays via
$\staui \nu_{\tau}$) in particular needing to be optimised.  Possible
methods for selecting the correct lepton mapping used to calculate the
$\chioi$ momentum also deserve further study. With these improvements
and/or more integrated luminosity it should then be possible both to
increase the accuracy of the chargino mass measurement and to study
quantities such as the helicity of the $\chipm$ through measurement of
the invariant mass distribution of the $W^{\pm}$ and the hard jet
produced alongside the $\chipm$ in the decay of the parent $\sql$.

\section{CONCLUSIONS}

A study of the identification and measurement of charginos decaying to
$W^{\pm} \chioi$ produced at the LHC has been performed. The results
indicate that for one particular mSUGRA model the mass of the $\chipm$
can be measured with a statistical precision $\sim$ 11 \% for 100
fb$^{-1}$ of integrated luminosity.

\vskip1cm
\noindent

\section*{ACKNOWLEDGEMENTS}

This work was performed in the framework of the workshop:
Les Houches 2003: Physics at TeV Scale Colliders.
We wish to thank the staff and organisers for all their hard work 
before, during and after the workshop.
We thank members of the ATLAS Collaboration for helpful
discussions. We have made use of ATLAS physics analysis and
simulation tools which are the result of collaboration-wide
efforts.  DRT wishes to acknowledge PPARC and the
University of Sheffield for support.

\setcounter{figure}{0}
\setcounter{table}{0}
\setcounter{section}{0}
\setcounter{equation}{0}
\clearpage

\def\nn {\nonumber}

\def\ti    {\tilde}

\def\sel   {{\ti e}_L}
\def\ser   {{\ti e}_R}
\def\sne  {{\ti\nu}_e}
\def\el   {{\ti e}^-_L}
\def\er   {{\ti e}^-_R}
\def\nt   {{\ti \chi}^0}

\def\stauone  {{\ti\tau}_1}
\def\stautwo  {{\ti\tau}_2}
\def\snt      {{\ti\nu}_\tau}

\def\anue {{\bar\nu}_e}
\def\anum {{\bar\nu}_\mu}
\def\anut {{\bar\nu}_\tau}

\def\cL   {\cos\Phi_L}
\def\sL   {\sin\Phi_L}
\def\ctL  {{\cos2\Phi_L}}
\def\stL  {{\sin2\Phi_L}}
\def\cR   {\cos\Phi_R}
\def\sR   {\sin\Phi_R}
\def\ctR   {\cos2\Phi_R}
\def\stR   {\sin2\Phi_R}

\newcommand{\mse}[1]   {m_{\ti e_{#1} }}
\newcommand{\mstau}[1] {m_{\ti \tau_{#1} }}
\newcommand{\mnt}[1]   {m_{\ti \chi^0_{#1} }}
\newcommand{\mch}[1]   {m_{\ti \chi^\pm_{#1} }}
\newcommand{\msnue}     {m_{\ti \nu_e}}

\newcommand{\gsim}{\;\raisebox{-0.9ex}
           {$\textstyle\stackrel{\textstyle>}{\sim}$}\;}

\newcommand{\lsim}{\;\raisebox{-0.9ex}{$\textstyle\stackrel{\textstyle<}
          {\sim}$}\;}

\newcommand{\eq}[1]{eq.~(\ref{#1})}
\newcommand{\fig}[1]{fig.~\ref{#1}}
\newcommand{\tab}[1]{table~\ref{#1}}

 \part{Chargino/Neutralino Sector In Combined Analyses At LHC/LC \label{leshouches_fin}}
{\it K. Desch, J. Kalinowski, G. Moortgat-Pick,
M.M. Nojiri and
G. Polesello}
\maketitle
\begin{abstract}
We demonstrate how the interplay of a future $e^+e^-$  LC at its first
stage with $\sqrt{s} \lsim 500$~GeV and of  the LHC
could lead to a precise determination  of the
fundamental SUSY parameters in the gaugino/higgsino sector without
assuming a specific supersymmetry breaking scheme.
The results are shown for  the benchmark
scenario SPS1a, taking into account
realistic errors for the masses and cross sections
measured at the LC with polarised beams and mass measurements at
the LHC. 
\end{abstract}

\section{INTRODUCTION}
The unconstrained MSSM has 105 new parameters and
SUSY analyses at future experiments, at the LHC and at a future Linear
Collider (LC), will have
to focus on the determination of these parameters 
\cite{Tsukamoto:1995gt,Feng:1995zd}.
An interesting possibility
to explore SUSY is to start with the
gaugino/higgsino particles which are expected to be among
the lightest SUSY particles.
At tree level, this sector depends
only on 4 parameters: $M_1$, $M_2$, $\mu$ and $\tan\beta$ -- the U(1)
and SU(2) gaugino masses, the higgsino mass parameter and the ratio of
the vacuum expectations of the two Higgs fields, respectively.

Some strategies have been worked out for the determination of
the parameters $M_2$, $M_1$, $\mu$, $\tan\beta$ even if only the
light gaugino/higgsino particles, $\tilde{\chi}^0_1$,
$\tilde{\chi}^0_2$ and $\tilde{\chi}^\pm_1$ were kinematically
accessible at the first stage of the LC \cite{Choi:2001ww,Choi:2002mc}.  In this
contribution we demonstrate how such an LC analysis could be
strengthened if in addition some information on the mass of the
heaviest neutralino from the LHC is available.  We consider the cases: 
(i) stand alone LC data and (ii) joint analysis of the 
LC and the LHC data.  The results in the
last scenario will clearly demonstrate the essentiality of the LHC and
LC and the benefit from the joint analysis of their data.

We take the SPS1a as a
working benchmark \cite{Allanach:2002nj,Ghodbane:2002kg} 
and assume that only the first
phase of a LC with a tunable energy up to $\sqrt{s}=500$~GeV
would overlap with the LHC running. Furthermore, we assume an
integrated luminosity of  $\int {\cal L} \sim 500$~fb$^{-1}$
and polarised beams with $P(e^-)=\pm 80\%$, $P(e^+)=\pm 60\%$.
In the following $\sigma_L$ will refer to cross sections obtained
with $P(e^-)=- 80\%$, $P(e^+)= + 60\%$, and $\sigma_R$ with $P(e^-)=+
80\%$, $P(e^+)= - 60\%$.

\section{THE GAUGINO/HIGGSINO SECTOR}
The  mass matrix ${\cal M}_C$
of the charged gaugino $\ti W^\pm$ and higgsino $\ti
H^\pm$ depends on $M_2$, $\mu$, $\tan\beta$. The mass eigenstates
are the two charginos $\tilde{\chi}^\pm_{1,2}$. For real ${\cal M}_C$
the two unitary diagonalisation
matrices can be parameterised with two mixing angles
$\Phi_{L,R}$. The 
 mass eigenvalues $m^2_{\tilde{\chi}^\pm_{1,2}}$ and the mixing
angles are analytically given by the Susy parameters (see e.g. 
\cite{Choi:2000ta,Choi:1998ut}).
The cross section 
$\sigma^\pm\{ij\}=\sigma(e^+e^-\to\tilde{\chi}^{\pm}_i
\tilde{\chi}^{\mp}_j)$ can be expressed as a function of 
$(\cos 2 \Phi_{L,R},m^2_{\tilde{\chi}^\pm_{i}})$; the coefficients for
$\sigma^\pm\{11\}$ are explicitly given in \cite{DKMNP}.

The neutralino mixing matrix ${\cal M}_N$ depends on $M_1$, $M_2$, $\mu$ and 
$\tan\beta$. Analytic expressions for the mass eigenvalues  
$m^2_{\tilde{\chi}^0_{1,\ldots,4}}$ and the eigenvectors are e.g. given in
\cite{Choi:2001ww,Choi:2002mc}. The characteristic equation of the mass matrix squared,
${\cal M}_N {\cal M}^{\dagger}_N$, is written explicitly
as a quadratic equation for the parameter $M_1$ \cite{DKMNP}. 

\section{STRATEGY FOR THE DETERMINATION OF THE SUSY PARAMETERS}
At the initial phase of future $e^+e^-$
linear--collider operations with polarised beams, the collision energy
may only be sufficient to reach the production thresholds of the light
chargino $\tilde{\chi}^\pm_1$ and the two lightest neutralinos
$\tilde{\chi}^0_1,\, \tilde{\chi}^0_2$. Nevertheless the entire
tree level structure of the gaugino/higgsino sector can be 
unraveled~\cite{Choi:2001ww,Choi:2002mc,Choi:2000ta}.

Chargino  cross sections measured at  
$\sqrt{s}=400$~GeV and 500 GeV with polarised beams 
and the lightest chargino mass 
are sufficient to determine unambiguously the 
mixing angles $\cos 2 \Phi_{L,R}$. Then the $M_1$ can be obtained from
the  quadratic equation   
${\cal M}_N {\cal M}^{\dagger}_N$. However,  using the kinematically
accessible cross sections for
the neutralino production $\sigma^0_{L,R}\{12\}$ and $\sigma^0_{L,R}\{22\}$
leads to a precise determination of the fundamental Susy
parameters \cite{DKMNP}.

In the following we perform this strategy for the benchmark scenario
SPS1a \cite{Allanach:2002nj,Ghodbane:2002kg} 
defined at the electroweak scale:
$M_1=99.13~\mbox{GeV}$, $M_2=192.7~\mbox{GeV}$, 
$\mu=352.4~\mbox{GeV}$, $\tan\beta=10$; 
the resulting masses are given in \tab{tab_mass_LC}.
\subsection{SUSY PARAMETERS FROM THE LC DATA}
We use the light chargino and neutralino masses
$m_{\tilde{\chi}^{\pm}_1}$, $m_{\tilde{\chi}^{0}_{1,2}}$  
and the polarised cross sections for the processes
$e^+e^-\to\tilde{\chi}^+_1\tilde{\chi}^-_1$, 
$\tilde{\chi}^0_1\tilde{\chi}^0_2$, $\tilde{\chi}^0_2\tilde{\chi}^0_2$
at $\sqrt{s}=400$, $500$~GeV as experimental input.

In our scenario the light chargino $\ti \chi^\pm_1$ and also the
neutralino $\tilde{\chi}^0_2$ decay mainly via $\tilde{\tau}_1$
chains producing the final states similar to that of stau pair
production, however with different topology.  Therefore, we assume that the
contamination of stau production events can be subtracted from the
chargino and neutralino production \cite{DKMNP}.

In our analysis we take the production cross sections with statistical
errors induced by the following
uncertainties:
\begin{itemize}
\item
The chargino mass measurement has
been simulated and the expected error is 0.55~GeV,
see \tab{tab_mass_LC}.
\item With $\int {\cal L}=500$~fb$^{-1}$
at the LC,  we assume 100~fb$^{-1}$ per each polarisation configuration
and we take into account 1$\sigma$ statistical error.
\item Since the chargino (neutralino) production is sensitive to 
$m_{\tilde{\nu}_e}$ ($m_{\tilde{e}_{L,R}}$),
we include the experimental error of their mass determination of
0.7~GeV (0.2~GeV, 0.05~GeV), 
see \tab{tab_mass_LC}.
\item Concerning the neutralino cross sections we 
estimate the statistical error based on an experimental 
simulation\footnote{M. Ball, diploma thesis, 
University of Hamburg, January 2003,
http://www-flc.desy.de/thesis/diplom.2002.ball.ps.gz.}
yielding an efficiency of 25\% and include
an additional systematic error ($\delta\sigma_{\mbox{bg}}$) which takes 
into account the uncertainty in the background subtraction, 
for details see \cite{DKMNP}.
\item The beam polarisation measurement is assumed
with an uncertainty
of $\Delta P(e^{\pm})/P(e^ {\pm})=0.5\%$.
\end{itemize}
\begin{table}
\begin{center}
\begin{tabular}{|c|cc|cccc|ccc|}
\hline & ${\tilde{\chi}^{\pm}_1}$ & ${\tilde{\chi}^{\pm}_2}$ &
${\tilde{\chi}^0_1}$ & ${\tilde{\chi}^0_2}$ &${\tilde{\chi}^0_3}$ &
${\tilde{\chi}^0_4}$ & ${\tilde{e}_R}$ & ${\tilde{e}_L}$ &
${\tilde{\nu}_e}$ \\ \hline mass [GeV] & 176.03 & 378.50 & 96.17 & 176.59 &
358.81 & 377.87 & 143.0 & 202.1 & 186.0 \\ error [GeV] & 0.55 & & 0.05 & 1.2
& & & 0.05 & 0.2 & 0.7 \\ \hline
\end{tabular}
\caption{ Chargino, neutralino and slepton  masses in SPS1a, and
the simulated experimental errors at the 
LC$^{2}$.
It is assumed that the heavy
chargino and neutralinos are not observed at the first phase of the
LC operating at $\sqrt{s}\le 500$~GeV.
\label{tab_mass_LC}}
\end{center}
\end{table}
\footnotetext[2]{H.U.~Martyn, LC-note LC-PHSM-2003-071.}
\begin{table}
\begin{tabular}{|l|cc|cc|}
\hline
\phantom{++++}$\sqrt{s}$ &\multicolumn{2}{|c|}{400~GeV} &
\multicolumn{2}{|c|}{500~GeV} \\
($P(e^-)$, $P(e^+)$)
&$(-80\%,+60\%)$ &$(+80\%,-60\%)$ &
 $(-80\%,+60\%)$ & $(+80\%,-60\%)$\\ \hline
$\sigma(e^+e^-\to\tilde{\chi}^+_1\tilde{\chi}^-_1)$
& 215.84 & 6.38 & 504.87 &  15.07\\ \hline
$\delta\sigma_{\mbox{total}}$ & 7.27 & 0.35 & 5.28 & 0.51 \\ \hline\hline
$\sigma(e^+e^-\to \tilde{\chi}^0_1\tilde{\chi}^0_2)$
& 148.38 & 20.06 & 168.42 &  20.81\\ \hline
$\delta\sigma_{\mbox{total}}$ & 3.0 & 1.58 & 3.52 & 1.57 \\ \hline\hline
$\sigma(e^+e^-\to\tilde{\chi}^0_2\tilde{\chi}^0_2)$
& 85.84 & 2.42 & 217.24 &  6.10\\ \hline
$\delta\sigma_{\mbox{total}}$ & 3.6 & 0.41 & 4.3 & 0.62 \\ \hline
\end{tabular}
\caption{ Cross sections
$\sigma_{L,R}^\pm\{11\}=
\sigma_{L,R}(e^+ e^-\to \tilde{\chi}^+_1\tilde{\chi}^-_1)$,
$\sigma^0_{L,R}\{12\}=
\sigma_{L,R}(e^+ e^-\to \tilde{\chi}^0_1\tilde{\chi}^0_2)$
 and
$\sigma^0_{L,R}\{22\}=
\sigma_{L,R}(e^+ e^-\to \tilde{\chi}^0_2\tilde{\chi}^0_2)$
with polarised beams $P(e^-)=\mp 80\%$, $P(e^+)=\pm 60\%$
at $\sqrt{s}=400$ and 500~GeV and assumed errors (in fb) corresponding to
 100~fb$^{-1}$ for each polarisation configuration. \label{tab_sig11}}
\end{table}
The resulting errors are listed in \tab{tab_sig11}.

From $\sigma_{L}^{\pm}(\tilde{\chi}^{+}_1\tilde{\chi}^-_1)$ at 
$\sqrt{s}=500$, 400~GeV and
$\sigma_{R}^{\pm}(\tilde{\chi}^{+}_1\tilde{\chi}^-_1)$  at
$\sqrt{s}=500$~GeV 
exploiting the relation 
$\cos 2 \Phi_R=f(\cos 2 \Phi_L,\sigma^\pm _{L,R}\{11\})$
we first predetermine chargino mixing angles as 
\begin{equation}
\cos 2 \Phi_L= [0.62,0.72],\quad
\cos 2 \Phi_R= [0.87,0.91] \label{eq_c2rrange}
\end{equation}
Then using the neutralino cross sections
$\sigma_{L}^{\pm}(\tilde{\chi}^{0}_1\tilde{\chi}^0_2)$,
$\sigma_{L}^{\pm}(\tilde{\chi}^{0}_2\tilde{\chi}^0_2)$ at
$\sqrt{s}=500$, 400~GeV and  light
neutralino masses $m_{\tilde{\chi}^0_{1,2}}$ within their experimental
errors, a rather accurate determination of 
the SUSY parameters can be obtained  from the $\Delta \chi^2$ test defined as
$\Delta \chi^2  =\sum_i |(O_i -\bar O_i)/\delta O_i|^2$.
The sum over physical observables $O_i$  includes
$m_{\nt_1},m_{\nt_2}$ and
neutralino production cross sections
$ \sigma^0_{L,R}\{12\},\sigma^0_{L,R}\{22\}$ measured
at both energies of 400 and 500 GeV, $\bar O_i$ stands for the
physical observables taken at the input values of all
parameters, and $\delta O_i$ are the corresponding errors.
The $\Delta \chi^2$ is a function of unknown $M_1,\ctL,\ctR$ with
$\ctL,\ctR$ restricted to the ranges given in
eqn.~(\ref{eq_c2rrange}) as
predetermined from the chargino sector.
\begin{figure}[htb]
\setlength{\unitlength}{1cm}
\begin{center}
{\epsfig{file=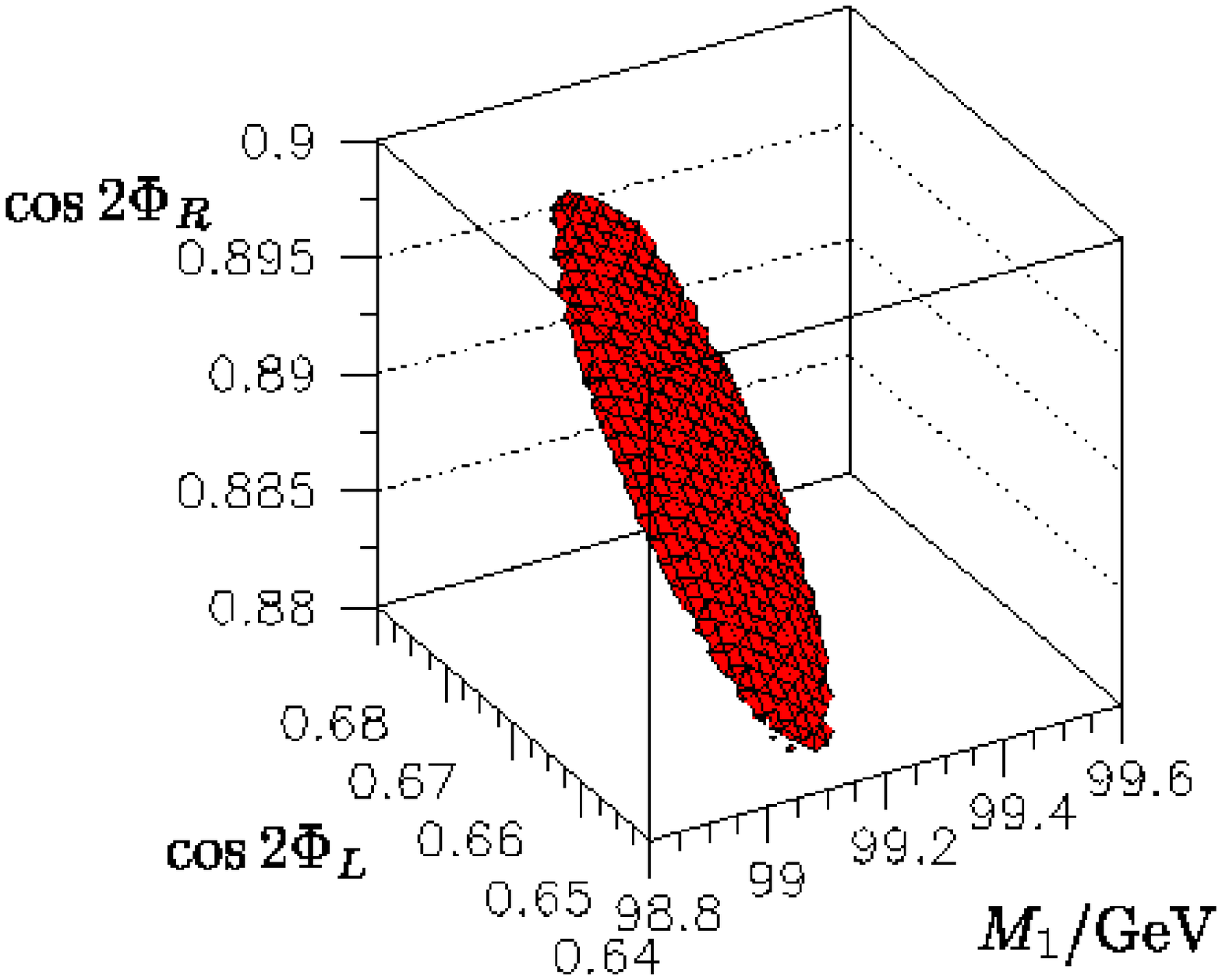, scale=.35}\epsfig{file=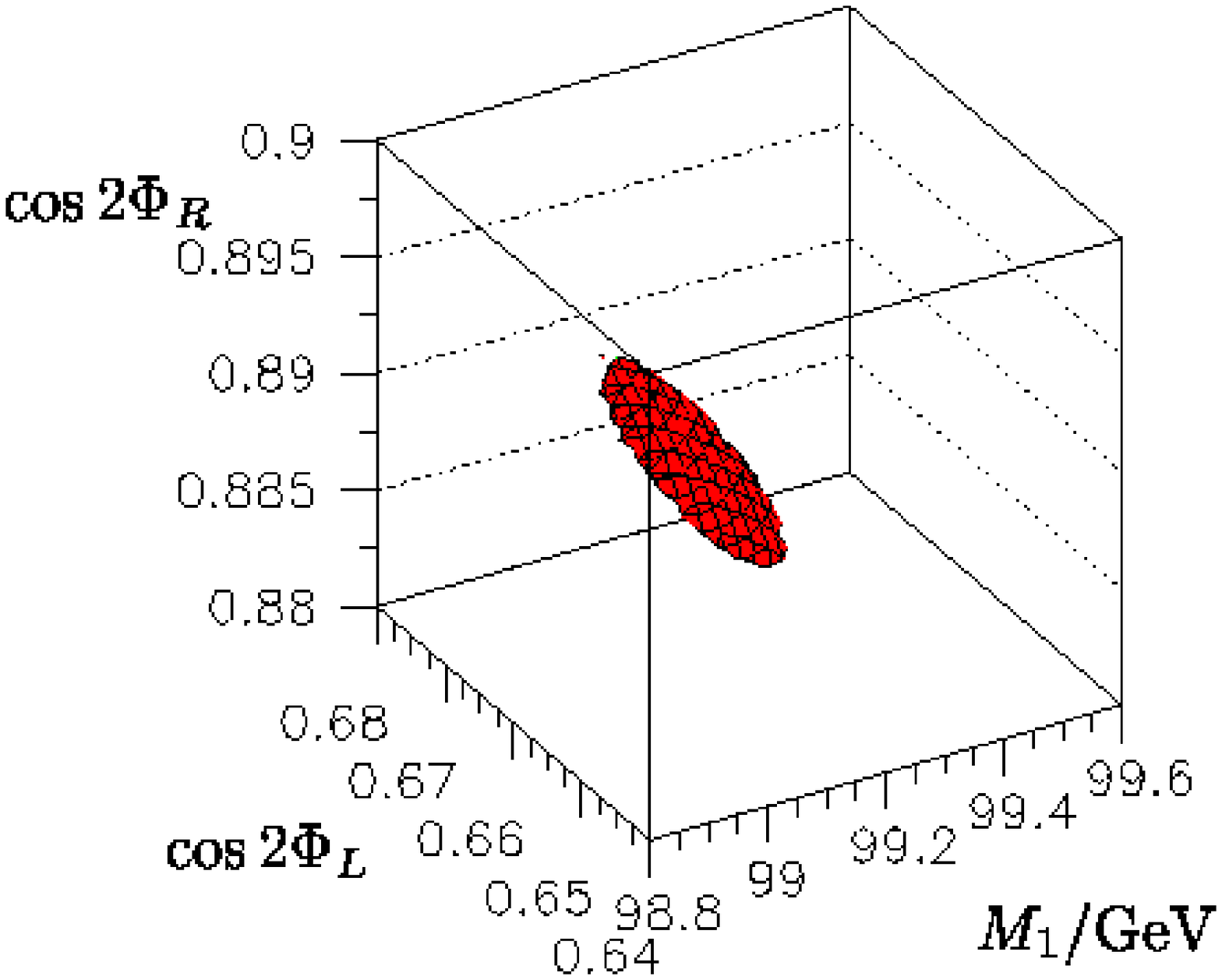,scale=0.35}}
\end{center}
\caption{\it 
The $\Delta \chi^2=1$ contour in the \{$M_1,\ctL,\ctR$\} parameter space
derived a) from the LC data 
and b) from the joint analysis of the LC
data and LHC data \cite{DKMNP}.}
\label{fig_4cygLC}
\end{figure}
In \fig{fig_4cygLC}a the contour of $\Delta \chi^2=1$ is shown in the
$M_1,\ctL,\ctR$ parameter space along with
its three 2dim projections. The projection of the contours onto the
axes determines 1$\sigma$ errors for each parameter.

Values obtained for $M_1,\ctL,\ctR$ together with $\mch{1}$ can be
inverted to derive the fundamental parameters $M_2$, $\mu$ and
$\tan\beta$. At the same time the masses of the heavy chargino and neutralinos
are predicted, see \tab{tab_lc}. As can be seen
in \tab{tab_lc}, the parameters $M_1$ and $M_2$ are
determined at the level of a few per-mil. The $\mu$ is  reconstructed
within a few per-cent, while for $\tan\beta$ the error is of order 15\%.

\begin{table}
\hspace*{-1cm}
\begin{tabular}{|c|cccc|ccc|}
\hline
&\multicolumn{4}{|c|}{SUSY Parameters}&
\multicolumn{3}{|c|}{Mass Predictions}\\
 & $M_1$ & $M_2$ & $\mu$ & $\tan\beta$ & $m_{\tilde{\chi}^{\pm}_2}$ &
$m_{\tilde{\chi}^0_3}$ & $m_{\tilde{\chi}^0_4}$ \\[2mm] \hline
LC& $99.1\pm 0.2$ & $192.7\pm 0.6$ & $352.8\pm 8.9$ & $10.3\pm 1.5$
& $378.8 \pm 7.8$ & $359.2\pm 8.6$ & $378.2 \pm 8.1$\\ \hline\hline
LC/LHC&$99.1\pm 0.1$ & $192.7\pm 0.3$ & $352.4\pm 2.1$ & $ 10.2 \pm 0.6$
& $378.5 \pm 2.0$ & $358.8\pm 2.1$ & --\\ \hline
\end{tabular}
\caption{ SUSY parameters with 1$\sigma$ errors
derived from the  analysis of  the LC data and 
from the combined analysis of the LHC and LC data
(with $\delta m_{\tilde{\chi}^0_2}=0.08$~GeV and
$\delta m_{\tilde{\chi}^0_4}=2.23$~GeV derived from the LHC when using
the LC input of $\delta m_{\tilde{\chi}^0_1}=0.05$~GeV)
  collected at the first phase of operation.
Shown are also the mass predictions of the heavier chargino/neutralinos.
\label{tab_lc}}
\end{table}

\subsection{COMBINED STRATEGY FOR THE LHC AND LC}
A rather large error of the $\mu$ parameter   derived above stems from
the gaugino-dominated character of the light charginos/neutralinos 
in the SPS1a scenario. A significant improvement for the $\mu$ (and 
$\tan\beta$) is expected with additional information on 
heavy neutralinos available from the LHC.

The LHC will provide a first measurement of
the masses of $\nt_1$, $\nt_2$ and $\nt_4$, see\footnote[3]{B.K. Gjelsten, 
J.~Hisano, K.~Kawagoe, E. Lytken, 
D. Miller, M.~Nojiri, P. Osland, G. Polesello,
contribution in the LHC/LC working group document, see also
http://www.ippp.dur.ac.uk/$\tilde{}$ georg/lhclc/.}.
The measurements of $\nt_2$ and $\nt_4$ will be 
achieved through the study of the processes
$\tilde\chi^0_i\rightarrow\tilde\ell\ell\rightarrow\ell\ell\nt_1$, 
(with $i=2,4$) in which  
the invariant mass of the two leptons in the final state shows
an abrupt edge at $(m^{{\rm max}}_{l^+l^-})^2=\mnt{i}^2 (1-m^2_{\ti
  \ell}/\mnt{i}) (1- \mnt{1}/m^2_{\ti \ell})$.

With the LHC data, the achievable precision on
$m_{\nt_2}$ and $m_{\nt_4}$ is expected be respectively  4.5 and 5.1 GeV
for an integrated luminosity of 300~fb$^{-1}$.
However, since the uncertainty on the $\mnt{2}$ and $\mnt{4}$ depends
both on the endpoint determination and also on $\mnt{1}$ and
$m_{\tilde\ell}$, a much higher precision can be achieved with 
$\mnt{1}$, $m_{\tilde e_R}$  and
$m_{\tilde e_L}$ measured at the LC with
precisions respectively of 0.05, 0.05 and 0.2  GeV,
\tab{tab_mass_LC}.
With this input the precisions on the LHC+LC measurements of
$m_{\tilde{\chi}^0_2}$ and $m_{\tilde{\chi}^0_4}$
become: $\delta m_{\tilde{\chi}^0_2}=0.08$~GeV and
$\delta m_{\tilde{\chi}^{0}_4}=2.23$~GeV.
Performing 
again
the $\Delta\chi^2$ test with this additional input one gets a  
significant improvement in the accuracy, see \fig{fig_4cygLC}b and 
\tab{tab_lc} for the final results. The accuracy for the parameters $\mu$ and
particularly $\tan\beta$ is now much better, better than from  other
SUSY sectors 
\cite{Boos:2003vf,Barger:2000fi,Gunion:2002ip} (and references therein).
\section{SUMMARY}
We have worked out in a specific example, an mSUGRA scenario with rather high
$\tan\beta=10$, how the combination of the results from
the two accelerators, LHC and LC,  allows
a precise determination of  the fundamental
SUSY parameters without assuming a specific supersymmetry breaking scheme.
We have shown that a promising hand-in-hand procedure consists
of feeding the LSP and slepton masses from the LC 
to the LHC analyses and injecting back a precise experimental determination
of the $\nt_2$ and $\nt_4$ masses. It provides a determination
of $M_1$, $M_2$, $\mu$ at the $\le O(1\%)$ level and of (rather high)
$\tan\beta$ of the 
order of $\le$ 10\%,  reaching a 
stage where radiative corrections become relevant in the electroweak
sector and which will have to be taken into account in  future fits.
    
\section*{ACKNOWLEDGEMENTS}
The authors would like to thank G. Weiglein for motivating this
project. We thank the organisers for the
wonderful atmosphere at the Les Houches workshop.
We are indebted to
G. Blair, U. Martyn  and W. Porod for providing the errors for simulated mass
measurements at the LC.
The work is supported in part by the European Commission 5-th
Framework Contract HPRN-CT-2000-00149.
JK was supported by the KBN Grant
2 P03B 040 24 (2003-2005).

\setcounter{figure}{0}
\setcounter{table}{0}
\setcounter{section}{0}
\setcounter{equation}{0}
\clearpage

\part{Proposal For A New Reconstruction Technique For Susy Processes
At The LHC \label{new6}}
{\it M.M. Nojiri, G. Polesello and D.R. Tovey}
\maketitle
\begin{abstract}
When several sparticle masses are known, the kinematics of SUSY decay
processes observed at the LHC can be solved if the cascade decays
contain sufficient steps. We demonstrate four examples of this full
reconstruction technique applied to channels involving leptons, namely
a) gluino mass determination, b) sbottom mass determination, c) LSP
momentum reconstruction, and d) heavy higgs mass determination.
\end{abstract}

\section{INTRODUCTION}

The potential of the LHC for SUSY parameter determination has been
studied in great detail for the past seven years. One of the most
promising methods involves the selection of events from a single decay
chain near the kinematic endpoint. Information on the masses involved
in the cascade decay can be extracted from the endpoint
measurements. It has been established that one can achieve a few
percent accuracy for sparticle mass reconstruction using this
technique with sufficient statistics.

In this paper we propose a new method for reconstructing SUSY events
which does not rely only on events near the endpoint. Instead one
kinematically solves for the neutralino momenta and masses of heavier
sparticles using measured jet and lepton momenta and a few mass
inputs.

To illustrate the idea we take the following cascade decay chain
\begin{equation}
\tilde{g}\rightarrow \tilde{b}b\rightarrow \tilde{\chi}^0_2 bb
\rightarrow \tilde{\ell}bb\ell \rightarrow \tilde{\chi}^0_1bb\ell\ell.
\label{eq1}
\end{equation}
This decay chain is approximately free from SM background with
appropriate cuts. The five SUSY particles which are involved in the
cascade decay have five mass shell conditions;
\begin{eqnarray}
m^2_{\tilde{\chi}^0_1}&=& p^2_{\tilde{\chi}^0_1},\cr
m^2_{\tilde{\ell}}&=& (p_{\tilde{\chi}^0_1}+ p_{\ell_1})^2,\cr
m^2_{\tilde{\chi}^0_2}&=& (p_{\tilde{\chi}^0_1}+
p_{\ell_1}+p_{\ell_2})^2,\cr
m^2_{\tilde{b}}&=& (p_{\tilde{\chi}^0_1}+
p_{\ell_1}+p_{\ell_2}+p_{b_1})^2,\cr
m^2_{\tilde{g}}&=& (p_{\tilde{\chi}^0_1}+ 
p_{\ell_1}+p_{\ell_2}+p_{b_1}+p_{b_2})^2.
\label{gluino}
\end{eqnarray}
Of these five masses, $m_{\tilde{\chi}^0_1}$,$m_{\tilde{\chi}^0_2}$
and $m_{\tilde{\ell}}$ can be measured at the LHC using first
generation squark cascade decays with an accuracy of $\sim$ 10\% (the
mass difference is measured more precisely).  Moreover, with input
from a future high energy Linear Collider these masses might be
determined with an accuracy $\sim O(1\%)$.
We therefore assume for the present work that the masses of the two
lighter neutralinos and of the right handed slepton are known, and we
ignore the corresponding errors.

For a $bb\ell\ell$ event, the equations contain six unknowns
($m_{\tilde{g}}$, $m_{\tilde{b}}$ and $p_{\tilde{\chi}^0_1}$) which
satisfy five equations. For two $bb\ell\ell$ events, we have ten
equations while we only have ten unknowns (two neutralino four
momenta, $m_{\tilde{g}}$ and $m_{\tilde{\chi}^0_1}$ ). Mathematically,
one can obtain the sbottom and gluino masses and all neutralino
momenta if there are more than two $bb\ell\ell$ events.

We call this technique the ``mass relation method'' as one uses the
fact that sparticle masses are common for events which go though the
same cascade decay chain.  Note events need not be near the endpoint
of the decay distribution to be relevant to the mass determination. In
the next section we demonstrate the practical application of this
method to measurement of the masses of the gluino and sbottom.

As a byproduct of the technique, once the mass of the squark and of
all the sparticles involved in the decay are known, the momentum of
the lighter neutralino can be fully reconstructed,
and this further constrains the event.

In SUSY events sparticles are always pair produced and there are two
lightest neutralinos in the event. If squark decays via
$\tilde{q}\rightarrow \tilde{\chi}^0_2\rightarrow
\tilde{\ell}\rightarrow \tilde{\chi}^0_1$ can be identified on one
side of the event then the neutralino momentum can be reconstructed as
described above.  The transverse momentum of the lightest neutralino
in the other cascade decay can then be obtained using the following
equation
\begin{equation}
{\bf p_T}(\tilde{\chi}^0_1(2)) = {\bf p_T}({\rm miss}) +
{\bf p_T}(\tilde{\chi}^0_1(1)), 
\label{ptmiss}
\end{equation}
provided that there are no hard neutrinos involved in the decay.  This
transverse momentum can be used to constrain the cascade decay of the
other sparticle.

For the case where the other squark decays via $\tilde{q}\rightarrow
\tilde{\chi}^+_1q \rightarrow \tilde{\chi}^0_1q W$ followed by
$W\rightarrow q'q''$, the chargino mass can be determined by using
Eq. (\ref{ptmiss}) and the following relations,
\begin{eqnarray}
p_{\tilde{q}}&=& p_{\tilde{\chi}^0_1(2)}+p_j+p_W,
\cr
p^2_{\tilde{q}}&=&m^2_{\tilde{q}},
\end{eqnarray}
where $p_j$ is the momentum of the selected high $p_T$ jet which comes
from the squark decay and $p_W$ is the momentum of the two jet system
consistent with the $W$ interpretation.  The neutralino momentum
resolution is important for the chargino mass reconstruction and we
discuss this in section 3. The reconstruction will be discussed in a
separate contribution\cite{tovey}.

The full reconstruction technique can be extended for higgs mass
reconstruction.  In section 4, we discuss the heavy higgs mass
determination from the process $H\rightarrow
\tilde{\chi}^0_2\tilde{\chi}^0_2$ followed by $\tilde{\chi}^0_2
\rightarrow \tilde{\ell} \ell\rightarrow ll\tilde{\chi}^0_1$. This
process is also useful for discovery of heavy higgs bosons. The four
lepton momenta and missing momentum can be used to reconstruct the
higgs mass assuming that the $p_T$ of the higgs boson is very small.

\section{GLUINO CASCADE DECAY}

We first discuss the results of a simulation study of the process
where a gluino cascade decays into a sbottom at model point
SPS1a\cite{Allanach:2002nj}. The relevant sparticle masses for this
study are listed in Table 1. The events were generated using the
HERWIG 6.4 generator \cite{Corcella:2000bw} \cite{Moretti:2002eu} and
passed through ATLFAST~\cite{atlfast}, a parametrised simulation of
the ATLAS detector.

\begin{table}
\begin{center}
\begin{tabular}{|c|c|c|c|c|}
\hline
$m_{\tilde{g}}$ & $m_{\tilde{b}_1(2)}$ & $m_{\tilde{\chi}^0_2}$ & 
$m_{\tilde{\ell}_R}$ 
&$m_{\tilde{\chi}^0_1}$ \cr
\hline
595.2& 491.9(524.6)& 176.8&  136.2& 96.0\cr
\hline
\end{tabular}
\caption{Some sparticle masses in GeV at SPS1a. }
\end{center}
\end{table}

We study only events which contain the cascade decay shown in
Eq.(\ref{eq1}). We then apply the following preselections to reduce
backgrounds:
\begin{itemize}
\item $p_T^{miss}>100$~GeV
\item $M_{\rm eff}>600$~GeV
\item at least 3 jets with $p_{T1}>150$~GeV, $p_{T2}>100$~GeV and
$p_{T3}>50$~GeV. 
\item exactly two jets with $p_T>50$~GeV tagged as $b$-jets
\item exactly two OS-SF leptons with $p_{Tl1}>20$~GeV, $p_{Tl2}>10$~GeV, and
invariant mass $40$GeV$<m_{ll}<78$~GeV.
\end{itemize}

The solution of Eq. (\ref{gluino}) can be written in the following
form:
\begin{eqnarray}
m^2_{\tilde{g}}&=& F_0+ F_1 m^2_{\tilde{b}}\pm  F_2 D, \cr
{\rm where} \ \ 
D^2 &\equiv& D_0+ D_1 m^2_{\tilde{b}} +D_2 m^4_{\tilde{b}}.
\label{sol}
\end{eqnarray}
Here $F_i$ and $D_i$ depend upon $p_{\ell_i}$ and $p_{b_i}$ and the
neutralino and slepton masses.  In the event, there are two $b$ jets
and we assume that the $b$ jet with larger $p_T$ originates from the
$\tilde{b}$ decay.  The two leptons must come from $\tilde{\chi}^0_2$
and $\tilde{\ell}$ decay.  There are maximally four sets of gluino and
sbottom mass solutions together with two lepton assignments for each
decay, because we cannot determine from which decay the lepton
originates. To reduce combinatorics we take the event pair which
satisfies the following conditions:
\begin{itemize}
\item  Only one lepton assignment has a solution to the Eq. (\ref{sol}) 
\item For a pair of events there are only two solutions and there is a
difference of more than 100~GeV between the two gluino mass solutions.
\end{itemize}

\begin{figure}[thb]
\begin{center}
\includegraphics[width=0.5\textwidth]{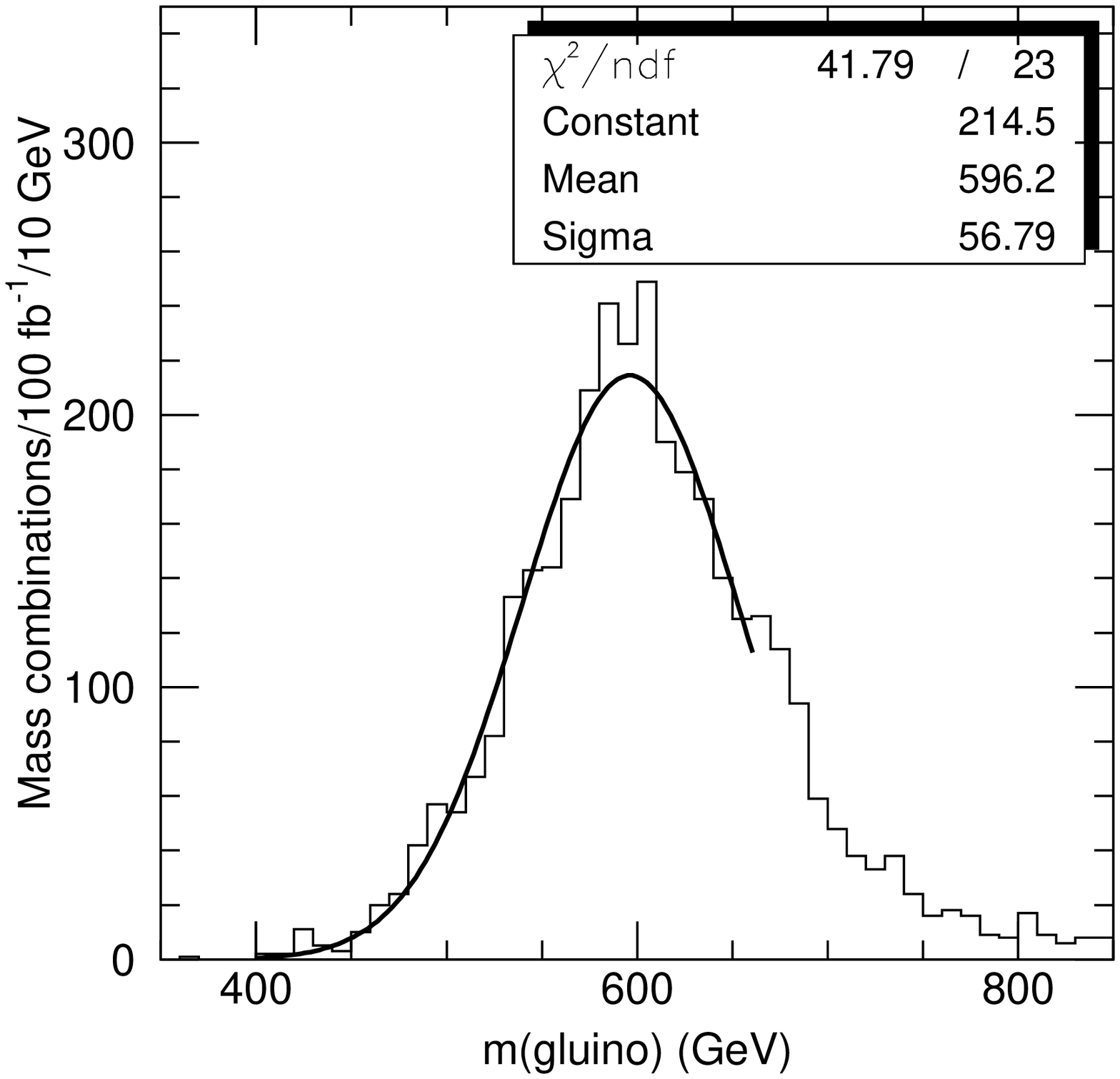}
\caption{$m_{\tilde{g}}$ obtained by using Eq. (\ref{sol})
for two $bb\ell\ell$ events.}
\end{center}
\end{figure}

In Fig 1, we plot the minimum $m_{\tilde{g}}$ solution which satisfies
the conditions given above. The peak position is consistent with the
gluino mass, and the error on the peak position obtained by a Gaussian
fit is around 1.7~GeV for 100 fb$^{-1}$.  For the events used in the
reconstruction, each event is used on average five times.  Note that
the $\sigma$ of the Gaussian fit is large ($\sim 56.7$~GeV) and is
determined by the resolution on the momentum measurement of the four
$b$-jets.  It is worth stressing that the results presented here were
produced by using a parametrised simulation of the response of the
ATLAS detector to jets, based on the results of a detailed simulation.
Results which crucially depend on the detailed features of the
detector response, such as the possibility of discriminating the two
sbottom squarks (see below) need to be validated by an explicit
detailed simulation of the detector performed on the physics channel
of interest. We only attempt here to evaluate the impact of the new
technique on sparticle reconstruction.

Once the gluino mass has been determined one can reconstruct the
sbottom mass by fixing the gluino mass to the measured value.  Here
one need only solve Eq.(\ref{sol}), which involves only two $b$-jets
in the fit, and therefore errors due to the jet resolution are
expected to be less than those for the gluino mass reconstruction.

For each event, there are two sbottom mass solutions
$m_{\tilde{b}}$(sol1) and $m_{\tilde{b}}$(sol2), each sensitive to the
gluino mass input.  The difference between the gluino and sbottom mass
solutions is however stable against variation in the assumed gluino
mass. The mass itself may have a large error in the absolute scale,
but the mass differences are obtained rather precisely, as is the case
in the endpoint method.

In Fig. 2 (left), we plot the solutions for all possible lepton
combinations in the $m_{\tilde g}-m_{\tilde{b}}$ (sol1) $m_{\tilde{g}}
-m_{\tilde{b}}$ (sol2) plane. Here we use the $b$-parton momentum
obtained from generator information.  One of the solutions tends to be
consistent with the input sbottom mass. Moreover the two decay modes
$\tilde{g}\rightarrow \tilde{b}_1b$ and $\tilde{b}_2b$ are clearly
separated.

We can compare the results from the previous analysis with those from
the endpoint analysis\cite{LHCLC}, where one uses approximate the
formula
\begin{equation}
{\bf p}_{\tilde{\chi}^0_2}=
\left(1-\frac{m_{\tilde{\chi}^0_1}}{m_{\ell\ell}}\right){\bf p}_{\ell\ell}.
\label{wrong}
\end{equation}
This formula is correct only at the endpoint of the three body decay
$\tilde{\chi}^0_2\rightarrow\chi^0_1 \ell\ell$, but is nevertheless
approximately correct near the edge of $\tilde{\chi}^0_2\rightarrow
\tilde{\ell}\ell \rightarrow \ell\ell\tilde{\chi}^0_1$ for SPS1a.  The
sbottom mass obtained by using Eq.(\ref{wrong}) is shown in
Fig. 2(right). For this case, the $\tilde{b}_2$ peak at 70.6~GeV is
not separated from the $\tilde{b}_1$ peak at 103~GeV.

\begin{figure}[thb]
\begin{center}
\includegraphics[width=0.45\textwidth]{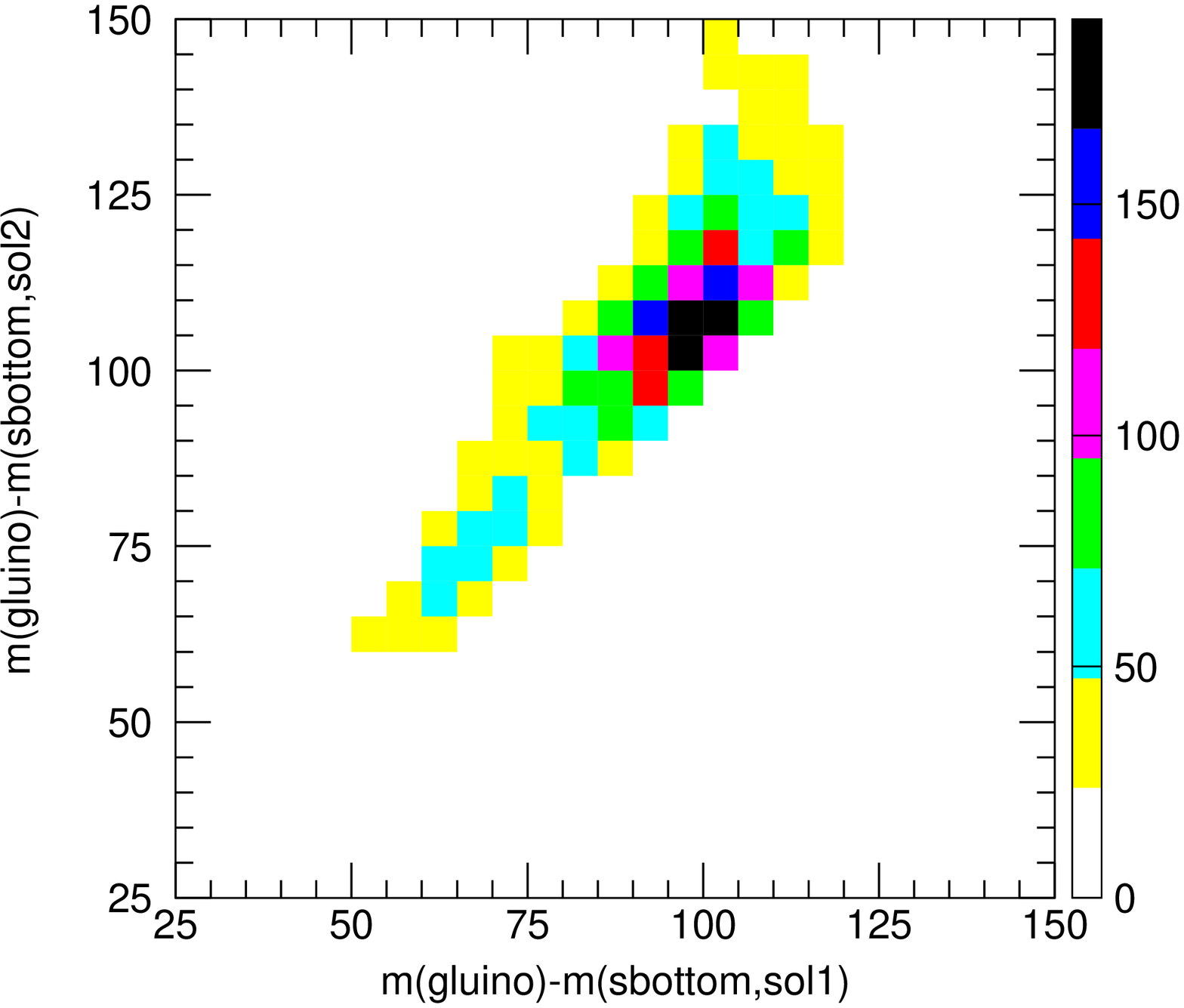}
\includegraphics[width=0.45\textwidth]{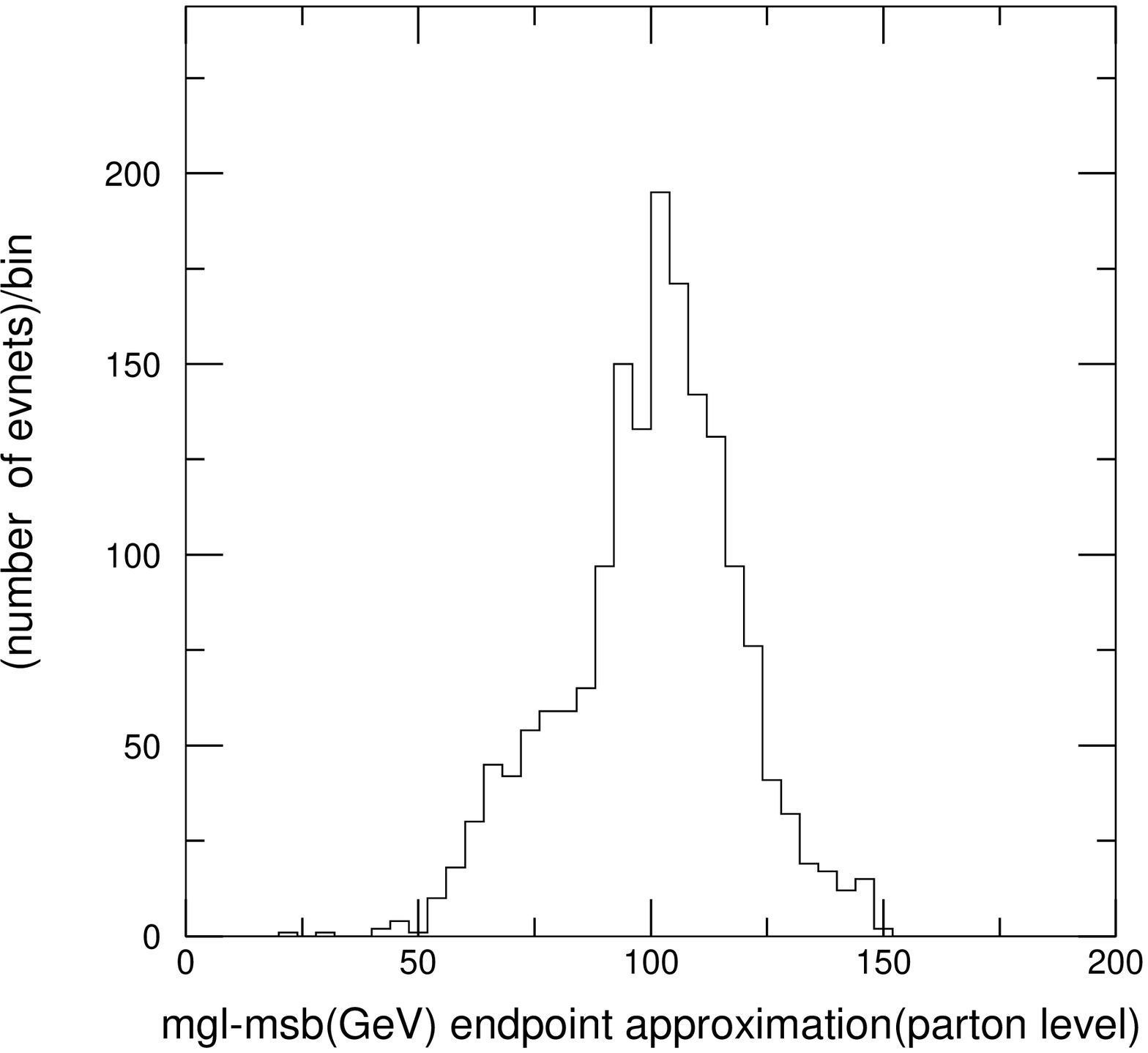}
\caption{The distribution of $m_{\tilde{g}}- m_{\tilde{b}}$ calculated
using the parton level $b$ momentum by solving Eq.(\ref{gluino})
(left) and using the approximate relation Eq. \ref{wrong}(right).}
\end{center}
\end{figure}

\begin{figure}[thb]
\begin{center}
\includegraphics[width=0.45\textwidth]{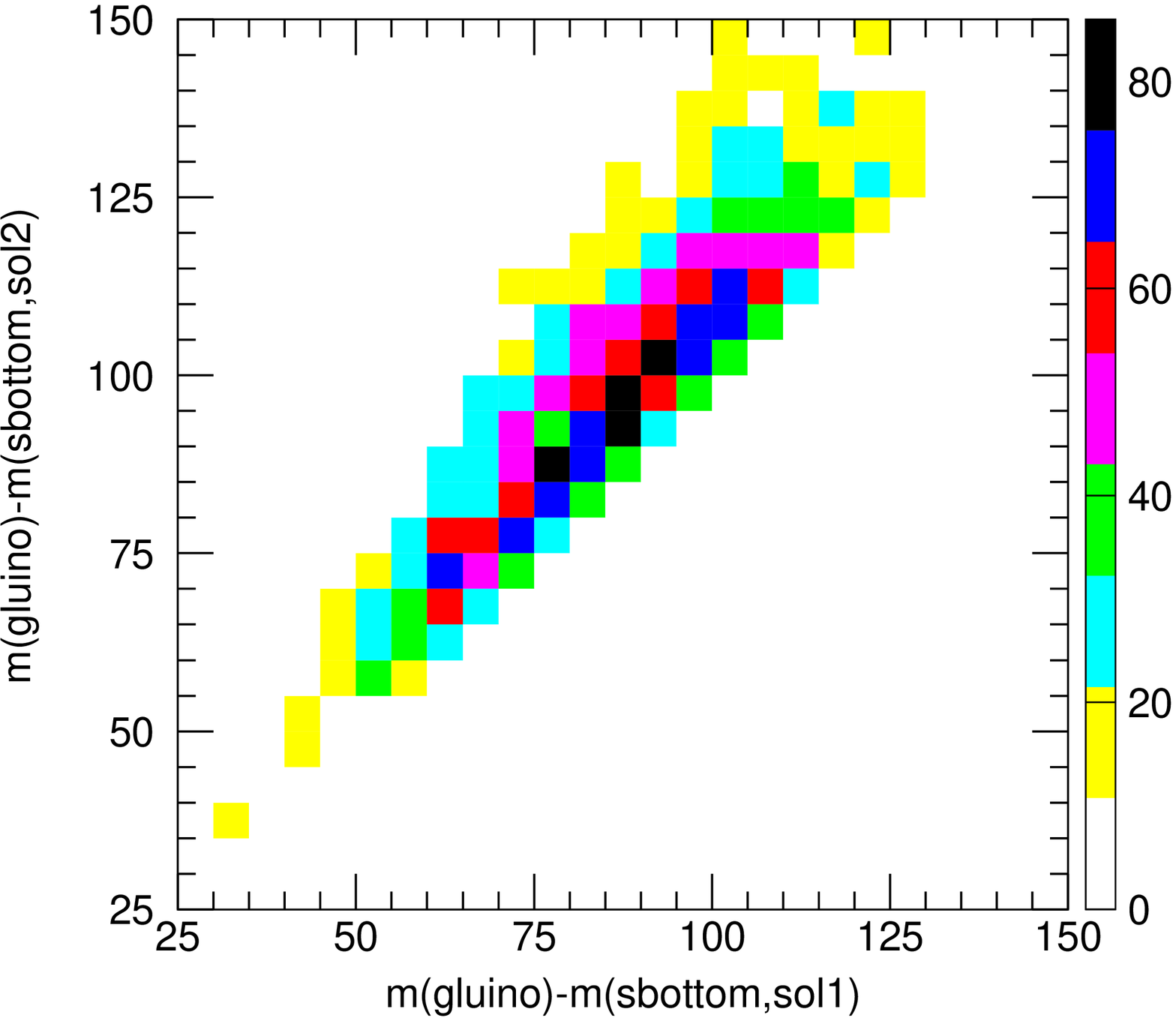}
\hskip 1cm 
\includegraphics[width=0.45\textwidth]{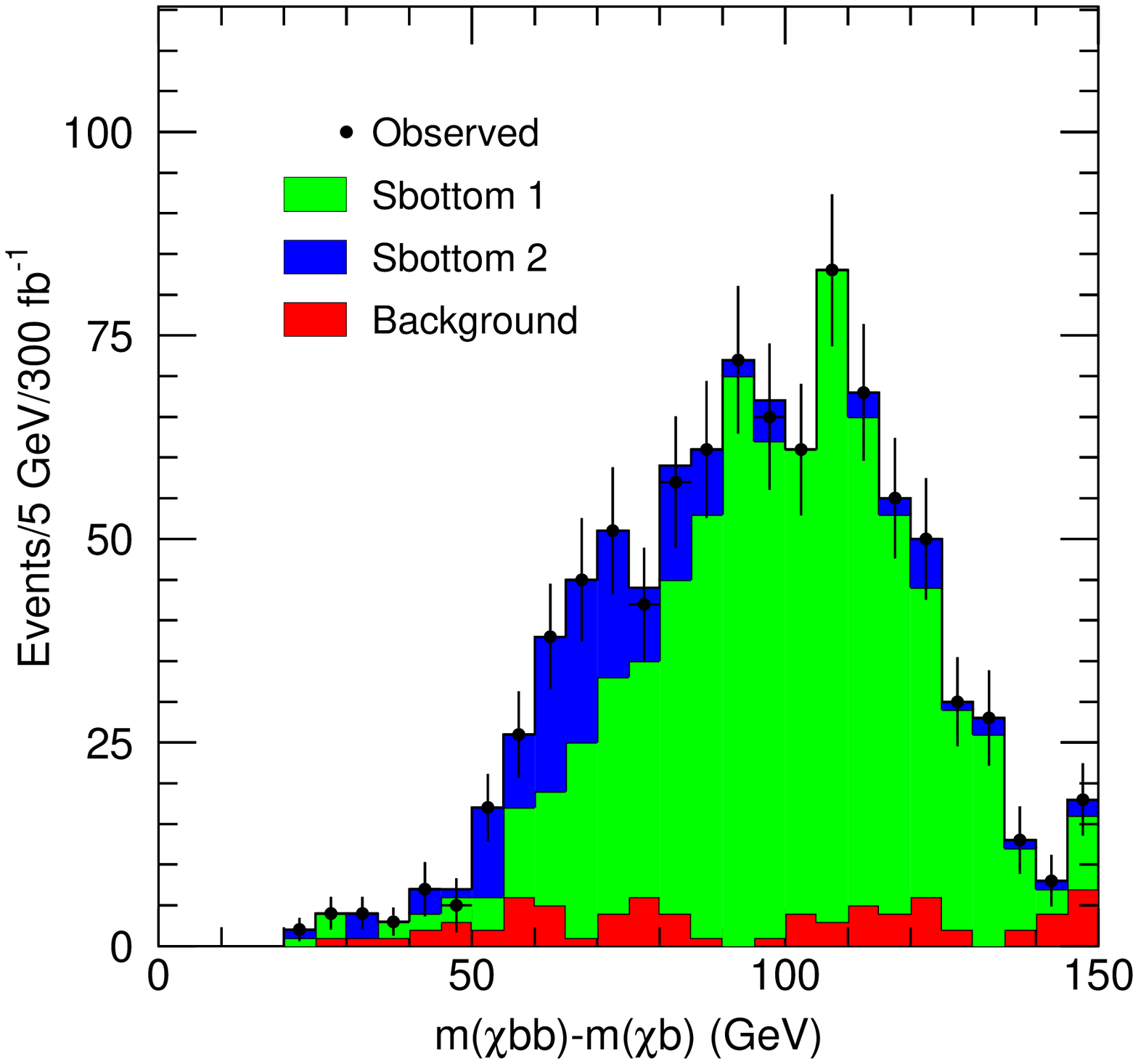}
\caption{As for Fig. 2 but with the $b$ jet momentum used instead of
the $b$ parton momentum.}
\end{center}
\end{figure}

The $\tilde{b}_1$ mass, or the weighted average of the sbottom masses,
is easily obtained.  The $b$ jet resolution is not sufficient however
to clearly separate the $\tilde{b}_1$ and $\tilde{b}_2$.  This can be
seen in Fig. 3 where the plots show the distributions corresponding to
Fig.2(left) and (right) but now with the $b$ parton momenta replaced
by $b$ jet momenta.  For the endpoint analysis (Fig.3 right), 
a correct evaluation of the sbottom masses would require 
a fit taking into account the shape of the response of ATLAS to b-jets.
In order to approximately evaluate the achievable statistical precision,
a naive double gaussian fit was performed on the distribution
shown in Fig.3 right, which corresponds to $\int dt L=300$ fb$^{-1}$.
The resulting statistical uncertainties are 
$\pm 1$~GeV ($ \pm 2.5$~GeV) for the
$m_{\tilde{g}}-m_{\tilde{b}_{1}}$($m_{\tilde{g}}-m_{\tilde{b}_{2}}$ )
peak positions respectively. Additional systematic uncertainties, 
not yet evaluated, as well a 1\% error due to the uncertainty on the
jet energy scale should also be considered.
These numbers are obtained assuming the presence of two gaussian peaks
in the data.

For the mass relation method the number of events available for the
study is larger by a factor of 2 because events away from the
endpoints can be used.  We also use the exact formula for the mass
relation method.  Although the analysis is more complicated due to the
multiple solutions, we believe it to be a worthwhile technique for use
when attempting to reconstruct the $\tilde{b}_1$ and $\tilde{b_2}$
masses.

\section{NEUTRALINO MOMENTUM RECONSTRUCTION} 

In this section, we discuss the reconstruction of the momentum of the
lightest neutralino. As we have discussed already, the mass shell
condition can be solved for long decay cascades, such as
$\tilde{q}\rightarrow \tilde{\chi}^0_2 q\rightarrow
\tilde{\ell}q\ell\rightarrow \tilde{\chi}^0_1q\ell\ell$.  For this
process we have two neutralino momentum solutions for each lepton
assignment.  One may wonder if the solutions for the neutralino
momentum might be smeared significantly, because of the worse jet energy
resolution as compared to leptons, and the jet $p_T$
is generally much larger than the neutralino momentum for the cascade
decay.  In Fig.  4(left) we show the distribution of
$p_T$(reco)/$p_T$(truth) for the point studied in \cite{tovey}.  Here
we choose the correct lepton combination using generator information,
and take the solution which minimizes $\vert
p_T$(reco)/$p_T$(truth)$-1\vert$.  Except for the case where we took
the wrong jet as input the reconstructed $p_T$ is within 20\% of the
true neutralino momentum. The result for the gluino cascade decay into
sbottom Eq.(1) is similar.
 
In Fig. 4(right) we show a similar reconstruction for the gluino
cascade decay, but unlike Fig.4(left), we use both lepton
combinations.  We fix the gluino mass to the input value\footnote {Here
we adopt an event selection which makes use of the true (input) gluino
and sbottom mass values, although in practice fitted values would be
used.}  and take events where one of the four sbottom mass solutions
is consistent with the input sbottom mass such that $\vert
m_{\tilde{b}_1}-m_{\tilde{b}}({\rm best})\vert<10$~GeV.  We then take
the solution where the sbottom mass is closest
to the input $m_{\tilde{b}_1}$.  There are still two
$p_{\tilde{\chi}^0_1}$ solutions, and we choose the one which 
minimize $\vert \min(p_{T}({\rm reco})/p_T({\rm truth}), 
p_{T}({\rm truth})/p_T({\rm reco}))
-1\vert$.   
The neutralino momentum resolution is
worse than that obtained using the correct lepton assignments only.
Nevertheless a significant fraction of events are reconstructed with
$0.8<\vert {\bf p}({\rm reco})/ {\bf p}({\rm truth})\vert <1.2$.

\begin{figure}[thb]
\begin{center}
\includegraphics[width=0.45 \textwidth]{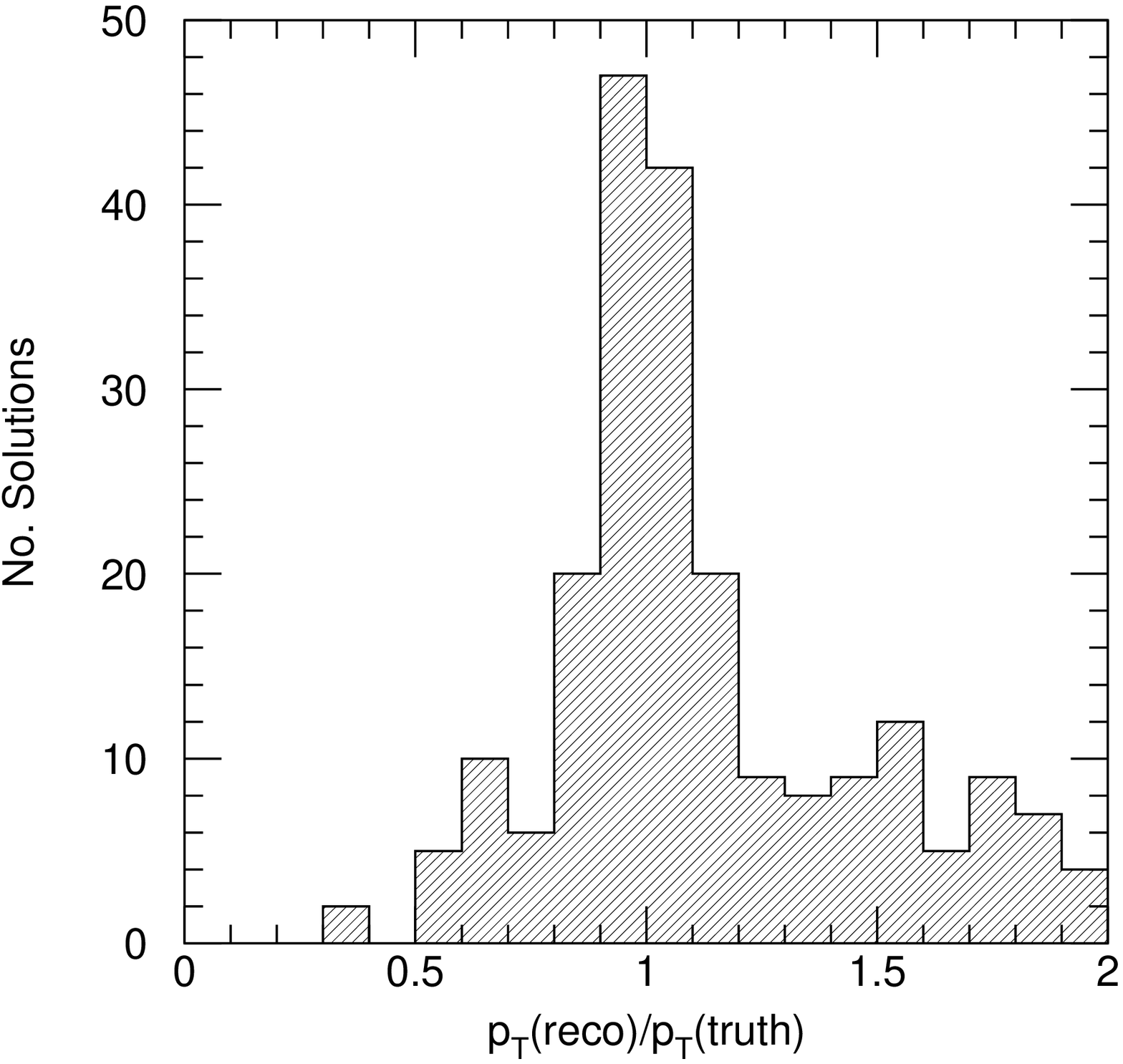}
\includegraphics[width=0.45 \textwidth]{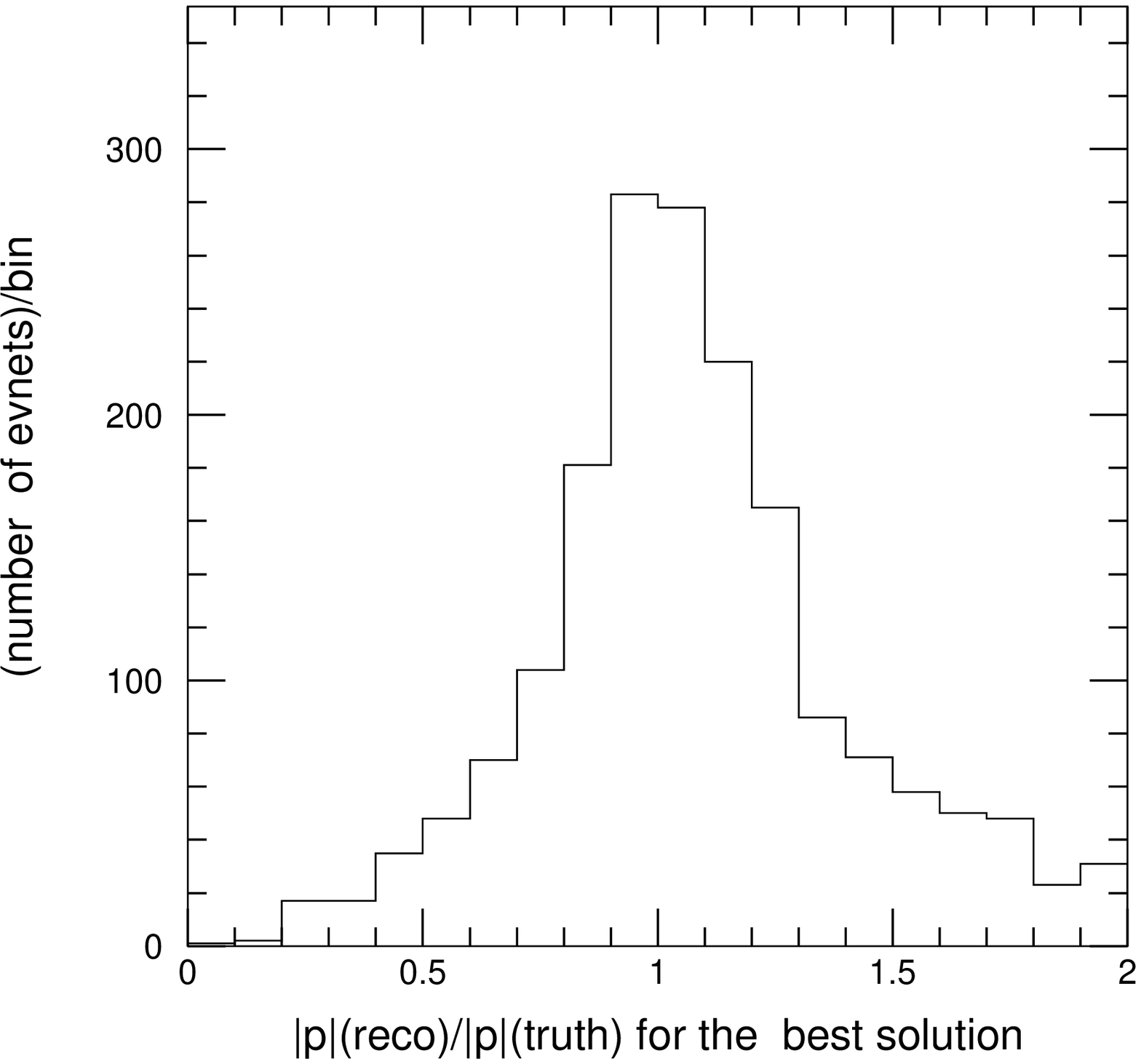}
\caption{Left: The calculated $\tilde{\chi}^0_1$ transverse momentum
divided by the true transverse momentum.  The decay
$\tilde{q}_L\rightarrow \tilde{\chi}^0_2 q \rightarrow \tilde{\ell}
q\ell \rightarrow\tilde{\chi}^0_1q \ell\ell$ is studied for the model
point used for the chargino study: $m_0=100$~GeV, $m_{1/2}=300$~GeV,
$A_0=-300$~GeV, $\tan\beta=6$, and $\mu>0$.  Only the correct lepton
choice is used.  Right: $\vert {\bf p}$(reco)$/{\bf p}$(truth)$\vert$
for the decay chain Eq.(\ref{eq1}) for SPS1a.}
\end{center}
\end{figure}

\section{HIGGS MASS RECONSTRUCTION}
\def\tchi{\tilde{\chi}} \def\tl{\tilde{\ell}} \def\etmiss{E_T^{miss}}
A promising decay for the observation of heavy and pseudo-scalar higgs
bosons in the difficult region with intermediate $\tan\beta$ is the
decay into two neutralinos. When both neutralinos decay through the
chain
$$
\tchi^0_2\rightarrow\tl_R\ell\rightarrow\ell\ell\tchi^0_1
$$
the resulting signature consists of events with four isolated leptons
(paired in opposite-sign same-flavour pairs) and no jet activity. The
main SM backgrounds to this signature are $t\bar{t}$ production, where
both the $b$-jets and the $W$s decay into leptons and $Zbb$
production. The key element for the rejection of these backgrounds is
the fact that the leptons from $b$ decays are not isolated. A detailed
study of the performance of lepton isolation in the detector is needed
to assess the visibility of the signal.  Additionally there is an
important SUSY background, including irreducible backgrounds from
direct slepton and gaugino decay.  Full background
studies as a function of the SUSY parameters were  performed by the
ATLAS and CMS Collaborations \cite{atlasTDR,filip}.  
We propose here, along the lines of
the previous sections, a technique for the complete reconstruction of
the higgs peak, based on the knowledge of the masses of $\tchi_2^0$,
$\tl_R$ and $\tchi_1^0$.  In this case one has 8 unknown quantities:
the 4-momenta of the two LSP's, and 8 constraints: six on-shell mass
constraints (3 for each leg), and the two $\etmiss$ components.

To demonstrate the power of the method, we apply it to Point SPS1a,
for which the mass of the $A$ and of the $H$ is $\sim394$~GeV.  The BR
into $\tchi^0_2\tchi^0_2$ is 6\% (1\%) for the $A$($H$).  We perform
the study on 1000 events for
$$A\rightarrow\tchi^0_2\tchi^0_2\rightarrow
\ell\ell\ell\ell\tchi^0_1\tchi^0_1$$ corresponding approximately to
the expected statistics for 300 fb$^{-1}$.  We simply require 2
isolated leptons with $p_T>20$~GeV and 2 further isolated leptons
with $p_T>10$~GeV, all within $|\eta|<2.5$.  The efficiency of these
cuts is $\sim 60\%$.

\begin{figure}[thb]
\begin{center}
\includegraphics[width=0.45 \textwidth]{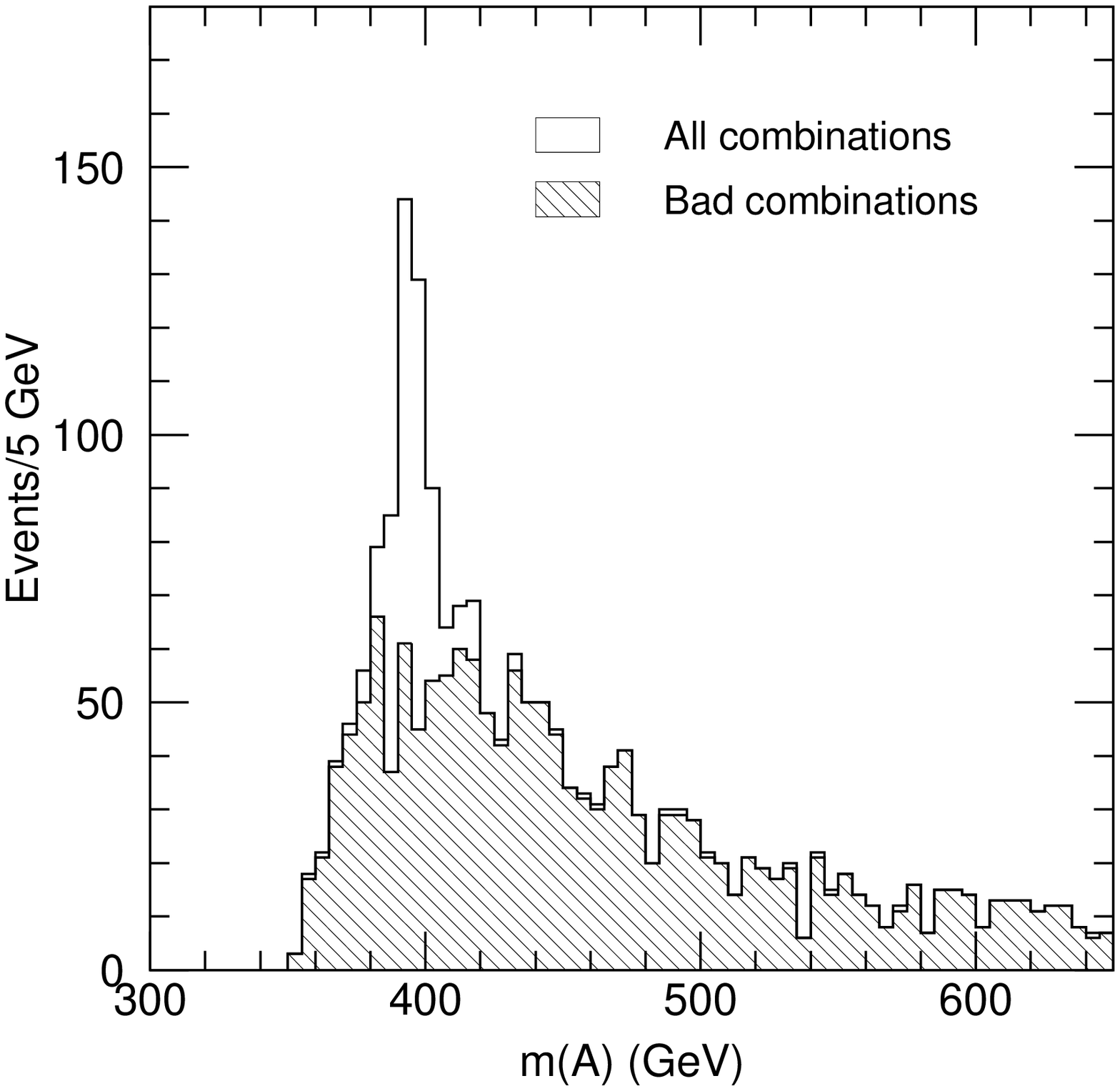}
\caption{Reconstructed mass of the $A$.} 
\label{fig:higgs}
\end{center}
\end{figure}

We have not performed any background simulations because at this stage
we only wish to explore the viability of the full reconstruction
technique.  The main problem for the reconstruction is the correct
assignment of the leptons to the appropriate decay chain.  The first
selection is based on requiring a unique identification of the lepton
pairs coming from the decays of the two $\tchi^0_2$s.  We therefore
require that either of the following two criteria is satisfied:
\begin{itemize}
\item
the flavour configuration of the leptons is $e^+e^+\mu^+\mu^-$
\item
the lepton configuration is either $e^+e^-e^+e^-$ or
$\mu^+\mu^-\mu^+\mu^-$, but for one of the two possible pairings the
invariant mass of one of the pairs is larger than 78~GeV, i.e.  above
the lepton-lepton edge for the $\tchi^0_2$ decay.
\end{itemize}
The total efficiency after these cuts is $\sim30\%$.  At this point,
on each of the two legs there is still an ambiguity due to the fact
that each lepton can be either the product of the first or of the
second step in the decay chain. This gives 4 possible combinations.
Furthermore, the full reconstruction results in a quartic equation
which can have zero, two or four solutions.  We show in
Fig.~(\ref{fig:higgs}) the distribution of the calculated $A$ mass for
all of the retained combinations as a full line. The dashed line shows
the combinations with the wrong lepton assignment.  A clear and narrow
peak emerges over the combinatorial background.  The width is
approximately 6~GeV, determined by the resolution of the measurement
of the momentum of the leptons.

\section{CONCLUSIONS}
In this contribution, we have described a novel technique for
reconstructing the mass and momenta of SUSY particles.  This technique
does not rely on any approximate formulae nor on endpoint
measurements. All events contribute to the sparticle mass
determination and decay kinematics reconstruction, even if they are
away from the endpoint of the distribution.  The method may be
particularly useful when the SUSY mass scale is large. In that case
the statistics can be so low that the endpoint cannot be seen clearly
while the SUSY sample itself is very clean.

The method applies most effectively when we know some of the
sparticles' masses exactly, because the number of unknown parameters
in e.g.  Eq.(\ref{gluino}) is reduced.  In the particular case where
some of the sparticle masses are measured at a LC the sparticle
cascades may be solved completely and study of the decay distributions
and higher mass determination becomes possible at LHC.

When all the sparticle masses are known the neutralino momentum can be
reconstructed if four sparticles are involved in the cascade
decay. The sparticles would be pair produced, and if we can identify
both of the cascade decay chains in the events then we only need six
sparticles in the cascade decay to solve both of the neutralino
momenta on account of the missing momentum constraint. The
reconstruction of sparticle momenta provides us with an interesting
possibility for studying the decay distribution at the LHC.

On the other hand, our method is not valid when some of the particles
in the cascade produce hard neutrinos.  This is unfortunately the case
when the chargino decays into (s)leptons, when a $\tilde{\tau}$ is
involved in the decay, or when a $W$ is produced and decays
leptonically.  If such SUSY decay processes dominate then this method
may not be useful.

\section*{Acknowledgments}

This work was performed in the framework of the workshop:
Les Houches 2003: Physics at TeV Scale Colliders.
We wish to thank the staff and organisers for all their hard work 
before, during and after the workshop.
We thank members of the ATLAS Collaboration for helpful
discussions. We have made use of ATLAS physics analysis and
simulation tools which are the result of collaboration-wide
efforts.  DRT wishes to acknowledge PPARC and the
University of Sheffield for support.

\setcounter{figure}{0}
\setcounter{table}{0}
\setcounter{section}{0}
\setcounter{equation}{0}
\clearpage

 \part{Building on a Proposal for a New Reconstruction Technique for
   SUSY Processes at the LHC \label{chris}}
{\it C.G. Lester}
\begin{abstract}
There has recently been interest in ``a new reconstruction technique
for SUSY processes at the LHC''.  The primary intention of this note
is to describe a modification to the way the technique is used.  This
modification suggests the method is much more powerful than originally
proposed.  We demonstrate that, in principle at least, the method does
not need to rely on input from other experiments.
We show that the method is capable of
standing on its own, and is able to measure the masses of all
the sparticles participating in the relevant decay chains.  Results
from other experiments such as a future linear collider may easily be
incorporated if desired.
\end{abstract}

\def\slashchar#1{\setbox0=\hbox{$#1$}           
   \dimen0=\wd0                                 
   \setbox1=\hbox{/} \dimen1=\wd1               
   \ifdim\dimen0>\dimen1                        
      \rlap{\hbox to \dimen0{\hfil/\hfil}}#1 
   \else                                        
      \rlap{\hbox to \dimen1{\hfil$#1$\hfil}}/                                    \fi}          
\def\etmiss{\slashchar{E}_T}
\newcommand\pmiss{{{\slashchar{p}}}}
\newcommand\ptmiss{\slashchar{p}_T}
\newcommand\Ptmiss{\slashchar{{\bf p}}_T}
\renewcommand{\topfraction}     {0.95}
\renewcommand{\bottomfraction}  {0.95}
\renewcommand{\textfraction}    {0.05}
\def\ie{i.e.}
\def\eg{e.g.}
\def\etal{{\it et. al.}}
\def\etc{{\it etc.}}
\def\ptwo{{\bf p}}
\def\ptlepOne{{p_T^{l_1}}}
\def\ptlepTwo{{p_T^{l_2}}}
\def\PtlepOne{{{\bf p}_T^{l_1}}}
\def\PtlepTwo{{{\bf p}_T^{l_2}}}
\def\Ptalpha{{{\bf p}_T^{\alpha}}}
\def\Ptbeta{{{\bf p}_T^{\beta}}}
\def\mttwo{{m_{T2}}}
\def\mttwomax{{m_{T2}^{\rm{max}}}}
\def\chginooneplus{{\ntlino^+_1}}
\def\chginoonepm{{\ntlino^\pm_1}}
\def\chginooneminus{{\ntlino^-_1}}
\def\chginotwoplus{{\ntlino^+_2}}
\def\chginotwominus{{\ntlino^-_2}}
\def\gluino{{\tilde g}}
\def\sbottom{{\tilde b}}
\def\slepton{{\tilde l}}
\def\ntlino{{\tilde \chi}}
\def\ntlinoone{{\ntlino^0_1}}
\def\ntlinotwo{{\ntlino^0_2}}
\def\ntlinothree{{\ntlino^0_3}}
\def\ntlinofour{{\ntlino^0_4}}
\def\ntlone{\ntlinoone}
\def\chgone{\chginooneplus}
\def\mtthree{{m_{T3}}}
\def\mtfour{{m_{T4}}}
\def\mttwosq{{m_{T2}^2}}
\def\mttwopm{{m_{T2\pm}}}
\def\mttwosqpm{{m_{T2\pm}^2}}
\def\mttwosqp{{m_{T2+}^2}}
\def\mttwosqm{{m_{T2-}^2}}
\def\denom{{\Delta}}
\def\denompm{{\Delta_\pm}}
\def\denomp{{\Delta_+}}
\def\denomm{{\Delta_-}}
\def\mpi{{\pi}}
\def\mchi{{\chi}}
\def\half{{\frac 1 2}}
\def\sixth{{\frac 1 6}}
\def\third{{\frac 1 3}}
\def\twoThirds{{\frac 2 3}}
\def\fourThirds{{\frac 4 3}}
\def\T{{T}}
\def\W{{W}}
\def\LM{{E_{\rm 1}}}
\def\DE{{E_{\rm 2}}}
\def\deltaE{{\delta_E}}
\def\mylambda{{\lambda}}
\def\mymu{{\mu}}
\def\mydelta{{\delta}}
\def\Z{{Z}}
\def\mynabla{{\Delta}}
\def\mybigq{{Q}}
\def\tbar{{\bar t}}
\def\ttbar{{t \tbar}}
\def\ifb{{{\mathrm {fb}}^{-1}}}
\def\eV{{{\mathrm {eV}}}} 
\def\keV{{{\mathrm {keV}}}} 
\def\MeV{{{\mathrm {MeV}}}} 
\def\GeV{{{\mathrm {GeV}}}} 
\def\TeV{{{\mathrm {TeV}}}} 
\newcommand{\slptwo}{{{\slashchar{{\bf p}}}}}
\newcommand{\twoMatrix}[4]{{\renewcommand\arraystretch{1.0} \begin{array}[c]{cc} 
    #1   &   #2  \\ #3   &   #4    
\end{array}}}
\newcommand{\smallTwoVec}[2]{{\renewcommand\arraystretch{0.6} \begin{array}[c]{c} 
\!\!\!    #1    \!\!\!\!
\\ 
\!\!\!    #2    \!\!\!\!
\end{array} }}
\newcommand{\twoVec}[2]{{\renewcommand\arraystretch{1.0} \begin{array}[c]{c} 
\!\!\!    #1    \!\!\!\!
\\ 
\!\!\!    #2    \!\!\!\!
\end{array} }}
\newcommand{\smallThreeVec}[3]{{\renewcommand\arraystretch{0.6} \begin{array}[c]{c} 
\!\!\!    #1    \!\!\!\!
\\ 
\!\!\!    #2    \!\!\!\!
\\ 
\!\!\!    #3    \!\!\!\!
\end{array} }}
\newcommand{\threeVec}[3]{{\renewcommand\arraystretch{1.0} \begin{array}[c]{c} 
\!\!\!    #1    \!\!\!\!
\\ 
\!\!\!    #2    \!\!\!\!
\\ 
\!\!\!    #3    \!\!\!\!
\end{array} }}
\newcommand{\capbox}[2]{\parbox{0.85\textwidth}{\caption[#1]{\textit{#2}}}}
\def\sus{supersymmetry}
\def\Sus{Supersymmetry}
\def\susic{supersymmetric}
\def\Susic{Supersymmetric}
\newcommand{\mysecref}[1]{Section~\ref{#1}}
\newcommand{\myeqref}[1]{Equation~(\ref{#1})}
\newcommand{\mytabref}[1]{Table~\ref{#1}}
\newcommand{\myfigrefatstartofsentence}[1]{Figure~\ref{#1}}
\newcommand{\myfigref}[1]{Figure~\ref{#1}}
\newcommand{\definmath}[2] {\def#1{\ifmmode#2\else$#2$\fi}}
\definmath\amin{\mathrm{min}}
\definmath{\cht}{{\tilde{\chi}}}
\definmath{\DeltaMChi}{{\Delta M_{\cht_1}}}
\def\mtx{{m_{TX}}}
\def\mtxsq{{m_{TX}^2}}
\definmath{\squark} {{\tilde{q}}}


\section{INTRODUCTION\label{sec:myintrintor} }
The authors of \cite{Nojiri:2003tu} propose ``a new reconstruction
technique for SUSY processes at the LHC''.  Their method is described
in detail in their article elsewhere in these proceedings, and so only
an outline of their method will be provided here.  The reader is
strongly encouraged to read their article before reading this one.

\par

The authors of \cite{Nojiri:2003tu} refer to their technique as the ``mass
relation method''.  I wish to narrow the meaning of this phrase,
as I want to draw a distinction between (1) the {\em idea} that makes
the whole method work, and (2) any particular {\em implementation} of
that idea.  I will use the phrase ``mass relation method'' to describe any method
which, for its success, is forced to rely on the extraction of
information from {\em two or more independent events} which are
related only by their sharing {\em similar or identical particle content}.

%

\section{OUTLINE OF ORIGINAL IMPLEMENTATION}

In \cite{Nojiri:2003tu} it is suggested that ``when several sparticle
masses are known, the kinematics of SUSY decay processes observed at
the LHC can be solved if the cascade decay contains sufficient
steps''.  Therein, an original implementation is described which
accomplishes this solution.  The efficacy of this implementation is
demonstrated using the decay chain
\begin{eqnarray}
  \gluino \rightarrow \sbottom b \rightarrow \ntlinotwo b b \rightarrow
  \slepton b b l \rightarrow \ntlinoone b b l l. \label{eq:myoinlyeeeq}
\end{eqnarray}
In order to complete the demonstration, the authors make the
assumption that ``... the masses of the two lighter neutralinos and
the right handed slepton are known, and [they] ignore the
corresponding errors''.  By ignoring the corresponding errors, they in
effect demonstrate their method in a scenario in which the slepton and
neutralino masses are already known to something like the one-percent
level -- something not usually assumed to be possible with LHC data
alone.  Having made these assumptions, the authors set out to measure
the masses of the two remaining sparticles.  Firstly, on an event-pair
by event-pair basis, they obtain an estimate of their gluino mass with
an accuracy of about $\pm 60~\GeV$ (approximately 10\%).  By
histogramming the results from all these event-pairs an overall
measurement of the gluino mass is obtained with an error of a few
$GeV$.  This measurement is then fed back into the events which are
analysed in a second pass in order to make a measurement of the
sbottom mass.\footnote{In practice, as in most of these analyses, the
mass difference between the sbottom and the gluino is measured more
accurately than the absolute value of either of their masses.}

\subsection{A small part of the original implementation in more detail\label{sec:idectheirmethod}}





\label{sec:originlimplmntatnojrri}

As already mentioned in \mysecref{sec:myintrintor}, the authors begin
by assuming prior knowledge of $m_\ntlinoone$, $m_\slepton$ and
$m_\ntlinotwo$; that is to say the masses of the two lightest
neutralinos and the mass of the slepton participating in the decay
chain shown in \myeqref{eq:myoinlyeeeq}.  This leaves in any one event
the six unknown real quantities comprising: $m_\gluino$, $m_\sbottom$
and the four components of $p_\ntlinoone$.  These six unknowns, are
however constrained to satisfy the five mass constraints:
\begin{eqnarray}
  m_\ntlinoone^2 & = & p_\ntlinoone^2 , \nonumber \\
  m_\slepton^2 & = & (p_\ntlinoone+p_{l_1} )^2 , \nonumber \\
  m_\ntlinotwo^2 & = & (p_\ntlinoone+p_{l_1}+p_{l_2} )^2 , \label{eq:myfiveconsdrtrinats}\\
  m_\sbottom^2 & = & (p_\ntlinoone+p_{l_1}+p_{l_2}+p_{b_1} )^2 ,
  \mbox{\qquad and} \nonumber \\
  m_\gluino^2 & = & (p_\ntlinoone+p_{l_1}+p_{l_2}+p_{b_1}+p_{b_2} )^2,\nonumber 
\end{eqnarray}
in which $p_{l_1}$, $p_{l_2}$, $p_{b_1}$ and $p_{b_2}$ are the
four-momenta of the emitted standard model
particles.\footnote{Strictly speaking these mass constraints apply
only to the {\em true} rather than the {\em measured} momenta of the
emitted standard model particles.  However, for the purposes of the
``original implementation'' this distinction did not need to be drawn,
and the measured momenta were used ``as if'' true.  The resulting
smearing of the answer was accepted as a source of reconstruction
error.  There are differences between this method and that of my
proposal.}  Since the number of unknowns (six) exceeds the number of
constraints (five) it is not possible to conclude much from one
event.

\par

Taking a second event together, however, the number of unknowns rises
by only four, namely the four components of the $\ntlinoone$-momentum
in the new event.  As ever, the new event, like the old, will satisfy
another five mass constraints of the form shown in
\myeqref{eq:myfiveconsdrtrinats}.  With two events, then, the number
of unknowns and number of constraints have each risen to ten.  So in
principle, with only two non-degenerate events, it is now possible to
determine all the unknowns.  This amounts to full reconstruction of
both events and determination of $m_\gluino$ and $m_\sbottom$.  The
interested reader is directed to \cite{Nojiri:2003tu} to see how the
authors handle choice-ambiguities that arise from (a) the solution of
simultaneous quartic and quadratic equations, and (b) lack of
knowledge of which of the two observed leptons is $l_1$ and which of
the two observed $b$-tagged jets came from $b_1$.
\label{sec:chouceambigueitieiasredisccssd} 

\par


The implementation of \cite{Nojiri:2003tu} proposes that one should do
exactly as described above: namely consider events in pairs.

\section{MOTIVATIONS FOR BUILDING ON THE ABOVE}

A natural reaction on seeing the original implementation is to ask:
\begin{quote}
Since two events are better than one, why not consider even more?
\end{quote}
The majority of the rest of this note tries to address the above
question.  The motivation for building on the above is that every new
event adds five more constraints but only four more unknowns.  Put
another way, for every additional event that is acquired, one can
either answer one new question, or else better constrain any answers
that one already has.  This note concentrates on the last of these two
possibilities.

%

\section{PROPOSAL FOR A ``NEW IMPLEMENTATION'' OF A MASS RELATION METHOD}
\subsection{General comments}

The new proposal is to do nothing more than consider {\em all} the
relevant events simultaneously.

\par

In this note, we will {\em not} address the important question of
whether, in a real LHC experiment, it would be possible to satisfy the
preconditions for the success of {\em any} mass relation method,
namely the requirements that it be possible to construct {\em
sufficiently pure} samples of the appropriate standard model samples
from chains of {\em sufficiently similar or identical} particle
content.  This needs to be addressed in further papers, and has
already been considered in part by \cite{Nojiri:2003tu}.  The
intention of this note is only to look at what may be achieved {\em
if} such selection were possible.\label{sec:lestercavvvesats}

\subsection{Detailed description of proposal}

\def\mvec{{\bf m}} \def\data{{\mbox{data}}} Ideally we would like to
know the masses of the sparticles in our events.  Realistically, we can
only expect to find the masses within some finite precision or error.
Bearing correlations in mind, the best we can expect to
determine is the relative probability of any particular combination of
the five masses $\mvec=(m_\ntlinoone, m_\slepton, m_\ntlinotwo,
m_\sbottom, m_\gluino)$ given the data.  In short, we would like to
plot $p(\mvec | \data)$.

\par

By Bayes' theorem, $p(\mvec|\data) \propto p(\data|\mvec) p_0(\mvec)$.
The first factor, the likelihood, will be determined purely from the
events considered and the mass relation method itself and is thus the
objective ``result'' of this experiment.  The last factor, the prior,
incorporates all existing knowledge gained from other experiments (if
you should wish to include them) and any subjective preferences you
might have.  Because we choose here to use a non-informative prior
(uniform in the hierarchical sparticle masses\footnote{Additionally,
the cosmetic constraints $30~\GeV < m_\ntlinoone < 300~\GeV$ and
$m_\gluino<1000~\TeV$ are incorporated into our prior in order to
frame all plots nicely.}) the reader may view the results at his or
her discretion as either (a) simple plots of the objective likelihood
distribution $p(\data|\mvec)$, or else, (b) indicative of the results
which a single experiment would provide $p(\mvec|\data)$ in the
absence of data from other experiments.


\par


\par

\def\event#1{{\mbox{event}_#1}}

%

The only thing remaining to be defined is $p(\data|\mvec)$. 
As the data consists of many independent events, we have
$
p(\data|\mvec) = p( \event 1 | \mvec) . p( \event 2 | \mvec) .
\cdot\cdot\cdot . p(\event n | \mvec),
$
and we are left needing to evaluate $p( \event i | \mvec)$
for a given event $i$.  

\par

Evaluating the event-likelihood $p( \event i | \mvec)$ properly and
efficiently is the hard part.  In principle there is only one right
answer, which you would obtain by taking the square of the matrix
element for the observed final state and integrating it over all the
unknowns in the problem, namely the measurement errors and the
momentum distribution of the unobserved chain progenitor.  The answer
you obtain will thus depend on which model assumptions you wish to
make, \eg\ whether you would choose to model the differing spins of
the sparticles.

\par

In order to meet the time constraints imposed by the submission of
this note, however, it was necessary to implement the event-likelihood
using an ad-hoc approximation.\footnote{A more in-depth paper
currently in preparation deals with the evaluation of the full form of
$p( \event i | \mvec)$.}  It is hoped that the approximation to the
event-likelihood described later is sufficiently similar to the full
form of $p( \event i | \mvec)$ that the basic features of the proposed
technique can be demonstrated.

\par

\par

We construct the approximation to the event-likelihood used in this
note as follows.  For a given chain momentum hypotheses $H$ consistent
with the given mass hypothesis $\mvec$, we can define a ``distance''
$\delta(H)$ between the observed and the true momenta of the visible
particles produced in that chain by
\begin{eqnarray}
\delta (H)= 
\epsilon_{l_1^H}^2/\sigma_l^2 +
\epsilon_{l_2^H}^2/\sigma_l^2 +
\epsilon_{b_1^H}^2/\sigma_b^2 +
\epsilon_{b_2^H}^2/\sigma_b^2 \label{eq:kfjhkfhbdmnr}
\end{eqnarray}
where (for example) $\epsilon_{l_1^H}^2/\sigma_l^2$ is square of the
number of standard deviations by which the measured momentum of $l_1$
differs from a hypothesised true value $p^H_{l_1}$.

\par

We can then perform a least-squares minimisation of $\delta(H)$ over
all possible chain momentum hypotheses $H$ consistent with the given
mass hypothesis $\mvec$.  We now make the assumption that this
least-squared minimisation will have provided us with the momenta
which are most consistent with the observation and the mass hypothesis
$\mvec$.  Finally, then, we can approximate the event-likelihood by
the simple process of evaluating the probability for the observed
momenta to deviate as far as their observed values, assuming that the
``true'' momenta are given by the result of the fit.  This
approximation thus depends crucially on a good understanding of the
measurement errors associated with the observed standard model
particles.

\par

The key features of this approximation to the event-likelihood are
that it will be large when $p( \event i | \mvec)$ is large and small
when $p( \event i | \mvec)$ is small.\footnote{Unlikely events (for a
given $\mvec$) are clearly those in which, no matter how hard you try,
you find a {\em huge disagreement} between the momenta of the
particles you see in the event and the momenta you would expect to
have seen considering {\em all} hypothesised chain momenta that would
have been consistent with the masses $\mvec$.  Put another way, you
will know you have a very unlikely event when you cannot hypothesise a
set of chain momenta in which the visible particles have momenta
``close'' to those observed in the event.  If then we discover that
$\delta (H)$ is large, no matter what hypothesis $H$ we choose
consistent with $\mvec$, then we know that that particular event is
unlikely given that particular $\mvec$.  Conversely, when $\mvec$ is
close to the right answer, the true chain momenta $H^{tr}$ will lead
to a small value for $\delta(H^{tr})$.}  This gives us confidence that
the maxima and minima of the event-likelihood will be well
approximated.  It will most probably not be the case, however, that
the widths of the resulting distributions or the finer shape details
(for a particular number of events) can be completely relied upon.
This, though
\label{sec:undesireableapproxm} undesirable, is not too great a
problem as the widths naturally scale with the number of events
analysed.  For this reason, the approximation may be thought of as
resulting in an uncertainty in how many events are necessary to
achieve a given reconstruction error, rather than an uncertainty in
the quality of the reconstruction itself.

\par

\section{RESULTS OF NEW PROPOSAL}

\def\mypthing{{$p(\mvec|\data)$}}

In order to demonstrate the potential of the above technique, it is
necessary to generate some events for analysis.  A toy montecarlo was
used to generate these events only.  For the reasons of
\mysecref{sec:lestercavvvesats} it simulated only relativistic
kinematics and decays were according to phase space only.  In effect
all particles were treated as scalars.  Furthermore, only the chain
described in \myeqref{eq:myoinlyeeeq} was simulated.  No extraneous
particles were produced, and nor is there an ``other side of the
event''.


\par

Events were simulated events using the following arbitrary values for
the sparticle masses: $m_\ntlinoone=150~\GeV$, $m_\slepton=200~\GeV$,
$m_\ntlinotwo=300~\GeV$, $m_\sbottom=500~\GeV$ and
$m_\gluino=650~\GeV$.

\par

Measurement errors were simulated by randomised rescaling of the
four momenta of the observed particles in a manner similar to that
described in \cite{atlfast20}, but without $\eta$ and $\phi$
dependence in the resolutions.  Lepton momenta were smeared by 1\% and
jet momenta by 5\%.  The reconstruction part of the analysis was
handed the two lepton momenta in a random order and the two $b$-jet
momenta in a random order so that it could not know which of them was
which.

\par

After generating 100 events of the form described above, the results
shown in Figures~\ref{fig:THE-moo-2D} 
were obtained.  Here we choose to plot $p(\mvec|\data)$ by sampling
from it using a Metropolis Markov-chain sampler
\cite{djminformationbook}, although this particular choice is
unimportant.\footnote{Finding good start points for the Markov-chain
  sampler
is not a difficult task, just a very
time consuming one due to the high dimensionality (five) of the space
to be probed.  I have avoided the computational overhead of doing such
pre-scans, and have opted to simply start the Markov-chain sampler off
in the vicinity of the ``correct'' answer.  It is possible that as a
result, the Markov-chain sampler may have failed to find other
islands of high probability separated from the correct area by
a deep ocean of improbability.  Experimentation with less well
constrained situations (\eg\ only 20 events instead of 100) suggests
that the other islands are not commonly a problem, but may be able to
admit extra solutions where two particles in the chain are close to
being mass degenerate.}


\par

\begin{figure}
\begin{center}
\subfigure[$m_\slepton$ against $m_\ntlinoone$]{
\includegraphics[width=6.5cm]{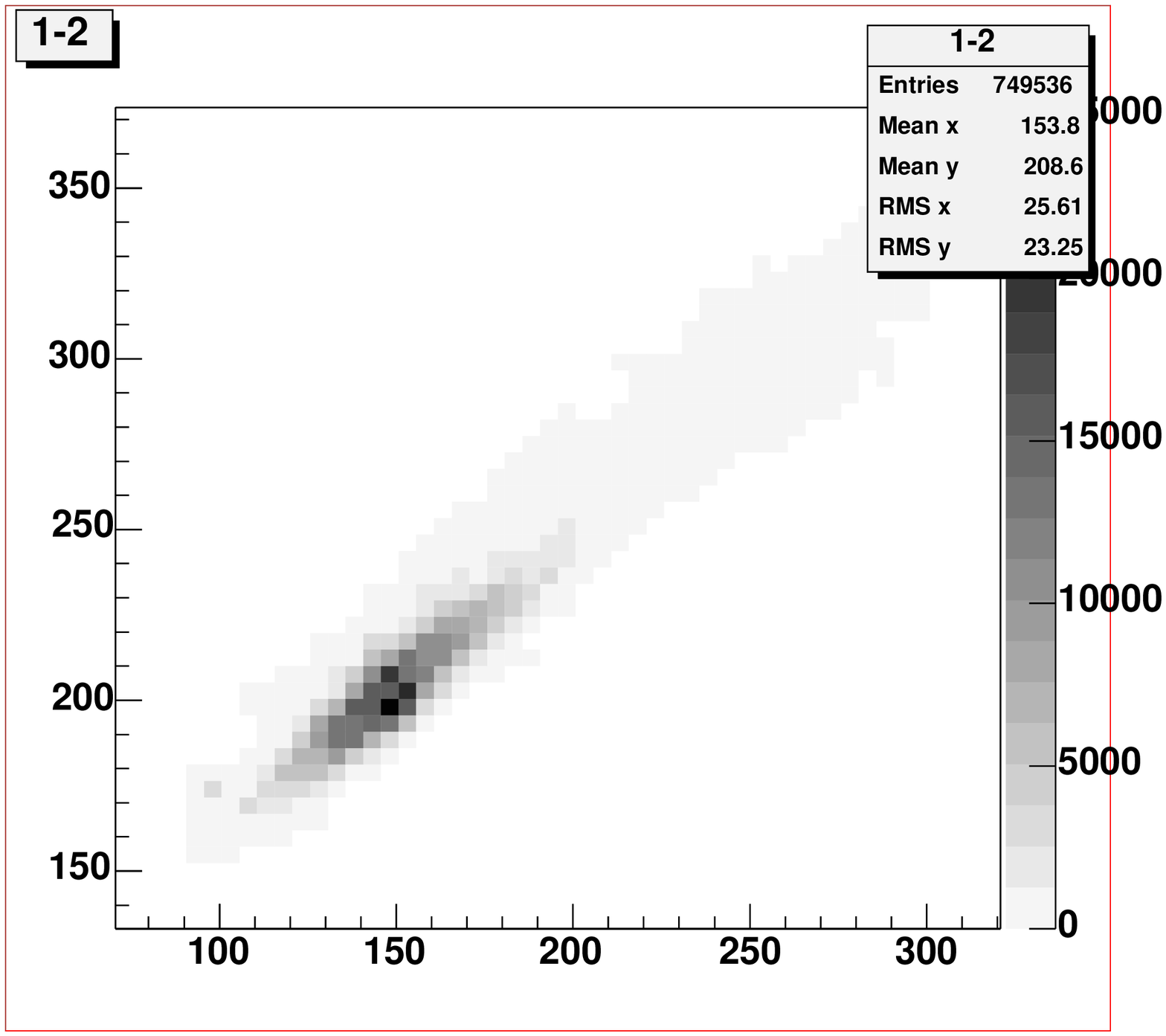} 
}
\subfigure[$m_\ntlinotwo$ against $m_\ntlinoone$]{
\includegraphics[width=6.5cm]{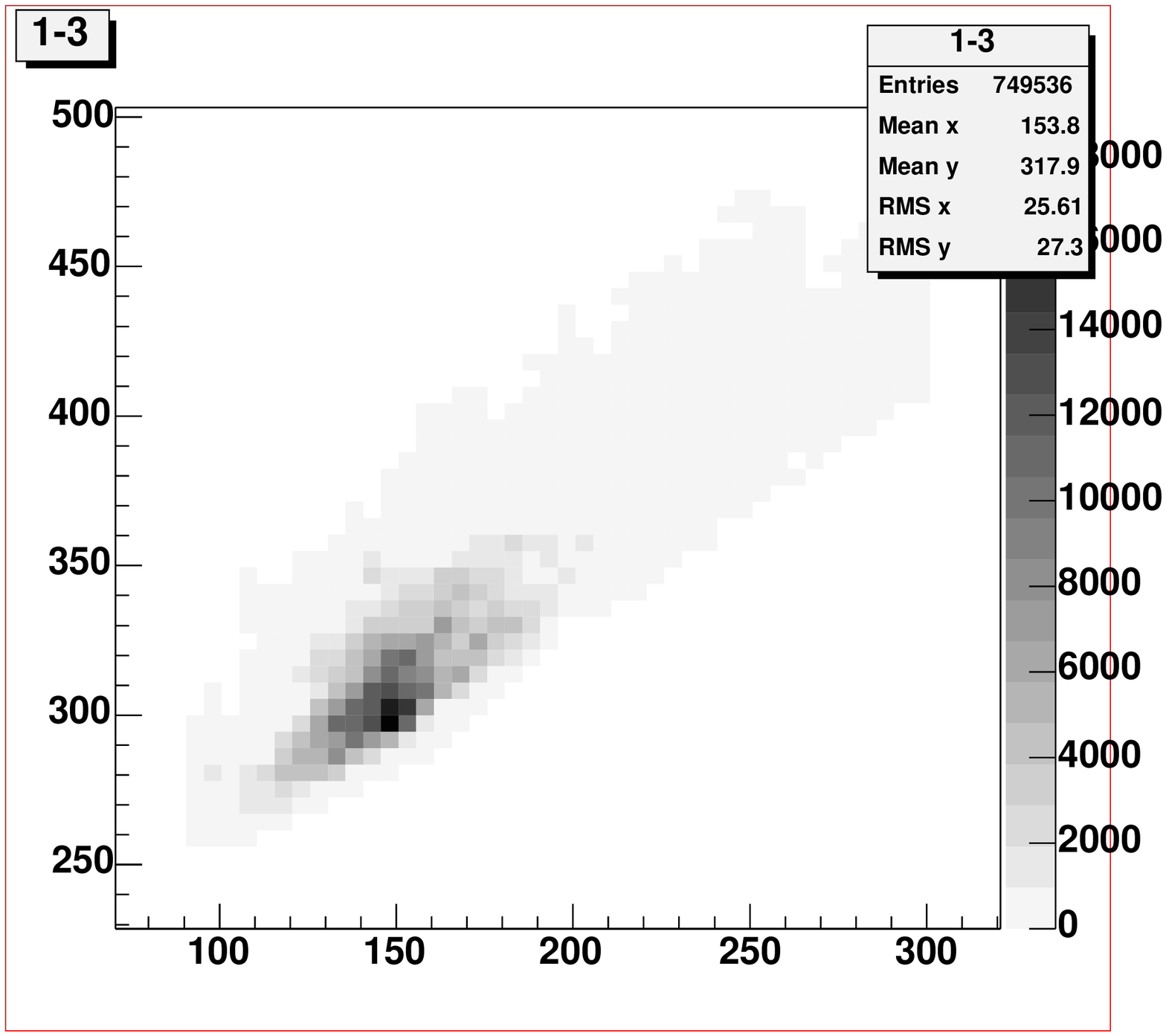} 
}
\subfigure[$m_\sbottom$ against $m_\ntlinoone$]{
\includegraphics[width=6.5cm]{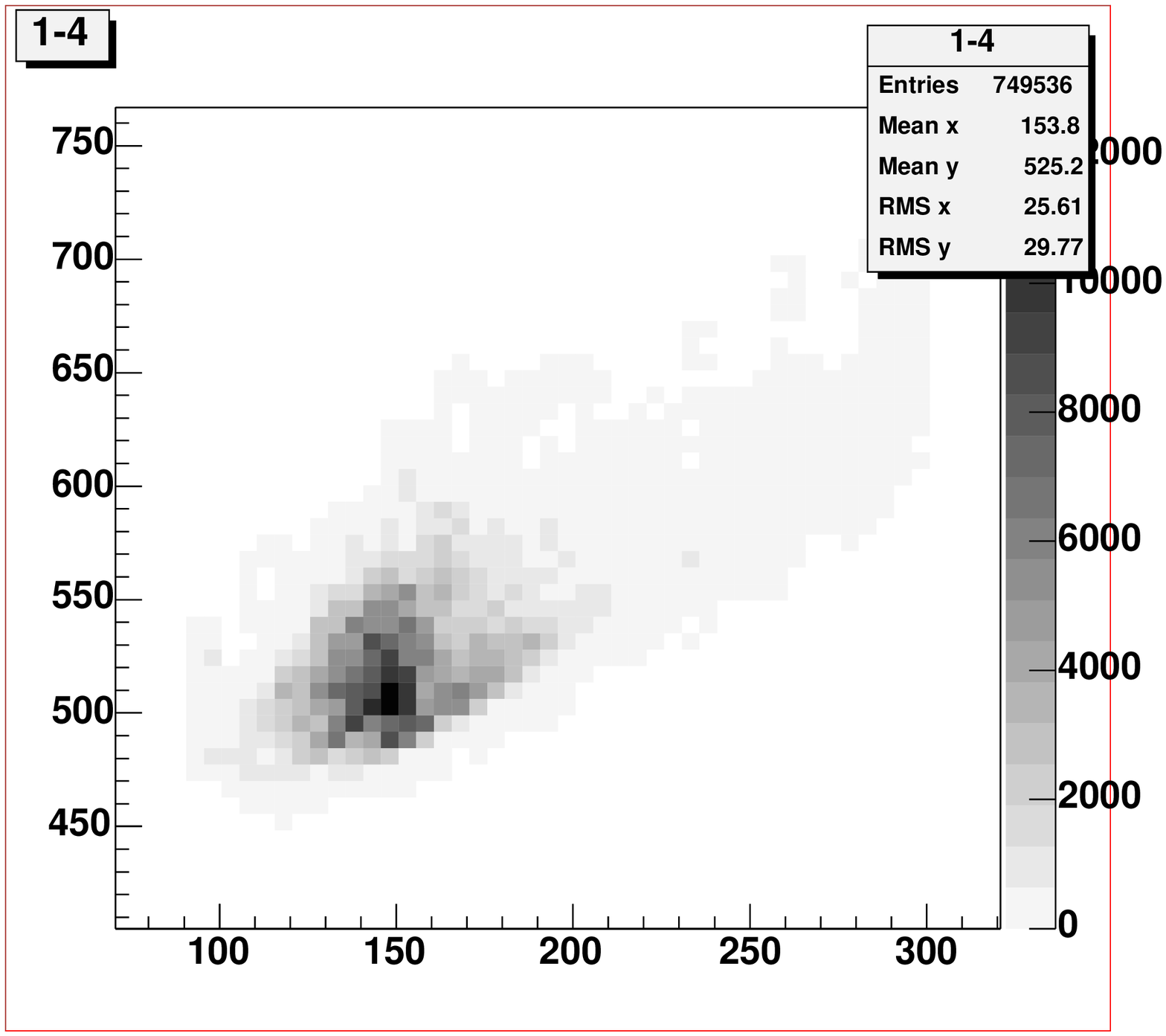} 
}
\subfigure[$m_\gluino$ against $m_\ntlinoone$]{
\includegraphics[width=6.5cm]{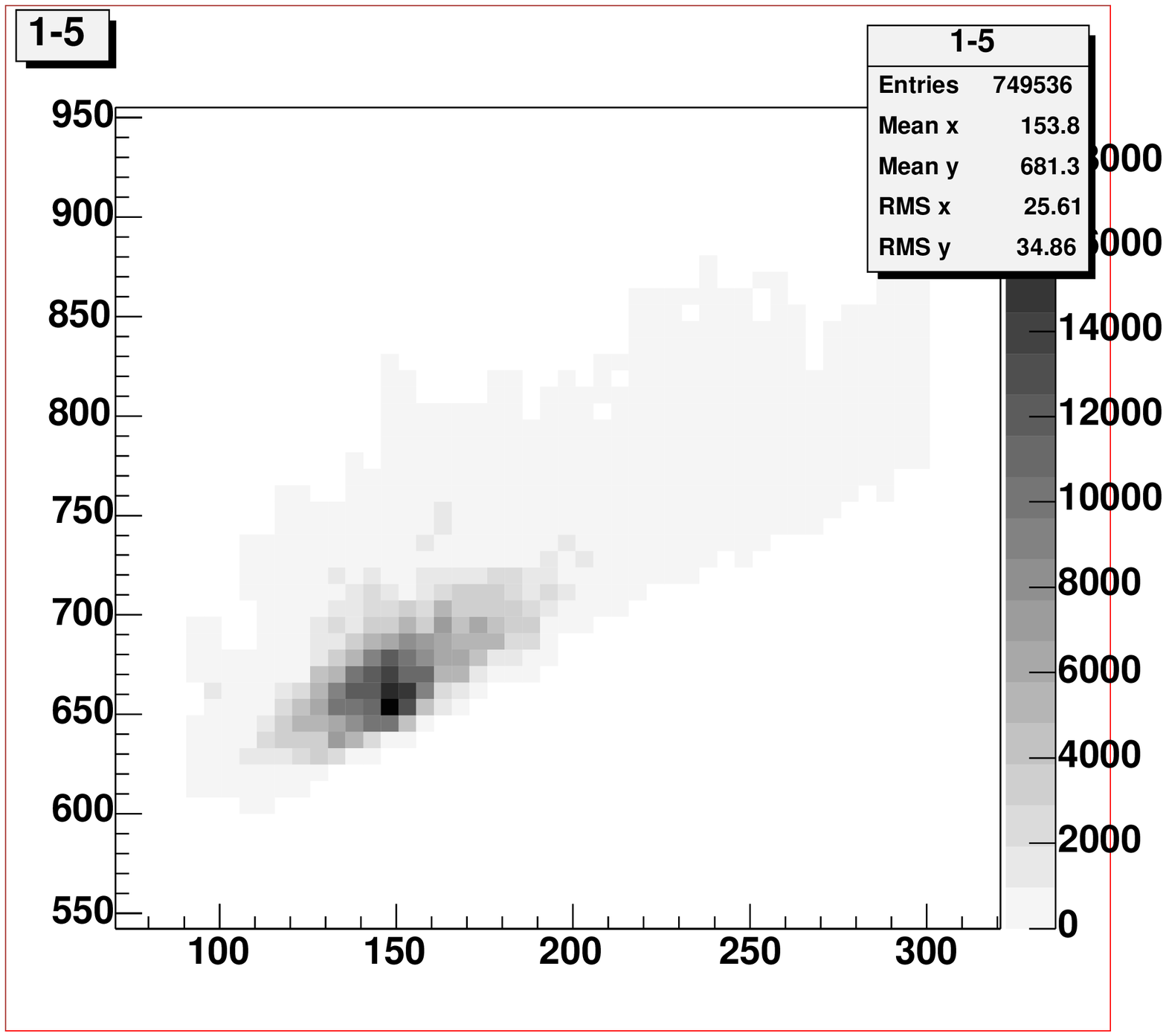} 
}
\caption{
  Samples drawn from \mypthing. 
  Axes in $\GeV$.
\label{fig:THE-moo-2D}}
\end{center}
\end{figure}





\par

Figure~\ref{fig:THE-moo-2D} shows how the reconstructed mass
distributions of the heavier sparticles correlate with the
reconstructed mass distribution of the lightest neutralino.  The
correlation show that at this point the mass differences tend to be
measured more accurately than the absolute masses.
%
%
%
The mass of the lightest neutralino is well reconstructed (input value
$150~\GeV$) with an error of order approximately
17\%.\footnote{Remember that there is an uncertainty associated with
the interpretation of this and the other reconstructed widths for the
reasons discussed in \mysecref{sec:undesireableapproxm}.}
%
%
The reconstructed mass distributions of the remaining sparticles are
similar to that of the lightest neutralino as the mass differences are
constrained better than the masses themselves.

%

\section{CONCLUSIONS}

This note aims to convey the message that {\em should it be possible}
at the LHC to isolate clean samples of the decay products of a hundred
or so sparticle decay chains containing enough sufficiently similar or
identical particles, then by using a mass relation method of the type
proposed, one should be able to reconstruct the masses of all the
sparticles in these decay chains to precisions of at least some tens
of percent, depending on the number of these events.

\par

Further work is needed to establish that the necessary sample purity
is achievable, and to ascertain the effect that a better model for the
event-likelihood would have on the widths of the reconstructed
distributions.  If the identified collections of outgoing particles
are sufficiently pure and large, it does not seem unreasonable to
believe that one might expect a precision on the reconstructed masses
which is competitive with any other independent method found so far.

\par

It seems likely that the usefulness of this method will be limited by
the ability to produce the necessary pure samples of decay-chain
products.\footnote{although there is no reason why a Bayesian
analysis could not include extra hypotheses for additional
simultaneous chains}

\section{ACKNOWLEDGEMENTS}

I would like to thank Mihoko Nojiri, Giacomo Polesello, Dan Tovey and
Andy Parker for helpful discussions.  I would also like to thank Ben
Allanach for his own helpful comments and for encouragement.  Finally
I would like to thank Alan Barr for his own nit-picks and for drawing
reference \cite{Nojiri:2003tu} to my attention.










\setcounter{figure}{0}
\setcounter{table}{0}
\setcounter{section}{0}
\setcounter{equation}{0}
\clearpage

 \part{Study of non-pointing photons at the CERN LHC \label{prieur}}
{\it K.~Kawagoe, M.M.~Nojiri,
G.~Polesello and D.~Prieur}
\maketitle
\begin{abstract}
Measurement of non-pointing photons is a key issue
to study the gauge mediation models at the CERN LHC. 
In this article we study the $\theta$ resolution of non-pointing photons
with the ATLAS electromagnetic calorimeter, and discuss the impacts to
the study of the gauge mediation models.

\end{abstract}

\section{GAUGE MEDIATION MODELS AND NON-POINTING PHOTONS}

Origin of the SUSY breaking in the hidden sector and its mediation to
the MSSM sector are key features of SUSY models.  When hidden sector
SUSY breaking is expressed by the order parameter $F$ and the scale
of the mediation to the MSSM sector by $M$, the mass scale of MSSM
sparticles $M_{SUSY}$ is of the order of $\lambda F/M$, where
$\lambda$ is the coupling of the hidden sector to the MSSM sector. The
SUSY breaking mediation may be due to renormalizable interactions,
such as the gauge interaction.  This is called ``gauge mediation''
(GM) models.  In the GM models $M$ and $F$ are arbitrary and we expect
$M\ll M_{\rm pl}$.

When $M\ll M_{\rm pl}$, the lightest SUSY particle (LSP) is the
gravitino ($\tilde{G}$) in the GM models.  The next lightest SUSY
particle (NLSP) is a particle in the MSSM sector which decays into a
gravitino.  If the lightest neutralino ($\tilde{\chi}^0_1$) is the
NLSP, the dominant decay mode is $\tilde{\chi}^0_1\rightarrow \gamma
\tilde{G}$.  The neutralino lifetime $c\tau$ is a function of $F_0$
and $m_{\tilde{\chi}^0_1}$, and the neutralino 
may be long-lived. Therefore it is an important subject to study
non-pointing photons at the CERN LHC.

In a paper \cite{Kawagoe:2003jv}, a procedure is proposed
to solve the gravitino
momentum and $\tilde{\chi}^0_1$ decay position for the cascade decay
$\tilde{\ell}\rightarrow \tilde{\chi}^0_1 \ell\rightarrow \tilde{G}
\ell\gamma$ using the ATLAS detector at the LHC. 
It is shown that one can determine 
the mass and lifetime of $\tilde{\chi}^0_1$.
To this purpose, one need to measure the photon momentum and arrival
time very precisely. A toy simulation is made under the following
assumptions to the photon momentum resolution;
\begin{itemize}
\item A good angular
resolution of $\sigma_{\theta} = 60\ {\rm mrad}/\sqrt{E}$ is
assumed, where $\theta$ is the polar angle of the photon momentum with
respect to the beam axis and $E$ is the photon energy 
measured in GeV.  
This resolution is based on a simulation for
pointing photons.
\item  The azimuthal angle of the photon momentum
$\phi$ is only poorly measured by the electromagnetic (EM) calorimeter. 
The $\phi$ resolution 
is good only  for the 
photons converted in the transition radiation tracker (TRT)
located in the EM calorimeter.
\end{itemize}

The analysis takes two steps. 
First, events with converted non-pointing photons
are selected. As the photon momentum is precisely measured, the
events can be used to determine the mass of $\tilde{\chi}^0_1$.  
In solving the kinematics, the direction of the gravitino momentum is
sensitive to the photon momentum. The mass resolution at GM
point ~G1 with $c\tau_{\tilde{\chi}^0_1}=100$cm is $\Delta
m_{\tilde{\chi}^0_1}=3.5$ GeV for $10^5$ generated SUSY events,
which corresponds to an integrated luminosity of 13.9~fb$^{-1}$.

Once the mass is determined precisely, the good
$\phi$ resolution of the photon momentum is not required
to solve the decay kinematics.
The decay kinematics can be solved for all $\ell\gamma$ events with
non-pointing photons, 
from 
the $\theta$ component of the photon momentum,
the arrival time and position at the ECAL, and the lepton
momentum. Therefore, all $\ell\gamma$
events with or without photon conversion  can be used to determine the
lifetime. The lifetime resolution for the GM point G1 is $\Delta
c\tau/c\tau=0.045$ for the $10^5$ generated SUSY events.

Although the result looks nice, 
some of the assumptions in the paper may be too optimistic,
and should be studied more realistically.
Among them, the $\theta$ resolution obtained for pointing photons  
are used to estimate that for non-pointing photons.
In
this article we study the $\theta$ resolution for non-pointing photons
by a full simulation
and discuss the impacts to the GM study.

\section{PARAMETRISATION OF ANGULAR RESOLUTION OF ATLAS BARREL ELECTROMAGNETIC CALORIMETER}

In this part we try to obtain a more refined parametrization of
the angular resolution for non-pointing photons 
with the ATLAS EM calorimeter.
So far the resolution has been studied
only for nearly pointing photons, i.e.
photons coming from the ATLAS interaction point, and it is
of the order of 60~${\rm mrad}/\sqrt{E}$
\cite{Airapetian:1996iv,atlasTDR}. 
In the following we
will try to see which resolution is achievable for non-pointing
photons using a full simulation of the ATLAS detector.

The EM calorimeter \cite{atlasTDR} is a projective calorimeter with
a good granularity to perform precision measurements of the shower
position. It is longitudinally divided into three compartments: strip,
middle and back compartment. The strip section is 
segmented along $\eta$ into very thin cells of $\Delta
\eta=0.003125$, leading to a resolution on $\eta$ position with
pointing photons of $0.30\times 10^{-3}$. The middle compartment
has a wider $\eta$ granularity of $\Delta \eta=0.025$ and it is
designed to contain most of the shower energy. It has a resolution
on $\eta$ position of $0.83\times 10^{-3}$. By combining the
measurement of the $\eta$ position in the first two compartments,
it is possible to determine the shower direction in $\eta$.

 For this study, different samples of single
photons have been generated. Each of these samples consist of
$20000$ photons of $p_t=60$~GeV, randomly triggered with $\eta$
from $-1.4$ to $1.4$. Pointing photons were generated at ATLAS
origin with a spread on the position of the generation vertex of
$5.6$~cm along Z axis and $15$ $\mu$m along radial axis, as it
should be in ATLAS final setup. For non-pointing photon samples,
the generation vertex has been shifted along ATLAS Z axis with
values from $10$~cm to $150$~cm. No spread on the generation
vertex position has been applied for these dataset. Finally all of
these events were fully simulated using Dice/Atlsim (v3.2.1), the
Geant3 ATLAS detector description \cite{Artamonov:1996}.

\noindent Here we focus only on the barrel part of the EM
calorimeter. The reconstruction of all events has been done using
ATLAS standard reconstruction software (Athena v6.5.0). No
electronic noise or pile-up have been added in the reconstruction.
The noise will certainly contribute to degrade the resolution,
however this has not been studied yet.

\noindent With the $\eta$ position, in each layer of the
calorimeter ($\eta_1$,$\eta_2$) and using a parametrization of the
shower depth for each layers ($R_1(\eta_1)$,$R_2(\eta_2)$)
\cite{Airapetian:1996iv}, we are able to reconstruct the shower
axis $\eta_{pointing}$ using the following relation:

\begin{equation}
\sinh(\eta_{pointing}) =
\frac{R_2(\eta_2)\sinh(\eta_2)-R_1(\eta_1)\sinh(\eta_1)}{R_2(\eta_2)-R_1(\eta_1)}
\end{equation}

The standard reconstruction uses a sliding window algorithm to
find regions of interest in the calorimeter, then $3 \times 3$
($\eta \times \phi$) clusters are made in order to compute the $\eta$
barycenter of the shower in each layer. This position is the
average of cells $\eta$ position weighted by energy in each cell:
\begin{equation}
\overline{\eta} = \frac{\sum{\eta_i E_i}}{\sum E_i}
\label{barycenter}
\end{equation}
The angular resolution achieved using the standard reconstruction
is shown in Fig. \ref{AngRes} as a function of the position of the
generation along ATLAS Z axis. The resolution is
significantly degraded from 60~mrad for $Z_{vertex}=0$~cm to about
800~mrad for $Z_{vertex}=100$~cm. This is due to several
effects. First, the S-shape corrections, that were tuned for
pointing photons, are no longer valid and tend to degrade the
resolution for large deviation from pointing. The S-shape effect
is a distortion of the reconstructed $\eta$ position due to the finite
cluster size. Then the $3 \times 3$ clusters are no longer sufficient
to contain all the shower and some energy leakage outside the
cluster is possible. Another point is that Eq. (\ref{barycenter})
is no more true for non-pointing photons and gives rise to
a systematic shift in computing $\eta$ position for each layer.
Finally, the shower depth parametrization, tuned from pointing
photons, is also no more valid for large deviation from pointing.

In order to improve the resolution, some changes have
been made to the standard reconstruction algorithm. First, the
cluster size has been extended to $5 \times 3$ and we do not apply
the S-shape corrections. For each layer the systematic shift observed
in the $\eta$ position reconstruction has been parameterized as a
function of the generation vertex position. Using this last
parametrization we have made an iterative algorithm which corrects
the $\eta$ position in each layer. The convergence is obtained
in about 3 iterations. Results of this correction on angular
resolution is show on Fig. \ref{AngRes}. For small vertex shift
($Z<30$~cm), the standard reconstruction gives the best
resolution. This is mainly due to the absence of S-shape
corrections. For larger vertex shift ($Z>30$~cm), the
reconstruction algorithm with iterative correction gives better
results than the standard one.

\begin{figure}[htbp]
\begin{center}
\includegraphics[width=0.95\linewidth ]{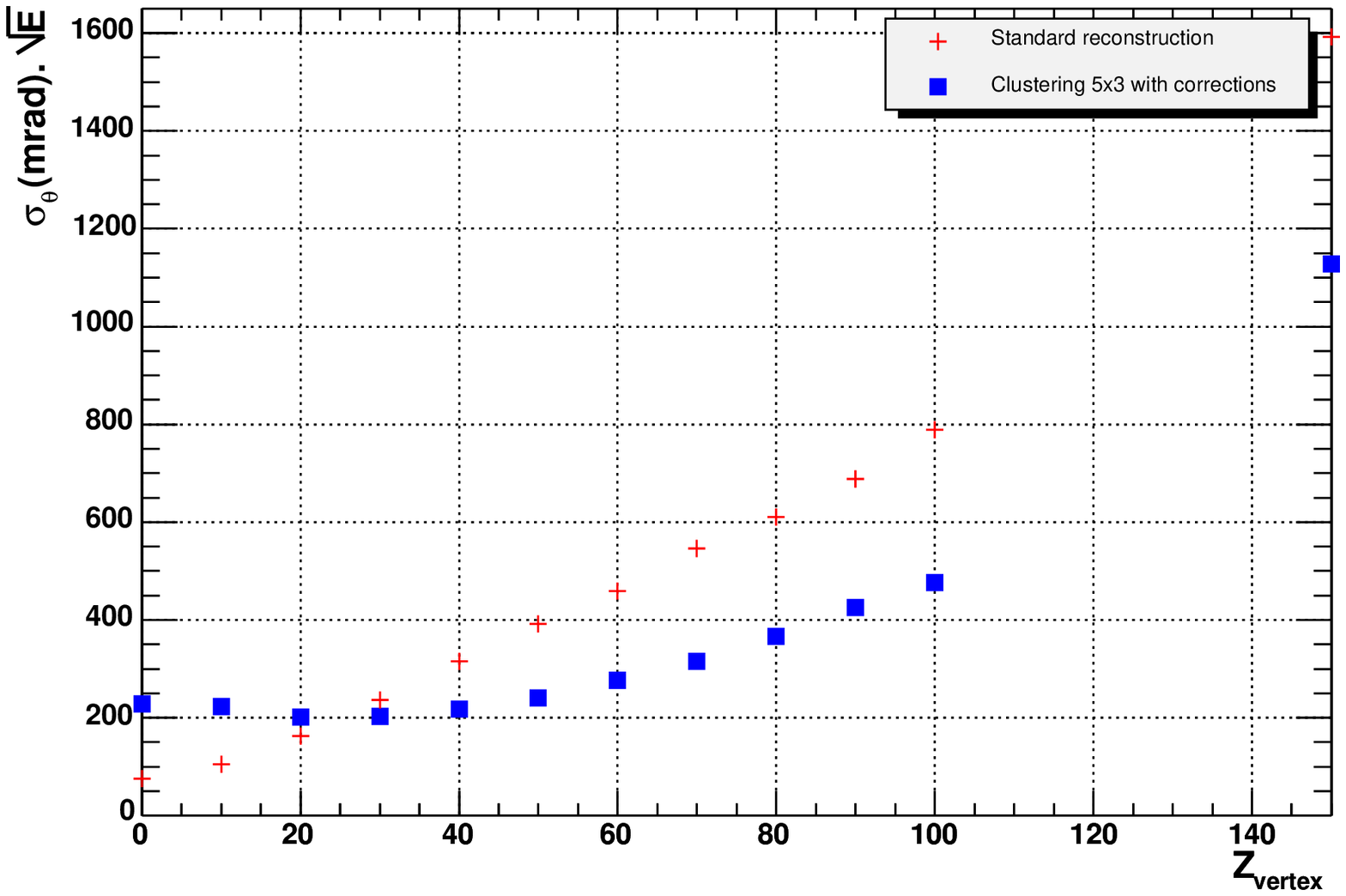}
\caption{Angular resolution of ATLAS barrel EM calorimeter as a
function of the position of the production vertex.} \label{AngRes}
\end{center}
\end{figure}

This study has shown that standard reconstruction algorithm is
not well suited for non-pointing photons and that a specific
treatment is necessary. The next step to improve the resolution
would be to try other clustering algorithms such as
NearestNeighbor. Another possible way would be to study the
dissymmetric shower profile of non-pointing photons and try to
extract an alternative method for computing
$\eta$ barycenter from this information.

\section{EFFECT ON THE GAUGE MEDIATION STUDY}

First of all, the $\theta$ resolution affects the selection efficiency of
non-pointing photons and background contamination from pointing photons.
Second, the resolution of the decay kinematics would be reduced if 
the photon angle resolution is reduced. 
We study this in terms of the angle $\psi$
defined as the opening angle of the gravitino direction and 
the photon direction. This  can be calculated by using 
the following formula
\begin{eqnarray}
\cos\psi &=& \frac{1-\xi^2}{1+\xi^2}\ , \cr
{\rm where }\ \ \xi  &\equiv& \frac{c t_{\gamma}-L\cos\alpha}{L\sin\alpha},
\label{kobayashi}
\end{eqnarray}
where 
$t_{\gamma}$ is the photon arrival time at 
the barrel EM calorimeter,  $L$ 
is the distance between the interaction point (O) and the point 
where the photon arrives at the EM calorimeter (A) , and $\alpha$ is the angle 
between the  photon momentum and the position vector \overrightarrow{OA}. 
Changing the $\theta$ resolution and 
keeping all the other resolutions same as those assumed in
Ref.~\cite{Kawagoe:2003jv}, 
we show in Table~\ref{reso}
the resolution of the angle $\psi$ for point G1 with 
$c\tau=100$ cm and $\int {\cal L}dt=13.9$~fb$^{-1}$. 
\begin{table} 
\begin{center}
\begin{tabular}{|c|rrrrrrr|}
\hline
$\sigma_{\theta} ({\rm mrad}) \times \sqrt{E({\rm GeV})}$ &0&60&100&200&300&400&500
\cr
\hline
$\sigma_{\psi}$(mrad) & 36.5& 38.2& 40.6& 46.9&  52.6 & 56.6&
60.2
\cr
\hline
RMS$_{\psi}$ (mrad)&46.2 & 48.0 & 50.9& 63.0&78.8 & 93.0& 105.5
\cr
\hline
\end{tabular}
\end{center}
\caption{Resolution of the angle $\psi$ for various asumptions of 
$\theta$ resolution.}
\label{reso}
\end{table}
Here, $\sigma_{\psi}$ is obtained by Gaussian fit using  center part of 
$\Delta \psi (\equiv \psi-\psi_{\rm true})$
distribution, while RMS$_{\psi}$ is obtained using the whole distribution.  
The error $\sigma_{\psi}$ increases as $\sigma_{\theta}$
increases.  This affects the error of the 
$\tilde{\chi}^0_1$ mass determination. 


\setcounter{figure}{0}
\setcounter{table}{0}
\setcounter{section}{0}
\setcounter{equation}{0}
\clearpage

 \renewcommand{\MeV}{\ensuremath{~\mathrm{MeV}}}
 \newcommand{\qloc}{\ensuremath{\mathrm{q}}}
 \newcommand{\qlocbar}{\ensuremath{\bar{\mathrm{q}}}}
 \renewcommand{\tbar}{\ensuremath{\bar{\mathrm{t}}}}
 \newcommand{\bbar}{\ensuremath{\bar{\mathrm{b}}}}
 \newcommand{\fb}{\ensuremath{~{\mathrm{fb}}}}
 \newcommand{\Wrm}{\ensuremath{{\mathrm{W}}}}

\part{Measuring Neutrino Mixing Angles At LHC \label{LH_BRPV}}
{\it W.~Porod and P.~Skands}
\maketitle
\begin{abstract}
We study an MSSM model with bilinear R-parity violation which is capable of
explaining neutrino data while leading to testable predictions for ratios of
LSP decay rates. Further, we estimate the precision with which such
measurements could be carried out at the LHC.
\end{abstract}

\section{INTRODUCTION}

Recent neutrino experiments 
\cite{Fukuda:1998mi,Fukuda:2001nj,Ahmad:2001an,Eguchi:2002dm}
clearly show that neutrinos are massive particles and that they mix.
In supersymmetric models these findings can be explained by the
usual seesaw mechanism 
\cite{Gell-Mann:1980vs,Yanagida:1980xy,Mohapatra:1980ia}. 
However, supersymmetry allows for an
alternative which is intrinsically supersymmetric, namely the breaking
of R-parity.
The simplest way to realize this idea
 is to add bilinear terms to the superpotential $W$:
\begin{eqnarray}
W = W_{\rm MSSM} + \epsilon_i \hat L_i \hat H_u
\label{eq:model}
\end{eqnarray}
For consistency one has also to add the corresponding 
bilinear terms to soft SUSY breaking which induce small vacuum expectation
values (vevs) for the sneutrinos. These vevs in turn induce a mixing between
neutrinos and neutralinos, giving mass to one neutrino
at tree level. The second neutrino
mass is induced by loop effects 
(see \cite{Romao:1999up,Hirsch:2000ef,Diaz:2003as} 
and references therein). The same parameters
that induce neutrino masses and mixings are also responsible for the
decay of the lightest supersymmetric particle (LSP). This implies that there
are correlations between neutrino physics and LSP decays 
\cite{Mukhopadhyaya:1998xj,Porod:2000hv,Hirsch:2002ys,Hirsch:2003fe}.    

In this note we investigate how well LHC can measure ratios of
LSP branching ratios that are correlated to 
 neutrino mixing angles in a scenario where
the lightest neutralino $\neut_1$ is the LSP.
In particular
we focus on the  semi-leptonic
final states $l_i \qloc' \qlocbar$ ($l_i=e,\mu,\tau$).
 There are several more
examples which are discussed in \cite{Porod:2000hv}.
In the model specified by Eq.~(\ref{eq:model})
the atmospheric mixing angle at tree level is given by
\begin{eqnarray}
 \tan \theta_{\rm atm} &=& \frac{\Lambda_2}{\Lambda_3} \\
\Lambda_i = \epsilon_i v_d + \mu v_i
\end{eqnarray}
where $v_i$ are the sneutrino vevs and $v_d$ is the vev of $H^0_d$. 
It turns out that the dominant part of the 
$\neut_1$-$\Wrm$-$l_i$ coupling $O^L_i$ is given by
\begin{eqnarray}
 O^L_i = \Lambda_i f(M_1,M_2,\mu,\tan \beta, v_d, v_u)
\end{eqnarray}
where the exact form of $f$ can be found in Eq.~(20) 
of ref.~\cite{Porod:2000hv}. The important point is that $f$ only depends on
MSSM parameters but not on the R-parity violating parameters. Putting 
everything together one finds:
\begin{eqnarray}
 \tan^2 \theta_{\rm atm} \simeq \left| \frac{\Lambda_2}{\Lambda_3} \right|^2
   \simeq \frac{BR(\neut_1 \to \mu^\pm \Wrm^\mp)}
               {BR(\neut_1 \to \tau^\pm \Wrm^\mp)}
   \simeq  
 \frac{BR(\neut_1 \to \mu^\pm \qlocbar \qloc')}
      {BR(\neut_1 \to \tau^\pm \qlocbar \qloc' )},
\label{eq:corr}
\end{eqnarray}
where the last equality is only approximate due to possible (small)
contributions from three body decays of intermediate sleptons and squarks.
The restriction to the hadronic final states of the $\Wrm$ is necessary
for the identification of the lepton flavour. Note that
Eq.~(\ref{eq:corr}) is a prediction of the bilinear model independent
of the R-parity conserving parameters.

\section{NUMERICAL RESULTS}

We take the SPS1a mSUGRA benchmark point \cite{Allanach:2002nj} as a
specific example, characterized by $m_0=100\GeV$, $m_\frac12=250\GeV$,
$A_0=-100\GeV$, $\tan\beta=10$, and $\mathrm{sign}(\mu)=1$\footnote{Strictly
  speaking, the SPS points should be defined by their low-energy parameters as
calculated with ISAJET 7.58.}. The
low--energy parameters were derived using \textsc{SPheno} 2.2
\cite{Porod:2003um} and passed to
\textsc{Pythia} 6.3 \cite{Sjostrand:2003wg} using the recently defined SUSY
Les Houches Accord \cite{Skands:2003cj}. The 
R-parity violating parameters (in \MeV) at the low scale are given by:
$\epsilon_1=43$, $\epsilon_2=100$, $\epsilon_3=10$, $v_1=-2.9$, $v_2=-6.7$
and $v_3=-0.5$. For the neutrino sector
we find $\Delta m^2_{\rm atm} = 3.8 \cdot 10^{-3}$~eV$^2$, 
$\tan^2 \theta_{\rm atm}= 0.91$, 
$\Delta m^2_{\rm sol} = 2.9 \cdot 10^{-5}$~eV$^2$,
$\tan^2 \theta_{\rm sol}= 0.31$. Moreover, we find that the following
neutralino branching ratios are larger than 1\%:\\
\begin{center}
\begin{tabular}{lll}
BR$(\Wrm^\pm \mu^\mp) = 2.2\%$, & BR$(\Wrm^\pm \tau^\mp) = 3.2\%$, &
BR$(\qlocbar \qloc' \mu^\mp) = 1.5\%$, 
 \\  BR$(\qlocbar \qloc' \tau^\mp) = 2.1\%$, &
BR$(\qloc \qlocbar \nu_i) = 4.7\%$, & BR$(\b \bbar \nu_i) = 15.6\%$, \\
BR$(e^\pm \tau^\mp \nu_i) = 5.9\%$, &BR$(\mu^\pm \tau^\mp \nu_i) = 30.3\%$, 
 &BR$(\tau^+ \tau^- \nu_i) = 37.3\%$, \\
\end{tabular}
\end{center}
where we have summed over the neutrino final states as well as over
the first two generations of quarks. Moreover, there are 0.2\% of neutralinos
decaying invisibly into three neutrinos. In the case that such events 
can be identified they can be used to distinguish this model from a model
with trilinear R-parity violating couplings because in the latter case they
are absent.

We now turn to the question to what extent
the ratio, Eq.~(\ref{eq:corr}), could be measurable at an LHC experiment. The
intention here is merely to illustrate the phenomenology and to give a rough idea of
the possibilities. For simplicity, we employ a number of shortcuts;
e.g.~detector energy resolution effects are ignored and events are only
generated at the parton level. Thus, we label a final-state quark or gluon
which has $p_\perp>15\GeV$ and which lies within the fiducial volume of the
calorimeter, $|\eta|<4.9$, simply as `a jet'. Charged leptons 
are required to lie within the inner detector coverage,
$|\eta|<2.5$, and to have $p_\perp>5\GeV$ (electrons), $p_\perp>6\GeV$
(muons), or $p_\perp>20\GeV$ (taus). 
The assumed efficiencies for such leptons are \cite{atlasTDR}
75\% for electrons, 95\% for
muons, and 85\% for taus decaying in the 3--prong modes (we do not use the
1--prong decays), independent of $p_\perp$. 

For SPS1a, the total SUSY cross section is 
$\sigma_{\mathrm{SUSY}}\sim 41~\mathrm{pb}$. This consists mainly of gluino and squark
pair production followed by subsequent cascades down to the LSP, the
$\neut_1$. With an integrated luminosity of $100\fb^{-1}$, approximately 8 million $\neut_1$ decays should thus 
have occurred in the detector. 

An important feature of the scenario considered here is that the  
$\neut_1$ width is sufficiently small to result in a potentially observable
displaced vertex. By comparing the decay length, $c\tau =0.5$~mm, 
with an estimated vertex resolution of about 20 microns in the transverse
plane and 0.5 mm along the beam axis, it is apparent that the two neutralino
decay
vertices should exhibit observable displacements in a fair
fraction of events. Specifically, we require that both neutralino
decays should occur outside an ellipsoid defined by 5 times the
resolution. For at least one of the vertices (the `signal' vertex),
all three decay products ($\mu\qloc\qlocbar'$ or $\tau\qloc\qlocbar'$) must be
reconstructed,  while we only require one
reconstructed decay product (jet in the inner detector or lepton in the inner
detector whose track does not intersect the 5$\sigma$ vertex resolution
ellipsoid) for the second vertex (the `tag' vertex).

Naturally, since the decay occurs within the detector, the standard SUSY
missing $E_\perp$ triggers are ineffective. Avoiding a discussion of detailed
trigger menus (cf.~\cite{tridas}), we have approached the issue by
requiring that each event contains either four jets, each with $p_\perp >
100\GeV$, or two jets with $p_\perp>100\GeV$ together with a
lepton (here meaning muon or electron) with $p_\perp>20\GeV$, or one jet with
$p_\perp>100\GeV$ together with two leptons with $p_\perp>20\GeV$.
Further,
since the Standard Model background will presumably be dominated by $\t\tbar$
events, we impose an additional parton--level $\b$ jet veto. 

To estimate the efficiency with which decays into each channel can be
reconstructed, a sample of 7.9 million SUSY events were generated with
\textsc{Pythia}, and the above trigger and reconstruction cuts were
imposed. To be conservative, we only include the resonant decay channels,
where the quark pair at the signal vertex has the invariant mass of the $\W$.
The number of generated decays into each channel, the fractions
remaining after cuts, and the expected total number of
reconstructed events scaled to an integrated luminosity of
$100\fb^{-1}$ are given in table \ref{tab:reconstruction}.
\begin{table}[tp]
\begin{center}
\begin{tabular}{lrr|r}
mode & $N_{gen}$ & $\epsilon_{\mathrm{rec}}$ & $N_{\mathrm{rec}}(100\fb^{-1})$\\ 
\toprule
$\neut_1\to \mu \W \to \mu\qloc\qlocbar'$ & 235000 & 0.10 & 12500  \\
$\neut_1\to \tau \W\to \tau_{\mathrm{3-prong}}\qloc\qlocbar'$ & 51600 & 0.054 &
1400 
\\ \bottomrule 
\end{tabular}
\caption{Statistical sample, estimated reconstruction efficiencies, and expected event numbers. 
\label{tab:reconstruction}}
\end{center}
\end{table}
The comparatively small efficiencies owe mainly to the
requirement that \emph{both} neutralino decays should pass the 5$\sigma$
vertex resolution cut. Nonetheless, using these numbers as a first estimate, 
the expected statistical accuracy of the ratio, $R={BR(\neut_1 \to
   \mu^\pm \W^\mp)} / {BR(\neut_1 \to \tau^\pm \W^\mp)}$, appearing in
  Eq.~(\ref{eq:corr}) becomes $\frac{\sigma(R)}{R} \simeq 0.028$.

\section{CONCLUSIONS}

We have studied neutralino decays in a model where bilinear R-parity violating
terms are added to the usual MSSM Lagrangian. This model can successfully
explain neutrino data and leads at the same time to {\it predictions}
for ratios of the LSP decay branching ratios. In particular we have considered
a scenario where the lightest neutralino is the LSP. In this case the  ratio 
${BR(\neut_1 \to \mu^\pm \W^\mp)}/{BR(\neut_1 \to \tau^\pm \W^\mp)}$ is
directly related to the  atmospheric neutrino mixing angle. 
Provided R-parity violating
 SUSY is discovered, the measurement of this ratio at colliders
would thus constitute an important test of the hypothesis of
a supersymmetric origin of neutrino masses. 

We have investigated the possibility of performing this measurement at
a `generic' LHC experiment, using \textsc{Pythia} to generate LHC SUSY
events at the parton level and imposing semi-realistic acceptance and
reconstruction cuts. Within this simplified framework, we find that
the LHC should be sensitive to a possible connection between 
 R-parity violating LSP decays  and
the atmospheric mixing angle, at least for scenarios with a fairly
light sparticle spectrum and where the neutralino decay length is
sufficiently large to give observable displaced vertices. Obviously,
the numbers presented here represent crude estimates and should not be
taken too literally. A more refined experimental analysis would be
necessary for more definitive conclusions to be drawn.

\vskip1cm
\noindent

\section*{ACKNOWLEDGEMENTS}

We thank A.~De Roeck and T.~Sj\"ostrand for useful discussions.
We are also grateful to the NorduGRID project for use of computing resources.
W.P.~is supported by the `Erwin Schr\"odinger fellowship No.~J2272'
of the `Fonds zur F\"orderung der wissenschaft\-lichen Forschung' of
Austria and partly by the Swiss `Nationalfonds'.

\setcounter{figure}{0}
\setcounter{table}{0}
\setcounter{section}{0}
\setcounter{equation}{0}
\clearpage

\newcommand{\mett}{{\not\!\!E}_{T}}
 \part{
Resonant slepton production at the LHC in models with an
ultralight gravitino \label{BMS_les}}

{\it B.C. Allanach, M. Guchait, K. Sridhar}

\maketitle

\begin{abstract}
We examine resonant slepton production at the LHC with 
gravitinos in the
final state. We investigate two cases: (i) where the slepton 
undergoes gauge
decay into neutralino and a lepton, followed by the neutralino 
decay into
a photon and a gravitino, and (ii) direct decays of a slepton 
into a lepton
and a gravitino. We show how to
accurately reconstruct both the slepton and neutralino masses in
the first case, and the slepton mass in the second case
for 300 fb$^{-1}$ of integrated luminosity at the LHC.
\end{abstract}

\section{INTRODUCTION}
This letter is devoted to the study of the signals at the Large Hadron 
Collider (LHC) due to a supersymmetric generalisation of the Standard 
Model (SM) which (a) violates $R$-parity, and (b) has an ultra-light 
gravitino in its spectrum. The anomalous events in the CDF experiment 
in the production rate of lepton-photon-$\mett$ in $p{\bar p}$ collisions
were explained~\cite{Allanach:2002ph,Allanach:2001hx,Allanach:2001mv} 
in the framework of a 
$R$-parity violating supersymmetric model with dominant 
$L$-violating $\lambda'_{211}$ coupling,
and an ultra-light gravitino of mass $\sim10^{-3}$ eV.

The resonant production of a smuon via the $R$-violating coupling,
its decay into neutralino and a muon and, finally, the decay of the
neutralino into a gravitino and a photon 
leads to the $\mu \gamma \mett$ final state studied in the CDF experiment.
The range of smuon and neutralino masses rel<evant to the explanation of 
these 
anomalous observations of the the CDF experiment is such that most of
this range will be explored at the Run II of the Tevatron. 
In the event
that this signal is not seen at Run II it will rule out the model at
the lower end of the neutralino and smuon masses. 
For heavier
smuon and neutralino masses (above 250 GeV, roughly), the aforementioned
Run I signal would be a statistical fluke and will probably disappear 
in Run II data. In that case, experiments at the LHC can be expected to 
discover and measure the sparticles. Here, we perform a 
study of the ability of the LHC to perform these two tasks, identifying 
the sensitive observables.

\section{THE MODEL}
We assume a single dominant $R$ violating coupling, $\lambda'_{211}$
for example. If the $R$-violating coupling is small, the existence 
of an ultralight gravitino in the mass range of $10^{-3}$~eV drastically 
alters the decay mode of the slepton. The slepton overwhelmingly decays
into a lepton and a (bino-dominated) neutralino, with the latter decaying 
into a photon and a gravitino resulting in a $l\gamma\mett$ final-state. 
The Feynman diagram for the process is shown in Fig.~\ref{feyn}. 
We should also expect signals from sneutrino production.
The background to the $\gamma\mett$ final-state that this would give rise
to depend crucially on cosmic ray events
which are difficult to estimate. Therefore we have not studied the signal
from sneutrino production. All sparticles except the neutralino,
gravitino and slepton are set to be arbitrarily heavy in our analysis.
\begin{figure}
\begin{center}
\begin{picture}(400,170)
\ArrowLine(60,10)(120,50)
\ArrowLine(120,50)(60,90)
\DashLine(120,50)(180,50){5}
\ArrowLine(180,50)(240,10)
\ArrowLine(240,90)(180,50)
\ArrowLine(300,60)(240,90)
\Photon(240,90)(300,130){5}{4}
\put(65,5){$q$}
\put(65,90){$\bar q'$}
\put(245,10){$l$}
\put(210,80){$\chi_0$}
\put(150,65){$\tilde l$}
\put(305,55){$\tilde G$}
\put(305,125){$\gamma$}
\end{picture}
\caption{Feynman diagram of resonant slepton production followed by 
neutralino decay.}
\label{feyn}
\end{center}
\end{figure}
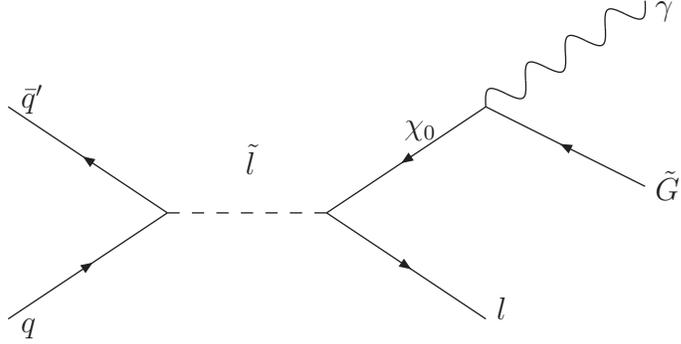

\section{SIMULATION RESULTS}
For our study of the process shown in Fig.~\ref{feyn} at the LHC ($pp$ 
collisions at $\sqrt s = 14$ TeV), we have chosen to work with the
following default set of model parameters (unless indicated otherwise):
\begin{itemize}
\item
Gravitino mass, $m_{\tilde G}=10^{-3}$~eV,
\item 
$R$-violating coupling $\lambda'\equiv\lambda'_{211}=0.01$,
\item
${\rm tan}\beta = 10$,
\item
sparticle masses $(m_{\chi_1^0}, m_{\tilde l})$=(120~GeV,200~GeV) or 
(200~GeV,500~GeV)
GeV (``low mass'' and ``high mass'' scenarios) respectively.
\end{itemize}
The choice of using $\lambda'_{211}$ rather than some other flavour
combination is arbitrary and can be easily generalised to other
$R$-violating couplings. We have checked that the chosen value for
$\lambda'_{211}$ is quite consistent with the existing bound~
\cite{Barger:1989rk,Allanach:1999ic}.
By selecting rather low values for $R$-parity
coupling and gravitino mass,we avoid
significant rates for the possible $R$-violating decays of $\chi_1^0 
\rightarrow \mu jj$ or $\chi_1^0 \rightarrow \nu jj$. $\chi_1^- \rightarrow
\gamma {\tilde G}$ is the dominant channel.
We use the {\small \tt ISASUSY}~\cite{Paige:1998xm} to generate the
SUSY spectrum, branching ratios and decays of the sparticles selecting
a representative point $\tan\beta$=10, $A_{t,\tau,b}=0$ along with
large values of  $\mu$ and other flavour diagonal soft 
supersymmetry breaking parameters.

The signals 
have been simulated using {\small \tt HERWIG6.4}~\cite{Corcella:2001pi} and
the $W \gamma$ SM background has been simulated
using {\small \tt PYTHIA}~\cite{Sjostrand:2003wg}.
In our simulations, both the signal and background, we have used
only selection cuts of $E_T, \mett > 25$ GeV on the
transverse energies of the muon, the photon and missing energy.
We have used the following cuts
on the rapidity of the photon and the muon: $|\eta_{\gamma,\mu}|<3$.
There is an isolation cut between the photon and other hard objects
$o$ in the
event of $\sqrt{(\eta_\gamma - \eta_o)^2 + (\phi_\gamma - \phi_o)^2}>0.7$.
Since the signal is hadronically quiet, we veto events with jets
reconstructed with $E_T>30$ GeV and $\eta_j<4$. Initial and final state
radiation effects, as well as fragmentation effects are included in the
background simulation.

\begin{figure}
\begin{center}
\unitlength=1in
\begin{picture}(5.4,2.7)
\put(2.7,0){\epsfig{file=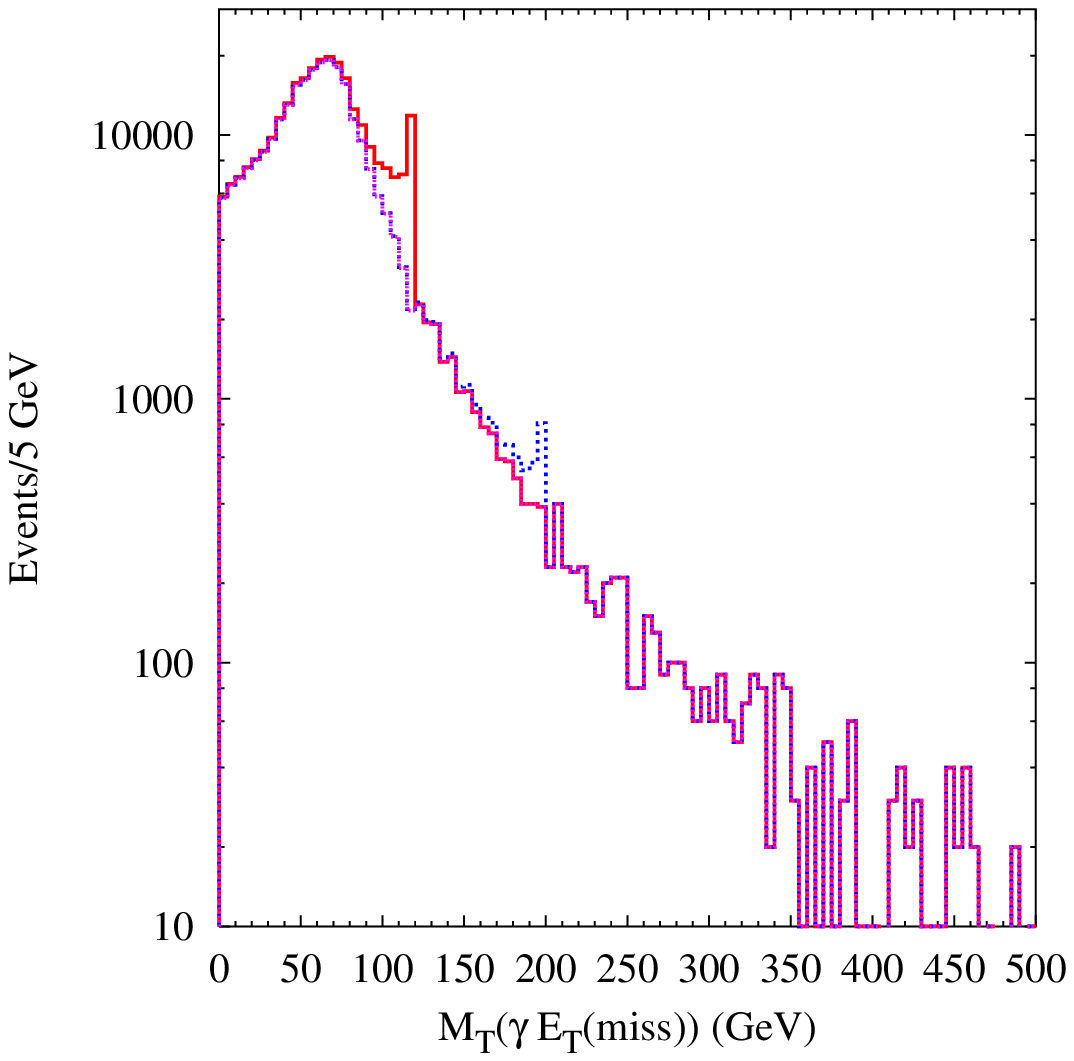, width=2.7in}}
\put(0,0){\epsfig{file=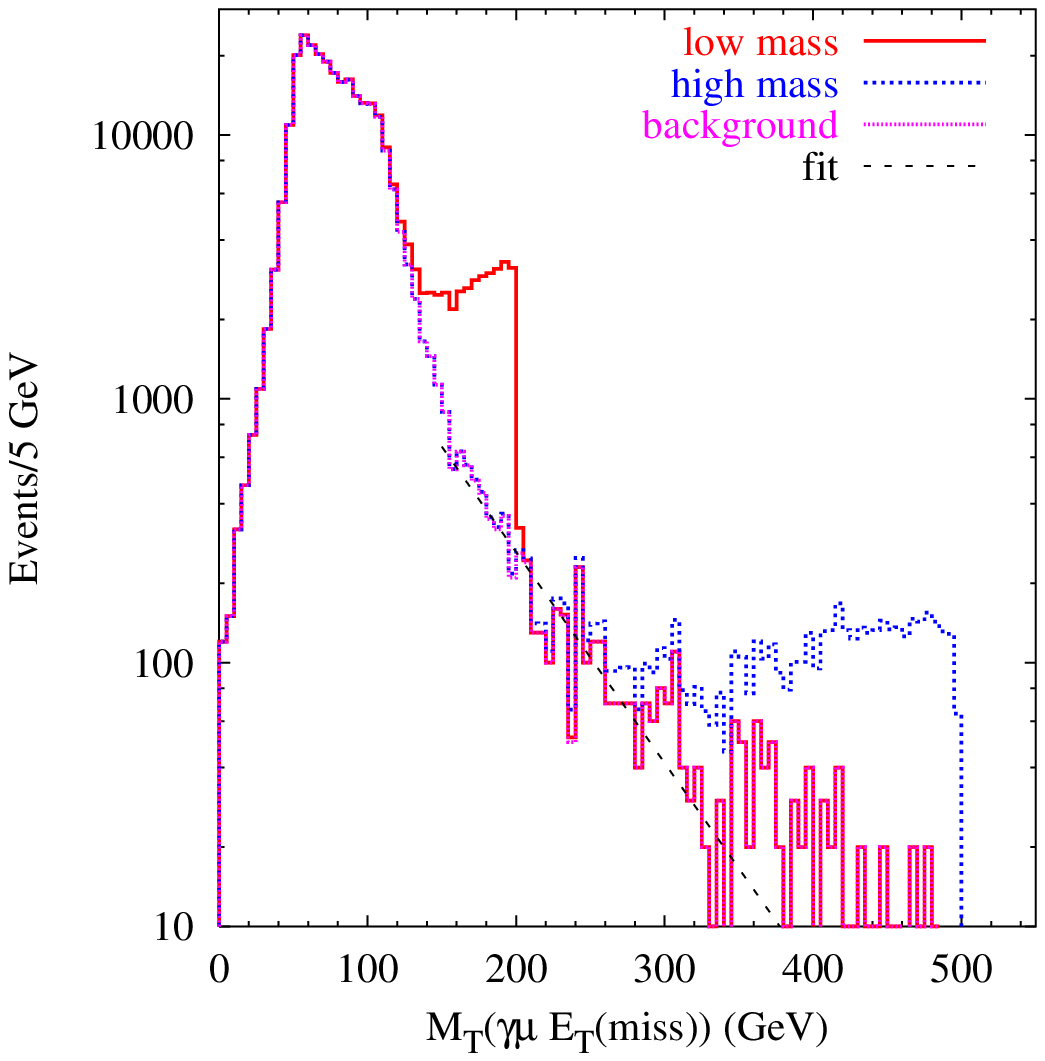, width=2.7in}}
\put(4.05,1.3){\epsfig{file=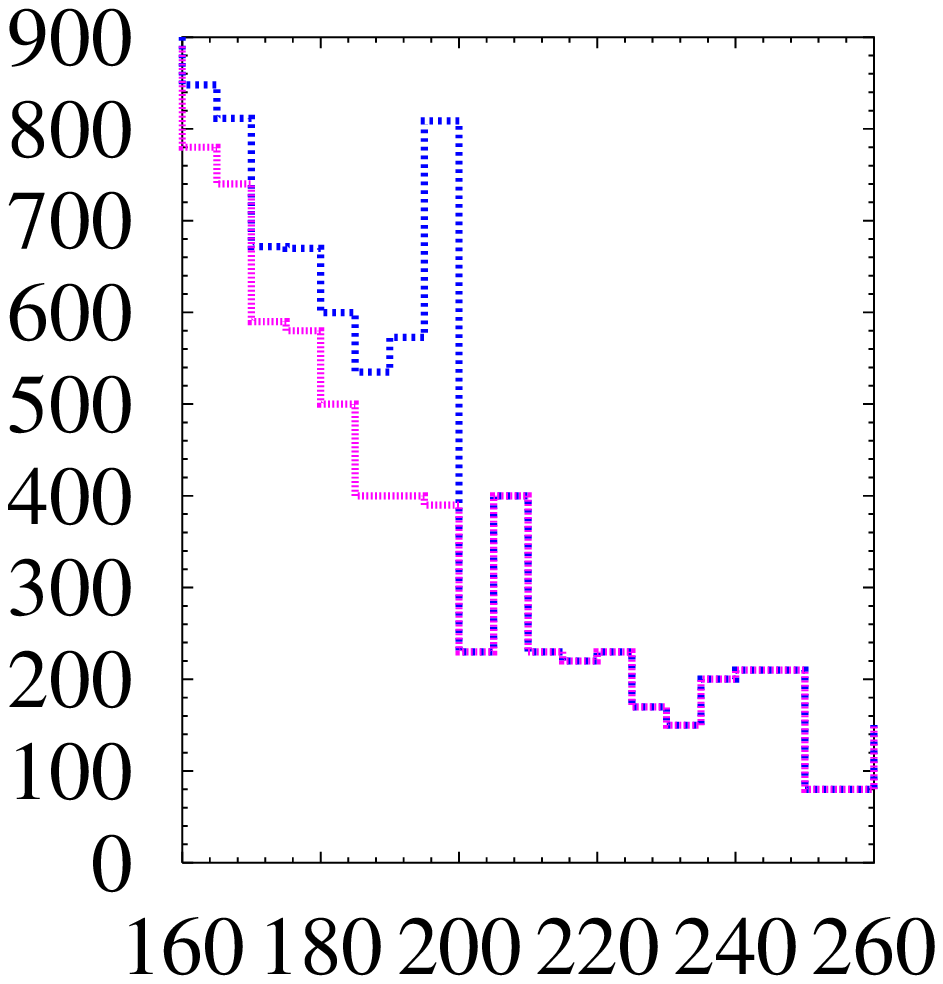, width=1.35in}}
\put(0,2.4){(a)}
\put(2.7,2.4){(b)}
\end{picture}
 \caption{$M_T$ distributions of (a) $\mu \gamma \mett$, (b) 
  $\gamma \mett$ for slepton.
  300 fb$^{-1}$ integrated luminosity  at the LHC is assumed.   
The purple (lighter) histograms display $W \gamma$ SM background, 
the red
  (darker) 
   histograms show signal plus background for $(m_{\chi_1^0}, m_{\tilde
  l})$=(120~GeV,200~GeV),  
 whereas the blue (dotted) histograms display the signal plus 
background distributions for
$(m_{\chi_1^0}, m_{\tilde l})$=(200~GeV,500~GeV).
In (a), the dashed black line displays a log-linear fit to the background
  distribution for $M_T=150-400$ GeV. In (b), the insert shows a linear 
scale
  magnification of 
  an area of the plot.}
\label{mtdist}
\end{center}
\end{figure}
The transverse mass distributions of
final state particles along with $\mett$ show a clear distinction between
signal and backgrounds. The $M_T(\mu \gamma \mett)$ and $M_T(\gamma \mett)$
distributions are displayed in Figs.~\ref{mtdist}a,b
for the simulated high and low
mass points and $W \gamma$ simulated SM background.
In Fig.~\ref{mtdist}a,
sharp peaks which are expected in the $M_T(\mu \gamma \mett)$ distributions
are clearly visible at values of the smuon mass and
will be detected above the SM $W\gamma$ background, leading to the
accurate measurement of the mas.
Fig.~\ref{mtdist}b shows that the signal peaks in $M_T(\gamma \mett)$
(predicted to be at the neutralino mass) should be able to
provide a measurement of the neutralino mass.

\begin{figure}
\begin{center}
\includegraphics[width=0.5\textwidth]{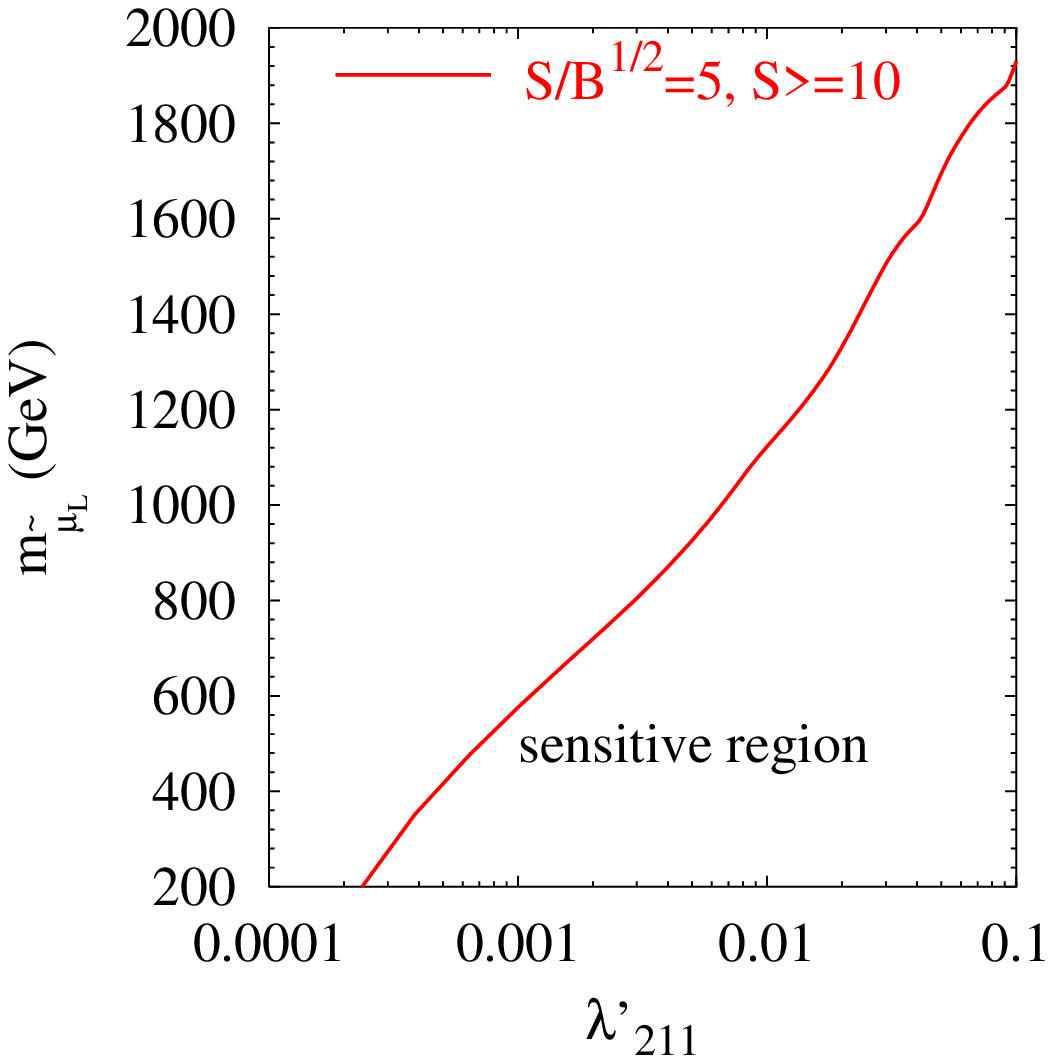}
 \caption{Search reach for the $\mu \gamma \mett$ signal 
(as defined in the
   text) for
   300 fb$^{-1}$ integrated luminosity  at the LHC. 
}
\label{search}
\end{center}
\end{figure}
In order to calculate the search reach, we use the signal $S$ in the
4 highest peak bins (covering 20 GeV) of the signal $M_T(\mu \gamma \mett)$
peak. The 
background distribution in these four bins is estimated by fitting a simple
function
B = 4 $\mbox{exp} \left[a M_T(\mu \gamma \mett) + b\right]$
to $M_T(\gamma \mu \mett)$ between 150-400 GeV in
Fig.~\ref{mtdist}a.
Using purely $\sqrt{B}$ statistical errors, we obtain
to $a=-0.018\pm0.001$, $b=9.25\pm0.22$.
$B$ is displayed in Fig.~\ref{mtdist}a
as the dashed black line.
We show the region of parameter space corresponding to
\footnote{The 
  statistical uncertainties
  on fitted $a$ and $b$ parameters make a negligible difference to the
final numerical results.} $S/\sqrt{B}>5$
and $S\geq10$
for 300 fb$^{-1}$ luminosity option, as a function of smuon mass
and R-parity conserving coupling in Fig.~\ref{search}a.

We now turn to the decay ${\tilde l} \rightarrow {\tilde G} l$. We ignore
sneutrino production in this case because it would lead to an invisible
final state. We have calculated the production matrix element and
the branching ratio and implemented in a parton-level Monte Carlo.
We have used cuts in our analysis on the muon and missing transverse
energy identical
to the $\gamma \mu \mett$ analysis, i.e. 
$\mett,E_T^\mu>25$ GeV and $|\eta_\mu|<3$.
\begin{figure}
\begin{center}
\includegraphics[width=0.5\textwidth]{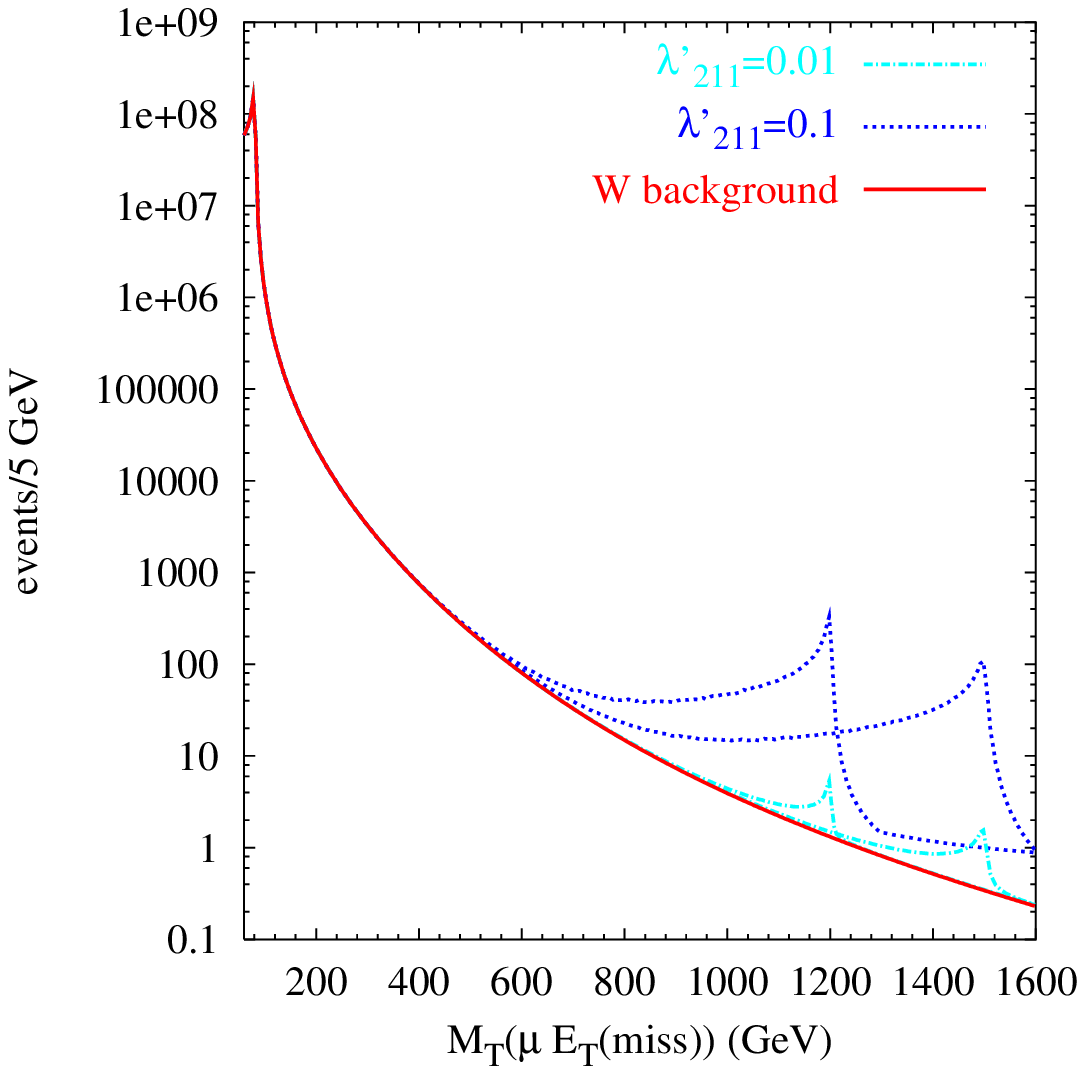}\includegraphics[width=0.5\textwidth]{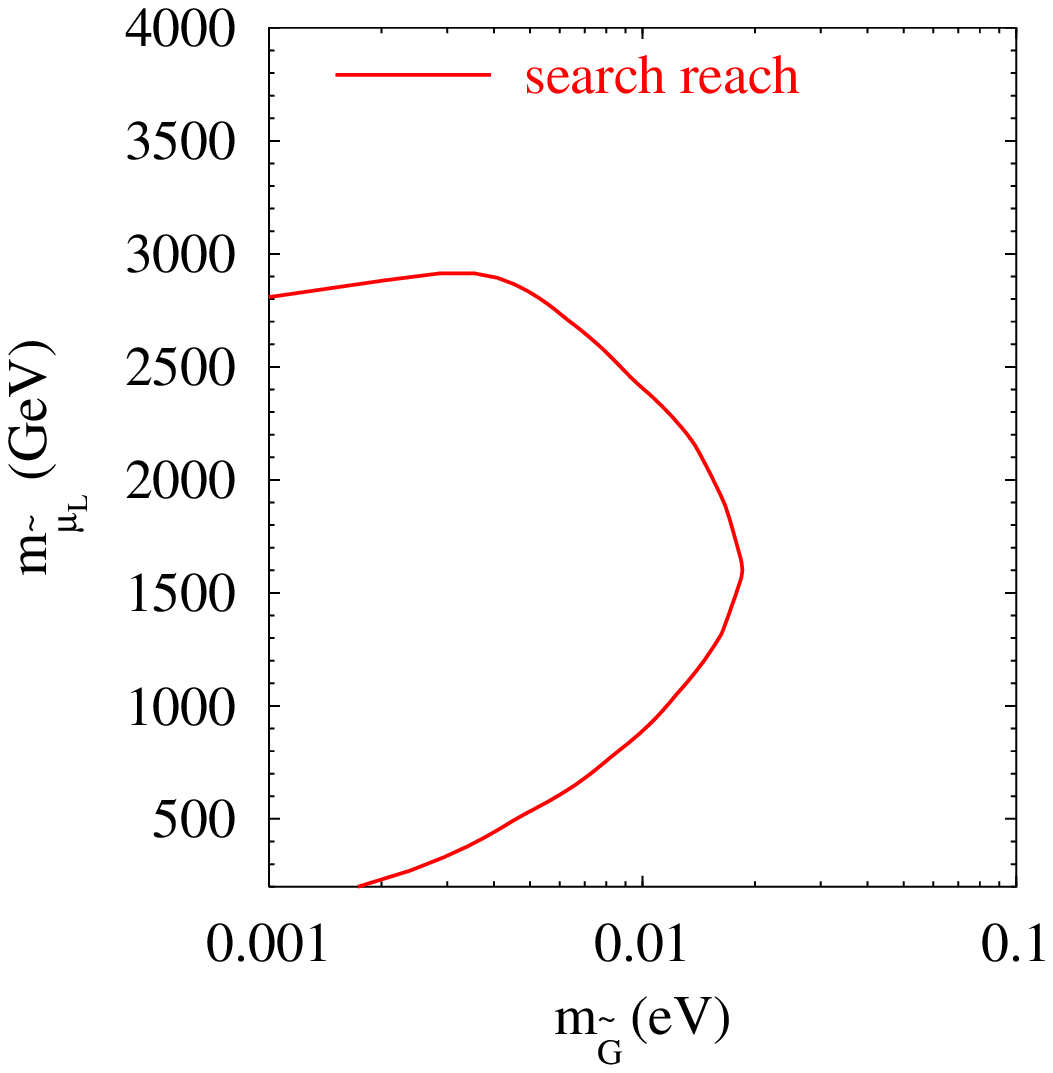}
 \caption{
(a) $M_T$ distribution of the $\mu {\tilde G}$ final state for two different
   values of the smuon mass and 300 fb$^{-1}$ integrated luminosity at the
   LHC. The left(right)-most three peaks are for smuon masses of 1.2 TeV and
   1.5 TeV respectively.
  (b) Search reach (as defined in the text) for the $\mu \tilde G$ final
  state
  and 300 fb$^{-1}$ integrated
  luminosity  at the LHC and $\lambda'_{211}=0.1$. 
  The search reach is contained to the left of the
  curve.
}
\label{new}
\end{center}
\end{figure}
Fig.~\ref{new}a displays the $M_T(\mu \mett)$ distribution for the $W$
background plus signal in the cases $m_{\tilde \mu}=1.2$ and 1.5 TeV
respectively and two different values of the $\lambda'$. For $\lambda'=0.01$
there are not enough signal events to be seen, but for higher values (eg
0.1), a clear mass peak should be seen in the tail of the $M_T$ distribution
of the $W$.
We define the search reach by the criteria that for $S$
signal and $B$
background events, $S/\sqrt{B}\geq 5$ and $S\geq 10$ in the 5 GeV
signal peak bin of $M_T(\mu \mett)$.
It is displayed for 
300  fb$^{-1}$ of integrated luminosity in Fig.~\ref{new}b.

\section*{CONCLUSIONS}
Resonant slepton production and its decays into $l \gamma {\tilde G}$ or $l
{\tilde G}$ can be discovered at the LHC for slepton masses into the multi-TeV
region, depending upon the $R_p$ violating coupling and provided that the gravitino is
ultra-light (with a mass less than 0.1 eV). Various $M_T$ distributions will allow the
accurate measurement of sparticle masses involved. For full details, the new
matrix element and additional
results (for example including the case of a smuon NLSP), see
ref.~\cite{Allanach:2003wz}.

\section*{ACKNOWLEDGEMENTS}
K Sridhar would like to thank CERN and LAPTH for hospitality offered 
during which some of the work contained herein was performed.

\setcounter{figure}{0}
\setcounter{table}{0}
\setcounter{section}{0}
\setcounter{equation}{0}
\clearpage


 \part{Radion Mixing Effects In The Two-Higgs-Doublet Model \label{hewett}}
{\it J.L. Hewett and T.G. Rizzo}
\begin{abstract}
We begin an examination of the effects of mixing between the radion of the 
Randall-Sundrum (RS) model and the Higgs fields of the Two-Higgs-Doublet 
model as would be motivated by, $e.g.$, supersymmetry. Preliminary results 
for the shifts in various particle masses and couplings are obtained. 
\end{abstract}

\section{INTRODUCTION}

The RS model{\cite {Randall:1999ee}} provides an interesting solution to the 
hierarchy problem which can be tested 
experimentally{\cite{Davoudiasl:1999jd,Davoudiasl:1999tf,Davoudiasl:2000wi}} 
at future colliders. One 
prediction of this model is the existence of a relatively light scalar radion 
which can mix with other scalars such as the Higgs boson of the Standard 
Model (SM). Such mixing can lead to substantial modifications in the expected 
properties of both the Higgs and the radion and has been extensively 
studied{\cite{Giudice:2000av,Csaki:2000zn,Hewett:2002nk,Dominici:2002jv}} 
in the literature. Here we extend this study to the case of two Higgs 
doublets as would be expected in a number of scenarios, $e.g.$, 
supersymmetry (SUSY). 
Although not necessary for solving the hierarchy problem within the RS 
scenario, SUSY may have other model 
building{\cite {Altendorfer:2000rr, Gherghetta:2000qt}} 
uses such as coupling constant 
unification or radius stabilisation{\cite {Goldberger:1999uk}}. 
The expectation from previous analyses of the single doublet model is that 
the properties of the mass eigenstate CP even neutral 
fields would substantially differ from those predicted in either the SM 
or the Minimal Supersymmetric Standard Model (MSSM). 
The preliminary discussion of our findings given here supports 
these expectations though further study is required to understand the 
breath of the possible modifications. 

\section{ANALYSIS}

With two Higgs doublets, mixing arises from the TeV brane action
\begin{equation}
S_{mix}=\int_{TeV} d^4x {\sqrt {det~g}}~R(g)~(\xi_1H_1^\dagger H_1
+\xi_2H_2^\dagger H_2+\xi_{12}H_1^\dagger H_2+h.c.)\,,
\label{actionmix}
\end{equation}
where $g$ symbolises the induced metric on the TeV brane, $R$, the induced 
curvature arising from $g$, 
$H_i$ are the two Higgs doublets and $\xi_i$ are dimensionless, order one 
parameters which we take to be real, 
thus assuming CP conservation for simplicity. The possible complexity of 
$\xi_{12}$ may lead to interesting phenomenology.  
(In what follows $\xi_1=\xi_2$ will be assumed since it is unlikely that 
gravitational interactions distinguish between these two Higgs doublets. 
This assumption, however, may be incorrect.) 
Thus, in the unitary gauge, 
the CP-odd field, $A$, as well as the charged scalars $H^\pm$ are not  
directly affected by modifications to the neutral CP even sector. (Of course, 
their couplings to the CP-even fields will be modified.) 
Before mixing with the radion, $r_0$,  
we denote the usual two CP-even Higgs by $h_0,H_0$ which have been 
obtained from the weak interaction eigenstate fields via rotations by 
the angles $\alpha,\beta$ as usual. Assuming $r_0$ obtains a mass from the 
stabilisation procedure, the above action can be expanded to quadratic order 
in the CP-even neutral 
fields from which one obtains the following effective Lagrangian 
generalising the notation of Csaki $etal.${\cite {Csaki:2000zn}}: 
\begin{equation}
{\cal L}=-{1\over {2}}h_0\partial^2 h_0-{1\over {2}}m_{h_0}^2h_0^2~+(h_0 
\rightarrow H_0,r_0) -3\sigma \gamma^2 r_0\partial^2 r_0+6\gamma(\tau_h h_0
\partial^2 r_0 +h_0 \rightarrow H_0)\,,
\label{efflang}
\end{equation}
where $\gamma=v/\sqrt{6} \Lambda_\pi \sim 0.05-0.10$, with $v$ the SM vev, and 
$\Lambda_\pi$, of 
order a few TeV or so, being the `TeV scale' of the RS model. The parameters 
$\sigma, \tau_{h,H}$ are functions of the $\xi_i$ and the usual mixing 
angles $\alpha,\beta$. The kinetic mixing in the above Lagrangian can be 
removed by a set of field redefinitions, $i.e.$, 
$h_0,H_0\rightarrow h',H'+6\gamma \tau_{h,H}r'/Z$ and $r_0\rightarrow r'/Z$. 
To obtain the mass eigenstate 
basis after the above redefinitions are employed 
a further orthogonal transformation, $(h',H'r')={\cal O}(h,H,r)$, must 
be performed. The elements of the orthogonal matrix, $\cal O$, as well as the 
corresponding mass eigenvalues can then be determined analytically. 

\begin{figure}[htbp]
\centerline{
\includegraphics[width=5.4cm,angle=90]{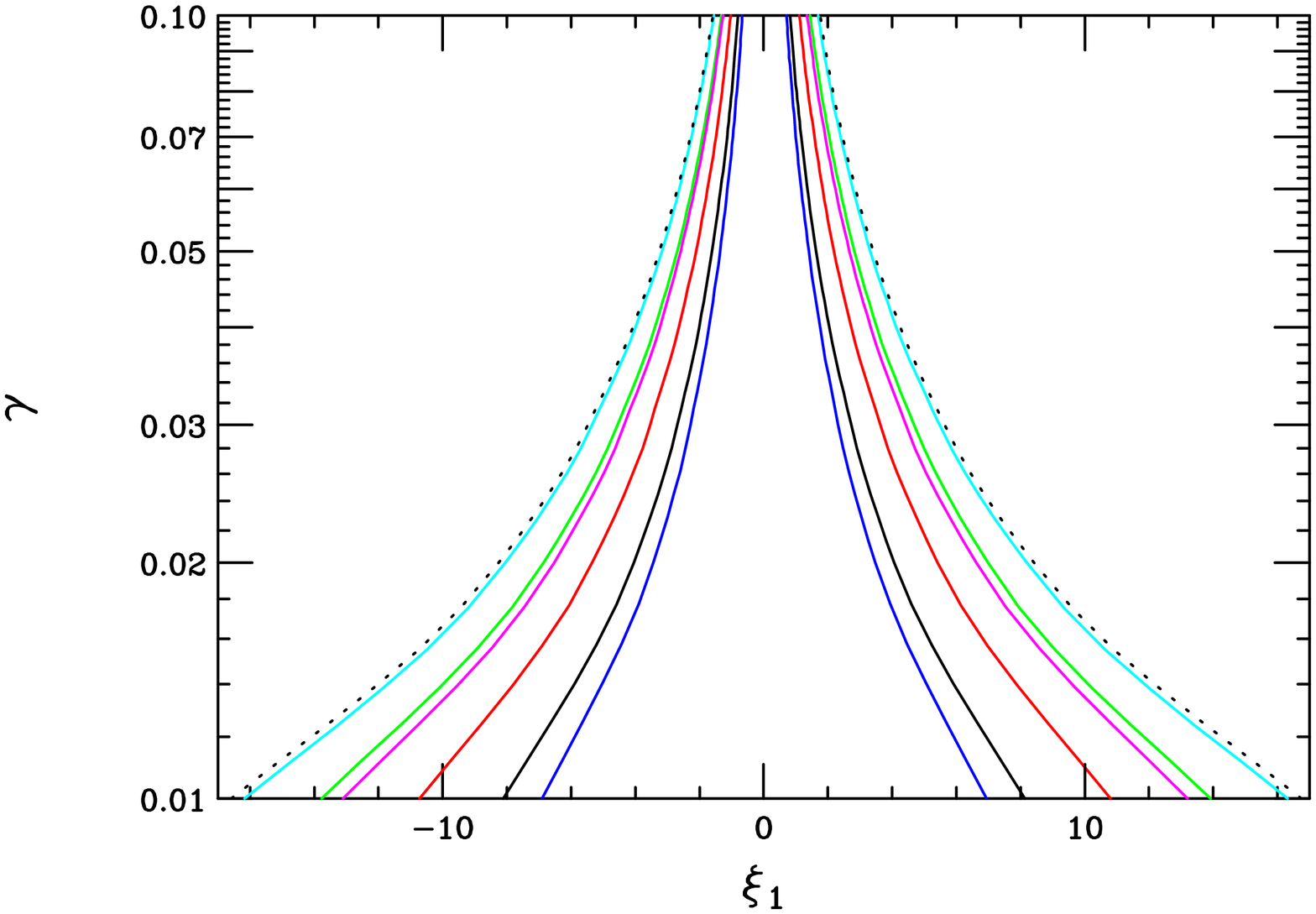}
\hspace*{1mm}
\includegraphics[width=5.4cm,angle=90]{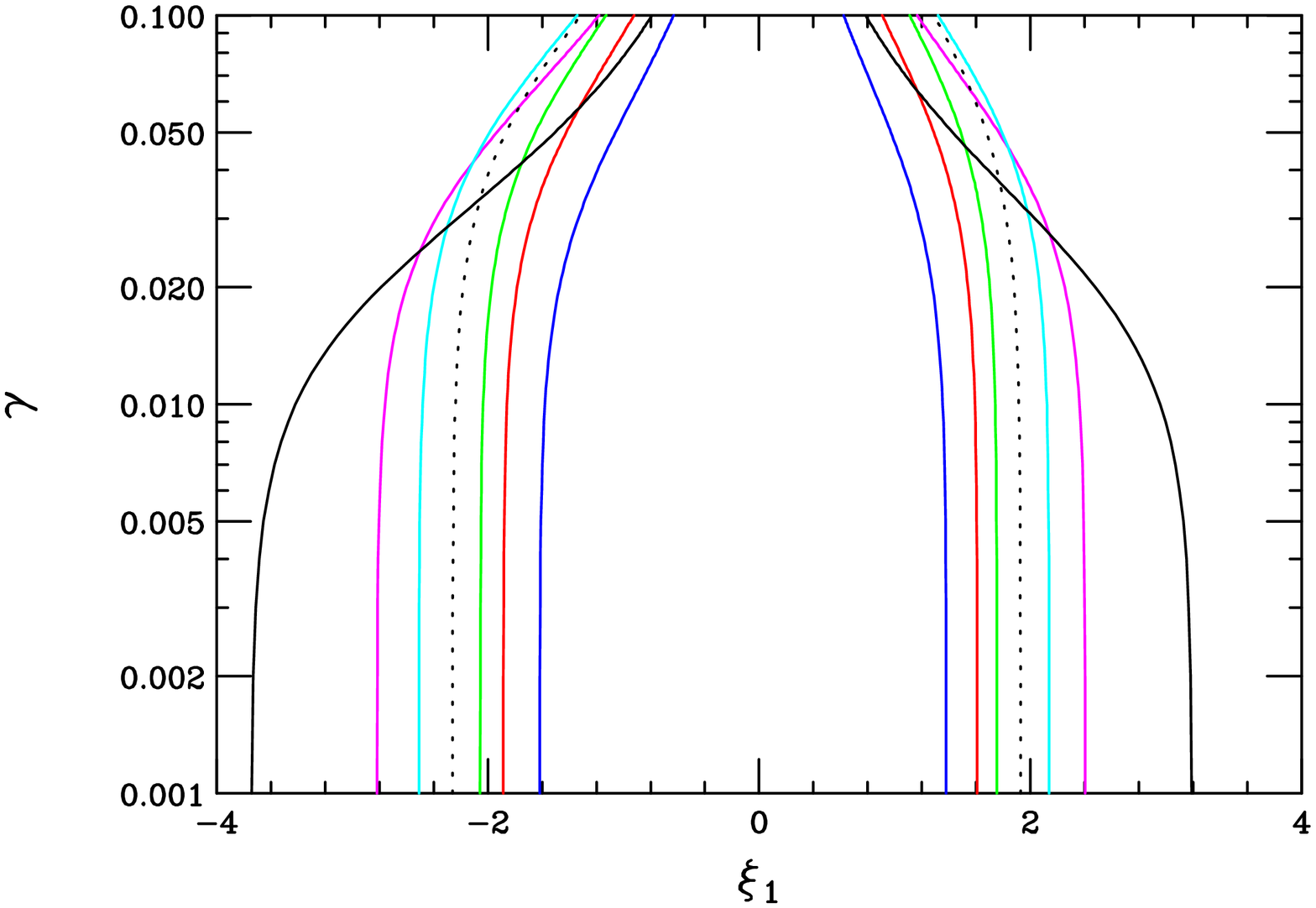}}
\vspace*{0.1cm}
\caption{Constraints from the absence of ghosts/tachyons(left) and 
perturbative unitarity(right) for different values of the ratio $R$ in the 
range between -2 and +2. The allowed region lies between the curves.}
\label{constraints}
\end{figure}

The number of parameters in this model is unfortunately rather large 
making it difficult to analyse in all generality. In order 
to show a specific example with somewhat fewer parameters we assume that 
the spectra, couplings, $etc.$, of the unmixed 
two-doublet-model Higgs sector to be that 
given by SUSY, including the effects of radiative 
corrections{\cite{Carena:2001bg}}, with $\tan \beta=10$, $M_A=500$ GeV, 
$A_t=A_b=-\mu$=1 TeV and $M_S^2=(M_{t_1}^2+M_{t_2}^2)/2=1$ TeV$^2$ where the 
$M_{t_i}$ are the stop masses. (The remaining parameters are $\xi_1, m_{r_0}, 
\gamma$ and the ratio $R=\xi_{12}/\xi_1$.) These parameter 
choices yield $m_h \simeq 125$ GeV. The resulting parameter space can be 
further restricted by noting that both $\xi_1$ and $R$ are of order unity, 
$\gamma$ is anticipated to be near the range described 
above and $m_{r_0}$ is expected to be of order the weak 
scale. Further numerical restrictions on the parameter ranges 
can be obtained by demanding that, 
$e.g.$, there are no ghosts or tachyons in the spectrum arising from the 
diagonalisation process and that $W_L^+W_L^-$ 
scattering satisfies perturbative unitarity{\cite {Han:2001xs}} 
up to the scale $\Lambda_\pi$; 
samples of such constraints can be seen in the figure above where we have 
assumed that $-2 \leq R \leq 2$. Note 
that for small values of $\gamma$ the constraints from unitarity are more 
restrictive than the requirement that no ghosts or tachyons be present; for 
larger $\gamma$ both constraints are found to be of comparable strength.

\section{PRELIMINARY RESULTS}

We now turn to a brief sample of our results which survey only a Small region 
of the allowed parameter space. 
The first thing to examine is the effect of mixing on the masses of the 
physical states $h,H,r$ as is shown in the next set of figures. (The curves 
are cut off at large values of $|\xi_1|$ 
by the no ghost/tachyon requirement.) Here we see the typical result that 
the mass `levels' of the three states repel each other due to mixing. 
In all of our examples, $m_{H_0}$ is the largest mass parameter and thus the 
mass of the $H$ is itself 
raised by mixing. Note that these upward shifts can be 
enormous near the parameter space boundaries. Though made heavier, $H$'s 
couplings will be seen to grow as well. When $m_{h_0}<(>)m_{r_0}$ we 
see that the $h(r)$ mass is pushed downward while that of $r(h)$ is pushed 
upward. These shifts are not as large as those experienced by the $H$. 
Thus while we might expect the light Higgs to have a mass $\leq 130$ GeV in 
the MSSM, mixing with the radion allows it to be larger provided the radion 
itself is less massive. 

\begin{figure}[htbp]
\centerline{
\includegraphics[width=5.4cm,angle=90]{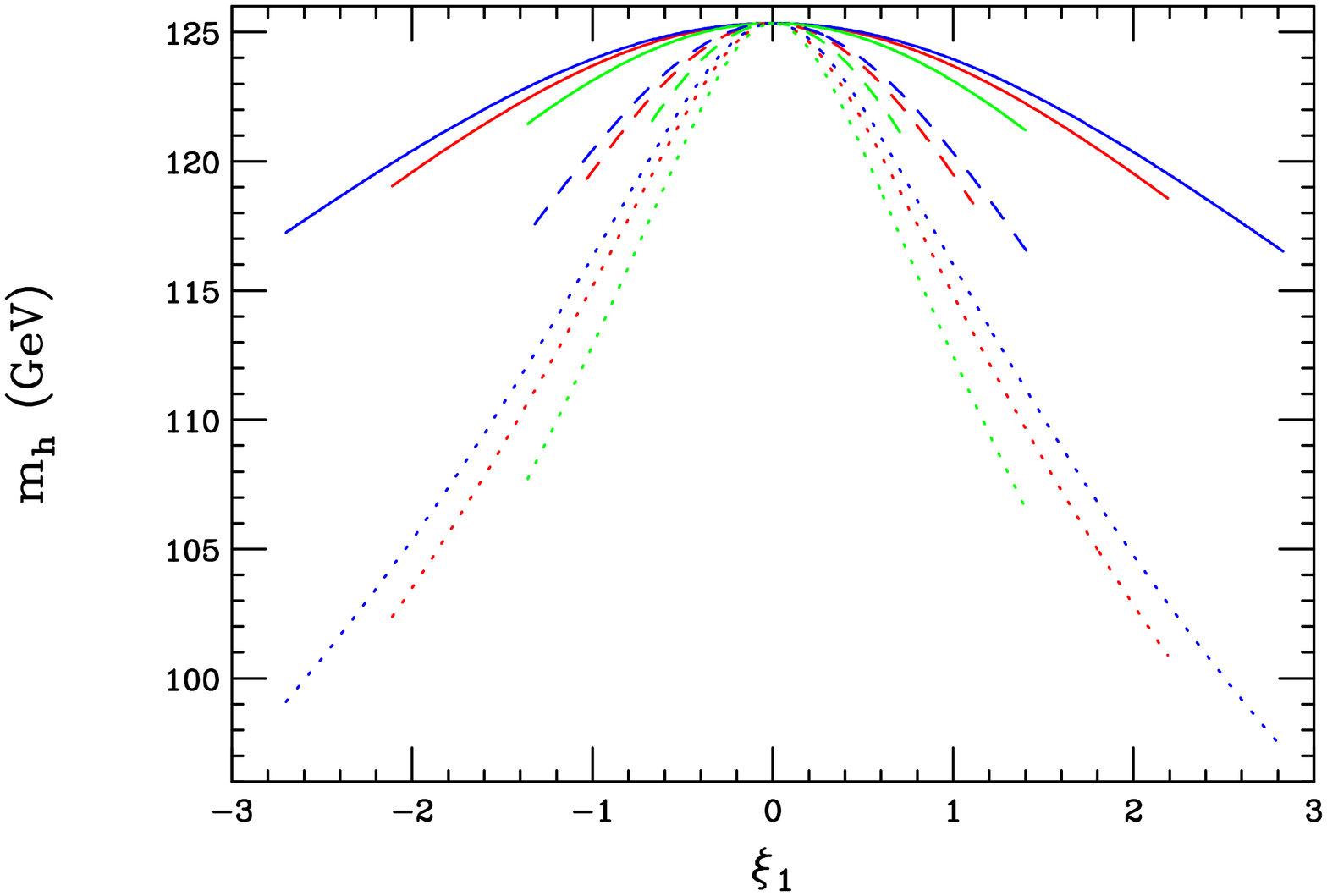}
\hspace*{1mm}
\includegraphics[width=5.4cm,angle=90]{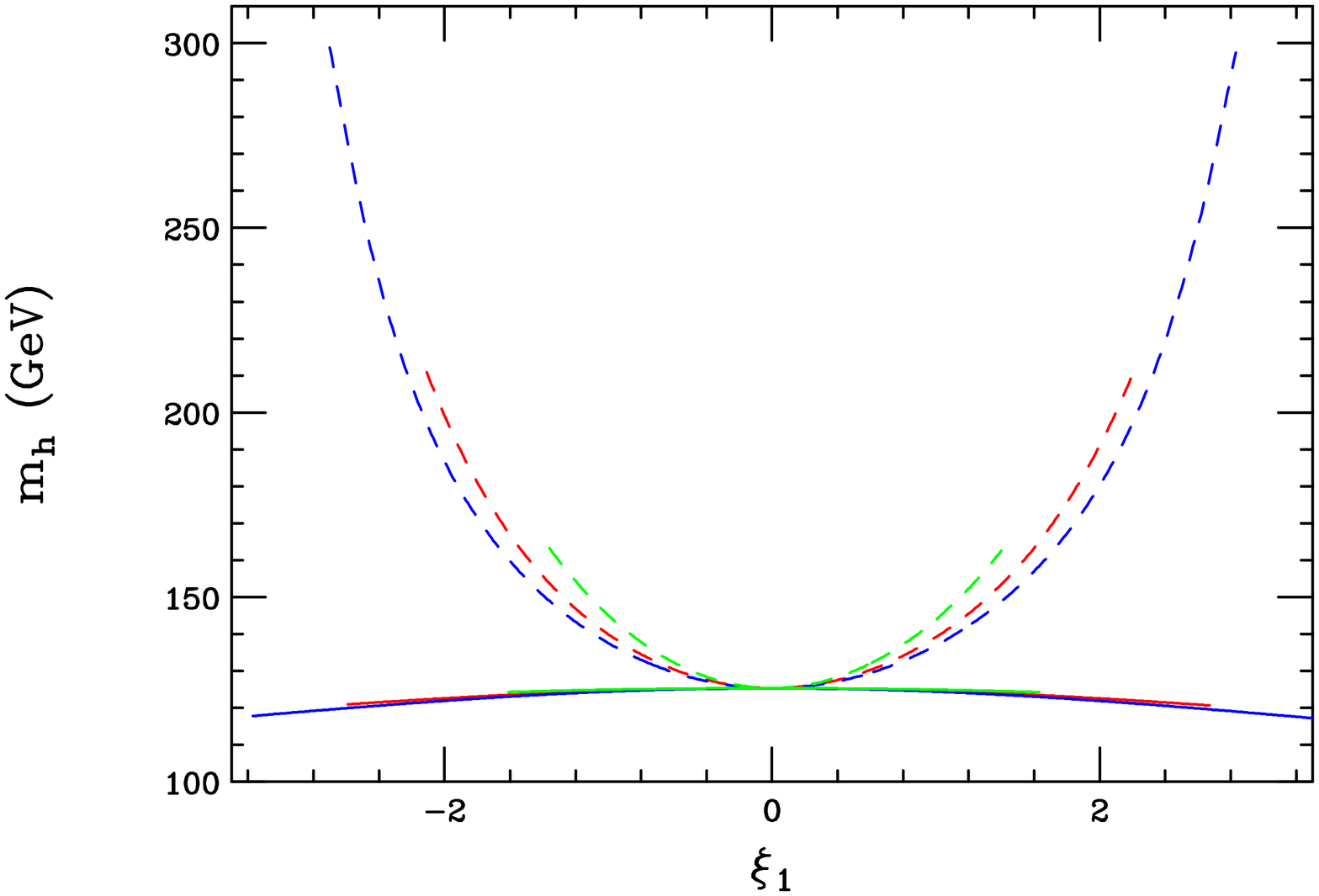}}
\vspace*{0.1cm}
\centerline{
\includegraphics[width=5.4cm,angle=90]{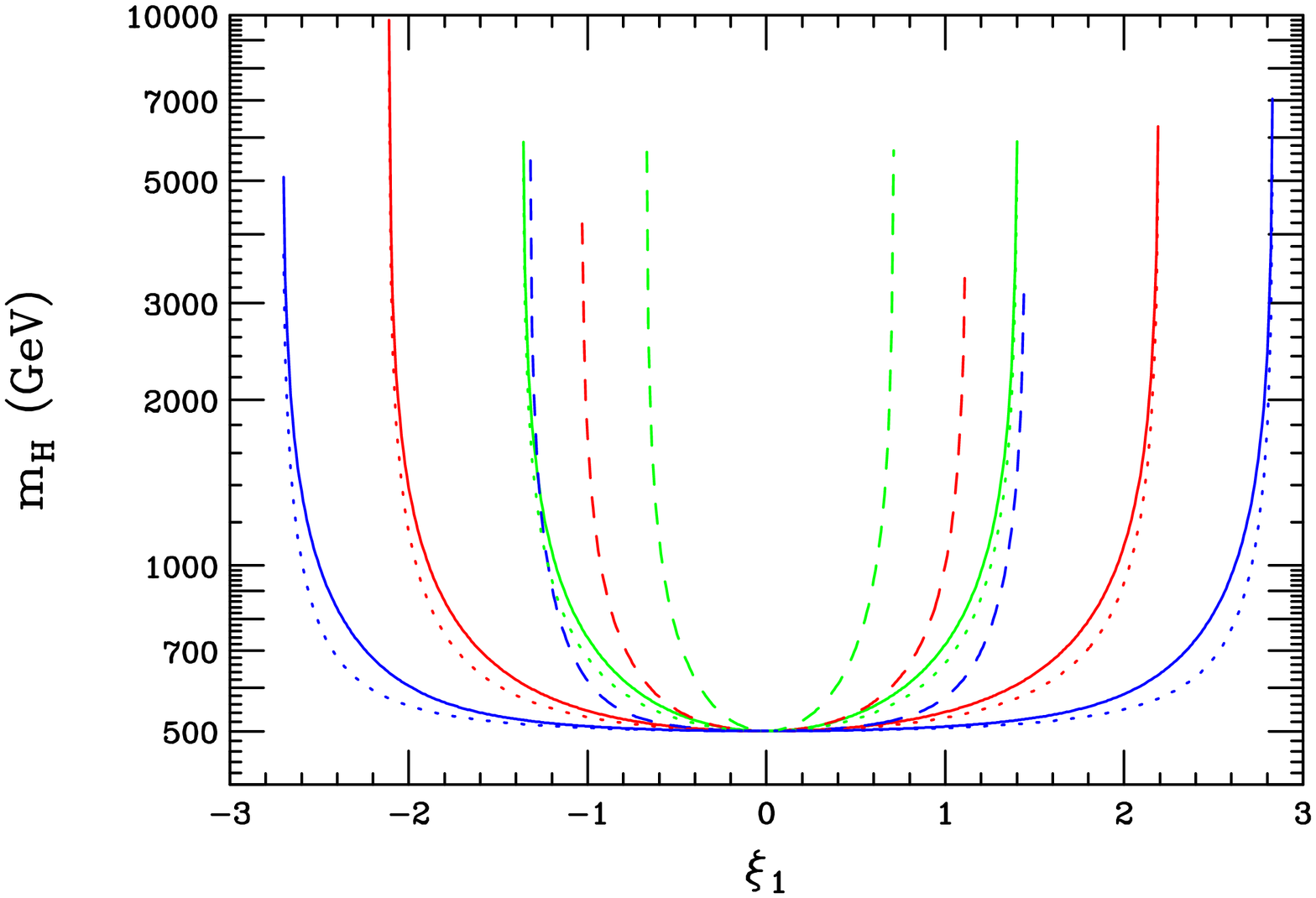}
\hspace*{1mm}
\includegraphics[width=5.4cm,angle=90]{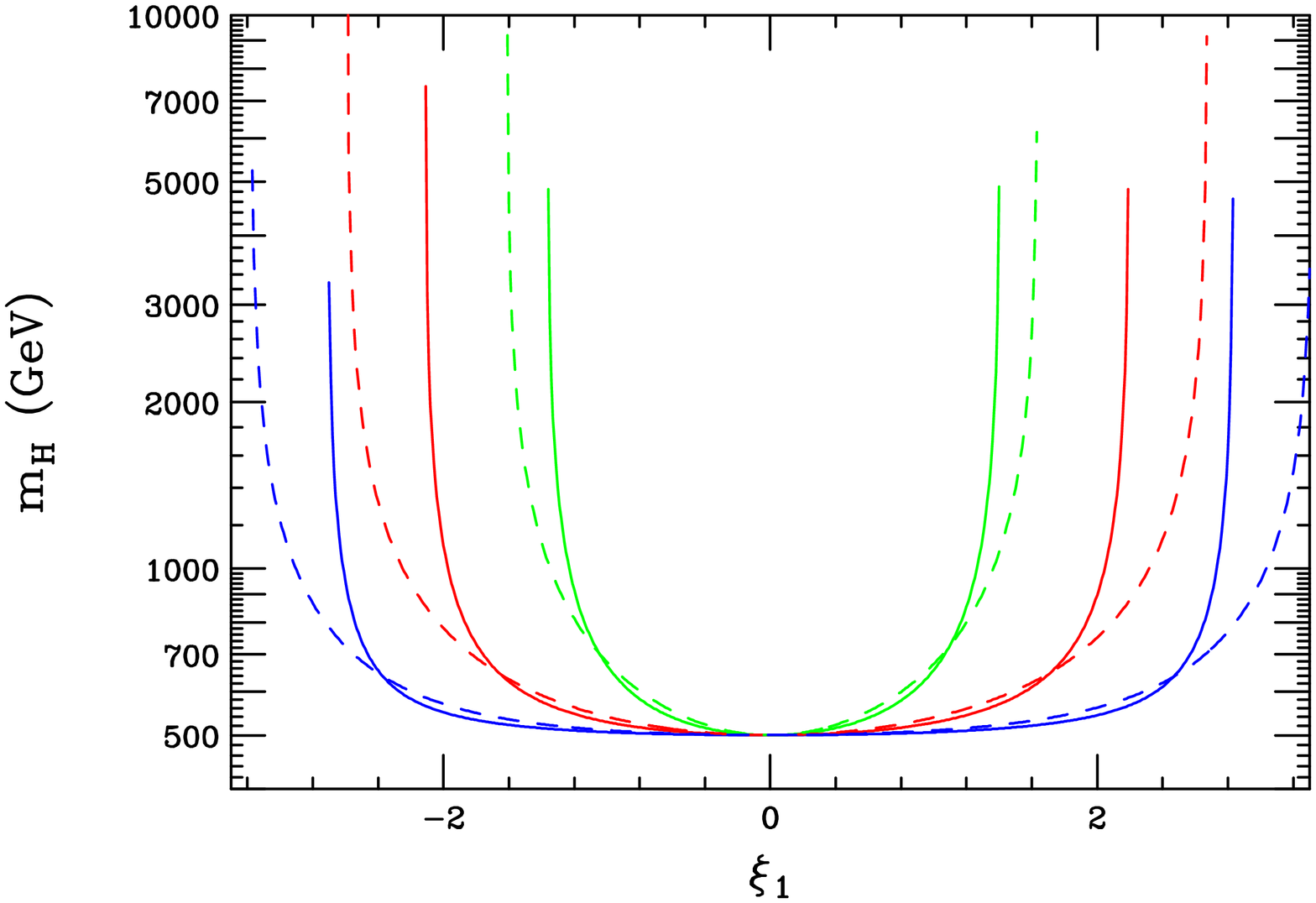}}
\vspace*{0.1cm}
\centerline{
\includegraphics[width=5.4cm,angle=90]{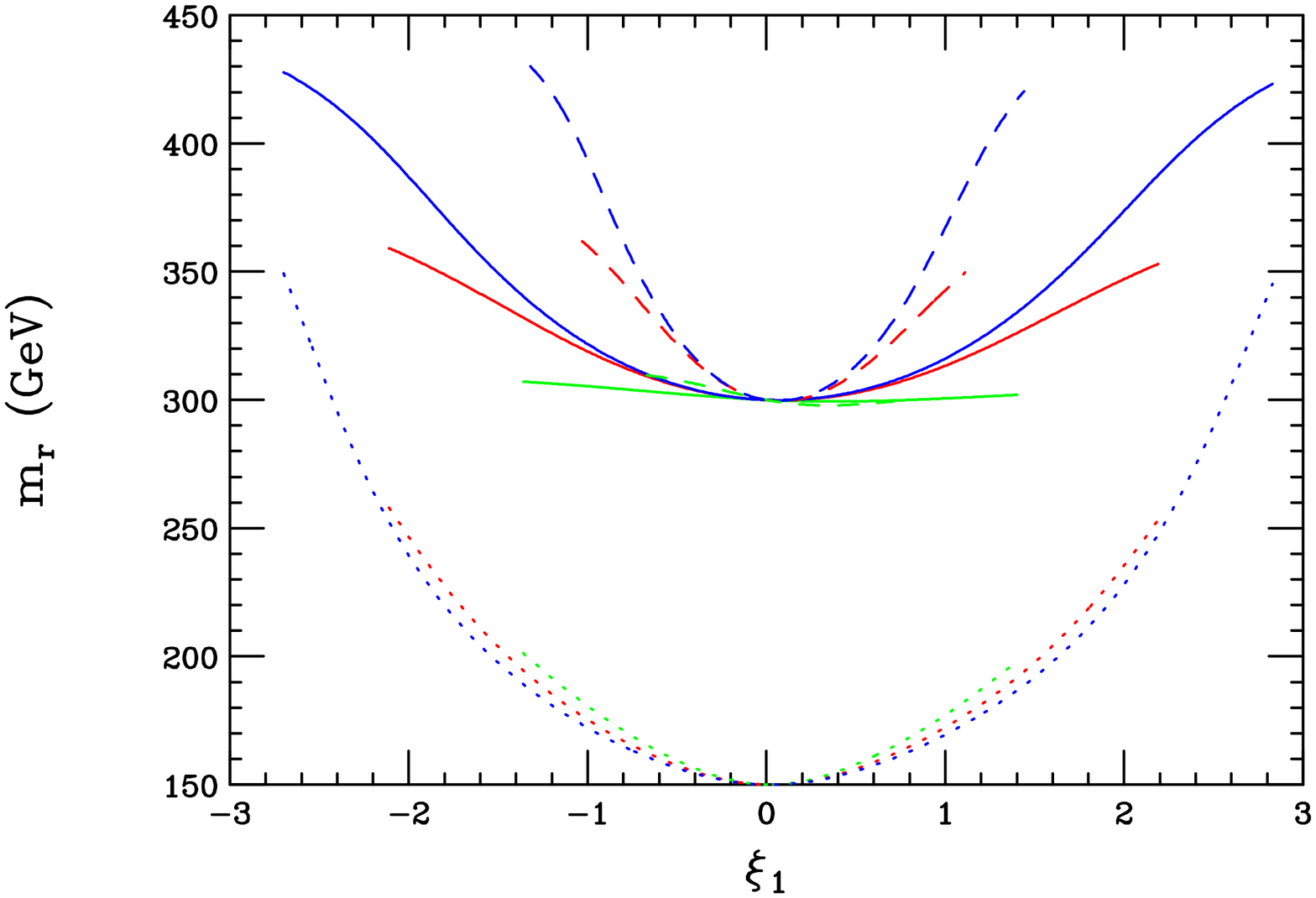}
\hspace*{1mm}
\includegraphics[width=5.4cm,angle=90]{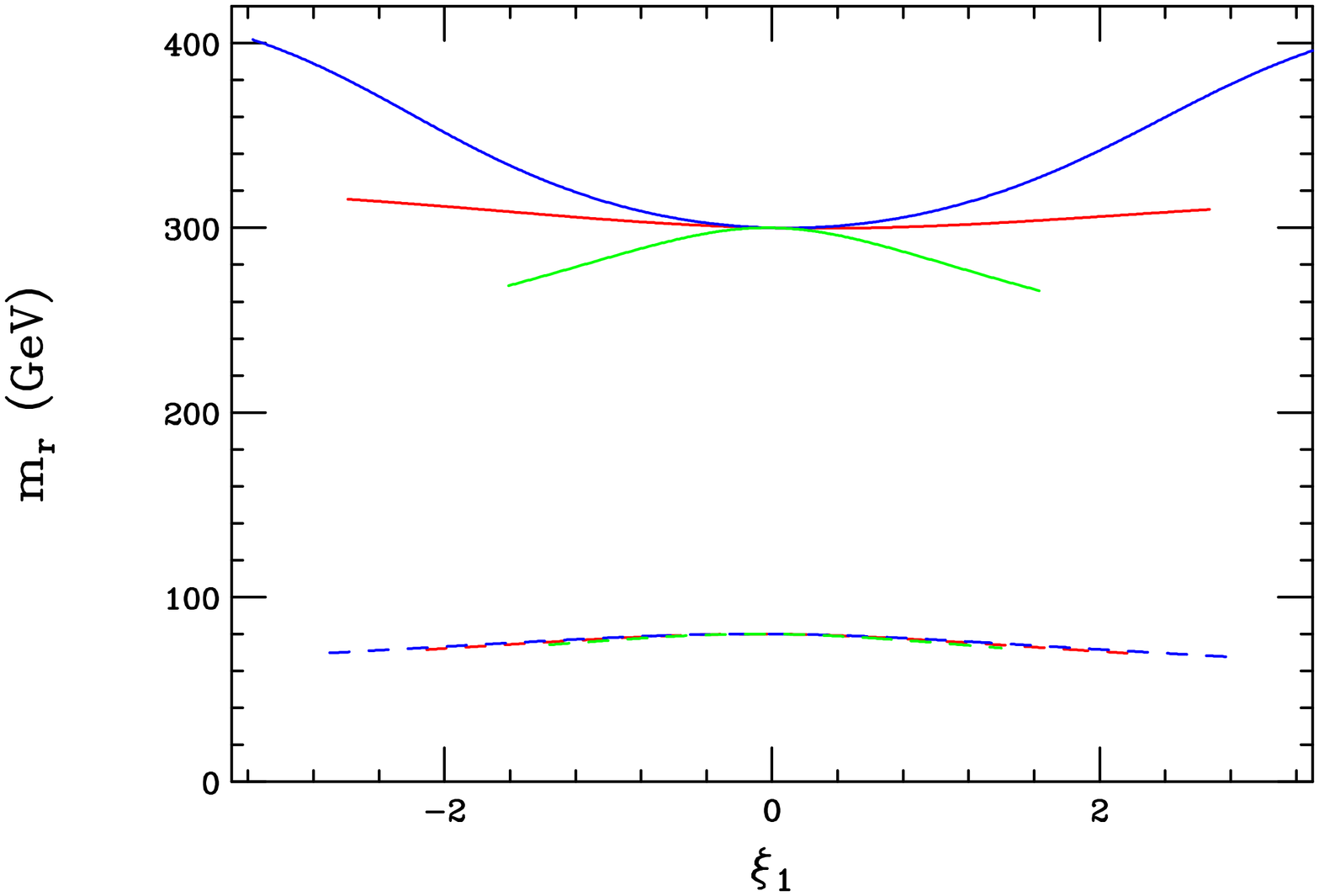}}
\vspace*{0.1cm}
\caption{Typical shifts in light Higgs(top), heavy Higgs(middle) and 
radion(bottom) masses due to mixing for different values of the model 
parameters. Figures in a given column are case correlated.}
\label{higgs1}
\end{figure}

Next we consider the shifts due to mixing in the squares of the couplings of 
the various fields to either $b\bar b$ or $t\bar t(c\bar c)$; it is important 
to note that the corresponding 
shifts for the couplings to $WW/ZZ$ are found to almost identical to those for 
$b\bar b$ in almost all cases so we do not present those results separately 
here. Note that we have scaled these couplings shown in the figures 
either by their value in 
the MSSM using the input parameters above, as in the case of $h,H$, or by  
their unmixed values, as in the case of $r$. The range of coupling shifts, 
especially for $H$ and $r$, are truly impressive being orders of magnitude 
in some cases. Not only are large enhancements seen for some parameter 
ranges, it is also important to note that there are regions of the 
parameter space, not necessarily near the boundaries, where couplings can 
completely vanish. For the light Higgs we see that while the mixing effects 
are not as large, for the cases at hand they 
always lead to a reduction in the 
coupling strengths in comparison with MSSM expectations.

\begin{figure}[htbp]
\centerline{
\includegraphics[width=5.4cm,angle=90]{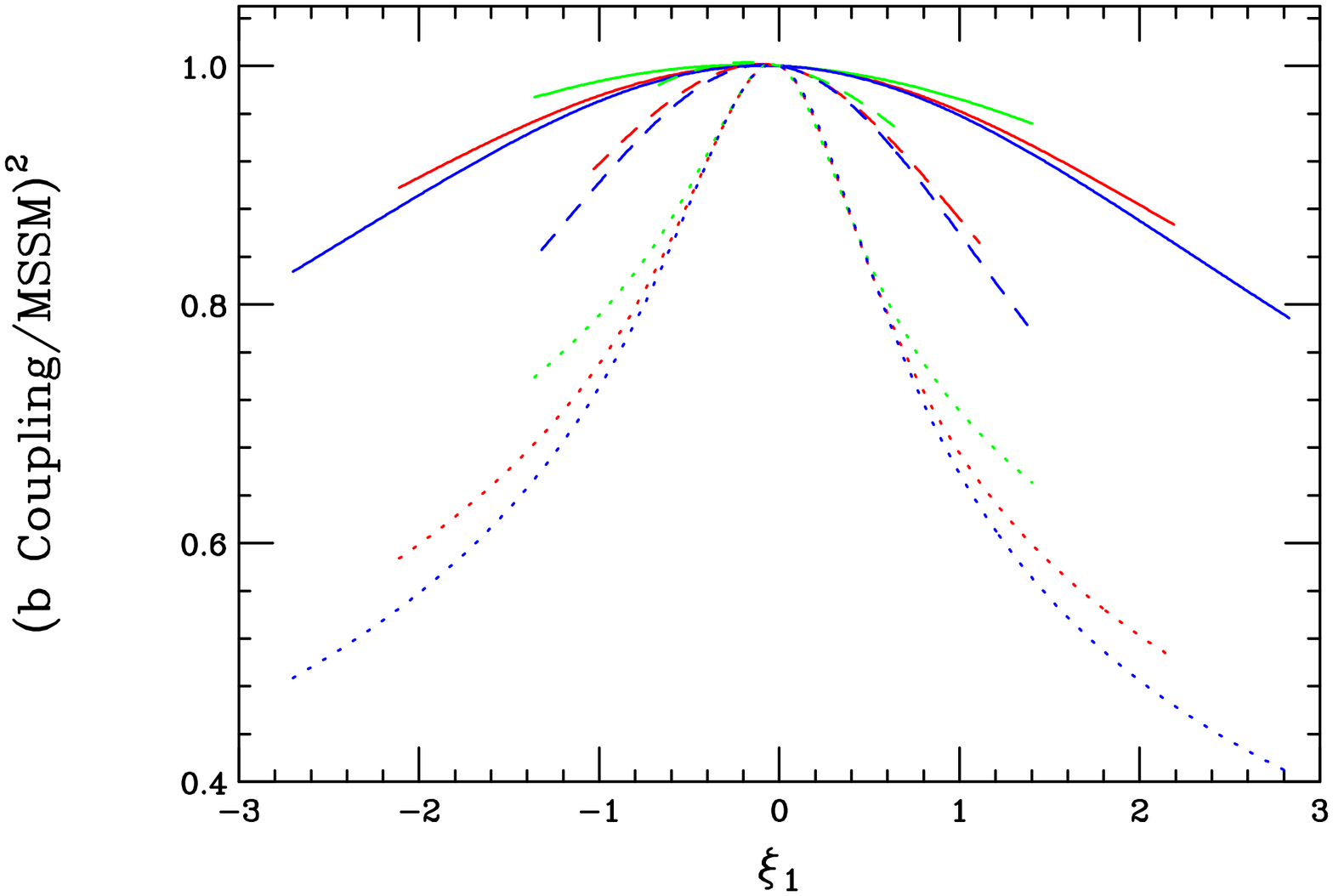}
\hspace*{1mm}
\includegraphics[width=5.4cm,angle=90]{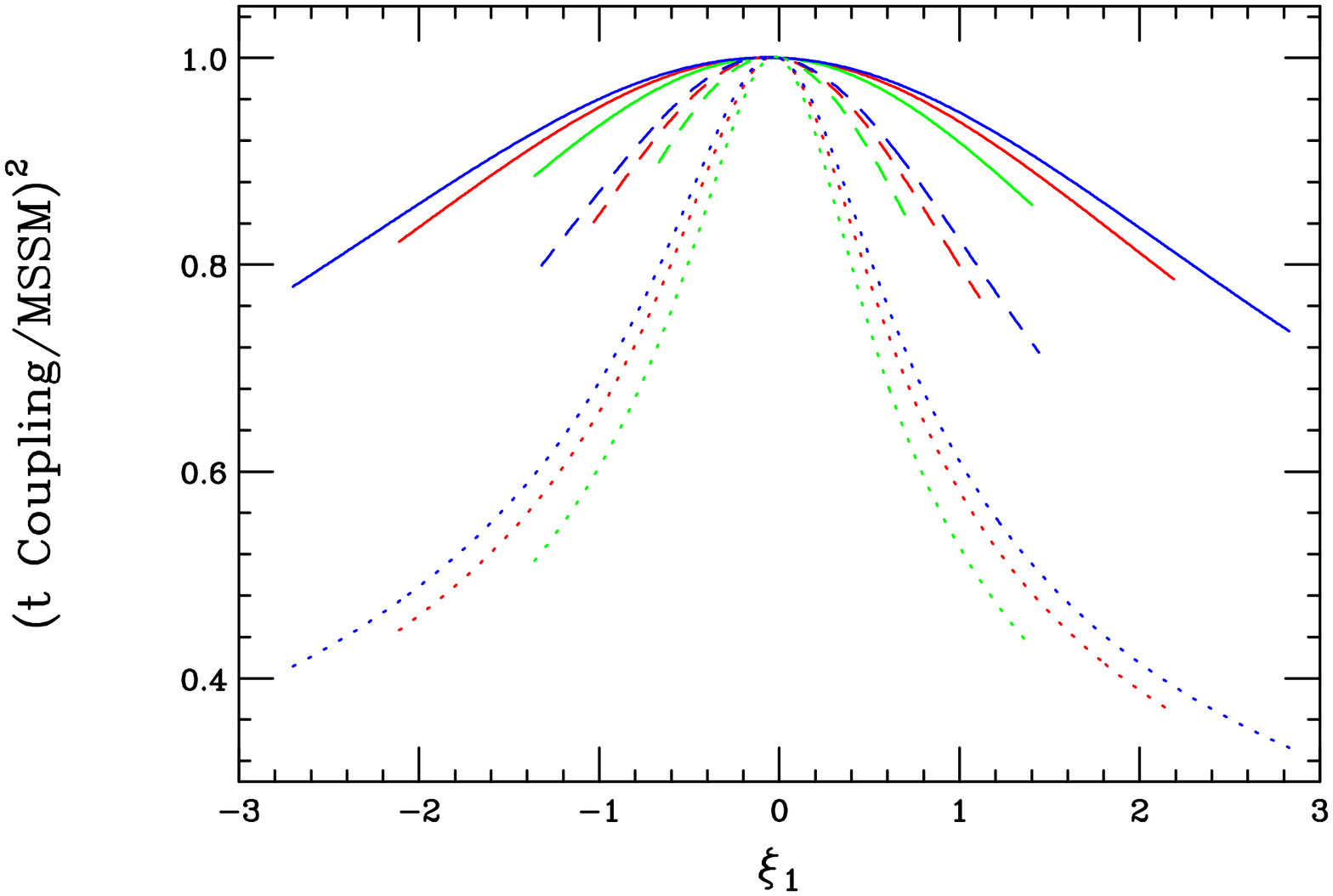}}
\vspace*{0.1cm}
\centerline{
\includegraphics[width=5.4cm,angle=90]{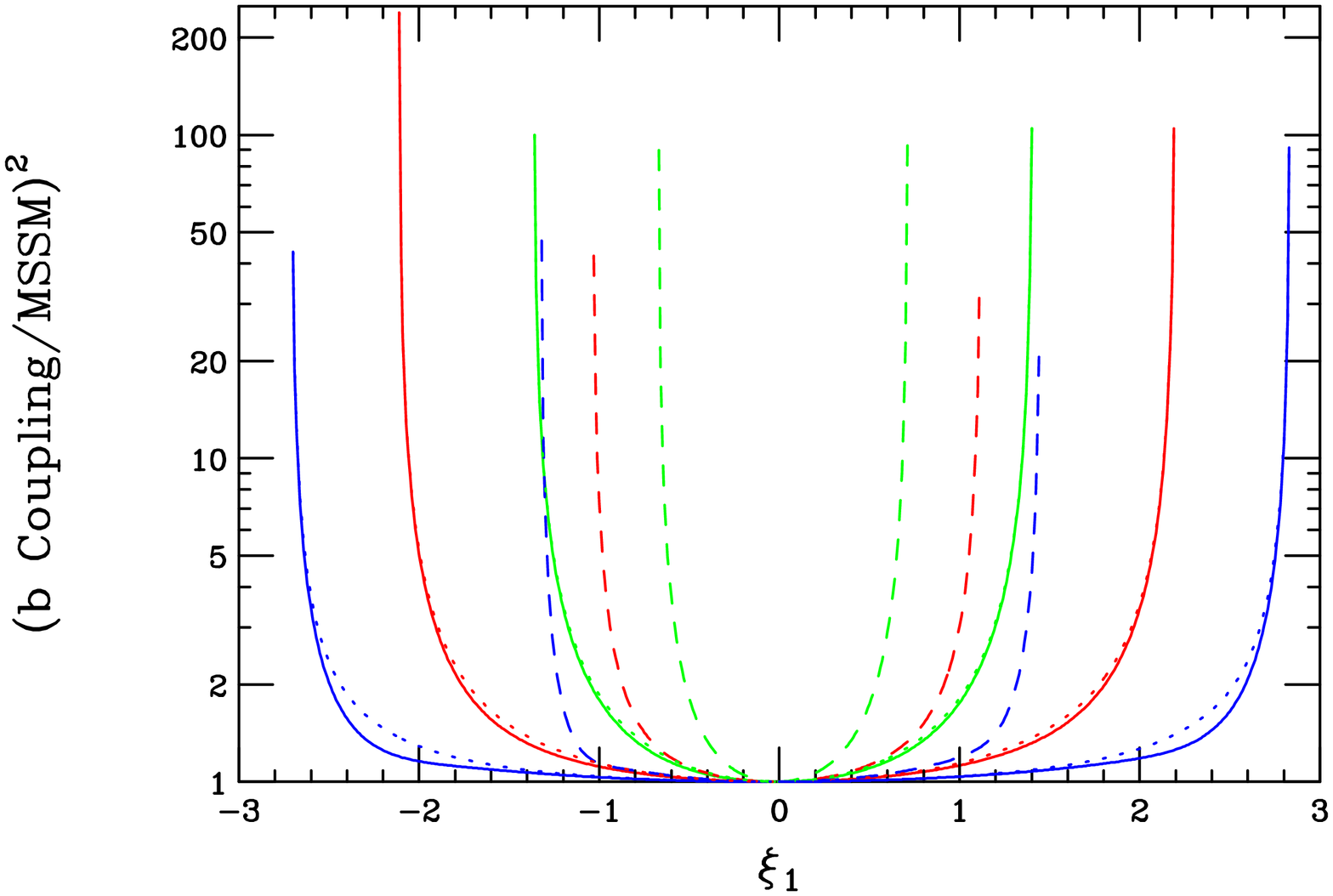}
\hspace*{1mm}
\includegraphics[width=5.4cm,angle=90]{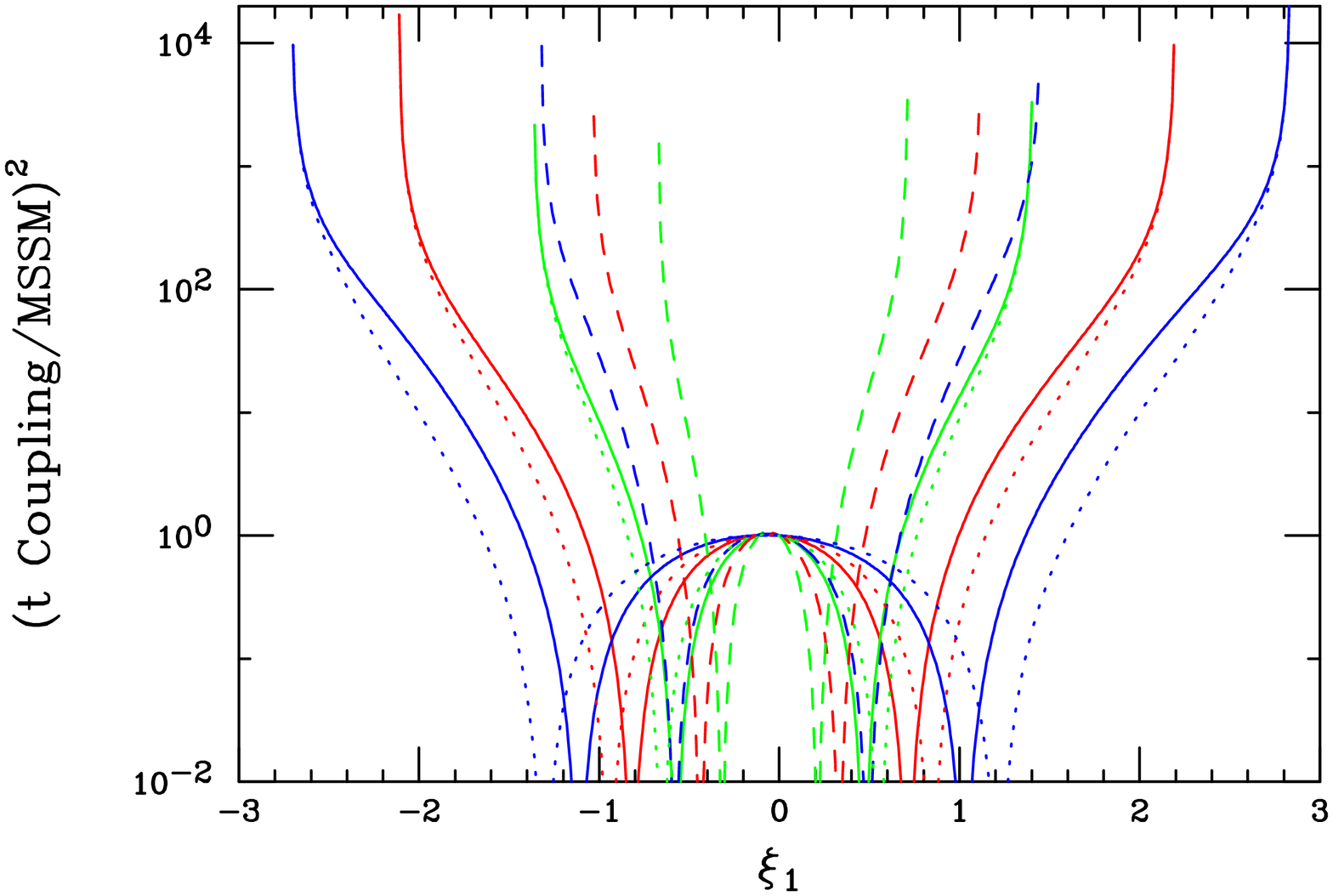}}
\vspace*{0.1cm}
\centerline{
\includegraphics[width=5.4cm,angle=90]{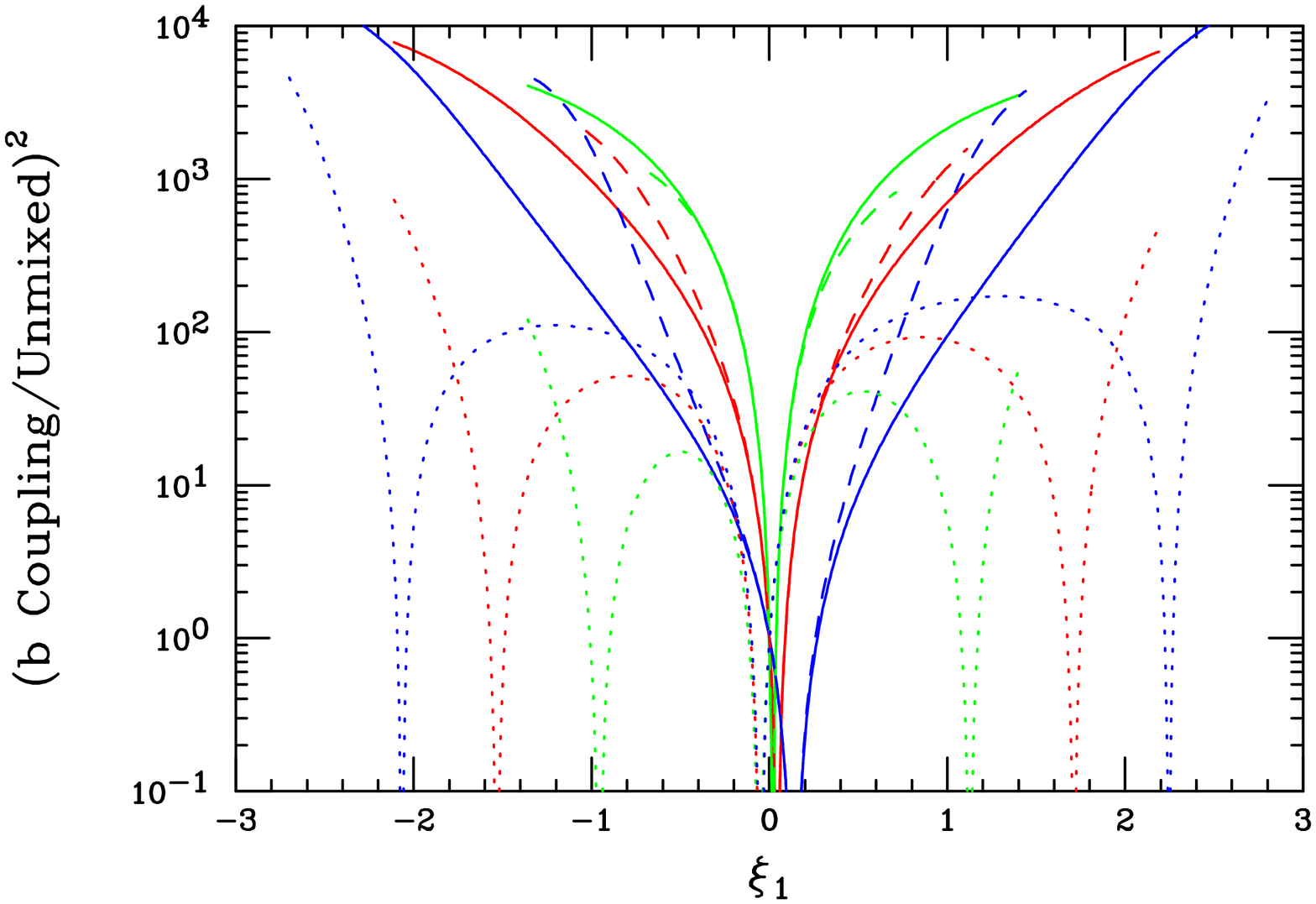}
\hspace*{1mm}
\includegraphics[width=5.4cm,angle=90]{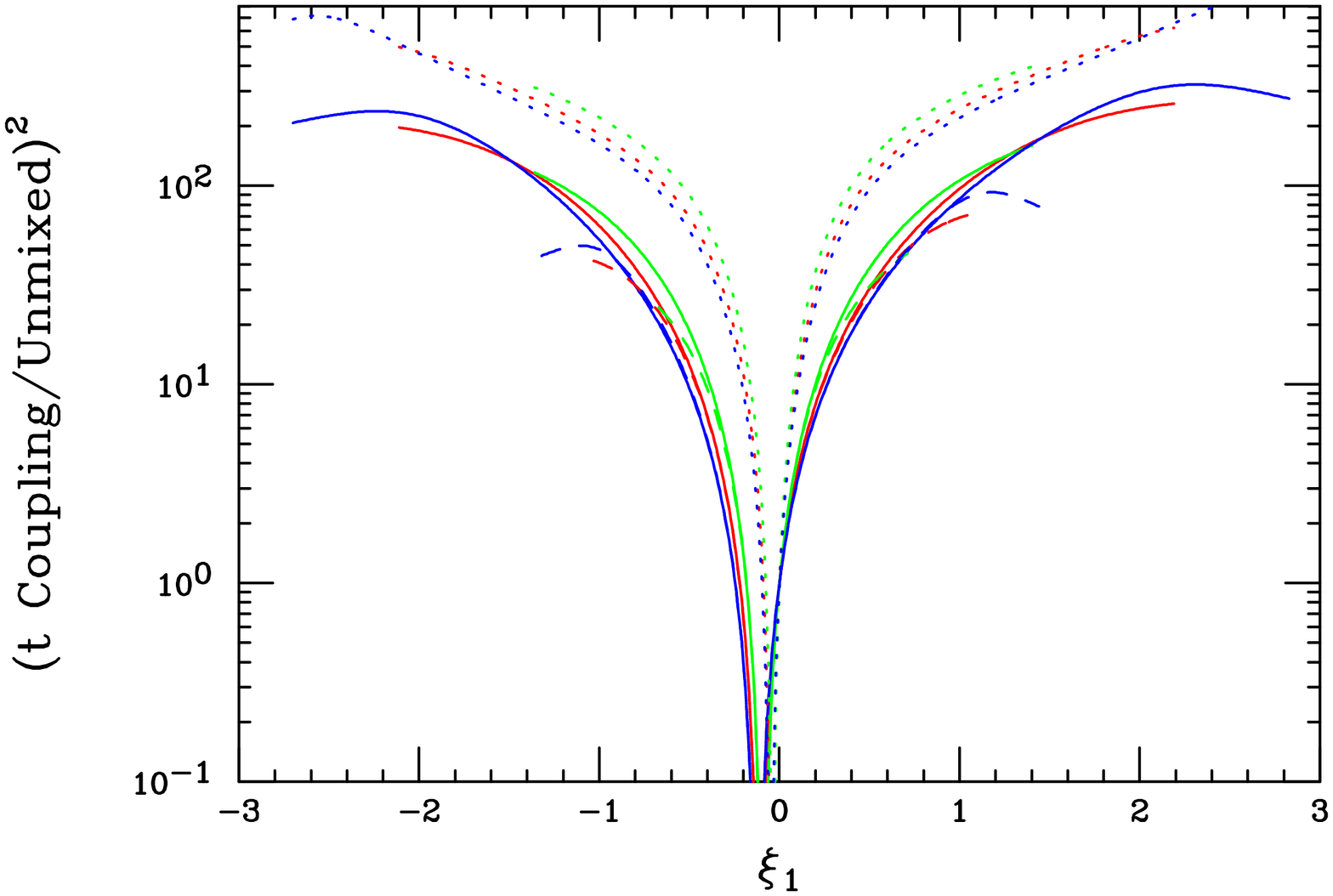}}
\vspace*{0.1cm}
\caption{Typical shifts in light Higgs(top), heavy Higgs(middle) and 
radion(bottom) couplings to $b\bar b$(left) and $t\bar t$(right) due to 
mixing for different values of the model parameters. Figures are correlated 
in a given column.}
\label{higgs2}
\end{figure}

For the light Higgs, $h$, it is particularly important to examine the 
variations in the couplings to the $gg$ and $\gamma \gamma$ final state since 
these control the dominant signal rate at the LHC; these are shown in the 
next set of figures. While the couplings of $h$ to quarks and massive vector 
bosons was generally reduced via mixing, we see here that the loop-induced 
processes can be either enhanced or suppressed depending upon the region of 
parameter space we happen to be sitting in. The shift in the 
couplings of $H,r$ to $gg$ and $\gamma \gamma$ will be presented elsewhere.

\begin{figure}[htbp]
\centerline{
\includegraphics[width=5.4cm,angle=90]{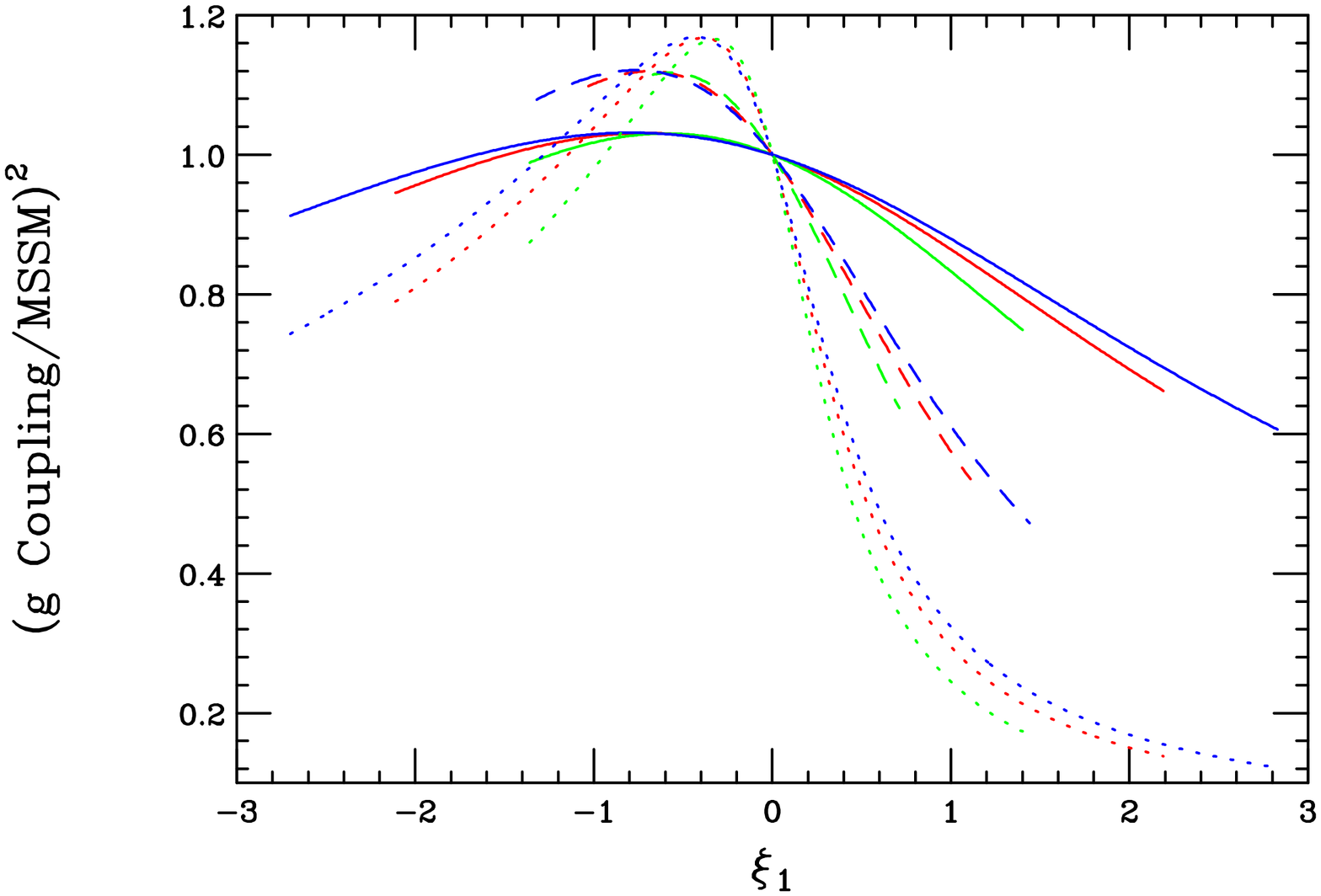}
\hspace*{1mm}
\includegraphics[width=5.4cm,angle=90]{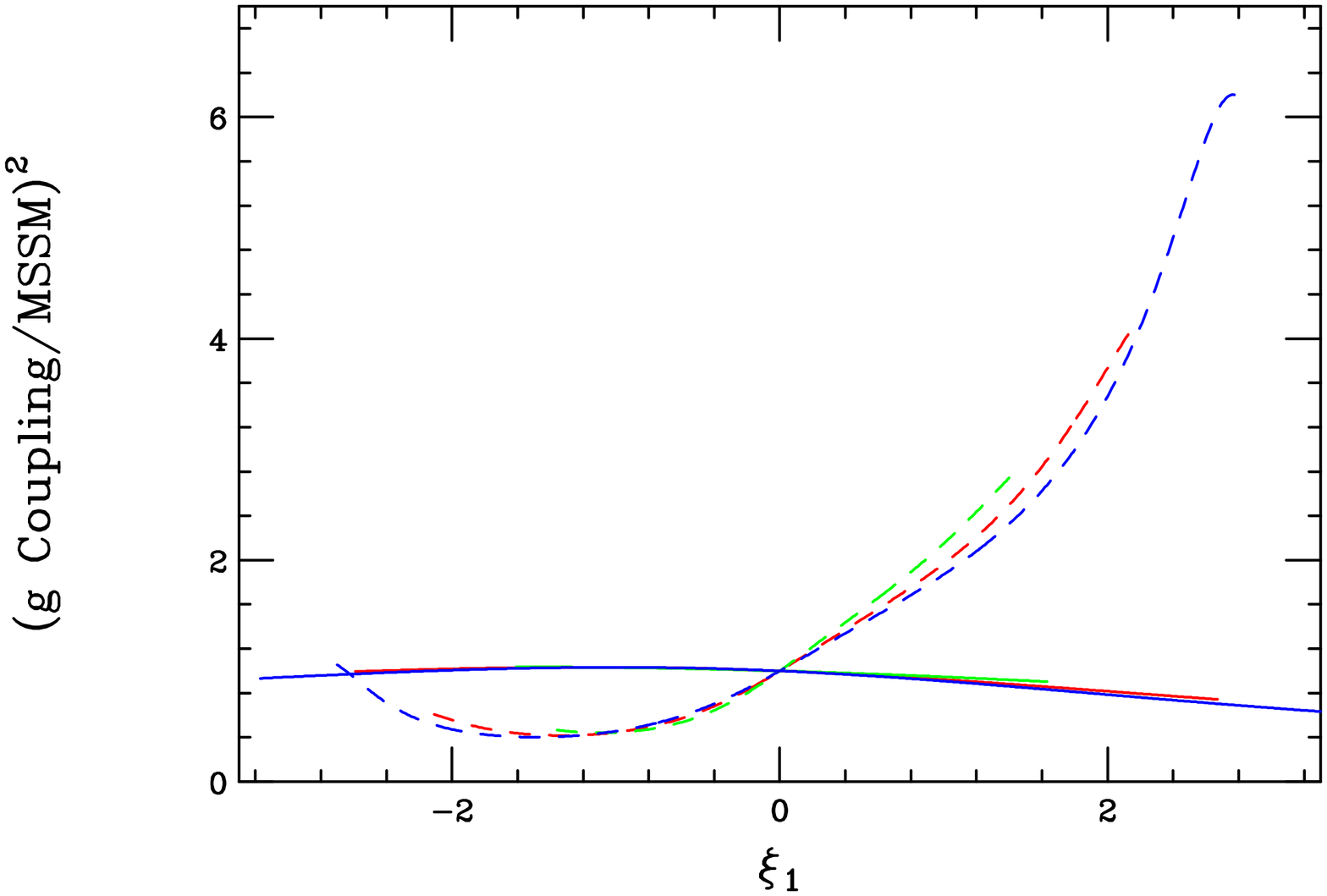}}
\vspace*{0.1cm}
\centerline{
\includegraphics[width=5.4cm,angle=90]{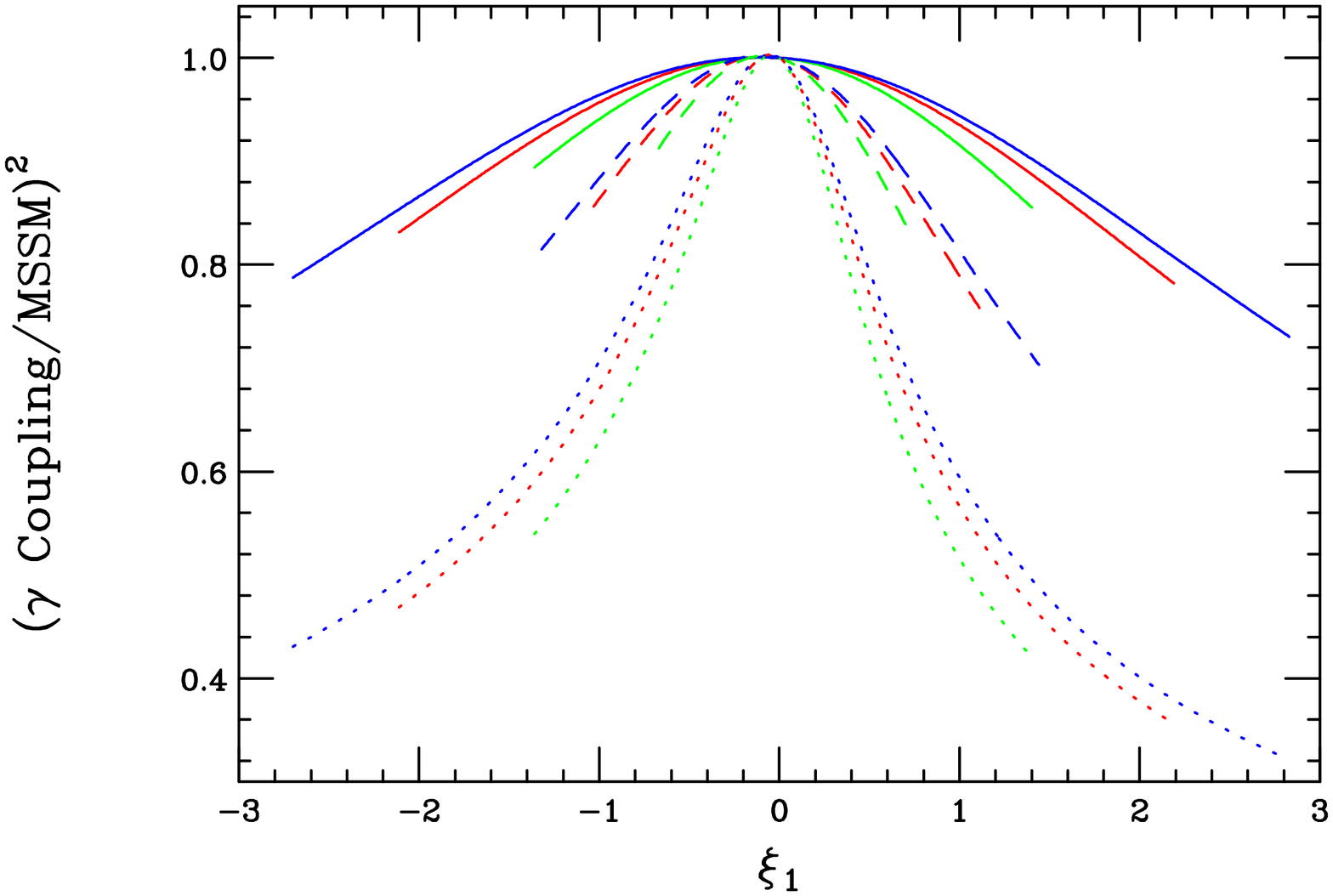}
\hspace*{1mm}
\includegraphics[width=5.4cm,angle=90]{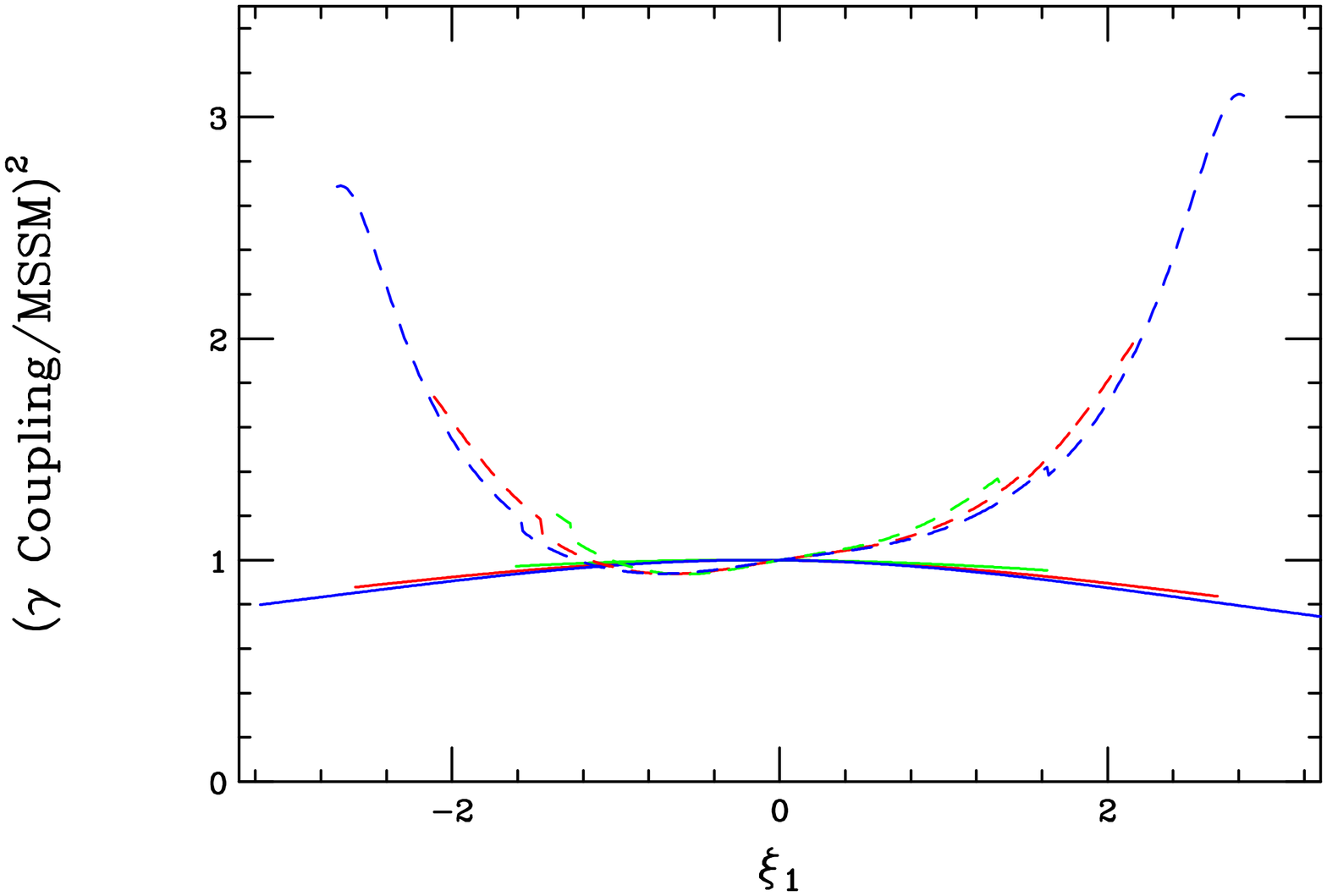}}
\vspace*{0.1cm}
\caption{Shifts in $gg$(top) and $\gamma \gamma$(bottom) couplings of the 
$h$ due to mixing. Figures in a given column are correlated.}
\label{higgs3}
\end{figure}

\section{CONCLUSIONS}

We have begun a preliminary 
examination of the effects of mixing between the radion and    
the two CP even Higgs fields present in the two-doublet-model. As a result of 
this mixing the masses and couplings of all of these fields are found to be  
substantially modified from the expectations of the MSSM. Further 
analysis into 
the details of these mixing effects and the corresponding shifts in the 
various particle widths is on-going.

\setcounter{figure}{0}
\setcounter{table}{0}
\setcounter{section}{0}
\setcounter{equation}{0}
\clearpage

\part{Search For The Radion Decay $\phi \rightarrow \rm h \rm h$ With
$\gamma \gamma$+$\rm b \bar{\rm b}$, $\tau \tau$+$\rm b \bar{\rm b}$ And  
$\rm b \bar{\rm b}$+$\rm b \bar{\rm b}$ Final States In CMS \label{paper}}

{\it D. Dominici, G. Dewhirst, S. Gennai, L. Fan$\grave{o}$, A. Nikitenko} 

\maketitle

\section{INTRODUCTION}
The Randall Sundrum model (RS) \cite{Randall:1999ee,Randall:1999vf}
 has recently received much attention
because it could provide a solution to the hierarchy
problem,  by means of an exponential
 factor in a five dimensional  non-factorizable metric.
 In the simplest version the RS model
is based on a five dimensional universe with two
four dimensional hypersurfaces (branes), located at the boundary of the
fifth coordinate $y$. 
By placing all the Standard Model
 fields on the visible brane at $y=1/2$
all the mass terms, which are of the
order of the Planck mass, are rescaled by the  exponential factor,
to a scale of the order of a TeV.
 The  fluctuations in the metric in the fifth dimension 
are described in terms of a scalar field, the radion
which  in general mixes with  the Higgs.
This scalar sector of the RS model is parametrized in terms
of a dimensionless parameter $\xi$, 
of the Higgs and radion masses $m_h$, $m_\phi$ 
 and  the vacuum expectation value
of the radion field $\Lambda_\phi$.
The  phenomenology of the Higgs and radion  at LHC
has been the object of several studies 
\cite{Giudice:2000av,Chaichian:2001rq,Hewett:2002nk,Dominici:2002jv,Battaglia:2003gb,Azeulos:2002xx} concentrating mainly on Higgs and radion
production. 
In general the Higgs and radion detection is not guaranteed 
in all the parameter space region.
The presence in the Higgs radion sector of trilinear
terms opens up the important possibility of $\phi\to hh$
decay and ${\rm h}\to\phi\phi$. For example for $\rm m_{\rm h}=120$ GeV/$c^2$,
$\Lambda_\phi=5$ TeV
and $\rm m_\phi\sim 250-350$ GeV/$c^2$ the $BR(\phi\to \rm hh)\sim 0.2-0.3$.
In this paper  we estimate the 
CMS discovery potential for the radion ($\phi$) in two Higgs decay
mode ($\phi \rightarrow \rm h \rm h$) with 
$\gamma \gamma$+$\rm b \bar{\rm b}$, $\tau \tau$+$\rm b \bar{\rm b}$ and
$\rm b \bar{\rm b}$+$\rm b \bar{\rm b}$ final states.
\section{ANALYSIS}

One point of 
$\rm m_{\phi}$=300 GeV/$c^2$ and $\rm m_{\rm h}$=125 GeV/$c^2$ was 
taken and the observability in the ($\xi$, $\Lambda_{\phi}$) plane was evaluated.
Signal events were simulated with the PYTHIA~\cite{Sjostrand:2001yu} 
MSSM $\rm g \rm g \rightarrow \rm H \rightarrow \rm h \rm h$ process when 
values of $\rm m_{\rm H}$ and $\rm m_{\rm h}$ were set to 300 and 
125 GeV/$c^2$. $\Gamma_{\rm H}$ was set to $\sim$ 1 GeV/$c^2$, thus the variation
of the radion width in the ($\xi$, $\Lambda_{\phi}$) plane was neglected.
However, the effect of changing of the radion width will be not visible
due to the fact that $\Gamma_{\phi}$ is small in comparison with the detector resolution
when the radion mass is reconstructed.

\subsection{$\gamma \gamma \rm b \bar{\rm b}$ final state}

Signal events were processed with the full detector simulation and 
reconstruction.
Events were required to pass the Level-1 and High Level Trigger (HLT)
selections for the di-photon stream 
~\cite{cms:daq_tdr,cms:daq_tdr2} with HLT thresholds
on photons of 40 and 25 GeV/$c$. Photons were required to be isolated with 
the tracker and the electromagnetic calorimeter. 
The two highest $\rm E_{\rm T}$ jets 
of $\rm E_{\rm T}>$ 30 GeV and $|\eta |<$2.4 were reconstructed with the calorimeter and 
were taken as b-jet candidates from $\rm h \rightarrow \rm b \bar{\rm b}$ 
decay. 
At least one of these two jets has to be tagged as a b jet. The efficiency
of single b tagging is 0.61 per event. Further selections require the
di-jet mass, $\rm M_{\rm b \rm j}$, to be in the window 
$\rm m_{\rm h} \pm$ 30 GeV/$c^{2}$ 
(efficiency 65 \%) and the di-photon mass, $\rm M_{\gamma \gamma}$, to be in the window 
$\rm m_{\rm h} \pm$ 2 GeV/$c^{2}$ (efficiency 78 \%). Finaly, the  
$\rm M_{\gamma \gamma \rm b \rm j}$ mass should be in the window of 
$\rm m_{\phi} \pm$ 50 GeV/$c^{2}$ (efficiency 95 \%). 
Figure~\ref{fig:radion_2g2b_mass} shows 
$\rm M_{\rm b \rm j}$ and $\rm M_{\gamma \gamma \rm b \rm j}$ distributions
with arbitrary normalization after selections. The signal efficiency of the
whole selection chain is 3.7 \%. For $\Lambda_{\phi}$ = 1 TeV and $\xi$ = 0 
the expected number of signal events with 30 fb$^{-1}$ is 41 
(with $\sigma$($\rm g \rm g \rightarrow \phi$) = 40.8 pb, 
Br($\phi \rightarrow \rm h \rm h$) = 0.33, 
Br($\rm h \rightarrow \rm b \bar{\rm b}$) = 0.61, 
Br($\rm h \rightarrow \gamma \gamma$) = 0.00225). 
\begin{figure}[ht!]
\begin{center}
\includegraphics[width=6cm]{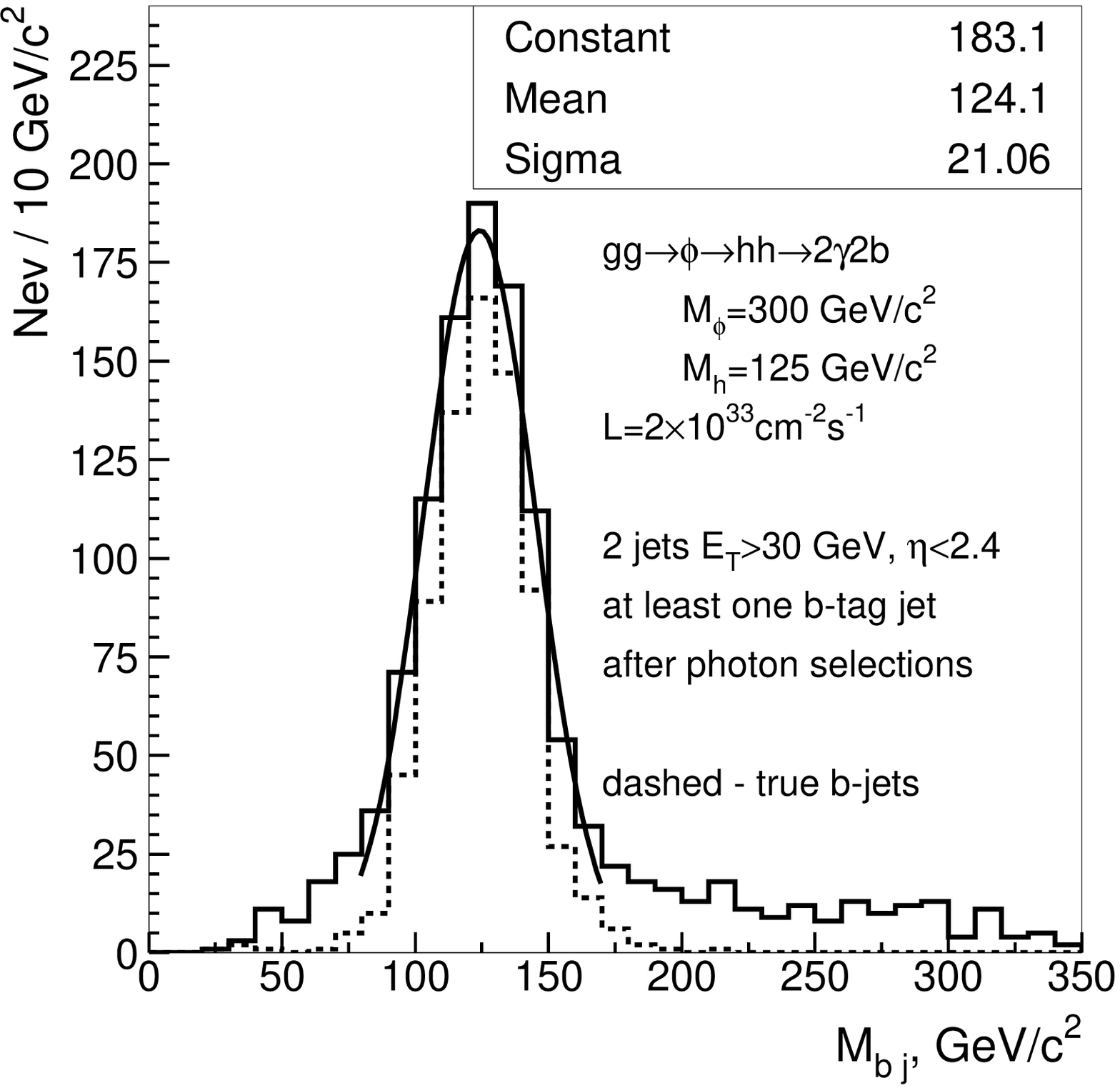}
\includegraphics[width=6cm]{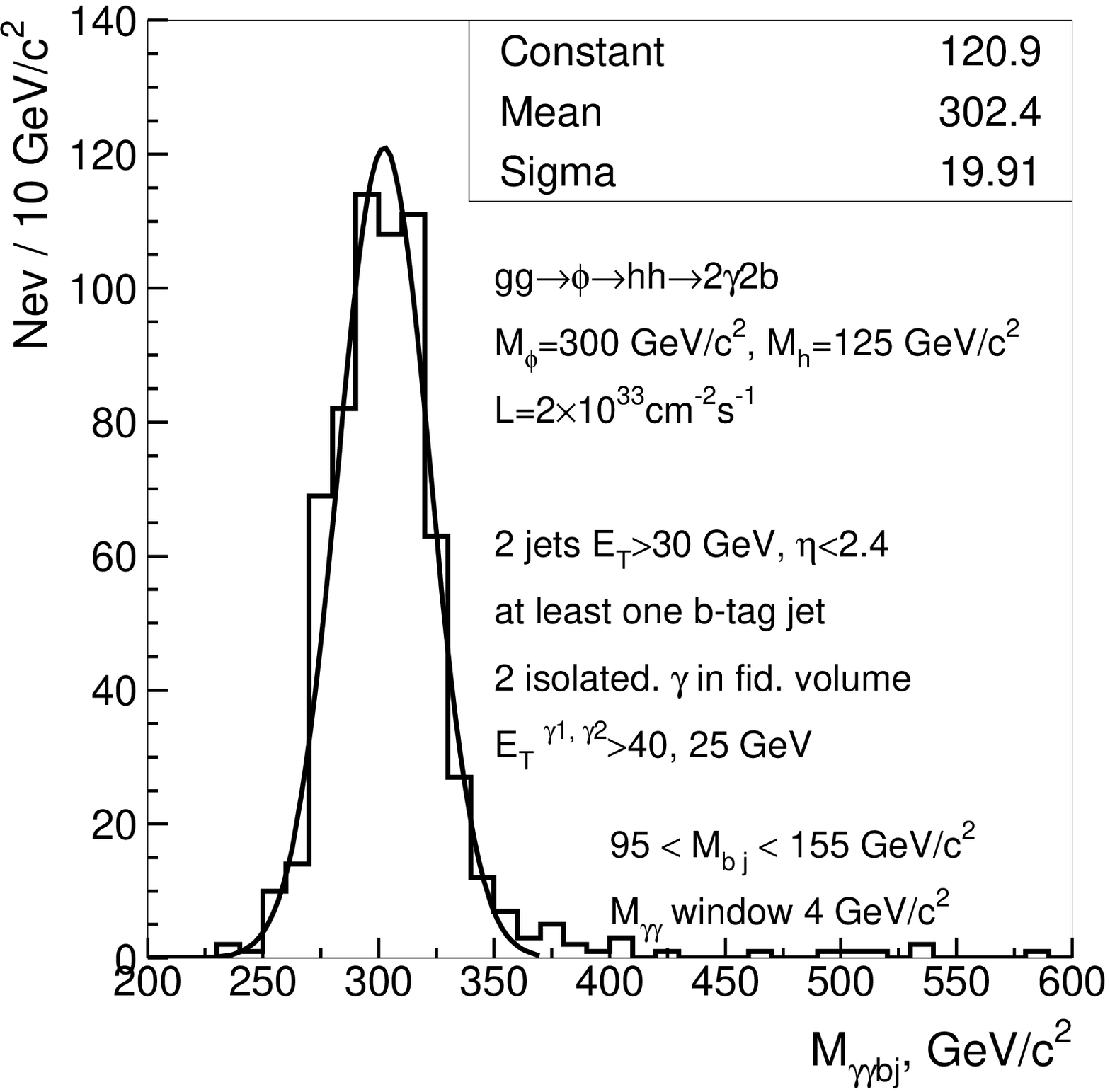}
\caption{Reconstructed b~j (left) and $\gamma \gamma \rm b \rm j$ mass for 
the signal.}
\label{fig:radion_2g2b_mass}
\end{center}
\end{figure}
\\
Irreducible di-photon backgrounds were generated with
CompHEP~\cite{Pukhov:1999gg} for the  $\gamma \gamma \rm j \rm j$ (j=u,d,s,g)
process and with MadGraph~\cite{Maltoni:2002qb,Stelzer:1994ta} 
for
the $\gamma \gamma \rm c \bar{\rm c}$ and $\gamma \gamma \rm b \bar{\rm b}$
processes with the factorization and the renormalization scales set to
$\rm M _{\rm Z}$ and CTEQ5L PDF. The generator level preselections are 
$\rm p_{\rm T,\rm max~(\rm min)}^{\gamma}>$35 (20) GeV/$c$,  
$\rm p_{\rm T}^{\rm j}>$20 GeV/$c$, $|\eta |<$2.5, 
$\Delta \rm R_{\gamma \gamma}>$0.3, 
$\Delta \rm R_{\gamma \rm j}>$0.3, $\Delta \rm R_{\rm j \rm j}>$0.3.
Cross sections are shown in Table~\ref{tab:radion_2g2b_bkg}. 
PYTHIA was used for the hadronization. Initial and final state
radiation in PYTHIA (ISR, FSR) were switched on.
A fast detector
simulation with the realistic resolution of the photon and jet energies and 
 track momentum was used. 
The track reconstruction efficiency of 0.9 was taken into 
account in the tracker isolation criteria. The efficiency of b tagging, 0.5, 
for b jets and the mistagging probability of 0.01 (0.1) for 
u,d,s,g (c) jets were used.
These numbers correspond to what was obtained with the full detector
simulation ~\cite{cms:btag} using the impact parameter tagging method. 
The efficiency of the selections and the expected number of the background events with
30 fb$^{-1}$  after all selections including b tagging are shown in 
Table~\ref{tab:radion_2g2b_bkg}. Statistical errors on the expected number of 
events are also shown. The number of background events was
then multiplied by 0.92 and by 0.90 to take into account of the Level-1 
e/$\gamma$ trigger and the calorimeter isolation efficiencies which were not 
taken into account in the fast simulation. These efficiencies were obtained 
from a full simulation of the signal events.
\begin{table}[ht]
\caption{Background cross sections, efficiency, number 
of events with 30 fb$^{-1}$  after all selections including b tagging.  
\label{tab:radion_2g2b_bkg}}
\centering
\begin{tabular}{|l|c|c|c|}
\hline
& 
$\gamma \gamma \rm j \rm j$ &
$\gamma \gamma \rm c \bar{\rm c}$ &
$\gamma \gamma \rm b \bar{\rm b}$ 
\\ \hline
cross-section, fb & 13310 & 778 & 76
\\ \hline\hline
selections & \multicolumn{3}{|c|} {efficiency}\\ \hline
$\rm E_{\rm T}^{\gamma _{1,2}}>$ 40, 25 GeV, $|\eta|<$ 2.5 
& 0.446 & 0.466 & 0.487
\\ \hline
tracker isolation in cone 0.3
& 0.328 & 0.345 & 0.379
\\ \hline
two jets $E_{\rm T}>$ 30 GeV, $|\eta|<$2.4
& 0.127 & 0.125 & 0.133
\\ \hline
$\rm M_{\gamma \gamma}$ window 4 GeV/$c^2$
& 0.00278 & 0.00263 & 0.00410
\\ \hline
$\rm M_{\rm j \rm j}$ window 60 GeV/$c^2$
& 0.00086 & 0.00096 & 0.00144
\\ \hline
$\rm M_{\gamma \gamma \rm j \rm j}$ window 100 GeV/$c^2$
& 0.00045 & 0.00061 & 0.00123
\\ \hline \hline
N events after all selections including b tagging & 
4.2 $\pm$ 0.8 & 2.0 $\pm$ 0.6 & 2.0 $\pm$ 0.6
\\ \hline 
\end{tabular}
\end{table}
\\
The CMS discovery reach was obtained in the ($\xi$, $\Lambda_{\phi}$) plane. 
Figure~\ref{fig:radion_2g2b_discovery} (left plot) shows the 
5 $\sigma$ discovery contour in the the ($\xi$, $\Lambda_{\phi}$) plane
when the irreducible background only (6.9 events with 30 fb$^{-1}$) was taken
into account. Theoretically excluded regions are also shown in the plot.
Dashed line contours present the discovery reaches when the irreducible 
background cross sections were calculated for the renormalization and 
factorization scales set to 0.5$\times \mu _{\rm 0}$ and to 
2$\times \mu _{\rm 0}$, where $\mu _{\rm 0}$ = $\rm M_{\rm Z}$. The background 
cross section uncertainty due to the scale variation found to be of 
$\simeq$ 40 \% for $\gamma \gamma \rm b \bar{\rm b}$. It was guessed
that the cross section variation for $\gamma \gamma \rm c \bar{\rm c}$ and
$\gamma \gamma \rm j \rm j$ production may be only slightly different, thus 
40 \% variation was applied to the cross sections of 
$\gamma \gamma \rm c \bar{\rm c}$ and $\gamma \gamma \rm j \rm j$ processes
as well.
\begin{figure}[ht!]
\begin{center}
\includegraphics[width=7cm]{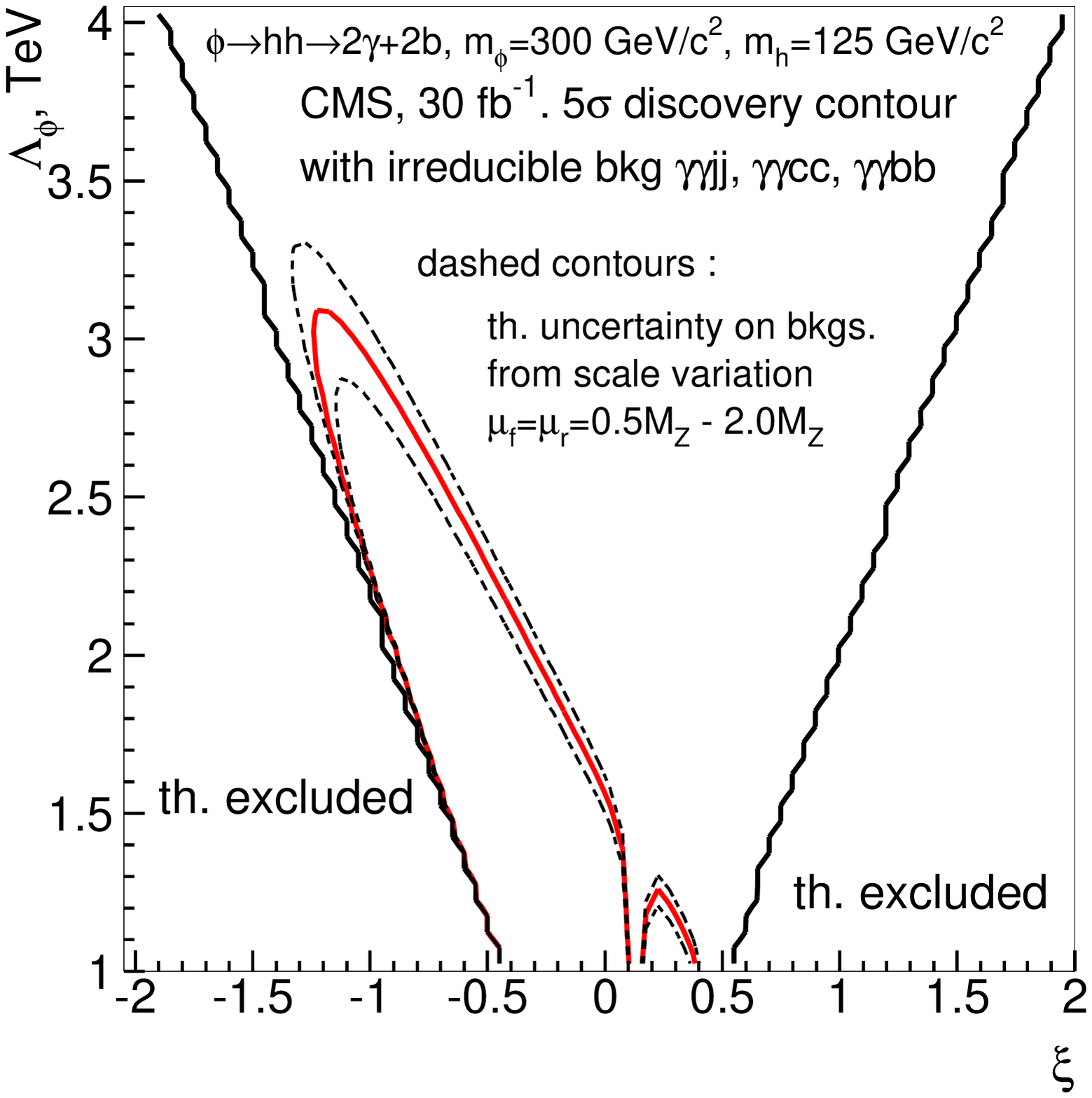}
\includegraphics[width=7cm]{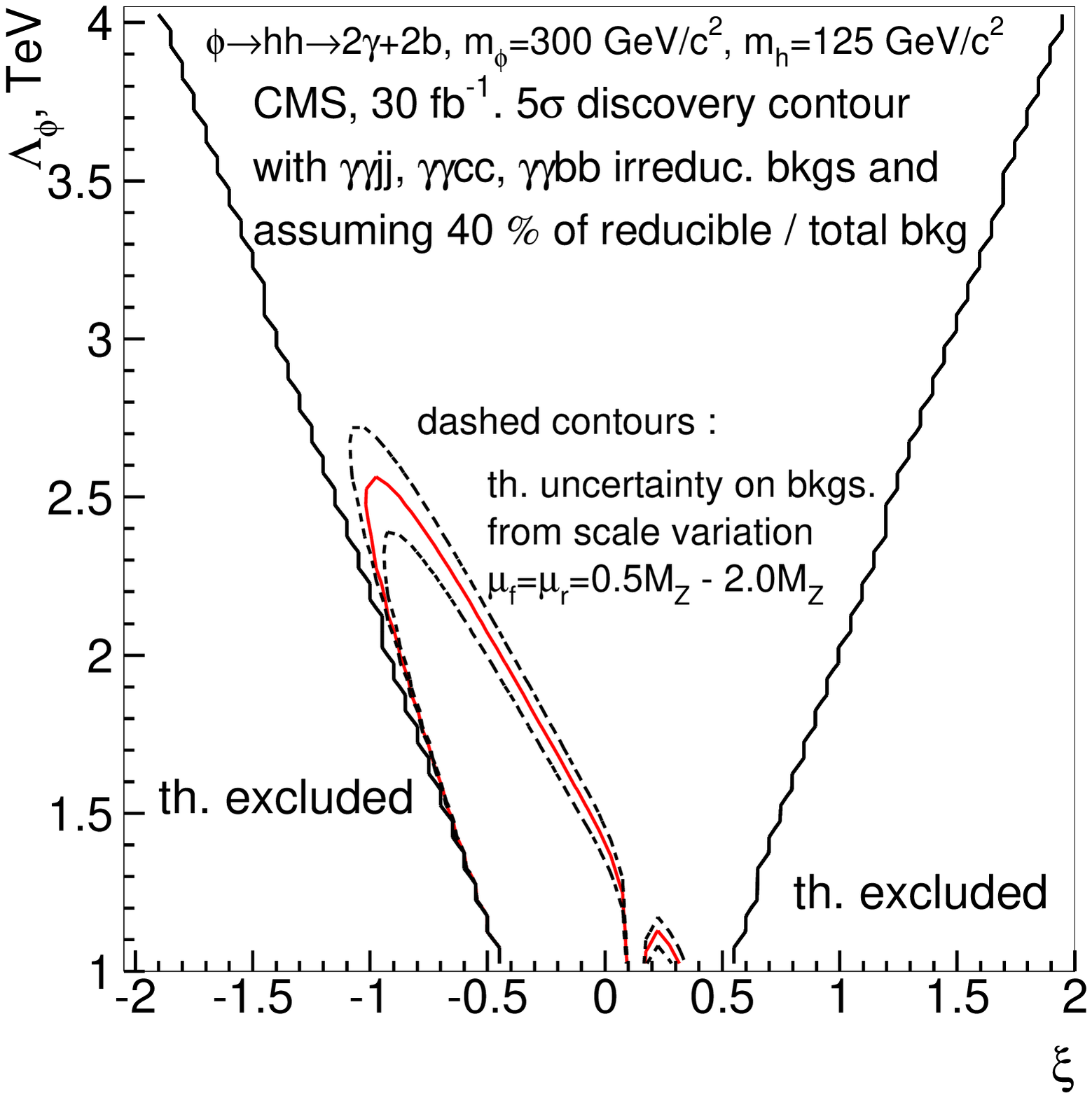}
\caption{5 $\sigma$ discovery contours for
$\phi \rightarrow \rm h \rm h \rightarrow \gamma \gamma$+$\rm b \bar{\rm b}$ 
channel ($\rm m_{\phi}$=300 GeV/$c^2$, $\rm m_{\rm h}$=125 GeV/$c^2$);
Left plot : with the irreducible background only; Right plot : with the total
background assuming the ratio of the reducible to the total background 
of 0.4. Dashed line contours present the discovery reaches when the 
irreducible background cross sections were calculated for the renormalization 
and factorization scales set to 0.5$\times \mu _{\rm 0}$ and to 
2$\times \mu _{\rm 0}$ where $\mu _{\rm 0}$ = $\rm M_{\rm Z}$.}
\label{fig:radion_2g2b_discovery}
\end{center}
\end{figure}
\\
Reducible backgrounds from $\gamma$ + three jets and four-jet processes 
still have to be evaluated. It is expected from the inclusive 
$\rm h \rightarrow \gamma \gamma$ studies that the reducible background 
will be about of 40 \% of the total background, thus the total background
is expected to be of 11.5 events with 30 fb$^{-1}$.
Figure~\ref{fig:radion_2g2b_discovery} (right plot) shows the 
5 $\sigma$ discovery contours in the the ($\xi$, $\Lambda_{\phi}$) plane
with the total background taken into account. The experimental systematics
uncertainty of the background estimated as $\simeq$ 5 \% hardly
affects the discovery reach due to the signal to background ratio in the
5 $\sigma$ region of the ($\xi$, $\Lambda_{\phi}$) plane is bigger than
two.

\subsection{$\tau \tau \rm b \overline{\rm b}$ final state}

The signature when one $\tau$ lepton decays leptonically and 
another $\tau$ lepton decays hadronically (producing a $\tau$ jet) was
considered.
The highest signal cross section times the branching ratios of 0.96 pb was 
obtained for $\xi$ = $-$0.35 and $\Lambda_{\phi}$ = 1 TeV.
The background processes considered in the analysis
are shown in Table~\ref{tab:trigger_eff} with the NLO cross sections taken 
from \cite{Campbell:2003hd,Chakraborty:2003iw, cern:smworkshop}.
\begin{table}[h!]
\caption{Trigger and total efficiency for the signal and the backgrounds; expected number of 
events with 30 fb$^{-1}$.}
\label{tab:trigger_eff}
\begin{center}             
\begin{tabular}{|l|c|c|c|c|} \hline
&&\multicolumn{2}{|c|}{efficiency (\%)}&number of\\\cline{3-4}
samples & $\sigma \times BR$ (pb)&trigger & trigger + off-line & events\\\hline
$\phi \rightarrow \tau\tau \rm b \overline{\rm b}$ &0.96 &6 $\pm$ 0.2&0.35 $\pm$0.06 & 102 \\\hline
$\rm t \overline{\rm t} \rightarrow \rm l + \nu + jets + \rm b  \overline{\rm b}$  
&180& 0.57 $\pm$ 0.02 &(1.6 $\pm$0.2)$\times 10^{-3}$& 111 \\\hline
$\rm t  \overline{\rm t}  \rightarrow \rm l + \nu + \tau ~ \rm jet + \rm b \overline{\rm b}$ & 15& 3.1 $\pm$ 0.2 &(7.7 $\pm$ 0.3)$\times 10^{-3}$&66 \\\hline
$\rm Z \rm b \overline{\rm b} \rightarrow \tau \tau + \rm b \overline{\rm b}$ & 5.4 &1.4 $\pm$ 0.2 &0.009$\pm$0.003& 21 \\\hline
$\rm Z + jets \rightarrow \tau \tau + jets$ & 306 &0.35 $\pm$ 0.02&(3.3 $\pm$ 0.5)$\times 10^{-4}$& 36 \\\hline
$\rm W + jets \rightarrow \rm l + \nu + jets$ &  175   & 0.039 $\pm$ 0.002&0& 0 \\\hline
\end{tabular}
\end{center}
\end{table}
Background Z+jets (W+jets) were
generated with $\hat{\rm p}_{\rm T}>$ 20 (80) GeV/$c$.
Signal events were processed with the full detector simulation and 
reconstruction, while the background was processed with the fast detector 
simulation package CMSJET~\cite{cms:cmsjet}.
The combined electron(muon)-plus-$\tau$ jet trigger~\cite{cms:daq_tdr2}
was used in this analysis. The Level-1 trigger threshold is 21 GeV for
electrons and 45 GeV for the $\tau$ jet.
The inclusive muon threshold is low enough (14 GeV/$c$) to allow a good 
efficiency 
however, to increase the background rejection a $\tau$ jet with 
$\rm E_{\rm T}>$35 GeV is required at the Level-1 trigger. The trigger efficiency is 
shown in 
Table~\ref{tab:trigger_eff} for the signal and background samples.
Missing transverse momentum and b tagging were used to reconstruct 
$\tau$ leptons and to identify $\rm b$ jets coming from the two Higgs bosons. 
In order to increase the signal statistics it was necessary to tag at least one jet. 
The off-line selections are the following:
\begin{itemize}
\item [--] $\Delta \phi$ between lepton and $\tau$ jet direction $>$0.1;
\item [--] $\rm E_{\rm T}$ of the $\rm b$-tagged jets $>$30 GeV, 
and $\rm E_{\rm T}$ of the most energetic jet $>$ 55 GeV;
\item [--] transverse mass of the lepton and missing momentum $<$ 
35 GeV/$c^2$;
\item [--] 75 $< \rm M_{\tau \tau}<$165 GeV/c$^2$, 100 $< \rm M_{\rm b \rm j}<$150 GeV/c$^2$, 
           265 $< \rm M_{\tau \tau \rm b \rm j}<$ 350 GeV/c$^2$.
\end{itemize}
Table~\ref{tab:trigger_eff} shows the signal and the background 
efficiencies for the off-line selections and the number 
of the expected events with 30 fb$^{-1}$.
Figure~\ref{fig:rad_mass} shows the reconstructed 
$\tau \tau \rm b \overline{\rm b}$ mass after all selections (left plot).
The total number of the background events after all selections is 
234 with 30 fb$^{-1}$. The 
biggest background is $\rm t \bar{\rm t}$ (177 events) while 
the $\rm W+jets$ background is negligible.
Estimating also the contribution of $\rm t \bar{\rm t}$ when both W bosons decay into $\tau$ the total
number of background events increases up to 254.
For the maximal signal cross section of 0.96 pb, ($\xi$= $-$0.35 and 
$\Lambda_{\phi}$ = 1 TeV) 
about 102 signal events are expected. Signal significance 
($\rm S/\sqrt{\rm B}$) at 
this point is 6.4.
Figure~\ref{fig:rad_mass} (right plot) shows a 5 $\sigma$ discovery contour in 
the ($\xi$, $\Lambda_{\phi}$) plane.The two contours correspond to the 
uncertainties of the background cross section values at NLO due to the
scale variation and different PDFs 
\cite{Campbell:2003hd,Chakraborty:2003iw, cern:smworkshop}. The experimental
systematics uncertainty of 3 \% for the total background was taken into 
account.  
\begin{figure}[ht!]
\begin{center}
\includegraphics[width=7cm]{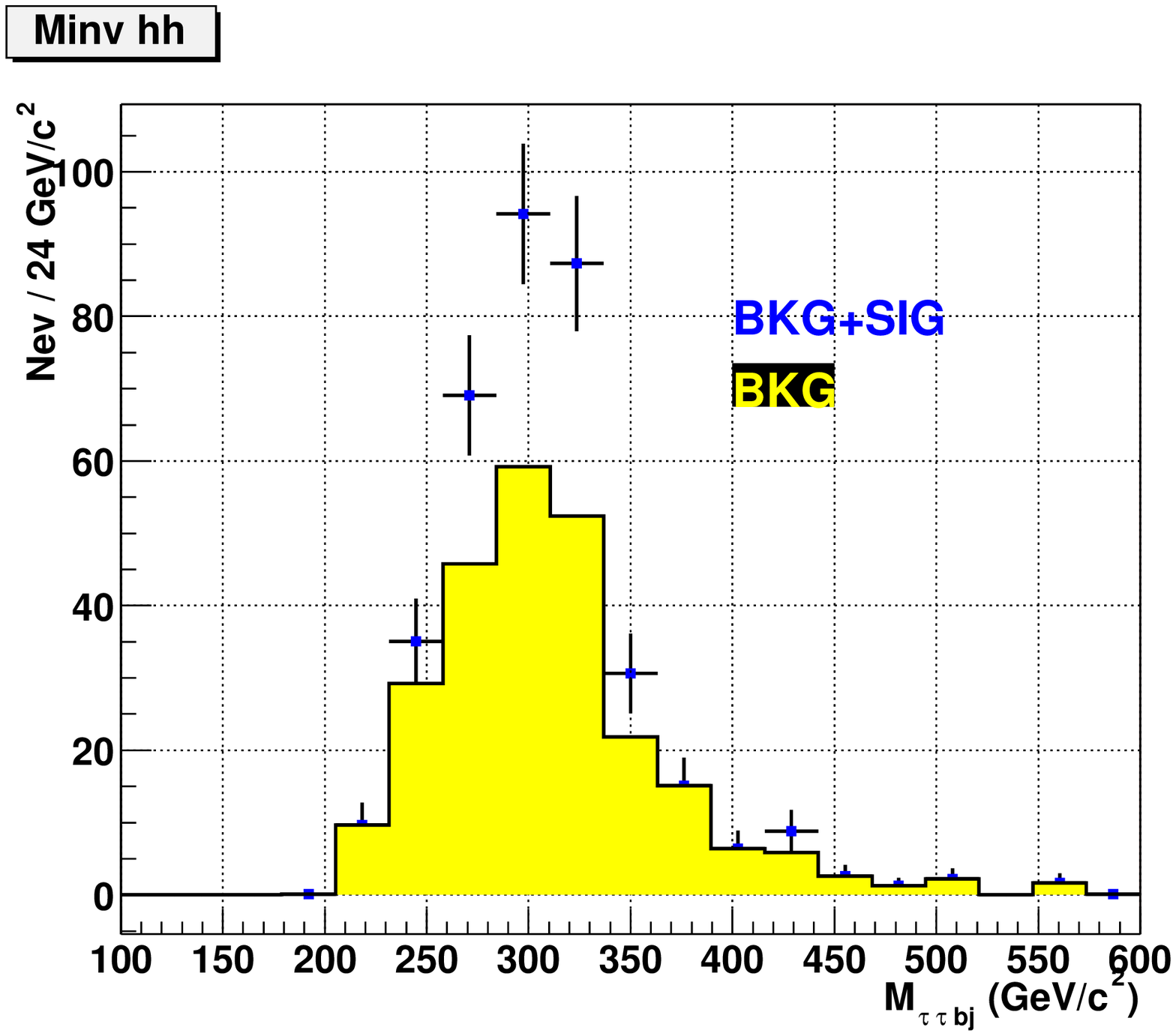}
\includegraphics[width=8cm]{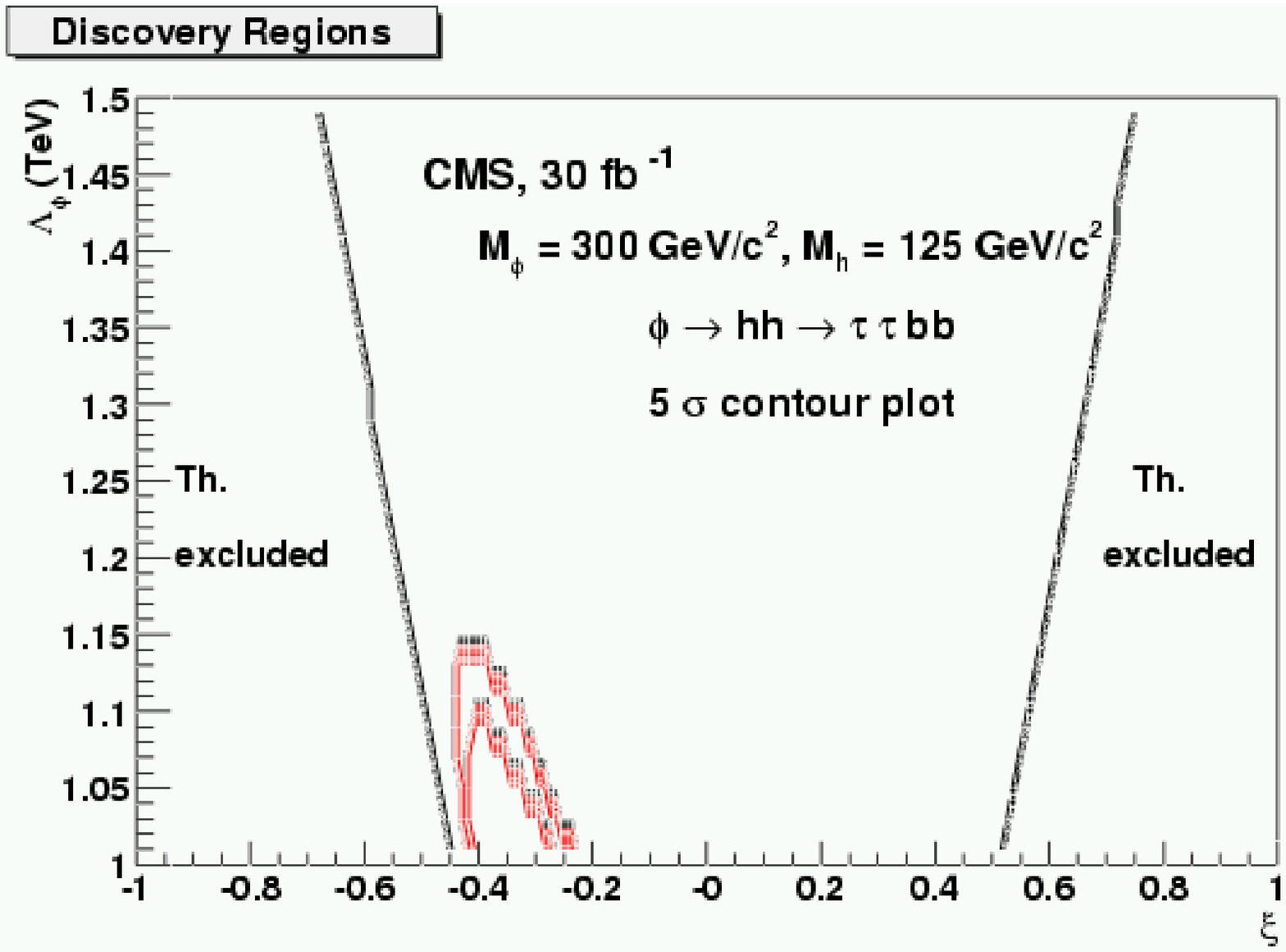}
\caption{Reconstructed $\tau \tau \rm b \rm j$ mass for the signal 
and background (left) 
and 5 $\sigma$ discovery contour (right).The two contours correspond to the maximum and minimum k-factor value
derives from  the NLO order calculation of background cross section.}
\label{fig:rad_mass}
\end{center}
\end{figure}

\subsection{$\rm b \bar{\rm b} \rm b \bar{\rm b}$ final state}

The signal cross section times branching ratios for the
$\rm g \rm g \rightarrow \phi \rightarrow \rm h \rm h \rightarrow \rm b \bar{\rm b} \rm b \bar{\rm b}$ process is 
10.3 pb for $\Lambda_{\phi}$ = 1 TeV and $\xi$ = $-$ 0.35 
The main QCD multi-jet background was generated by PYTHIA 
in different $\hat{\rm p}_{\rm T}$ bins. Other backgrounds considered are 
$\rm t\bar{\rm t}$, $\rm t\bar{\rm t} \rm j \rm j$ and $\rm Z \rm b\bar{\rm b}$. In Table~\ref{tabxsec} 
cross sections and expected numbers of events with 30 fb$^{-1}$ are summarized.
\begin{table}[ht]
\caption{Signal and background events with 30fb$^{-1}$.
\label{tabxsec}}
\centering
\begin{tabular}{|c|c|c|}\hline\
 &  cross section & events in 30 fb$^{-1}$
\\
\hline\hline
Signal & 10.3 pb & $3.1\times 10^{5}$
\\ \hline\hline
QCD$_{\hat{\rm p}_{\rm T}(30-170)}$ & 0.2257 mb & $6.79\times 10^{12}$ 
\\ \hline\hline
$\rm t\bar{\rm t}$ & 615 pb & $1.8\times 10^{7}$
\\ \hline
$\rm t\bar{\rm t}$jj & 507 pb & $1.5\times 10^{7}$
\\ \hline
$\rm Z \rm b\bar{\rm b}$ & 349 pb & $1.0\times 10^{7}$
\\ \hline
\end{tabular}
\end{table}
A rejection factor on background higher than $10^{6}$ is needed to reach a 
5$\sigma$ statistical evidence of the signal. A fast detector simulation with the
CMSJET package 
~\cite{cms:cmsjet} was used for both signal and background samples.
Dedicated trigger selections were developed to keep the QCD multi-jet 
background rate at the acceptable level whilst maintaining a high efficiency 
for the signal. At Level-1 multi-jet triggers were used with the thresholds 
taken from Table 15-13 presented in~\cite{cms:daq_tdr2} and 
restricted in 
pseudorapidity, $|\eta|<$0.8. At the High Level Trigger at least 4 jets were
required within the restricted pseudorapidity range, $|\eta|<$0.8. Two jets must be b-tagged 
with the impact parameter tagging method  
(2 associated tracks with significance on the transverse impact parameter $>$ 2). The output QCD rate 
after these selections is $\sim$ 5 Hz. In off-line selections all possible di-jet invariant masses were
calculated from the 4 highest $\rm E_{\rm T}$ jets. The two jet pairs were chosen minimizing the value 
of $\rm m_{i,j}$-$\rm m_{k,l}$, the same jets were then used to reconstruct the radion mass. 
The mean values (and $\sigma_{\rm fit}$) of the di-jet and four-jet effective masses reconstructed in this 
way are: 120 (39) GeV/$c^{2}$ and  313 (76) GeV/$c^{2}$. 
A 1.5 $\sigma$ window in mass around $\rm m_{\rm h}$ and $\rm m_{\phi}$ was used to select signal and 
background events. Efficiencies for the signal, background and the expected number of events with 
30 fb$^{-1}$ are summarized in Table~\ref{finale}. 
\begin{table}[ht]
\caption{Trigger and total selection efficiency and expected number of events with 30 fb$^{-1}$
\label{finale}}
\centering
\begin{tabular}{|c|c|c|c|}\hline\
 & $\epsilon_{\rm trigger}$  & $\epsilon_{\rm total}$  & events 
\\ \hline\hline
 signal & 0.038 & 0.031 & 9.57$^{.}10^{3}$  
\\ \hline\hline
 QCD $\hat{\rm p}_{\rm T}(80-120)$ & $1^{.}10^{-5}$ & $7^{.}10^{-6}$  & $7.5^{.}10^{5}$ 
\\ \hline
 QCD $\hat{\rm p}_{\rm T}(120-170)$ & $1^{.}10^{-4}$ & $6.6^{.}10^{-5}$  & $1.1^{.}10^{6}$ 
\\ \hline\hline
$\rm t\bar{\rm t}$ & 0.015 & 0.010 & $1.84^{.}10^{5}$
\\\hline
$\rm t\bar{\rm t}$jj & 0.056 & 0.026 & $1.8^{.}10^{5}$
\\\hline
$\rm Z \rm b\bar{\rm b} \rightarrow$ 4b  & 0.002 & 8$^{.}10^{-4}$ & $1.2^{.}10^{3}$
\\\hline
\end{tabular}
\end{table}
The di-jet and four-jet invariant mass for the background and the signal at 
$\Lambda_{\phi}$ = 1 TeV and $\xi$ = $-$0.35 point are shown respectively in 
the left and the right plots of Figure~\ref{rad_mass_bbbb}.
For this point it may be possible to achieve a signal significance 
($\rm S/\sqrt{\rm B}$) of 5.5 if the background shape
of the four-jet mass distribution is well understood (with $\sim0.1$\%
uncertainty). 
\begin{figure}[ht!]
\begin{center}
\includegraphics[width=7cm]{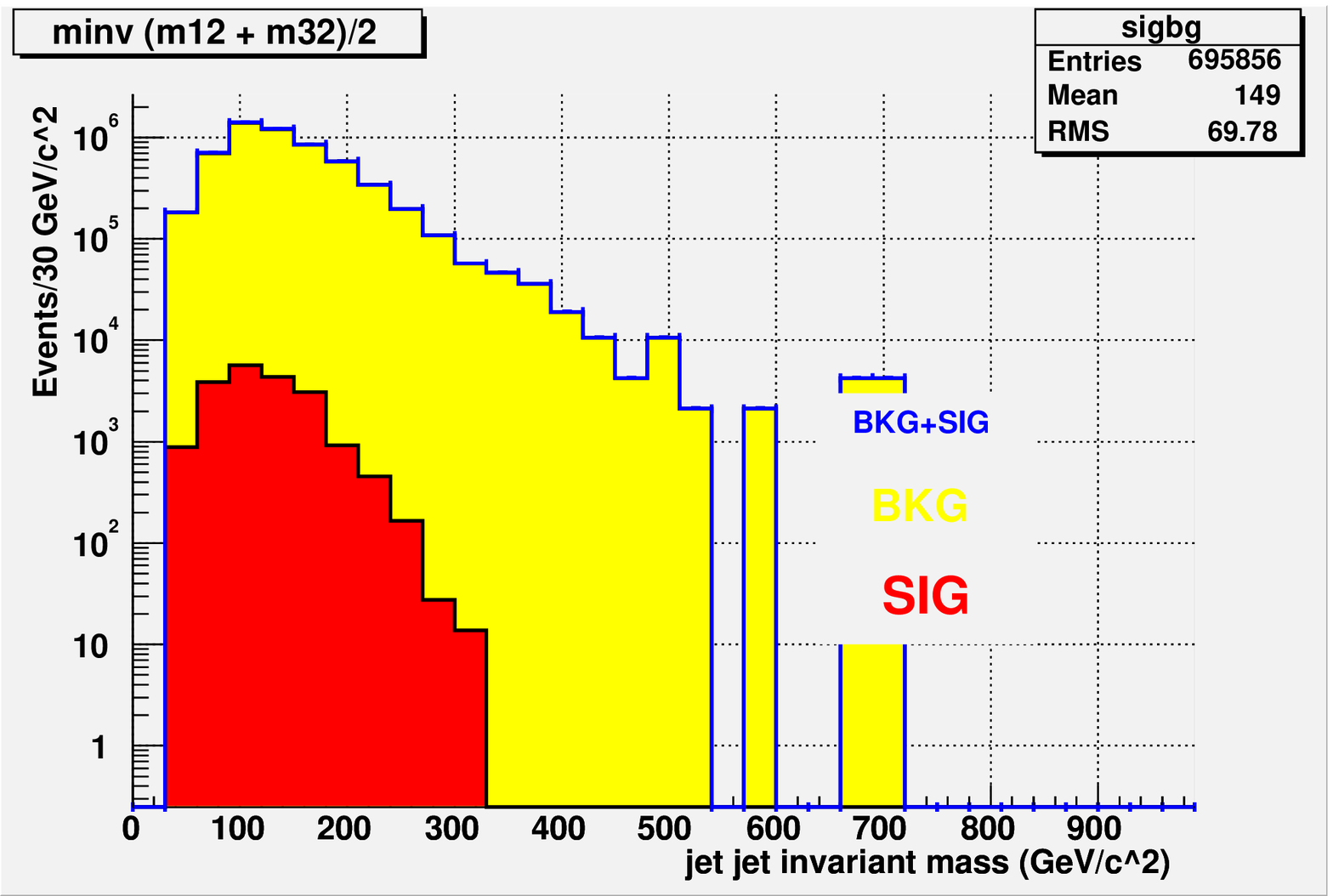}
\includegraphics[width=7cm]{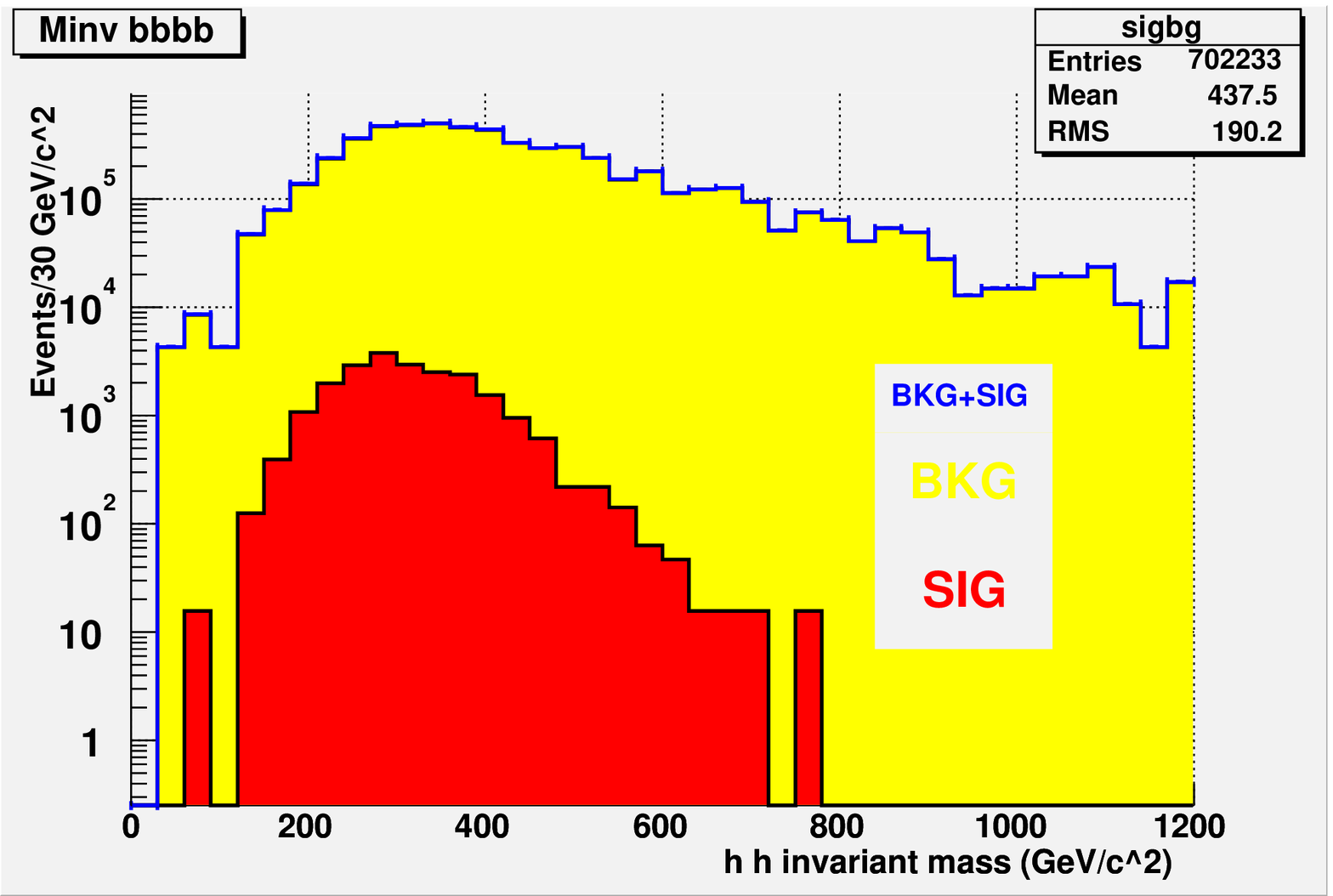}
\caption{Reconstructed di-jet (left) and four-jet mass (right) for signal and background.}
\label{rad_mass_bbbb}
\end{center}
\end{figure}

\section{CONCLUSION}

We estimated the CMS discovery potential for the radion into two Higgs decay
mode ($\phi \rightarrow \rm h \rm h$) with 
$\gamma \gamma$+$\rm b \bar{\rm b}$, $\tau \tau$+$\rm b \bar{\rm b}$ and
$\rm b \bar{\rm b}$+$\rm b \bar{\rm b}$ final states. One point of 
$\rm m_{\phi}$=300 GeV/$c^2$ and $\rm m_{\rm h}$=125 GeV/$c^2$ was 
taken and the observability in the ($\xi$, $\Lambda_{\phi}$) plane was 
evaluated. It was found that the $\gamma \gamma$+$\rm b \bar{\rm b}$ 
topology provides the best discovery potential.
The $\rm b \bar{\rm b}$+$\rm b \bar{\rm b}$ final state requires the 
dedicated High Level Trigger with the double b tagging and an excellent 
understanding of the background shape. 

\vskip1cm
\noindent

\section*{ACKNOWLEDGEMENTS}

A.N. would like to thank M.~Dubinin for CompHEP generation of
$\gamma \gamma \rm j \rm j$ events and F.~Maltoni for explanations  
about the MadGraph generator.

\setcounter{figure}{0}
\setcounter{table}{0}
\setcounter{section}{0}
\setcounter{equation}{0}
\clearpage

 \part{The Invisible Higgs Decay Width
In The  ADD Model At The LHC \label{inv_proc_3}}
{\it M. Battaglia, D. Dominici, J.F. Gunion and J.D. Wells}

\maketitle
\begin{abstract}
Assuming flat universal extra dimensions,
we demonstrate that for a light Higgs boson 
the process $pp\to W^*W^* +X \to {\rm Higgs, graviscalars} 
+X \to {\rm invisible}+X$
will be observable at the $5~\sigma$ level at the LHC 
over the
portion of the Higgs-graviscalar mixing ($\xi$) and effective
Planck mass ($M_D$) parameter space where channels
relying on visible Higgs decays fail to achieve
a $5~\sigma$ signal.  
Further, we show that for some values of $\delta$
and $\xi\sim O (1)$ the invisible decay signal can
probe values of $M_D$ up to and possibly above those probed by the 
($\xi$-independent) jets/$\gamma$ + missing energy signal
from graviton radiation. 
 We also discuss various
effects, such as Higgs decay to two graviscalars, that
could become important when $m_h/M_D$ is of order 1.
\end{abstract}
\def\to{\rightarrow}
\def\ptl{\partial}
\def\beq{\begin{equation}}
\def\eeq{\end{equation}}
\def\bea{\begin{eqnarray}}
\def\eea{\end{eqnarray}}
\def\nn{\nonumber}
\def\half{{1\over 2}}
\def\rhalf{{1\over \sqrt 2}}
\def\calo{{\cal O}}
\def\cala{{\cal A}}
\def\call{{\cal L}}
\def\calm{{\cal M}}
\def\del{\delta}
\def\eps{\epsilon}
\def\lam{\lambda}
\def\anti{\overline}
\def\delfac{\sqrt{{2(\del-1)\over 3(\del+2)}}}
\def\heff{h'}
\def\square{\boxxit{0.4pt}{\fillboxx{7pt}{7pt}}\hspace*{1pt}}
    \def\boxxit#1#2{\vbox{\hrule height #1 \hbox {\vrule width #1
             \vbox{#2}\vrule width #1 }\hrule height #1 } }
    \def\fillboxx#1#2{\hbox to #1{\vbox to #2{\vfil}\hfil}   }

\def\gev{~{\rm GeV}}
\def\mev{~{\rm MeV}}
\def\gam{\gamma}
\def\sn{s_{\vec n}}
\def\sm{s_{\vec m}}
\def\mm{m_{\vec m}}
\def\mn{m_{\vec n}}
\def\mh{m_h}
\def\sumn{\sum_{\vec n>0}}
\def\summ{\sum_{\vec m>0}}
\def\vl{\vec l}
\def\vk{\vec k}
\def\ml{m_{\vl}}
\def\mk{m_{\vk}}
\section{INTRODUCTION}
In several  extensions of the Standard Model (SM) there
exist mechanisms that  modify the Higgs production/decay 
rates in channels that are observable at the LHC.
One example is the Randall Sundrum model
where the   Higgs-radion mixing not only gives detectable 
reductions (or enhancements) in Higgs 
yields, but also allows the possibility of direct 
observation of radion production and 
decay~\cite{Dominici:2002jv,Battaglia:2003gb}. 
It is also possible for the Higgs rate 
in visible channels to be reduced
as a result of a substantial invisible width.
For example, this occurs in supersymmetric
models when the Higgs has a large 
branching ratio into the lightest gravitinos or neutralinos.
Invisible decay of the Higgs is also predicted 
in models with large extra dimensions felt by gravity (ADD)
\cite{Arkani-Hamed:1998rs,Antoniadis:1998ig}. 
In ADD models the presence of an
interaction between the Higgs $H$ and the 
Ricci scalar curvature of the induced 4-dimensional metric $g_{\rm ind}$, 
generates, after the usual shift $H=({v+ h\over \sqrt{2}},0)$,
the following mixing term \cite{Giudice:2000av}
\begin{equation}
{\cal L}_{\rm mix}=\epsilon  h \sum_{\vec n >0}s_{\vec n}
\end{equation}
with
\beq
\eps=-{2\sqrt 2\over M_P}\xi v \mh^2\sqrt{{3(\del-1)\over \del+2}}\,.
\eeq
Above, $M_P=(8\pi G_N)^{-1/2}$ is the Planck mass, $\delta$ is the number of extra
dimensions, $\xi$ is a dimensionless parameter and
$s_{\vec n}$ is a graviscalar KK excitation with
mass $m_{\vec n}^2=4\pi^2 \vec n^2/L^2$, $L$ being the
size of each of the extra dimensions.
(Note that with respect to \cite{Giudice:2000av}
our normalization is such that we have taken
only the real part of the $\phi_G^{\vec n}$ fields,
writing $\phi_G^{\vec n}={1\over\sqrt 2}(s_{\vec n}+i a_{\vec n})$ 
and using $\phi_G^{\vec n}=[\phi_G^{-\vec n}]^*$
to restrict  sums to $\vec n>0$, by which we mean that the first non-zero entry
of $\vec n$ is positive.) After diagonalization of the full mass-squared matrix
one finds that the physical eigenstate, $h'$,  acquires
admixtures of the graviscalar states and vice versa.
Dropping $\calo(\eps^2)$ terms and higher,
\beq
h'\sim \left[h-\sum_{\vec m>0}{\eps\over \mh^2-i \mh
\Gamma_{h}-m_{\vec m}^2}s_{\vec m}\right]\,,\quad
s'_{\vec m}\sim \left[ s_{\vec m}+{\eps\over \mh^2-i\mh\Gamma_{h} 
-m_{\vec m}^2} h\right]\,.
\label{eigenstate}
\eeq
In computing a process such as $WW\to h'+\sum_{\vec m>0}s_{\vec m}' \to F$,
normalization and admixture corrections of order $\eps^2$ that 
are present must be taken into account and the full coherent sum
over physical states must be performed. The result at the
amplitude level is 
\beq
\cala(WW\to F)(p^2)\sim {g_{WWh}g_{h F}\over  
p^2-\mh^2+i\mh\Gamma_h+iG(p^2)+F(p^2)}
\label{amplitude}
\eeq
where 
$F(p^2)\equiv -\eps^2 {\rm Re} \left[\sum_{\vec m>0}{1\over
    p^2-m_{\vec m}^2}\right]$ and $G(p^2)\equiv
-\eps^2{\rm Im}\left[\sum_{\vec m>0}{1\over p^2-m_{\vec m}^2}\right]$.
Taking the amplitude squared and
integrating over $dp^2$ in the narrow width approximation
gives the result
\beq
\sigma(WW\to h'+\sum_{\vec m>0}s'_{\vec m}\to F)=\sigma_{SM}(WW\to h \to F)
\left[{1\over 1+F'(m_{h\,{\rm ren}}^2)}\right]\left[{\Gamma_h\over \Gamma_h+\Gamma_{h\to {\rm graviscalar}}}\right]
\label{xsec}
\eeq
where $m_{h\,{\rm ren}}^2-\mh^2+F(m_{h\,{\rm ren}}^2)=0$ and we have defined
$\mh\Gamma_{h\to {\rm graviscalar}}\equiv G(m_{h\,{\rm ren}}^2)$. We will argue
that for a light Higgs boson both the wave function renormalization
and the mass renormalization effects will be small. In this case, 
the coherently summed
amplitude gives the Standard Model cross section suppressed by the
ratio of the SM Higgs width to the sum of the SM Higgs width and the
Higgs width arising from mixing with the graviscalars.

\section{INVISIBLE WIDTH}
As described, there is a decay of the Higgs arising from
the mixing (or oscillation) 
of the Higgs itself into the closest KK graviscalar
levels. These graviscalars are invisible since they
are weakly interacting and mainly reside in the 
extra dimensions whereas the Higgs resides on the brane.
The mixing width 
$\Gamma_{h\to {\rm graviscalar}}\sim G(\mh^2)/\mh$ thus corresponds to an invisible decay 
width. The equation for $G(\mh^2)$ 
below eq.~(\ref{amplitude}) 
shows that it is calculated by extracting the imaginary
part of the mixing contribution to the Higgs self energy.
The result is \cite{Giudice:2000av,Wells:2002gq}
\begin{eqnarray}
\Gamma(h\to {\rm graviscalar})&\equiv& \Gamma(h\to 
\sum_{\vec n>0}s_{\vec n})=
2\pi\xi^2 v^2 \frac {3(\delta -1)}
{\delta +2}
\frac {m_h^{1+\delta}}{M_D^{2+\delta}}{S_{\delta -1}}\nn\\
&\sim& (16\,\mev) 20^{\delta -2} \xi^2
S_{\delta-1}\frac {3(\delta -1)}
{\delta +2} \left ( \frac {m_h}{150\, \gev} \right )^{1+\delta}
\left ( \frac 
{3\, {\rm TeV}} {M_D}\right )^{2+\delta}
\label{invwidth}
\end{eqnarray}
where $S_{\delta-1}=2\pi^{\delta/2}/\Gamma(\delta/2)$ denotes the
surface of a unit radius sphere in $\delta$ dimensions while $M_D$ is
related to the $D$ dimensional reduced Planck constant ${\overline
  M}_D$ by $M_D= (2\pi)^{\delta/(2+\delta)}{\overline M}_D$.  Our eqs.
(\ref{invwidth}) are a factor of 2 larger than those presented in
refs.~\cite{Giudice:2000av,Wells:2002gq}.

\subsection{The wave function renormalization factor and mass renormalization}

A simple estimate of the quantity $F'(m_{h\,{\rm ren}}^2)$, appearing in the
wave function renormalization factor found in eq.~(\ref{xsec}),
suggests that it is of order $\xi^2{\mh^4\over \Lambda^4}$, where
$\Lambda$ is an unknown ultraviolet cutoff energy presumably of order
$\Lambda\sim M_D$ \cite{Giudice:1998ck}. 
Assuming this to be the case, $F'$ will
provide a correction to coherently computed LHC production cross
sections that is  very probably quite small for the
$\mh\ll M_D$ cases that we are about to explore.  However, one must
keep in mind that a precise calculation of $F'$ is not possible.
Similarly, the mass renormalization from $F(m_{h\,{\rm ren}}^2)$ should be
of order $\xi^2\mh^6/M_D^4$ and, therefore, small for $\mh\ll M_D$.
There are other incomputable sources of $v^4/M_D^4$ corrections
lurking in the theory beyond these sources, and the results presented
here are computed using the first, and perhaps only, calculable terms
in the perturbation series.

\subsection{Contribution to the invisible width from direct
two graviscalar decay}  

In addition to decay by mixing, one expects also a contribution to the
invisible width of the Higgs from its decays into two graviscalars.
This can be evaluated by using the transformation of
eq.~(\ref{eigenstate}) between the physical eigenstate $h'$ and the
unmixed $h$ to derive the relevant trilinear $h' s_k s_l$
vertices. These are used to compute the corresponding matrix element.
The final expression for $\Gamma(h'\to {\rm graviscalar~pairs})$ can be
written as
\begin{eqnarray}
\Gamma(h'\to {\rm graviscalar~pairs})
&=& {18\over \pi} {m_h^{3+2\del}v^2\over M_D^{4+2\del}}\xi^4
\left({\del-1\over\del+2}\right)^2 \left[{\pi^{\del/2}\over \Gamma(\del/2)}\right]^2 I\,,
\end{eqnarray}
where $I$ is an integral coming from the sum over all the possible
kinematically allowed $h'\to s_k s_l$ decays.  The integral $I$
decreases rapidly as $\del$ increases.  As a result, $\Gamma(h'\to
{\rm graviscalar~pairs})$ is only significant compared with $\Gamma(h\to
{\rm graviscalar})$ if $\del\leq 4$.  The ratio of the two widths is given
by:
\begin{equation}
{\Gamma(h'\to {\rm graviscalar~pairs})\over \Gamma(h\to {\rm graviscalar})}=
 {3(\del -1)\over 2\pi^2(\del+2)}\xi^2 \left({m_h\over M_D}\right)^{2+\del}
{\pi^{\del/2}\over \Gamma(\del/2)}I\,.
\end{equation}
From this result, we immediately see that even for small $\del$ the
pair invisible width will be smaller than the mixing invisible width
unless $m_h$ is comparable to $M_D$. 

To lowest order in $\xi^2(\mh/M_D)^{2+\del}$, 
decays of other states nearly degenerate with the $h'$
can be neglected in the computation of a 
cross section obtained by coherently summing over the 
$h'$ and the nearly degenerate $s'_{\vec m}$ states.
Thus, to this same order of approximation,
$\Gamma(h'\to {\rm graviscalar~pairs})$ should simply be
added to $\Gamma(h\to {\rm graviscalar})$ in the expression for
the narrow-width cross section of eq.~(\ref{xsec}). 

\begin{figure}[htbp]
\begin{center}
\includegraphics[width=7.0cm,height=5.0cm]{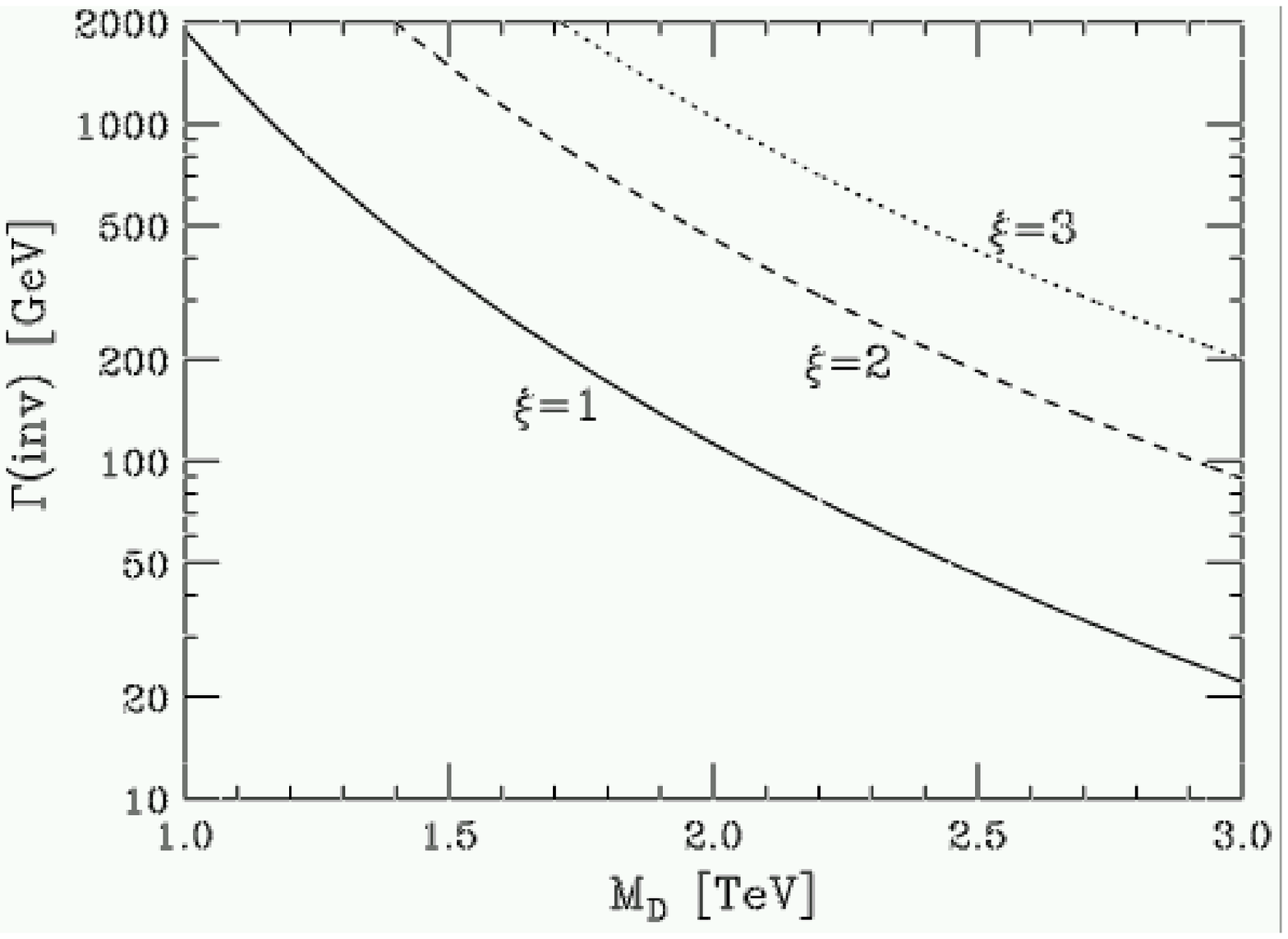} 
\includegraphics[width=7.0cm,height=5.0cm]{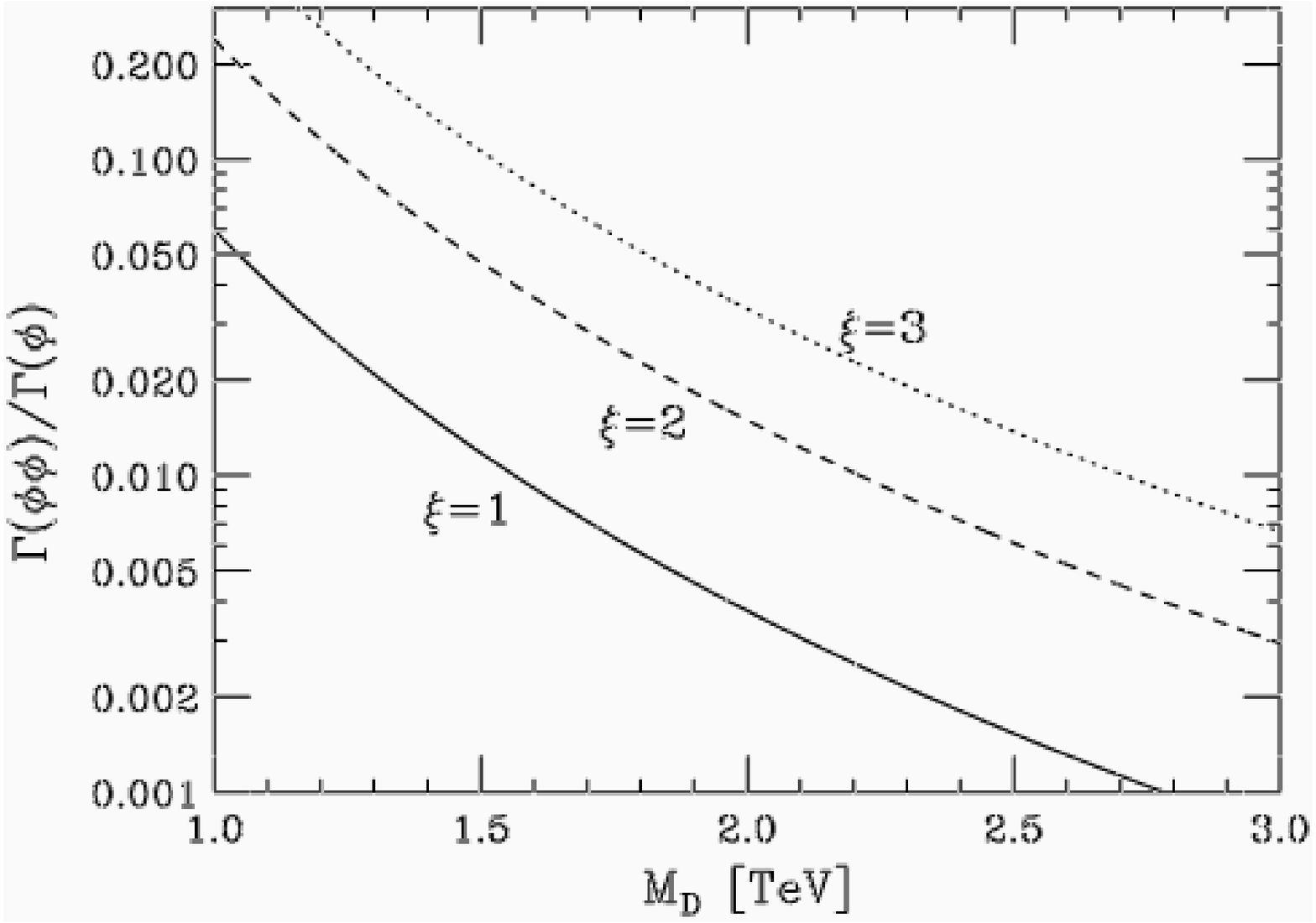}
\caption{In the left-hand plot, we display the total invisible width of a
  1~TeV Higgs boson into one and two graviscalars as a function of
  $M_D$ for various values of $\xi$ ($\xi=1$ solid, $\xi=2$ dashed,
  $\xi=3$ dotted).  For these plots we have fixed $\del=2$. The plot on
  the right shows the ratio of the two-graviscalars decay width 
  to the one-graviscalar decay width for the same choices of
  parameters.}
\label{paircomp}
\end{center}
\end{figure}
In Figure~\ref{paircomp}, we show an extreme case corresponding to
$\del=2$ and $m_h=1000\gev$. Depending on the values of the parameters
 $\xi$ and $M_D$, the pair
invisible width can be a significant correction to the invisible width
from direct mixing.  More generally, for $m_h>M_D$ the
graviscalar-pair invisible width can provide a 3\% to 20\% correction
to the direct-graviscalar-mixing invisible width. However, if $m_h$ is
substantially smaller than $M_D$, then the graviscalar pair width is
not important.  For example, for $\del=2$, $\mh=120\gev$ and
$M_D=500\gev$, $\Gamma(h'\to {\rm graviscalar~pairs})/\Gamma(h\to
{\rm graviscalar})<0.0015$ for $\xi<2$.  Therefore, in the following
analysis, where we will assume a light Higgs, we can safely neglect
the contribution to the invisible width from the decay into two
graviscalars and use the expression given by eq.~(\ref{invwidth}).

\section{MEASUREMENTS AT THE LHC}

For a Higgs boson with $m_h$ below the $WW$ threshold, the invisible
width causes a significant suppression of the LHC Higgs rate in the
standard visible channels.  For example, for $\delta=2$, $M_D=500\gev$ and
$\mh=120\gev$, $\Gamma(h\to {\rm graviscalar})$ is of order $50\gev$ already
by $\xi\sim 1$, {\it i.e.} far larger than the SM prediction of
$3.6\mev$.  Even when $m_h$ is greater than the $WW$ threshold,
Figure~\ref{paircomp} shows that the partial width into invisible states
can be substantial even for $M_D$ values of several TeV; therefore,
for any given value of the Higgs boson mass, there is a considerable
parameter space where the invisible decay width of the Higgs boson
could be the first measured phenomenological effect from extra
dimensions.

Detailed studies of the Higgs boson signal significance, with
inclusive production, have been carried out by the 
  ATLAS~\cite{atlasTDR} and CMS ~\cite{cmsnote}
experiments. If 115~GeV $< \mh <$ 130~GeV, the $h \to \gamma \gamma$
channel appears to be instrumental for obtaining a $\ge 5 \sigma$
signal at low luminosity.  The $t \bar{t} h$, $h \to b \bar{b}$ and $h
\to ZZ^{*} \to 4~\ell$ channels also contribute, with lower statistics
but a more favorable signal-to-background ratio. Preliminary results
indicate that Higgs boson production in association with forward jets
may also be considered as a discovery mode. However, here the
background reduction strongly relies on the detailed detector
response.

In the ADD model, these results are 
modified by the appearance of an invisible decay 
width suppressing the Higgs signal in the standard
visible channels. 
Here, we fix $\mh=120\gev$ and perform a full scan of the ADD parameter 
space by varying $M_D$ and $\xi$ for different values of the number of extra dimensions $\delta$ and 
demonstrate that there are regions at high $\xi$ 
where the significance of the 
Higgs boson signal in the canonical channels drops below the 5~$\sigma$ threshold. 
However, the LHC experiments will also be sensitive to an invisibly decaying Higgs 
boson through $WW$-fusion production, 
with tagged forward jets. A detailed CMS 
study has shown that, with only 10~fb$^{-1}$, an invisible 
channel rate of 
$\Gamma_{\rm inv}/\Gamma$=0.12-0.20 times the SM
$WW\to$Higgs production rate 
gives a signal exceeding the 95\% CL  significance for 
120~GeV $< \mh  <$ 400~GeV~\cite{DiGirolamo:2001yv,cmsnote}. 
Given that the effective (from the sum
over the $h$ state and nearby degenerate states)
$WWh$ coupling is of SM strength,
this defines the region in the ADD parameter space
where the Higgs boson signal can be recovered 
through its invisible decay .

Figure~ \ref{figure1} summarizes the results for specific choices of 
parameters. In the green (light grey) region, 
the Higgs signal in standard channels drops 
below the 5 $\sigma$ threshold  
with 30 fb$^{-1}$ of LHC data. But in the area above the
bold blue line the LHC search for invisible decays 
in the fusion channel yields a 
signal with an estimated significance exceeding 5 $\sigma$.
It is important to observe that, whenever the Higgs boson sensitivity is 
lost due to the suppression of the canonical decay modes, the invisible 
rate is large enough to still ensure detection through a dedicated analysis.
\begin{figure}[h]
\begin{center}
\begin{tabular}{c c}
\includegraphics[width=6.0cm,height=6.0cm]{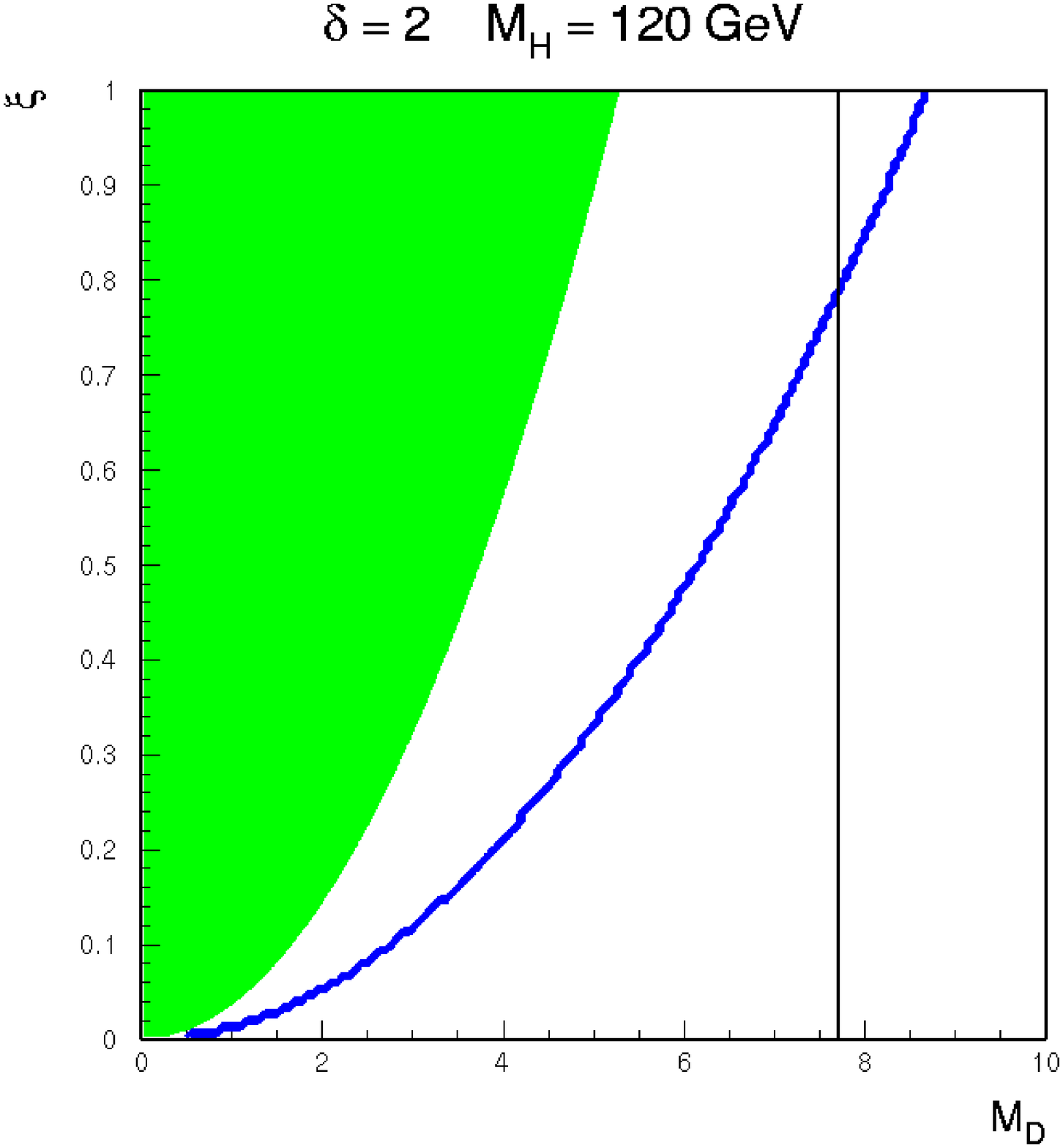} &
\includegraphics[width=6.0cm,height=6.0cm]{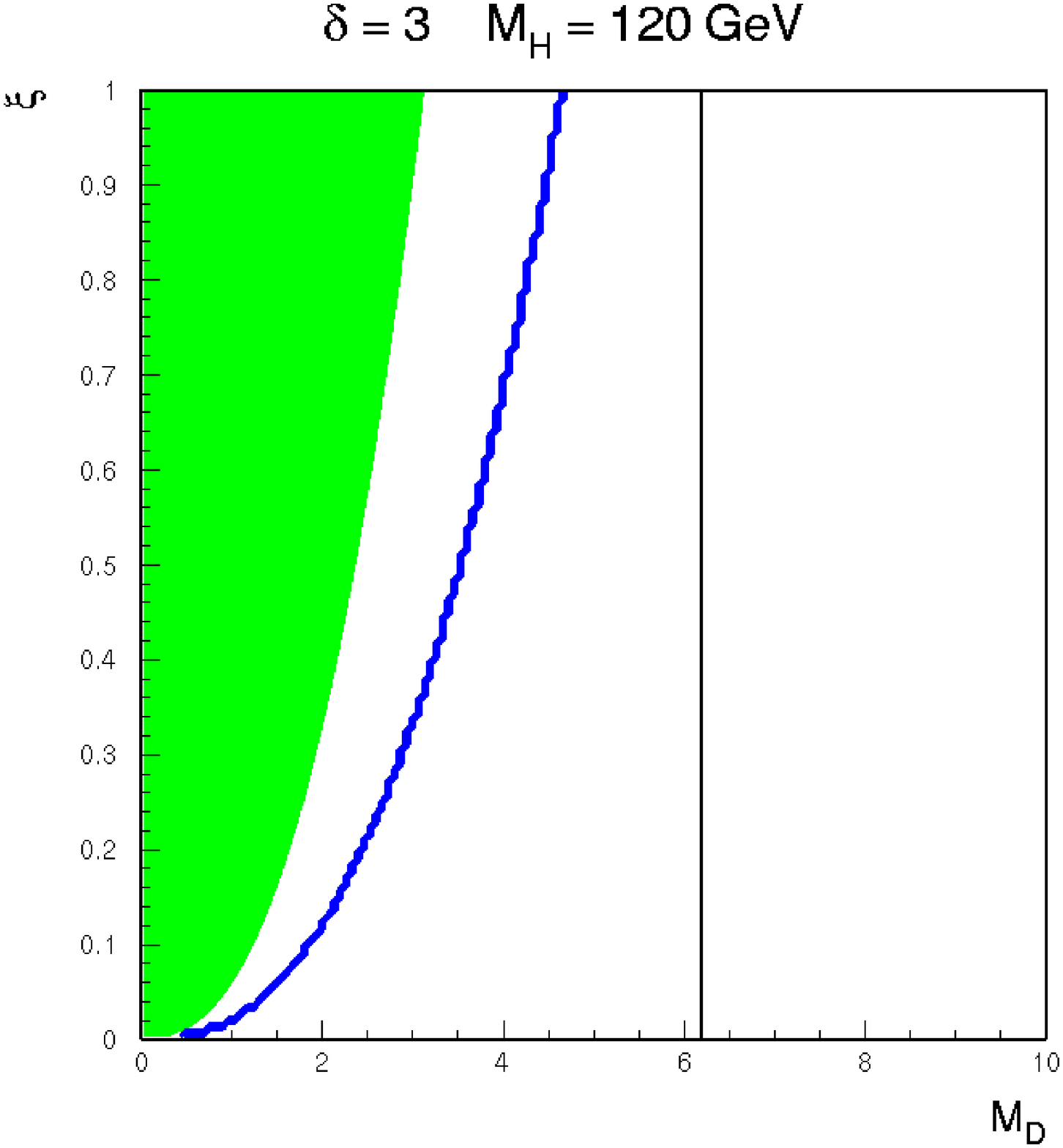}\\
\includegraphics[width=6.0cm,height=6.0cm]{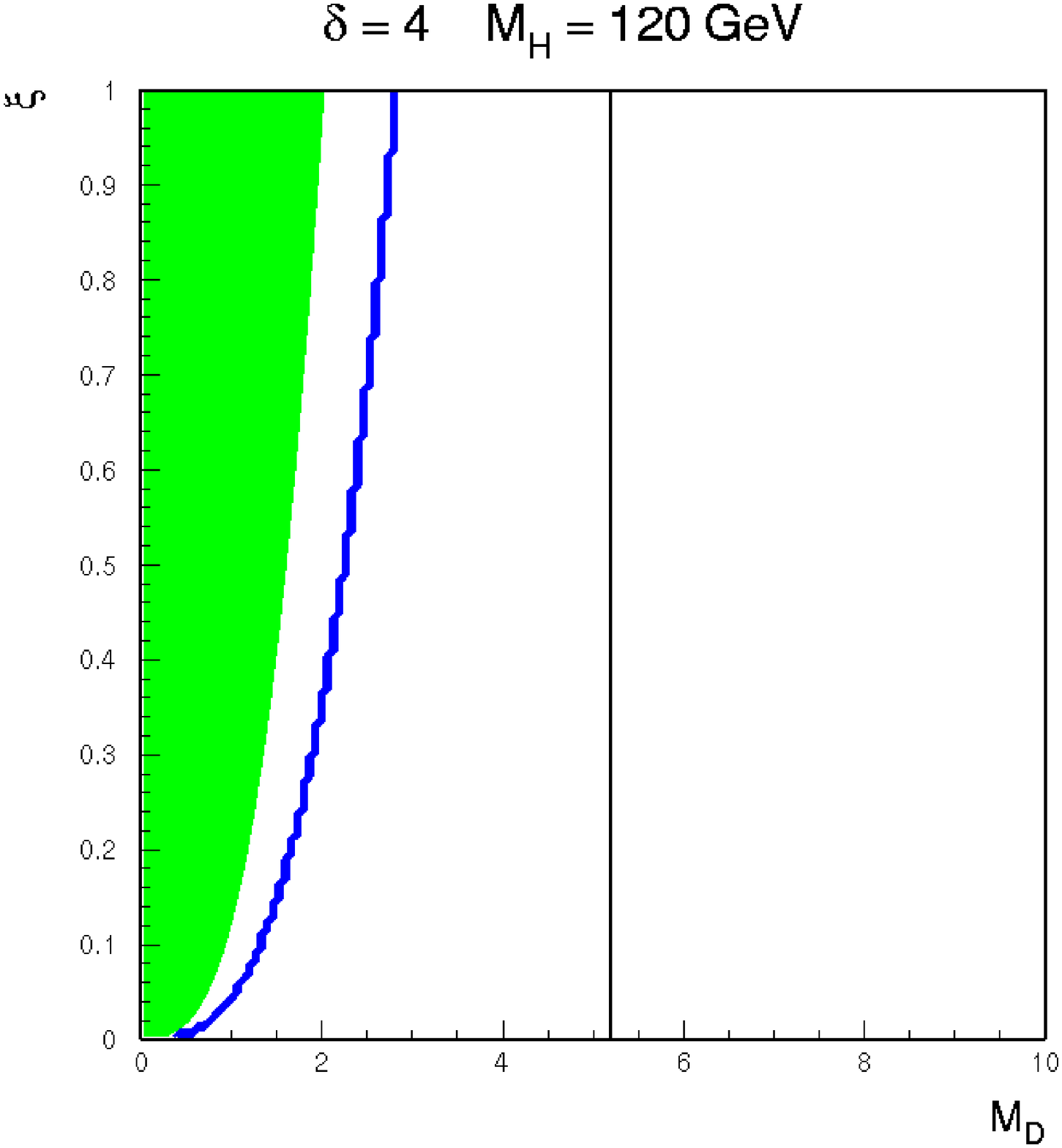}&
\includegraphics[width=6.0cm,height=6.0cm]{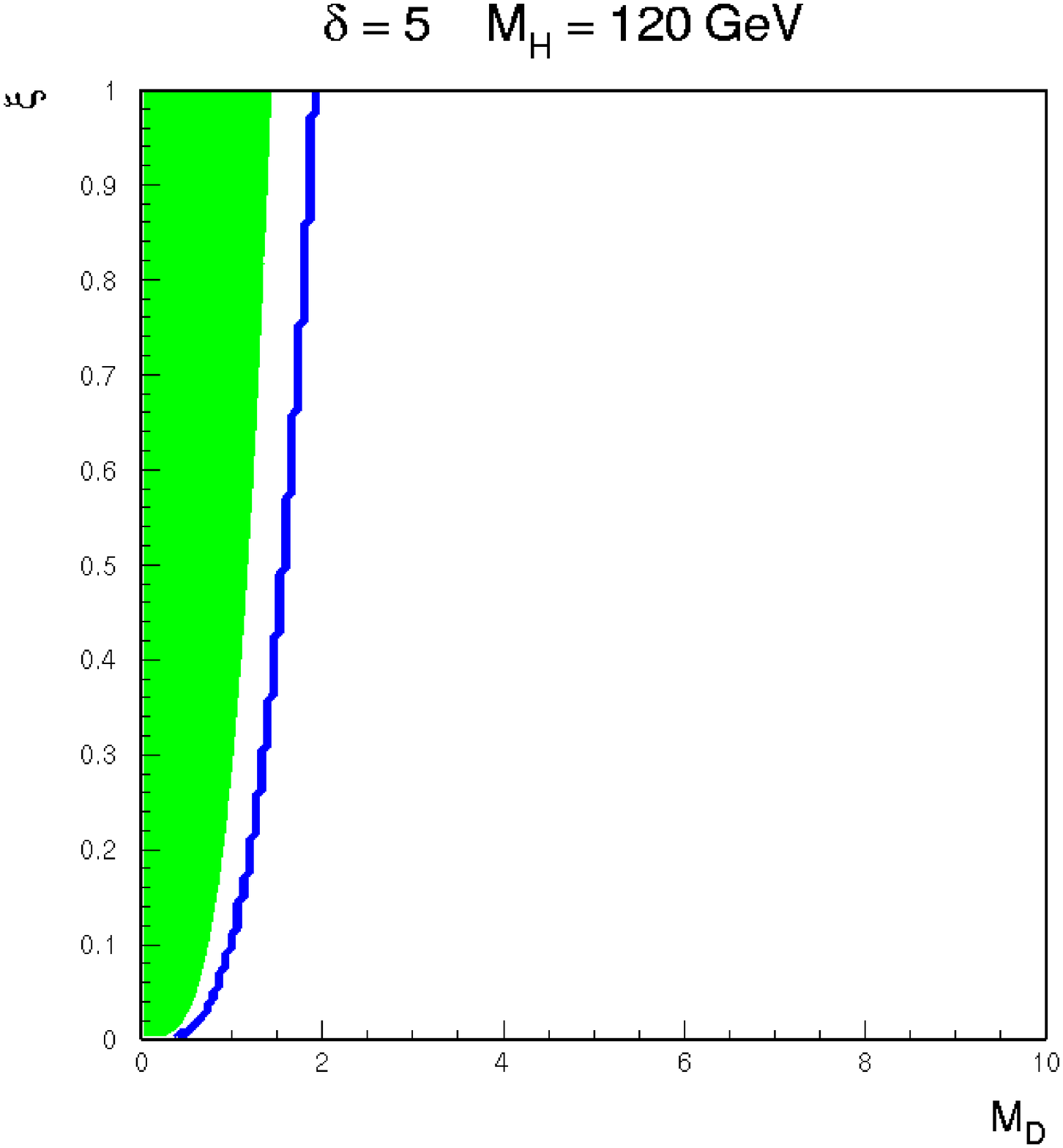}
\end{tabular}
\caption{Invisible decay width effects in the $\xi$ - $M_D$ plane for 
$M_H$ = 120~GeV. The green (grey) regions indicate where the Higgs signal 
at the LHC drops below the 5~$\sigma$ threshold for 30 fb$^{-1}$ of data. 
The regions above the blue (bold) line are the parts of the parameter space 
where the  invisible Higgs signal in the $WW$-fusion channel exceeds 5~$\sigma$
significance. The vertical lines show the upper limit on $M_D$  which can be 
probed by the analysis of jets/$\gamma$ with missing energy at the LHC. 
The plots are for different values of $\delta$: 2 (upper left), 3 (upper right), 4 (lower left), 5 (lower right).}
\label{figure1}
\end{center}
\end{figure}

The analysis of jet/$\gamma +$  missing energy is also sensitive to the ADD model 
over a range of the $M_D$ and $\delta$ parameters~\cite{Vacavant:2001sd}.
The invisible Higgs decay width appears to probe a parameter space up to, 
and possibly beyond, 
that accessible to these signatures (see Figure~\ref{figure1}). Further, the sensitivity 
of these channels decreases significantly faster with $\delta$ 
than  that of the 
invisible Higgs width if $\xi\sim 1$. 
Finally, it is interesting that, in the region where both 
signatures can be probed at the LHC, a combined analysis will provide a constraint 
on the fundamental theory parameters.
A TeV-class $e^+e^-$ linear collider will be able to further improve the determination 
of the Higgs invisible width. Extracting the branching fraction into invisible final 
states from the Higgsstrahlung cross section and the sum of visible decay modes affords 
an accuracy of order 0.2-0.03\% for values of the invisible branching fraction in the 
range 0.1-0.5. But the ultimate accuracy can be obtained with a dedicated analysis looking 
for an invisible system recoiling against a $Z$ boson in the $e^+e^- \to h Z$ process. 
A dedicated analysis has shown that an accuracy 
$0.025< \delta {\mathrm{BR}}/{\mathrm{BR}} < 0.04$ can be obtained 
for $0.1 < {\mathrm{BR}} < 0.5$~\cite{schumacher}.
This accuracy would establish an independent constraint on the $M_D$, $\xi$ and $\delta$ 
parameters.

\vskip1cm
\noindent

\section*{ACKNOWLEDGEMENTS}
JFG and JDW are supported by the U.S. Department of Energy.

\setcounter{figure}{0}
\setcounter{table}{0}
\setcounter{section}{0}
\setcounter{equation}{0}
\clearpage

\newcommand{\be}{\begin{equation}}
\newcommand{\ee}{\end{equation}}
\newcommand{\beas}{\begin{eqnarray*}}
\newcommand{\eeas}{\end{eqnarray*}} 
\newcommand{\ba}{\begin{array}}
\newcommand{\ea}{\end{array}}
\newcommand{\bi}{\begin{itemize}}
\newcommand{\ei}{\end{itemize}}
\newcommand{\ben}{\begin{enumerate}}
\newcommand{\een}{\end{enumerate}}
\newcommand{\gtsim}{\stackrel{\scriptscriptstyle>}{\scriptscriptstyle\sim}}

\part{Determining the extra-dimensional location of the Higgs boson \label{XDH}}

{\it 
A. Aranda, C. Bal\'azs, J. L. D\'{\i}az-Cruz
S. Gascon-Shotkin and O. Ravat
}

\maketitle

\begin{abstract}
In the context of a TeV$^{-1}$ size extra dimensional model, we consider the 
lightest Higgs boson as an admixture of brane and bulk scalar fields. We find 
that at the Tevatron Run 2 or at the LC the Higgs signal is suppressed.
Meanwhile, at the LHC or at CLIC one might find highly enhanced production 
rates. This will enable the latter experiments to distinguish between the extra 
dimensional and the SM for $M_c$ up to about 6 TeV and perhaps determine the 
extra-dimensional location of the lightest Higgs boson.
\end{abstract}

\section{INTRODUCTION}
\label{sec:introduction}

Extra dimensional models have been used recently to address a wide class of 
problems in particle physics, such as the hierarchy, unification and flavor 
problems~\cite{Arkani-Hamed:1998rs,Antoniadis:1998ig,Dienes:1998vh,Masip:1999mk}. 
In this work, we examine a TeV$^{-1}$ size extra dimensional model, in which the 
lightest Higgs boson emerges as an admixture of brane and bulk scalar fields.
This model predicts a suppression of the Higgs production cross section at LEP 
and the Tevatron, while it promises a significant enhancement of the signal at 
the CERN Large Hadron Collider (LHC) and possibly at a multi TeV linear collider 
(CLIC).
We present results for the cross section of the associated production of Higgs 
with gauge bosons at the LC and LHC.

\section{THE MODEL} 
\label{sec:model}

In this section we present the general features of the extra dimensional model 
(for a detailed description see~\cite{Aranda:2002dz}).
We work with a five dimensional (5D) extension of the SM that contains two Higgs doublets.
The SM fermions and one Higgs doublet ($\Phi_u$) live on a 4D boundary,
the brane, while the gauge bosons and the second Higgs doublet ($\Phi_d$),
are all allowed to propagate in the bulk. The constraints from electroweak
precision data~\cite{Masip:1999mk} show that the compactification scale
can be of ${\cal O}$(TeV) (3-4 TeV at 95 \% C.L.).
The relevant terms of the 5D $SU(2)\times U(1)$ gauge and Higgs Lagrangian are 
given by
\be \label{5dlagrangian}
{\cal L}^5 = -\frac{1}{4}\left(F^a_{MN}\right)^2
 -\frac{1}{4}\left(B_{MN}\right)^2
+ |D_M \Phi_d|^2 + |D_{\mu}\Phi_u|^2 \delta(x^5) \, ,
\ee
where the Lorentz indices $M$ and $N$ run from $0$ to $4$, and $\mu$ runs 
from $0$ to $3$.

After spontaneous breaking of electroweak symmetry one obtains the following 4D Lagrangian:
\bea \label{interactions}
\nonumber
{\cal L}^4 & \supset & \frac{g M_Z}{2c_W}\left(h\sin(\beta-\alpha)
+H\cos(\beta-\alpha)\right)Z_{\mu}Z^{\mu} \\ \nonumber
           &    +    &
\sqrt{2} \frac{g M_Z}{c_W} \left( h \sin\beta\cos\alpha
+H\sin\beta\sin\alpha \right)\sum_{n=1}^{\infty} Z_{\mu}^{(n)}Z^{\mu}
\\ \nonumber
           &    +    & g M_W \left(
h\sin(\beta -\alpha) +H\cos(\beta -\alpha)\right)
W^+_{\mu}W^{-\mu} \\  
           &    +    &
\sqrt{2}gM_W\left(h\sin\beta\cos\alpha + H\sin\beta\sin\alpha\right)
\sum_{n=1}^{\infty}\left(W^+_{\mu}W^{-(n)\mu} + 
W^-_{\mu}W^{+(n)\mu}\right) \, .
\eea
where $h$ and $H$ are the CP-even Higgses ($m_h < m_H$), $\alpha$ is the mixing angle
that appears in the diagonalization of the CP-even mass matrix, and $\tan\beta$ is the ratio
of vevs.

\section{HIGGS PRODUCTION AT FUTURE COLLIDERS}
\label{sec:results}

\begin{figure}
\vspace{-1cm}
\resizebox*{.32\textwidth}{.25\textheight}
{\includegraphics{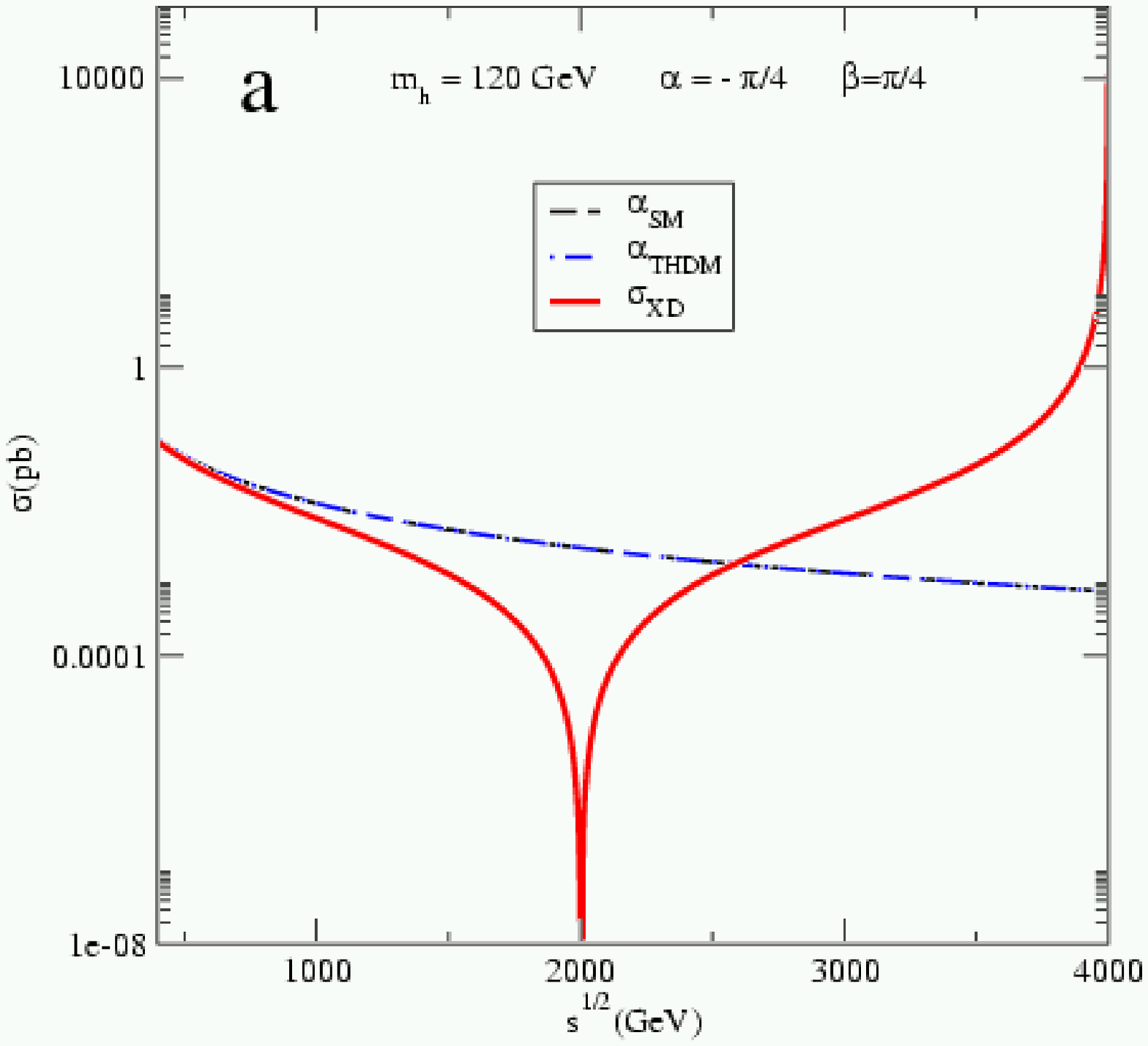}}
\resizebox*{.32\textwidth}{.25\textheight} 
{\includegraphics{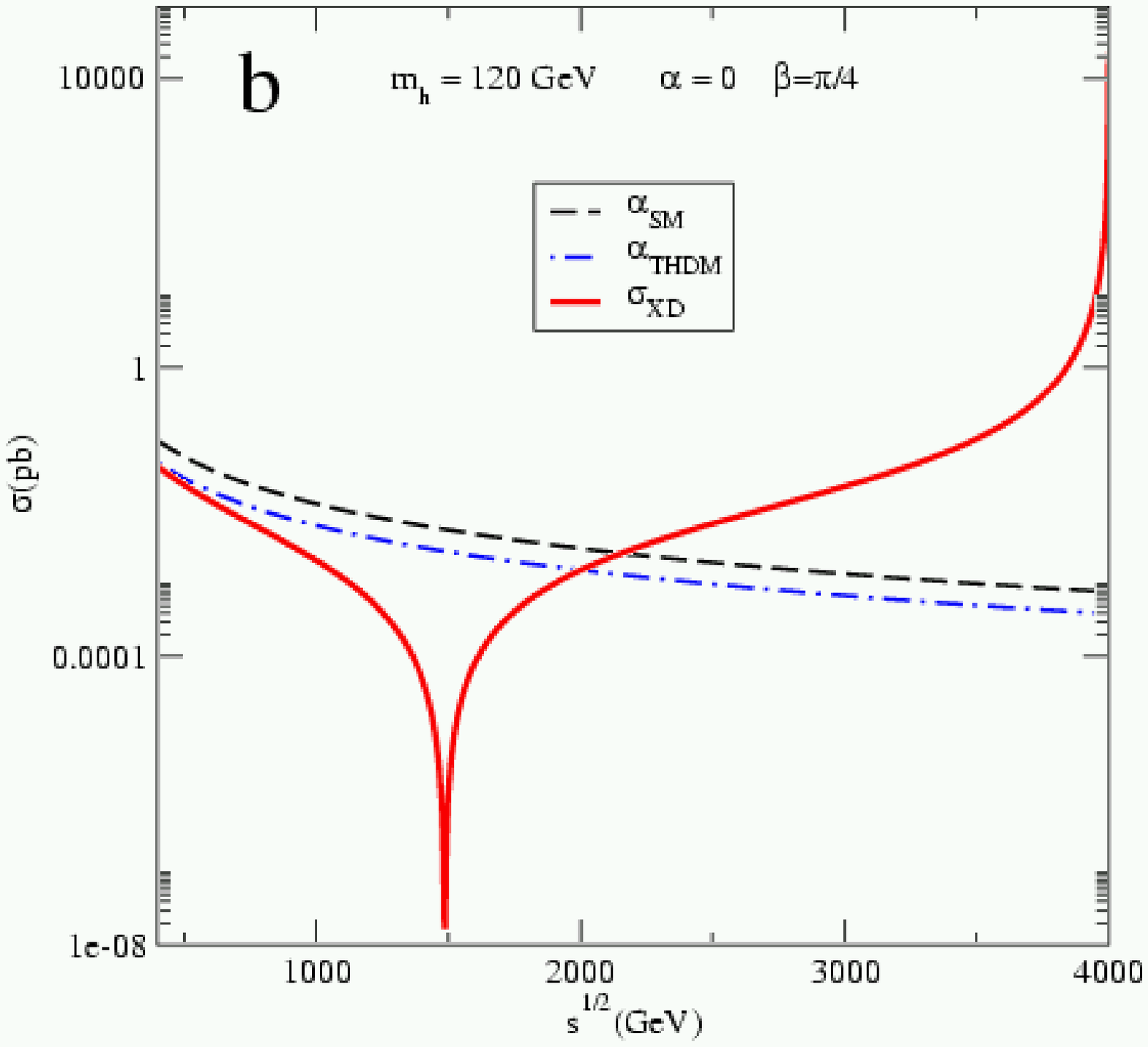}}
\resizebox*{.32\textwidth}{.25\textheight} 
{\includegraphics{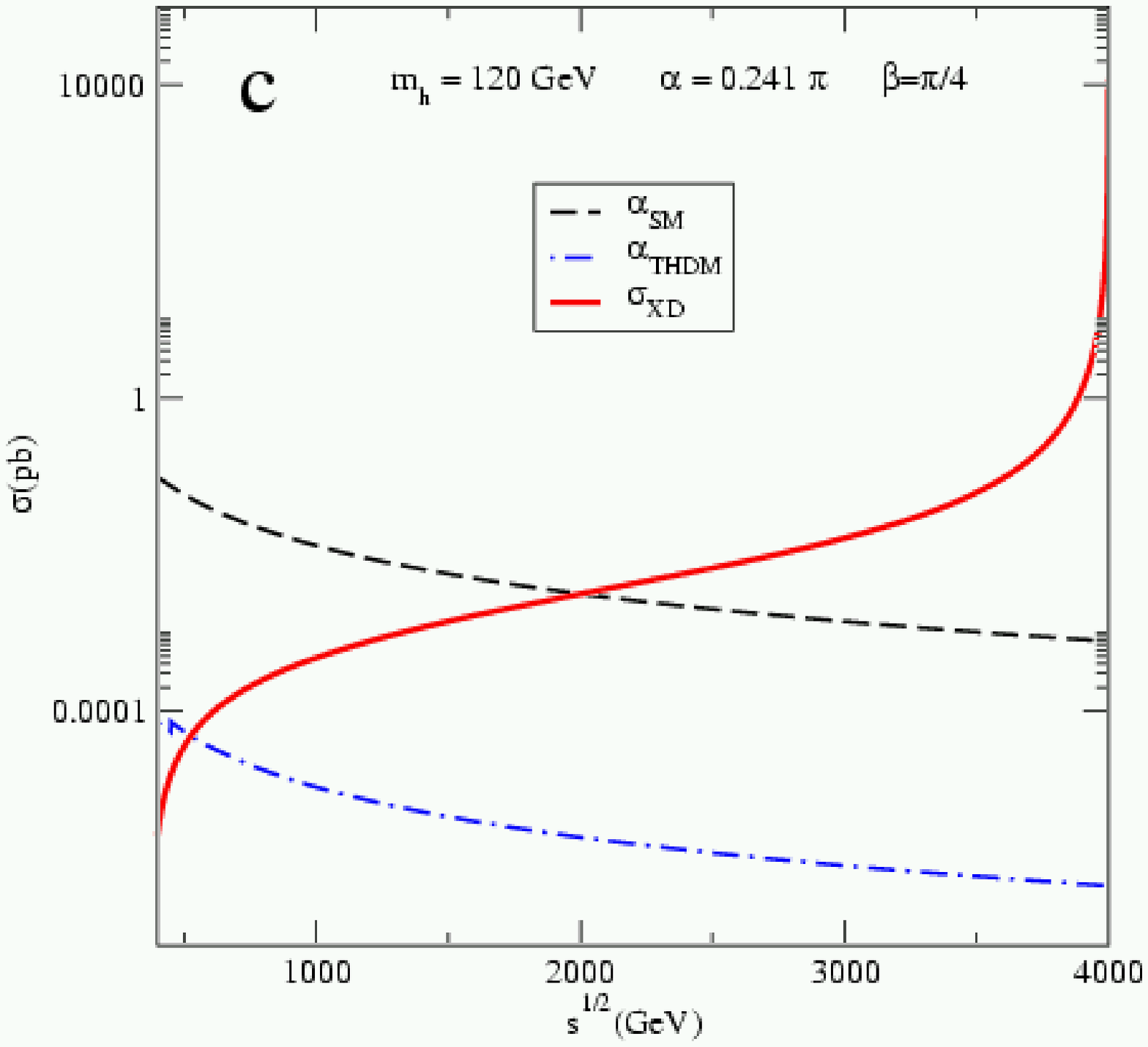}}
\caption{SM, THDM and XD cross sections for $e^+e^- \rightarrow Zh$. Each plot
corresponds to a different set of values for $\alpha$ and $\beta$ all with
$m_h = 120$~GeV and with a compactification scale $M_c = 4$~TeV.}
\label{figure1Blak}
\end{figure}

We present results for the associated $h+Z$ production cross section obtained from 
Eq.(\ref{interactions}) at linear colliders and at the LHC. 
Fig.~\ref{figure1Blak} shows the results for the $e^+e^- \to hZ$ cross section. 
$\sigma_{SM}$ stands for the standard model cross section, THDM labels the (4D) two 
Higgs doublet model, and the results from the extra dimensional model are denoted by 
XD. 
The three plots correspond to three different choices of the parameters $\alpha$ 
and $\beta$. It can be observed that the SM cross section dominates in all cases 
up to $\sqrt{s}\sim 2$~TeV. This is understood from the fact that the heavier KK 
modes, through their propagators, interfere destructively with the SM amplitude 
thus reducing the XD cross section. Moreover, as Fig.~\ref{figure1Blak} shows, once 
the center of mass energy approaches the threshold for the production of the 
first KK state, the cross section starts growing. For instance, with $M_c=4$ 
TeV,  $\sigma_{SM} \simeq \sigma_{XD}$ for $\sqrt{s} \simeq 2$ TeV. However, one 
would need higher energies in order to have a cross section larger than that of 
the SM, which may only be possible at CLIC~\cite{Assmann:2000hg}.
Based on this, we also conclude that at the Tevatron the luminosity required 
to find a light Higgs boson is higher than in the SM case.

The Higgs discovery potential in this model is more promising at the LHC. We 
illustrate this in Fig.~\ref{figLHC1}, showing the $pp\to hZ$ differential cross 
section as the function of the $hZ$ invariant mass $M_{hZ}$. 
The typical resonance structure displayed by Fig.\ref{figure1Blak} is preserved by 
the hadronic cross section. The resonance peak is well pronounced when $M_{hZ} 
\sim M_c$. This leads to a large enhancement over the SM (or THDM) cross 
section.
The singularity at $M_c = M_{hZ}$ is regulated by the width of the KK mode, 
which is included in our calculation (cf. Ref.\cite{Aranda:2002dz}). Depending 
on the particular values of $\alpha$ and $\beta$ the enhancement is more or less 
pronounced. For an optimistic set $\alpha = 0$ and $\beta = \pi/2$, the XD 
production cross section is considerably enhanced compared to the SM at $M_{hZ} 
= M_c$. This enhancement may be detectable up to about $M_c = 6$ TeV. We 
estimate that with 100 fb$^{-1}$ for $M_c = 6$ TeV there are about 20 $hZ$ 
events in the bins around $M_{hZ} \sim M_c$. As Fig.~\ref{figLHC1} shows, in the 
SM less than one event is expected in the same $M_{hZ}$ range. It is needless to 
say that similar results hold for $pp\to hW^\pm$, which further enhances the 
discovery prospects.

Based on these results, we conclude that in the Bjorken process alone the reach 
of the LHC may extend to about $M_c = 6$ TeV, depending on the values of 
$\alpha$ and $\beta$.  The $Zh$ (or $W^{\pm}h$) production cross section determines
only a specific combination of the mixing angles $\alpha$ and $\beta$.  In order
to determine the individual angles, one has to also measure the production 
cross sections of the heavier Higgs boson in association with a $Z$ (or $W^{\pm}$).
Fortunately, these cross sections are also enganced by the same amount as the ones
for the lightest Higgs boson.

\begin{figure}
\begin{center}
\vspace{-.7cm}
\includegraphics[width=11cm,height=6.5cm]{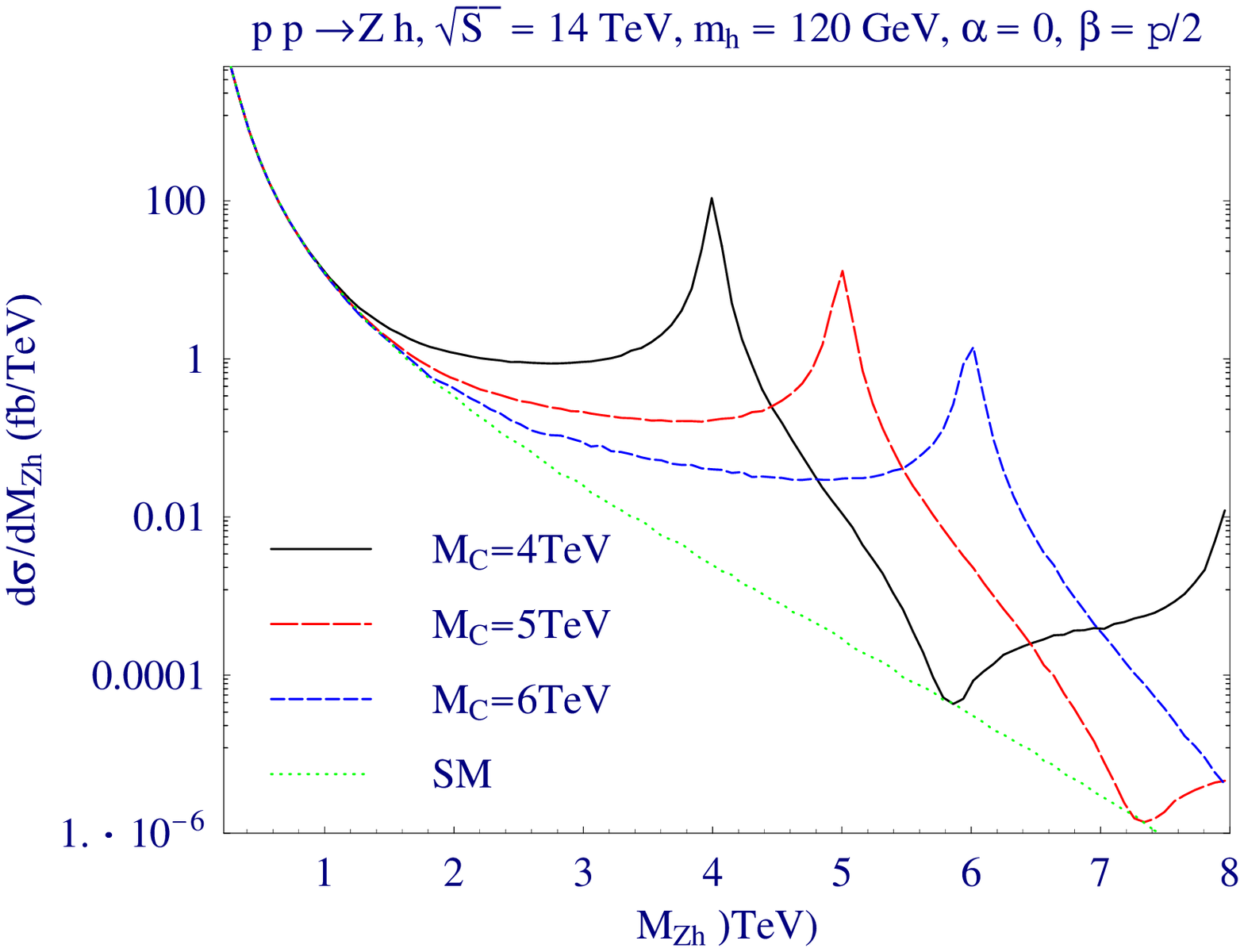}
\end{center}
\caption{Higgs production cross section in association with a $Z$ boson at the
LHC as a function of the compactification scale for selected values of the 
mixing parameters.}
\label{figLHC1}
\end{figure}

\section{CONCLUSIONS}
\label{sec:conclusions}

We presented Higgs production cross section calculations in the framework of an 
extra dimensional model that releaves the tension between the low mass 
predictions for the SM Higgs and the missing Higgs signal at LEP. We found a 
suppresson at the LC and at the Tevatron, but an enhanced signal at the LHC or 
CLIC. The fact that the lightest Higgs boson is an admixture of brane and bulk 
fields (that is it has a non-trivial bulk location) is the key ingredient in the 
suppression-enhancement mechanism for the signal. This may enable the LHC and CLIC
to determine this location.

\section*{ACKNOWLEDGEMENTS}
A.A. acknowledges support from the Alvarez-Buylla fund of the Universidad de
Colima. During this work, CB was supported by the U.S. Department
of Energy HEP Division under contracts DE-FG02-97ER41022 and
W-31-109-ENG-38,
and by LPNHE-Paris.
J.L. D.-C. was supported by CONACYT and SNI (M\'exico).

\setcounter{figure}{0}
\setcounter{table}{0}
\setcounter{section}{0}
\setcounter{equation}{0}
\clearpage

\part{The sensitivity of the LHC for TeV scale dimensions in dijet production \label{proceeding}}
{\it C. Bal\'azs, M. Escalier, S. Ferrag, B. Laforge and G. Polesello}
\maketitle

\begin{abstract}
In this work, we present results for dijet distributions at the LHC with 
the assumption of a TeV size extra dimension. In our calculation, we included 
the virtual effects of gluonic Kaluza-Klein state exchanges, as well as the 
modified running of the strong coupling constant (but restricted our numerical 
study to the case of standard $\alpha_S$ evolution). Computing the transverse 
momentum distribution of dijets, we found that the LHC is able to discover a 
single extra dimension up to $M_c \sim 15$ TeV.
\end{abstract}

\section{Introduction}


String theory is the most promising candidate for a unified framework of matter 
and interactions. Among the predictions of string theory are extra, compact 
space dimensions (XDs) which, depending on their sizes $R$, play a role in 
determining physics close to the weak scale. The string arguments of 
Refs.~\cite{Polchinski:1995mt,Horava:1996ma} do not prevent the standard gauge 
bosons from penetrating the bulk. If the compactification scale $M_c = 1/R$ is 
higher than ${\cal O}$(TeV), phenomenology and present experiments do not 
conflict with this scenario either~\cite{Pomarol:1998sd, Antoniadis:1998sd, 
Delgado:1998qr, Carone:1999nz, Dienes:1998vh, Dienes:1998vg}. These new 
dimensions, on the other hand, may be probed at near future particle 
accelerators, in particular at the CERN Large Hadron Collider (LHC).  

Since the LHC produces strongly interacting particles abundantly, if the gluons 
propagate in XDs then dijet production at the LHC is a sensitive discovery 
channel for TeV scale XDs. The main effect of these dimensions on the dijet 
production cross section is twofold. On the one hand, the gluonic Kaluza-Klein 
(KK) excitations enhance the dijet distributions in the high invariant mass 
(${\hat s}$) and transverse momentum ($p_T$) region \cite{Dicus:2000hm}. On the 
other, the modified evolution of the strong coupling ($\alpha_S$) further 
distorts these distributions \cite{Balazs:2001eu}. These competing effects are 
entangled and has to be taken into account simultaneously in order to predict 
the discovery potential of the LHC.

In this work, we computed dijet distributions for the LHC including both of 
these effects. We assumed that the standard model (SM) gauge bosons, especially 
gluons, propagate in a single TeV$^{-1}$ size compact dimension. We implemented 
gluonic KK excitations and the modified running of $\alpha_S$ in the Monte Carlo 
event generator PYTHIA \cite{Sjostrand:2000wi}, for dijets. Then, we used PYTHIA 
to calculate dijet $p_T$ distributions, determined the enhancement at large 
$p_T$, and estimated the significance of a potential discovery.

\section{Dijet production at the LHC}

\begin{floatingfigure}[r]{8.5cm}
\hspace{-1cm}
\includegraphics[width=8.5cm]{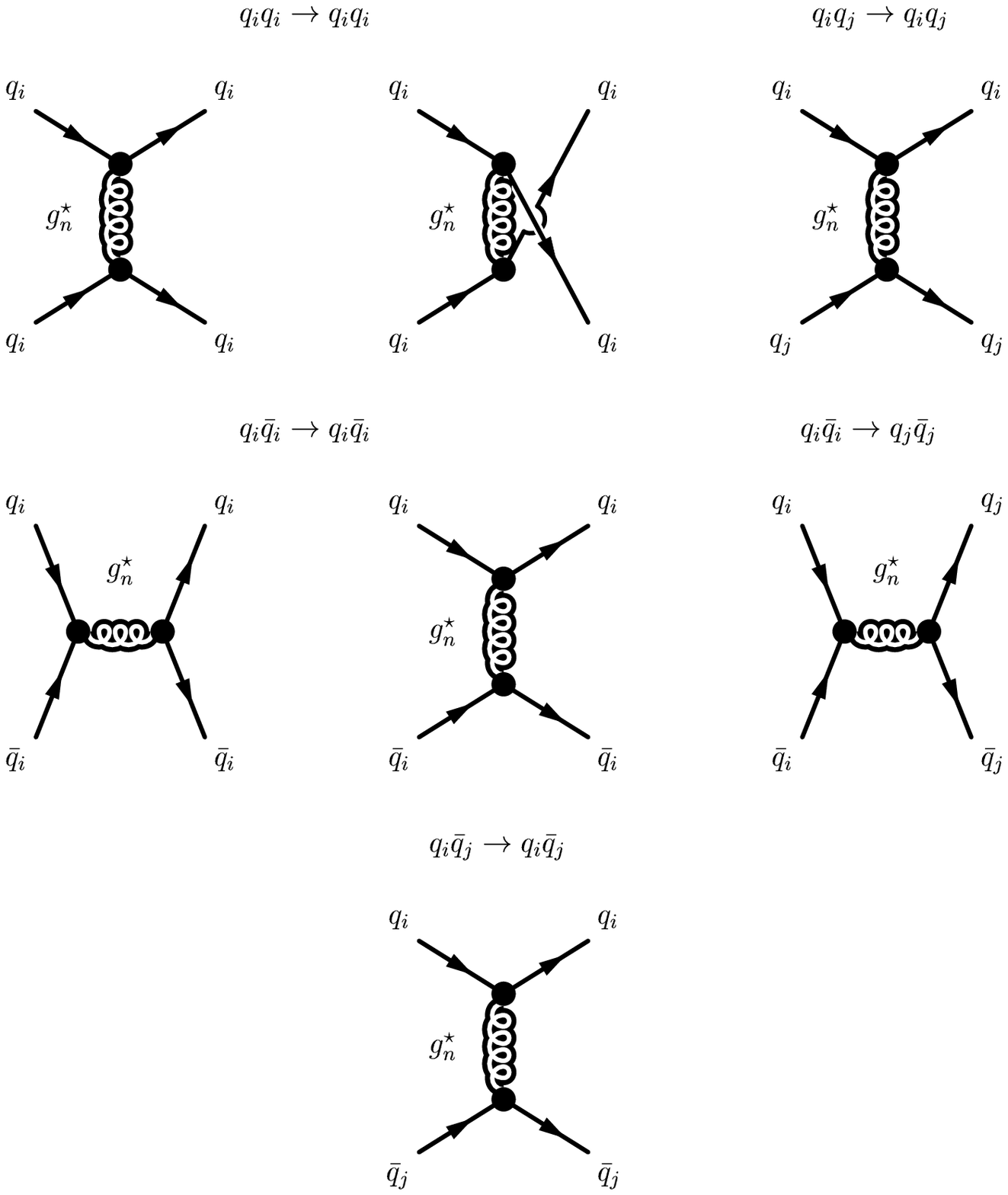}
\caption{Feynman diagrams for dijet production involving Kaluza-Klein 
excitations of the gluons. The indices $i$ and $j$ represent distinct 
($i \neq j$) quark flavors.}
\label{fig:diagrams}
\end{floatingfigure}
%
The formalism that we use is described in detail in Refs.\cite{Dicus:2000hm} and 
\cite{Balazs:2001eu}. When calculating the dijet production cross section, 
besides the SM ones, we include the tree-level diagrams shown in Fig. 
\ref{fig:diagrams}.\footnote{We note that five-momentum conservation forbids 
internal gluonic KK excitations in any tree-level dijet diagrams involving 
external gluons, that is the KK excitations do not affect the process $q \bar{q} 
\to gg$, for example.} In these diagrams $g_n^*$ signals that a KK tower of 
virtual gluons is exchanged. This means that in the SM diagrams we replace the 
gluon propagators by
\begin{eqnarray}
D_{\mathit{eff}}(p) = \sum_{n=0}^{N} c_n D_n(p) .
\end{eqnarray}
Here
\begin{eqnarray}
D_n(p) = \frac{c_n}{p_n'^2  +  i m_n \Gamma_n} ,
\end{eqnarray}
is the propagator of the $n^{th}$ gluon KK resonance with 
$p_n'^2=p^2-m_n^2$, $c_{n>0} = 2$, $m_n = n/R, \Gamma_n = 2 \alpha_s m_n$.
The SM gluon is identified with the zero mode and $c_0 = 1$.


When calculating the cross section, it is necessary to evaluate amplitude 
squares which will contain products of propagators of the form
\begin{eqnarray}
\frac{1}{2}\mbox{\raisebox{-.6ex}{\huge
$[$}}D_{\mathit{eff}}^{\star}( p )D_{\mathit{eff}}( q ) +
     D_{\mathit{eff}}( p )D_{\mathit{eff}}^{\star}( q )
     \mbox{\raisebox{-.6ex}{\huge $]$}}
 = \nonumber \\
\sum_{m,n=0}^{N} \frac{ c_m c_n (p_m'^2  q_n'^2 + m_m \Gamma_m m_n \Gamma_n)}
{((p_m'^2)^2 + m_m^2\Gamma_m^2) ((q_n'^2)^2 + m_n^2\Gamma_n^2)} \,
\label{eq:Deff}
\end{eqnarray}
(We note that in Eq.(\ref{eq:Deff}), we corrected a typo which is present 
in v.3 of hep-ph/0012259.)
As it was noted in Ref. \cite{Dicus:2000hm} the sum in Eq.~(\ref{eq:Deff}) 
converges rapidly. We checked that for the LHC (with $\sqrt{(S)} = 14$ TeV) and 
for $M_c > 1$ TeV, choosing $N = 50$ (or equivalently a cutoff scale of $M_s > 
50$ TeV) leads to a satisfactory numerical precision.
We implemented the effective propagators given by Eq.(\ref{eq:Deff}) in PYTHIA,
modifying the parton level processes represented by Fig.(\ref{fig:diagrams}).
Finally, we checked the implementation against the numerical results given
in Ref. \cite{Dicus:2000hm} and found a good agreement.

In our calculation, we also include the modified running of $\alpha_S$, as 
described in Refs.\cite{Dienes:1998vh,Dienes:1998vg,Balazs:2001eu}.
Above the compactification scale, we implemented the modified running of 
$\alpha_s$ as given by
\begin{eqnarray}
       \alpha_i^{-1}(\mu) = \alpha_i^{-1}(\mu_0) -
            {b_i-\tilde b_i\over 2\pi}\,\ln{\mu \over \mu_0} - 
          ~{\tilde b_i\over 4\pi}\,
             \int_{r\mu^{-2}}^{r\mu_0^{-2}} {dt\over t} \,
     \left[ \vartheta_3\left( {it\over \pi R^2} \right) \right]^\delta ,
\label{KKresult}
\end{eqnarray}
where $i = 1,2,3$ labels the gauge groups of the SM. The coefficients of
the usual one loop beta functions
\begin{eqnarray}
        (b_1,b_2,b_3) = (41/10,-19/6,-7)
\end{eqnarray}
are supplemented by new contributions from the KK towers
\begin{eqnarray}
      (\tilde b_1,\tilde b_2,\tilde b_3) = (3/5, -3, -6) + \eta \; (4, 4, 4)
\end{eqnarray}
(where, for simplicity, we set $\eta = 0$). In the last term of 
Eq. (\ref{KKresult}), $\vartheta_3$ denotes the elliptic Jacobi function and 
\begin{eqnarray}
      r = \pi \,(X_\delta)^{-2/\delta} ~~~ {\rm with} ~~~ 
      X_\delta = {2 \pi^{\delta/2} \over \delta \Gamma(\delta/2)}.
\end{eqnarray}
We note that in Refs. \cite{Dienes:1998vh,Dienes:1998vg} an approximate 
expression is used to calculate the running of the couplings, but we 
implemented the exact formula (\ref{KKresult}) in PYTHIA.

\section{Numerical results}


\begin{figure}
\includegraphics[width=7.5cm]{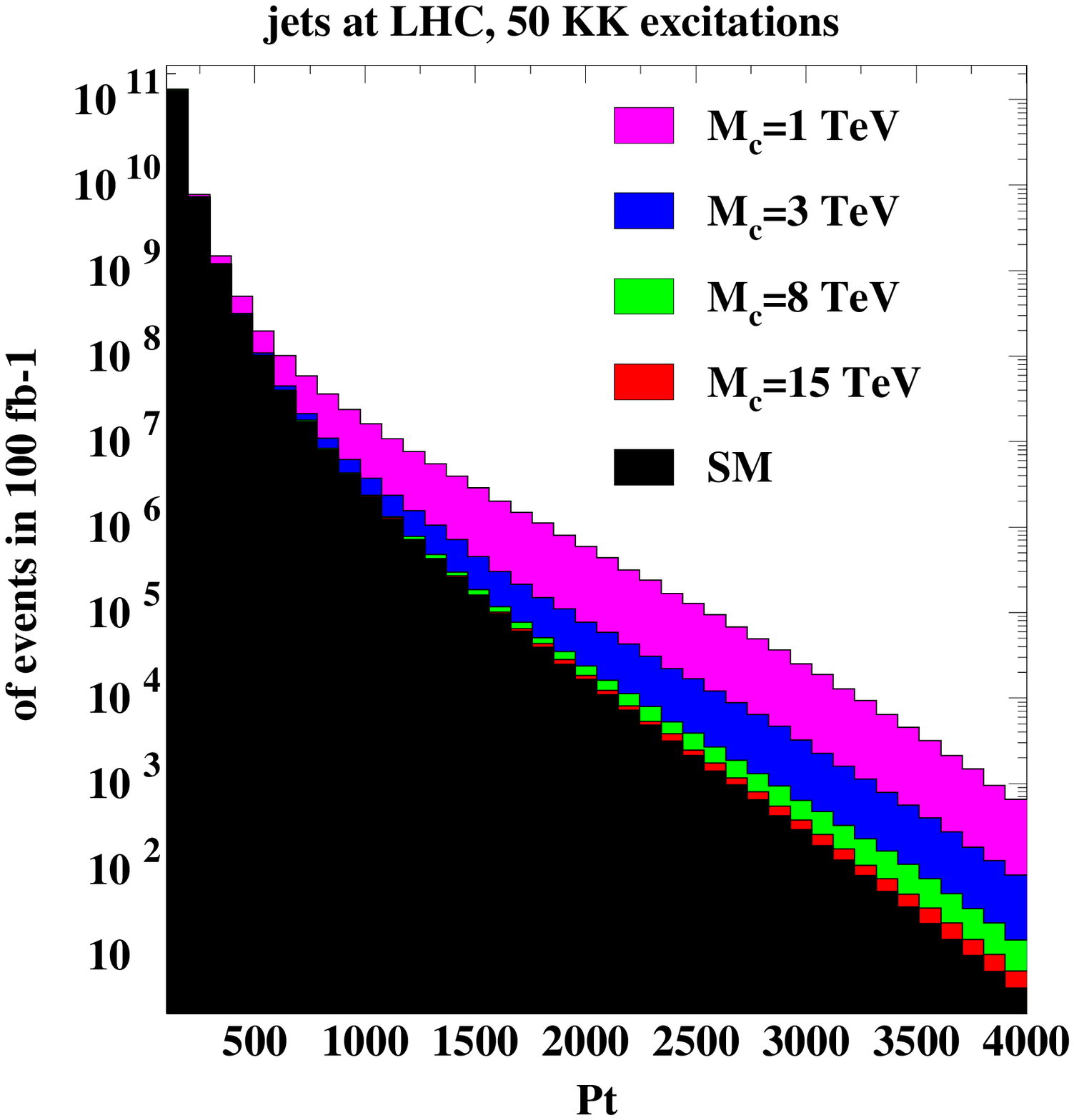}
\includegraphics[width=7.5cm]{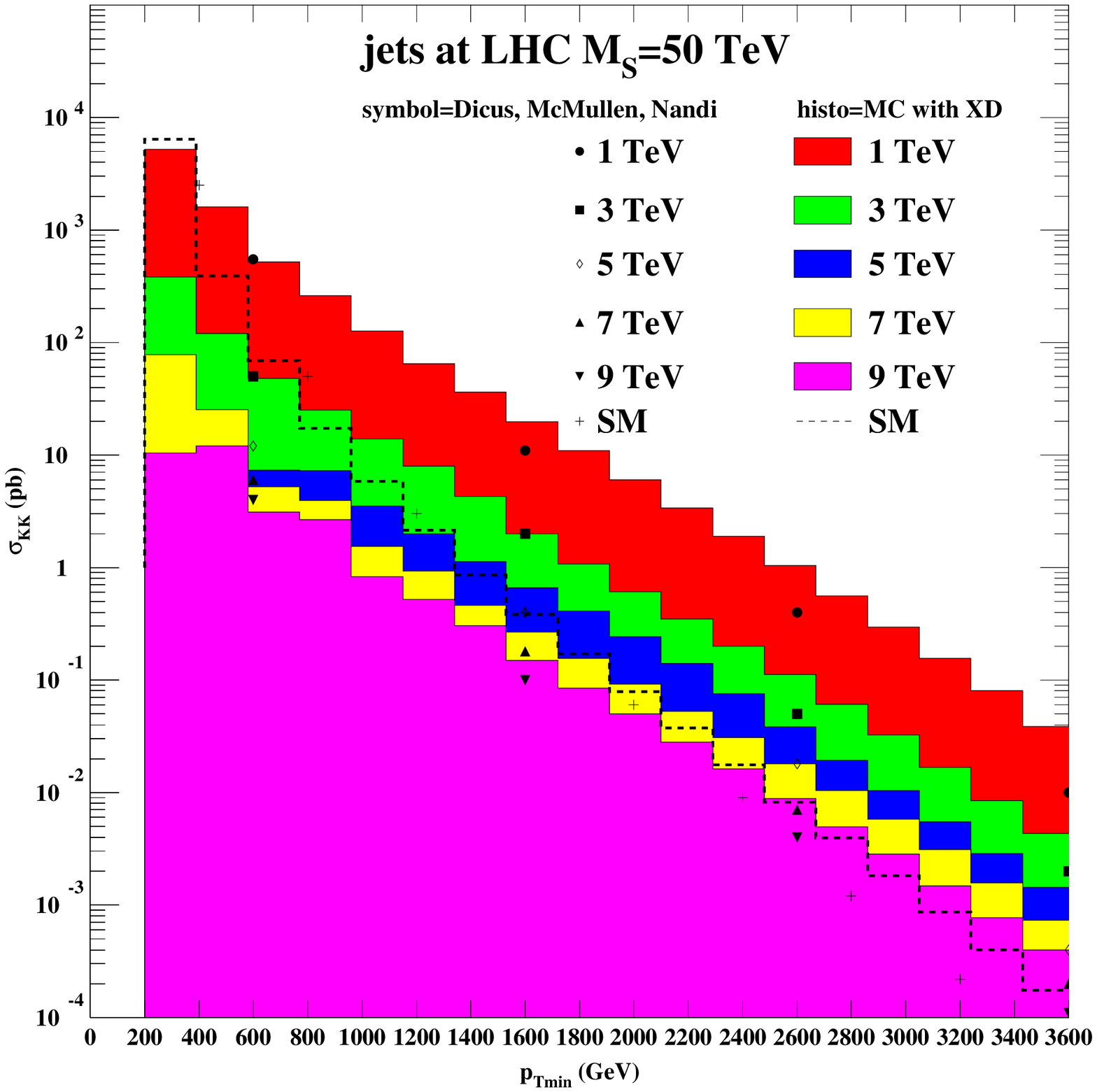}
\caption{
Left panel: number of dijet events vs. the dijet $p_T$ calculated in the SM
and for various values of $M_c$.
Right panel: KK contribution to the cross section as the function of $p_{T\min}$.}
\label{fig:pt}
\end{figure}

Our goal is to quantify the sensitivity of the LHC to a TeV size XD.  To this 
end, following Ref. \cite{Dicus:2000hm}, we compute the dijet transverse 
momentum ($p_T$) distribution. As a first step, we include 50 KK excitations of 
the gluons but keep the standard evolution of $\alpha_s$.  We use PYTHIA version 
6.210 with the modifications described in the previous section. On the final 
state, we apply the following kinematic cuts:
\begin{eqnarray}
p_T > p_{T\min}, ~~~ |y| < 2.5, ~~~ p_{Tjet} > 100 ~{\rm GeV~(on~each~jet)},
\end{eqnarray}
where $y$ is the rapidity of the two jet system. In PYTHIA, we also turn the 
initial and final state radiation (ISR and FSR) on.

The left panel of Fig. \ref{fig:pt} shows the results as the number of dijet 
events against the dijet $p_T$ assuming 100 fb$^{-1}$ integrated luminosity.  In 
this and in the subsequent computations we used the CTEQ6L1 parton distribution 
function (PDF) with the dijet invariant mass as the factorization scale. It is 
shown that there is a significant enhancement in the high $p_T$ for $M_c = 8$ 
TeV, and there is still a detectable excess for $M_c = 15$ TeV.
The right panel of Fig. \ref{fig:pt} shows the KK contribution to the cross 
section 
\begin{eqnarray}
\sigma_{\rm KK} = \sigma_{\rm total} - \sigma_{\rm SM},
\end{eqnarray}
as the function of $p_{T\min}$.  Considering the difference in the PDF, slightly
different cuts and the PYTHIA effects (ISR, FSR, etc.), our results still 
reasonably agree with Fig.3 of Ref. \cite{Dicus:2000hm}.

\begin{figure} 
\begin{center}
\includegraphics[width=7cm]{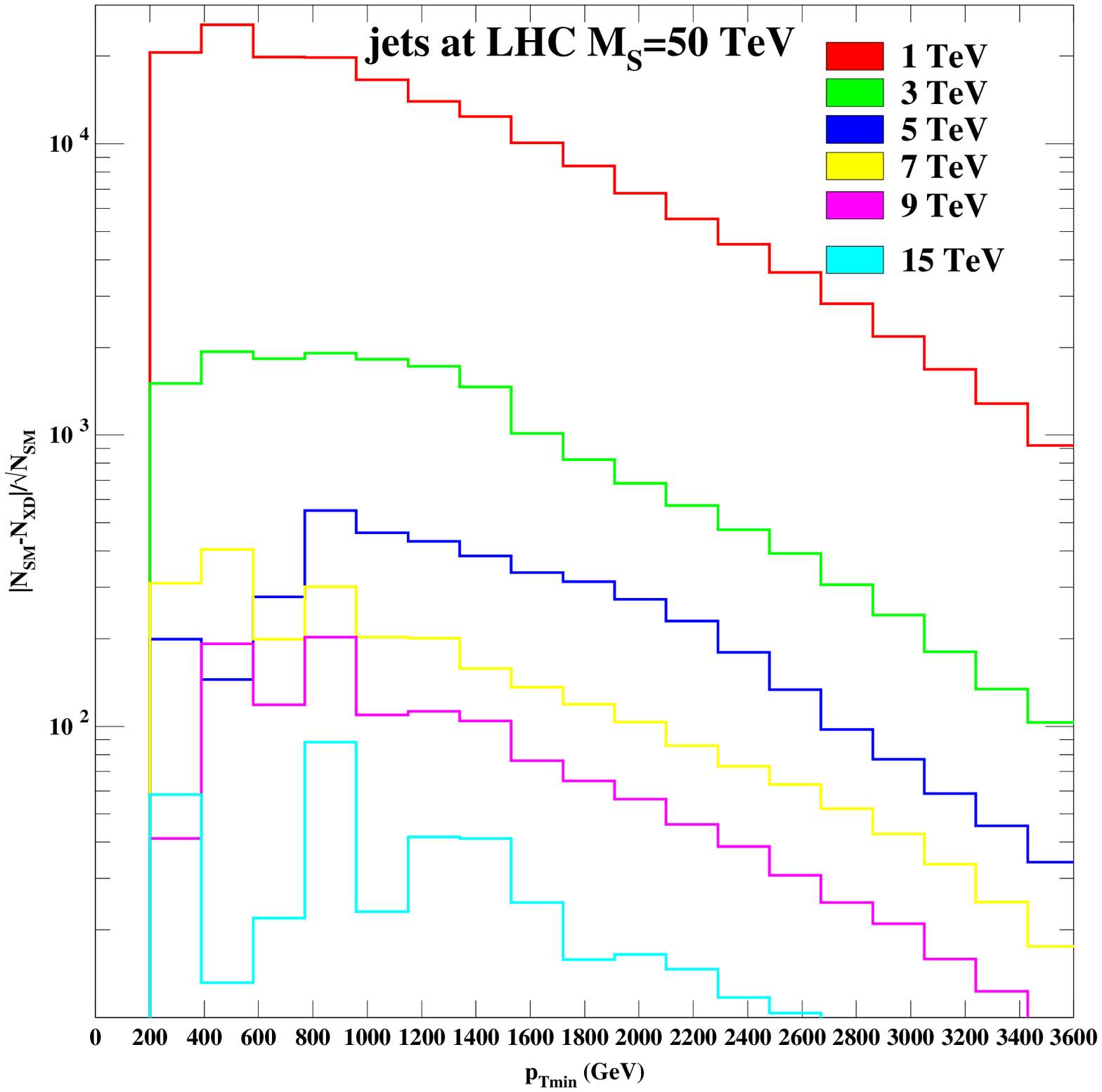}
\caption{Statistical significance of the XD dijet signal in units of $\sigma$.}
\label{fig:significance}
\end{center} 
\end{figure}

Finally, Fig. \ref{fig:significance} shows the statistical significance 
\begin{eqnarray}
S = |N_{SM}-N_{XD}|/\sqrt{N_{SM}},
\end{eqnarray}
of the XD dijet signal in units of $\sigma$ plotted against the dijet $p_T$. 
Here $N_{X}$ is the number of events predicted by model $X$. This plots shows 
that by measuring the dijet $p_T$ distribution in the 1-3 TeV region the new 
dimension can be easily discovered even if it is as small as 15 TeV$^{-1}$.


\section*{Acknowledgments}

The authors are grateful to the organizers of Les Houches 2003. During this 
work, CB was supported by the U.S. Department of Energy HEP Division under 
contracts DE-FG02-97ER41022 and W-31-109-ENG-38, and by Universites Paris VI 
\& VII.

\setcounter{figure}{0}
\setcounter{table}{0}
\setcounter{section}{0}
\setcounter{equation}{0}
\clearpage

 \part{Little Higgs Model: LHC Potential}
{\em K.Mazumdar}

\begin{abstract}
A recent idea of solving the 'little hierarchy' between physical mass of the 
Higgs boson and present limit of relatively low cut off scale of about 10 TeV
has a vector-like new heavy quark {\it Top} as one of the additional particles.
The potential of CMS experiment at LHC to discover it has been 
studied using the available  phenomenology for collider experiments.
\end{abstract}

\section{INTRODUCTION}

The {\it Little Higgs} model is an alternative solution for the hierarchy 
problem in the Standard Model (SM) of strong and electroweak interactions
which  
has passed stringent experimental verifications upto the electroweak scale. 
The precision measurements indicate that the physical mass of the
Higgs boson $m_{\rm H} \le 219$ GeV/c$^2$ to 95\% CL. 
Considering the SM as an effective theory valid 
upto an  energy scale $\sim $ few TeV, any new physics appearing below a
cutoff of about 10 TeV has to be weakly coupled to be consistent with 
precision measurements.
The gap between electroweak scale  and this cutoff value is called
the 'little hierarchy'. 

This hierarchy problem is solved in {\it Little Higgs} model by requiring 
$m_{\rm H}$
to be safe from only one-loop divergences. The Higgs fields are considered as 
Nambu-Goldstone-Bosons of a global symmetry which is spontaneously broken at
a higher scale, but the physical Higgs boson states continue to remain light
due to  
approximate global symmetry. In the minimal scenario both gauge and Yukawa
interactions are necessary to break all  the global symmetry which protect the 
Higgs boson mass and hence pushes the cutoff scale  to $\sim$ 10
TeV. The additional particles postulated in the model are three heavy gauge
bosons,   
a  heavy  top-like quark ({\it T}), a heavy Higgs triplet and a Higgs doublet. 
The lightest neutral scalar is identified with the SM Higgs boson.
Interestingly, the most important divergences are cancelled between loops
of particles with the same statistics.

The masses of the new particles need to be light enough to avoid fine tuning, 
whereas the precision electroweak constraints push them higher. Hence 
parameters of the model, {\it eg.} couplings, need to be tuned. Present
expectations of  
1-2 TeV mass 
range make these particles accessible at LHC and this motivates us to study
 their prospect of discovery.

\section{COLLIDER PHENOMENOLOGY}
The collider phenomenology has been worked out in ~\cite{han:2003a}. The 
heavy gauge bosons can be produced in LHC via Drell-Yan 
type process. The heavy  quark {\it T} can be produced via QCD 
(gluon-gluon fusion) in a model independent way but with increasing
mass the cross-section falls off rapidly. It can also be produced singly,
through t-channel fusion process $W^+ b \rightarrow T$, and the cross-section,
though  
larger, depends on  the 
parameters in the model, relating the {\it Top} mass 
$M_T$ to the scale, or equivalently to the Higgs condensate $f$ ($\sim$ TeV). 

The tree-level Feynman diagram for  the production of single  {\it T} and the
cross-section  at LHC as a function of its mass is given in
Fig.~\ref{fig:feyn}.  
The single {\it T} production cross-section is taken as 0.25 pb for $M_T=1$
TeV/c$^2$. 
 
\section{HEAVY TOP SEARCH IN CMS}

\begin{figure}[b]
\begin{center}
\unitlength=1cm
\begin{picture}(14,5)
\put(1,1){\includegraphics[width=7cm]{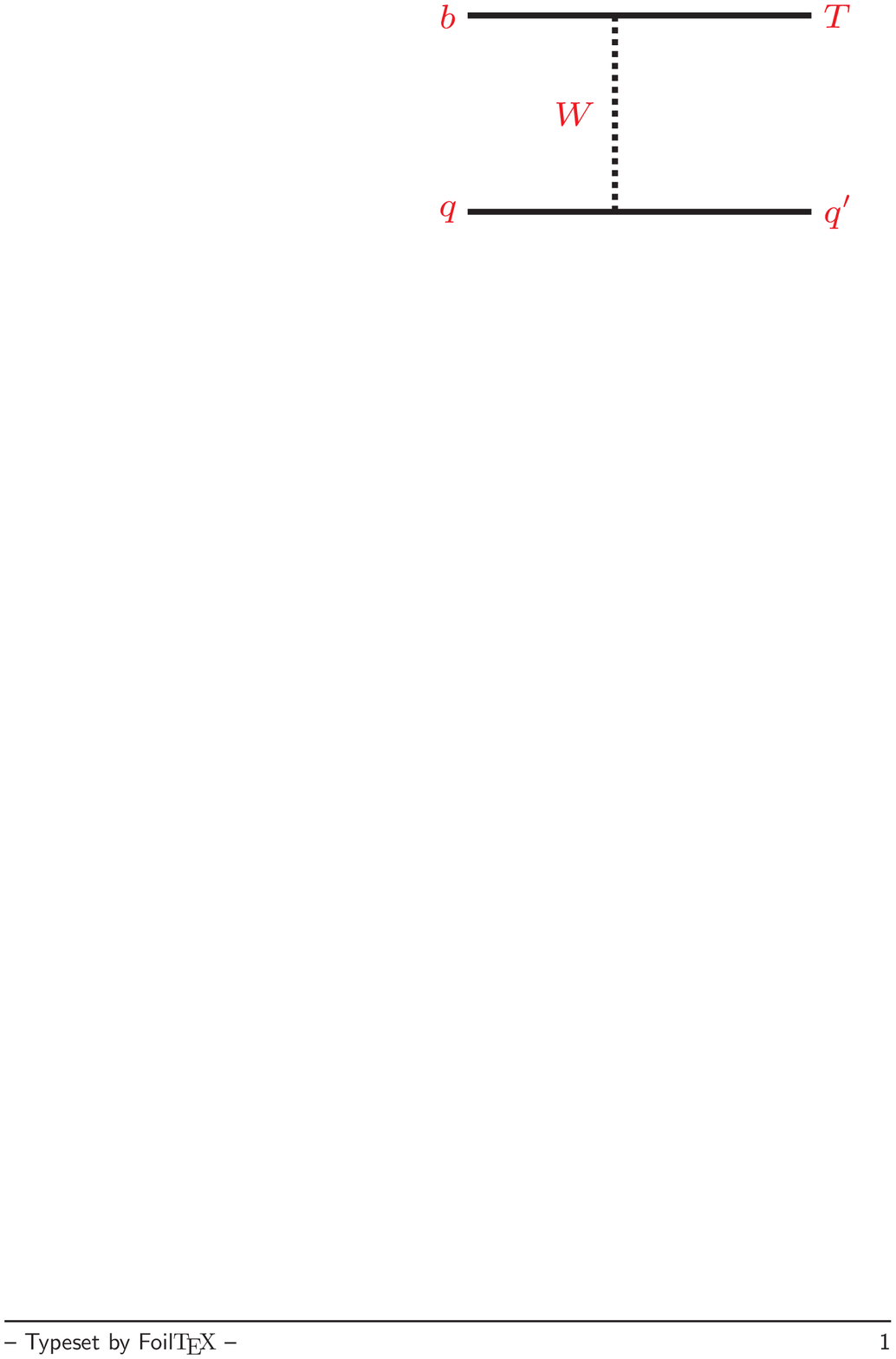}}
\put(7,0){\includegraphics[width=7cm]{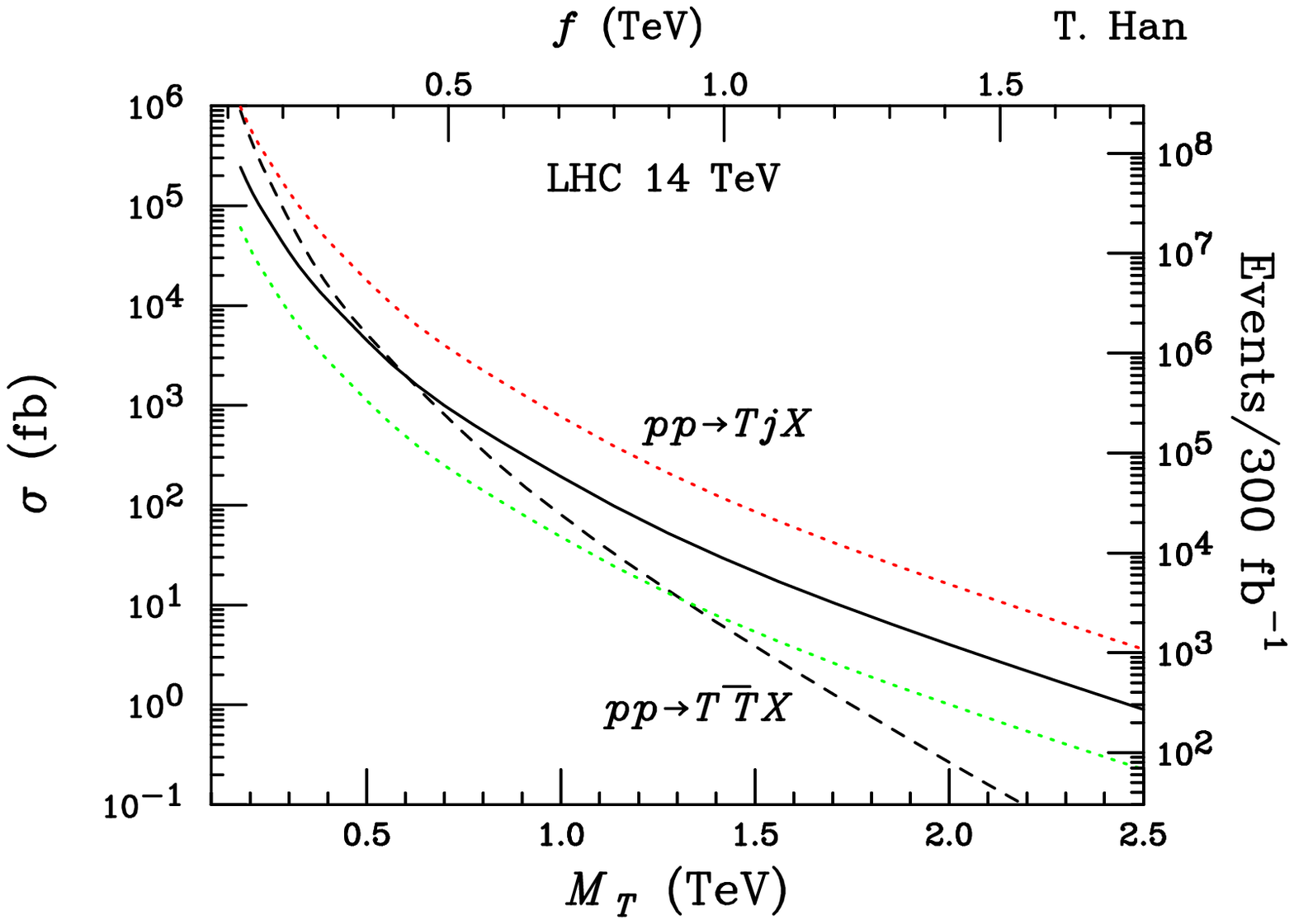}}
\end{picture}
\end{center}
\caption{The tree-level Feynman diagram for  the production of single  
{\it T} and the cross-section  at LHC as a function of its
mass. \label{fig:feyn}} 
\end{figure} 
We have investigated the potential of CMS experiment at LHC to discover some of these 
new particles predicted in the model. We report in the following on detection 
capability and possible discovery of 
singly-produced {\it T} decaying through channels $T\rightarrow th$ and 
$T\rightarrow tZ$, each with a branching fraction of 25\%, where $t, ~h, ~Z$ are the
top quark,  the Higgs scalar and the neutral vector boson in SM. The other mode 
$T \rightarrow bW$, though have 
50\% branching ratio, is 
expected to be overwhelmed at LHC by backgrounds and hence we have not studied it.  

\subsection{Simulation method}
We have  used Pythia subprocess no. 83 for 
$4^{th}$ generation heavy quark  production in t-channel 
process $q_i f_j \rightarrow Q_k f_l$ ($Q_k = T$). Pythia decay table is also modified 
according to model predictions. Background events are various SM processes and they are generated
using Pythia or Alpgen (multi-parton final states) generator. In all cases, CTEQ5L
structure function is used.

The detector response is emulated according to  CMS experimental characteristics 
and using a fast monte carlo package CMSJET where detector effects are 
parametrised using full detector simulation. The jet reconstruction and identification 
is assumed to be possible upto pseudorapidity $|\eta| \le 4.5$ and above transverse 
energy $E_T \ge 40$ GeV.
The b-tagging of jets plays a crucial role and it is effective upto $|\eta| \le 2.5$
with  an efficiency of about 60\%. While considering possible 
bakground events, we note that the signal may be mimicked due to the mis-identification 
of a light quark jet as a b-jet, though the probability is about 1\% only.

We have first done a preliminary study of the kinematics of the signal event at 
particle level after event generation through Pythia as exemplified in Fig.~\ref{fig:kine}.
 The source of the initial 
state b-quark is gluon splitting and being a t-channel process, both the initial 
state quarks are reasonably forward going.
In the final state the particles are highly energetic as expected due to the heavy mass involved; the rapidity distributions of the decay products from {\it T} are
central and the outgoing light quark, though not central, has $|\eta| \le 4.5$ for about 87\% of the cases. Hence this light jet can be used 
for tagging the event selection as will be discussed later. Note, since these are
preliminary studies our selections do not include any threshold on the missing transverse
energy in the event, though it is reasonably high for signal events.  
\begin{figure}
\begin{center}
\unitlength=1cm
{\includegraphics[width=10cm]{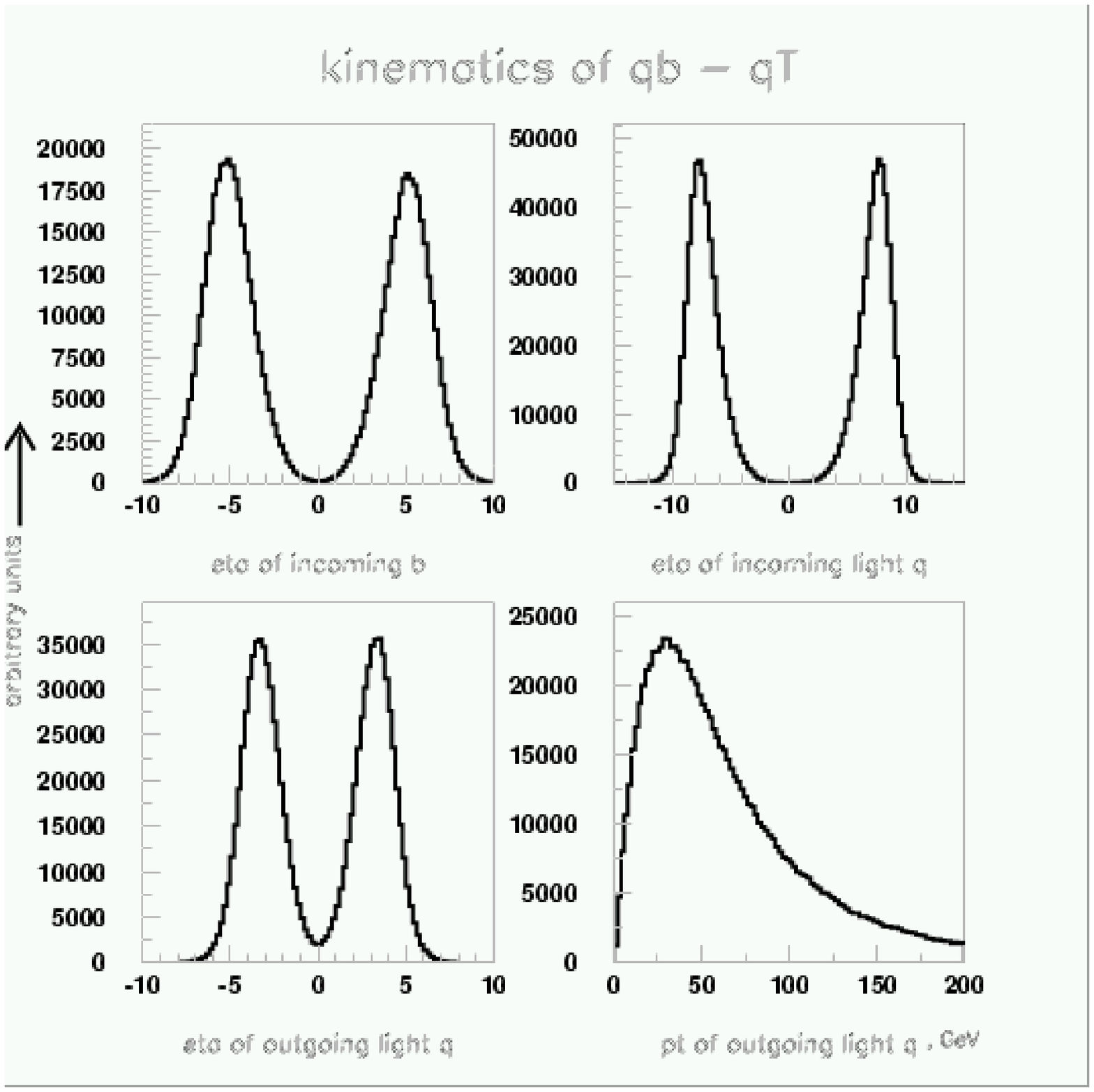}}
\end{center}  
\caption{Some of the kinematic distributions of the initial and final state
  particles in  
$qb\rightarrow q^\prime T$.
\label{fig:kine}}                                  
\end{figure} 

\section{HEAVY TOP SEARCH IN $T\rightarrow ht, ~h\rightarrow b{\bar b}$ MODE}

In this mode we have investigated the situation where the $W$ from top-quark decays 
leptonically, {\it i.e.,} $T\rightarrow ht$, $h\rightarrow b{\bar b}$, 
$t\rightarrow bW, ~~W\rightarrow \ell \nu$ where $\ell = e,~\mu$.
 For a higgs boson mass of 200 GeV/c$^2$, the 
decay branching to $b\bar b$ pair is about 65\%,
so the rate of the final state,  given by the 
signal production cross-section times the relevant branching ratios, is 8.9 fb.
The final state contains
 three b-jets, one isolated lepton, missing transverse energy  and one light-jet.
For background we have considered the following SM processes: 
\begin{itemize}
\item
$pp \rightarrow t{\bar t}$ production with subsequent decays of top-quarks to $bW$ mode. 
We consider here the leptonic decay of one W and the hadronic decay 
of the other. We have used Pythia to generate such events and the leading order 
 cross-section is 435 pb. 
\item
 $pp\rightarrow W b b j j$ with a total cross-section of
3.95 pb, including BR($W\rightarrow \ell\nu$).
\item
 $pp\rightarrow W + 4 jets$ with a 
total cross-section of 146 pb, including BR($W\rightarrow \ell\nu$). 
\end{itemize}
 For processes (2) and (3) we used Alpgen generator and are limited by statistics.

\subsection{Possibility of Trigger}
To judge the trigger viability of the signal events, we resort to CMS-designed  set of criteria for triggering interesting physics events where
the threshold values are decided by the DAQ capability. Triggering with an isolated
lepton (inclusive) is a suitable condition in this case due to the leptonic decay
mode of W. The 
reconstruction efficiencies for electron
and muon events are different and are taken into account to determine
the signal selection
efficiency according to the following set of kinematic criteria.
\begin{itemize}
\item
Electron: $p_T \ge 29$ GeV/c, $|\eta|\le 2.5$ with exclusion of
$1.442 \le |\eta|\le 1.566$
\item
Muon: $p_T \ge 14$ GeV/c, $|\eta|\le 2.1$
\item
Total selection efficiency for electron and muon events are 0.607 and 0.852 resp.ly. These
numbers are also independent of $m_{\rm H}$ in the interesting region.
\end{itemize}

\subsection{Event Selection}
We now discuss the main features of  signal 
{\it vis-a-vis} the main background of $t\bar t$ events after they have passed the 
 detector simulation and the trigger selection criteria. The momentum  distribution 
of the lepton from the
signal $T\rightarrow ht \rightarrow$ 3 b-jets + 1 lepton  + (light-jet) event
is harder compared to that from $t\bar t \rightarrow b\bar b,~W_1\rightarrow l\nu,~W_2\rightarrow jj$ event.
The b-tagged jets are both central and also more energetic. The 4-th jet is typically due to a light quark and is mostly in the higher pseudo-rapidity region as expected for signal events.
We have applied the following set of  criteria to select signal events;  the succesive 
efficiency factors are also quoted.
\begin{itemize}
\item
1 isolated lepton at trigger level, efficiency = 0.73 .
\item
Total no of jets with transverse energy $E_T \ge 60$ GeV in the event = 4, among which 
three most energetic jets  are b-tagged,  efficiency = 0.007 .
\item
First 3 jets should have minimum transverse energy $E_T \ge ~200, ~125,~80$ Gev, $|\eta| \le 2.3$ and
4th jet should be in the forward/backward region $|\eta| \ge 3.0$,
efficiency = 0.29 .
\item
Transverse momentum of the isolated lepton, ~$p_T^\ell \ge 40$ GeV/c, efficiency = 0.82 .
\end{itemize}
Combining all we get a grand efficiency of 0.12\%, whereas the signal cross-section is 
8.9 fb. Hence we conclude that the signal rate is too small to be observed even with large 
accumulated luminosity.

The above selection criteria could be relaxed and optimised by demanding, for example,
 less number of b-tagged jets. Evidently with a demand of three b-tagged jets very few events from 
the background processes could survive due to jet mistagging, but none survived all the
selections. This situation may change if we 
require only one or two b-tagged jets. We should also generate events of type $t\bar t$ +jets using Alepgen.

To make our exercise complete,  the mass of the higgs boson
from the two b-jets with minimal separation, $\Delta R = \sqrt(\Delta\eta ^2 + \Delta\phi ^2)$, is reconstructed and  subsequently 
the transverse mass of {\it T} also can be reconstructed. This can finally lead to 
the estimation of the model parameter. We note here that the distance 
 between the two b-quarks from the higgs boson decreases with increasing $m_{\rm H}$ due to higher boost.

\section{HEAVY TOP SEARCH IN $T\rightarrow Zt, ~Z\rightarrow \ell^+\ell^-$ ~($\ell =e,\mu$) MODE}
The complete final state considered is 
$q~b\rightarrow  q^\prime ~T, ~T\rightarrow Zt$, ~
$Z\rightarrow \ell^+\ell^-, t\rightarrow bW, ~~W\rightarrow \ell \nu$  and so there are 
3 isolated
charged leptons, one b-jet, one light jet and missing transverse energy. The cross-section is 0.92 fb. 
The main backgrounds are SM events of type WZ+jets, with a  cross-section of 91 fb, where
 both  ${\rm Z}$ and 
${\rm W}$ decay leptonically and the light-jet is mis-tagged as b-jet. 
As expected the leptons
from the signal event have harder momentum distribution.

\subsection{Trigger}
The signal can be triggered in di-lepton mode with the following criteria.
\begin{itemize}
\item
Electron: $p_T \ge 17$ GeV/c, $|\eta|\le 2.5$ with exclusion of
$1.442 \le |\eta|\le 1.566$
\item
Muon: $p_T \ge 3$ GeV/c, $|\eta|\le 2.1$
\item
For two leptons of same flavour and opposite signs, to satisfy the criteria,
 the efficiency for electron type events is 0.77 and for muon type it is 0.92 .
\end{itemize}                                                                  

\subsection{EVENT SELECTION}
The event selection criteria are discussed below with corresponding signal efficiencies 
for subsequent requirements.
\begin{itemize}
\item
3 isolated leptons with any two passing di-lepton trigger criteria = 0.83
\item
Maximum 2 jets in the event with $E_T \ge 40$ GeV, among which  one is b-tagged, efficiency = 0.049
\item
Highest $E_T$ jet has transverse energy  $E_T\ge 100$ GeV,  efficiency=0.86
\item
Invariant mass of 2 leptons, of same flavour and  opposite sign 
$M_{ll} \ge 80$ GeV/c$^2$, efficiency = 0.98
\end{itemize}

The WZ background, though large, is tamed by the requirement of accompanying hard b-jet and so no event finally survives the full set of requirements.
The grand efficiency is 0.035 and corresponding to a cross-section of 0.92 fb
we expect about 3.2 events with an integrated luminosity of 100 fb$^{-1}$.
This gives an encouraging result of signal-to-background ratio of 6.4. By
tuning the event selection criteria a better significance can be expected.

\section{CONCLUSION}
The {\it Little Higgs} model predicts the existence of a top-like heavy quark of mass
within the range of few TeV. We have studied the experimental signature of single 
production of this 
particle  at LHC and subsequent decays in modes 
$T\rightarrow ht, ~~h\rightarrow b{\bar b}$ and 
$T\rightarrow Zt, ~Z\rightarrow \ell^+\ell^-$ where $\rm W$ from top-quark decays leptonically.
Preliminary studies show that for the first channel, a requirement of 3 b-tagged jets lead to too low an efficiency for the signal. But the second channel is encouraging and could be even 
more promising when we consider the hadronic mode of $\rm W$ since the event will be triggered by the leptons from $\rm Z$ anyway. The study of the channel where $\rm W\rightarrow$ 2-jets is under progress. Of course  the background is expected to be larger. We plan to study
the potential of CMS experiment for the heavy bosons soon.

\section*{ACKNOWLEDGEMENTS}

I would like to thank the  organizers  of the workshop for the kind hospitalilty and the colleagues from ATLAS collaboration for their friendly help regarding
signal event generation.

\setcounter{figure}{0}
\setcounter{table}{0}
\setcounter{section}{0}
\setcounter{equation}{0}
\clearpage







\part{Z$'$ studies at the LHC: an update \label{zprime}}

{\it M. Dittmar, A. Djouadi, A.-S. Nicollerat}



\section{INTRODUCTION}

The LHC discovery potential for a $Z'$ in the  reaction $pp\rightarrow Z'
\rightarrow \ell^+\ell^-$ with $\ell=e,\mu$ is well known. As shown in previous
studies, a $Z'$ with a mass up to 5~TeV could be discovered at the LHC with
100~fb$^{-1}$. We make here a summary of the detailed work described
in~Ref.~\cite{Dittmar:2003ir} showing how, after a $Z'$ signal has been detected at
the LHC,  one could identify it. In contrast to previous studies,  where the
models were either analyzed from a more theoretical point of view or a
particular model was analyzed within a certain experimental frame, we combine
here different experimental observables in order to  investigate the realistic
potential of the LHC experiments to distinguish between models and determine
their parameters. In this study, two classes of $Z'$ models are considered:
$E_6$ models, parametrized with $\cos\beta$ and left--right (LR) models,
parametrized with $\alpha_{LR}$ (see~Ref.~\cite{Djouadi:1992sx}, that we will follow, 
for a theoretical account). 

\section{OBSERVABLES SENSITIVE TO $Z'$ PROPERTIES}

Future measurements of $Z'$ properties at the LHC can use the following
observables: 

\textit{The total decay width of the $Z'$} which is obtained from a fit to the
invariant mass distribution of the reconstructed dilepton system using a
non--relativistic Breit--Wigner function: $a_0/[(M_{\ell \ell}^2-M_{Z'}^2)^2+
a_1]$ with  $a_1 = \Gamma_{Z'}^2 M_{Z'}^2$. 

\textit{The $Z'$ cross section times leptonic branching ratio} which
is calculated from the number of reconstructed dilepton events lying within $\pm 3\Gamma$
around the observed peak. 

\textit{The leptonic forward-backward asymmetry}  $A_{FB}^{\ell}$, which is
defined from the lepton angular distribution with respect to the quark 
direction in the center of mass frame, as:
\begin{equation}
\mathrm{\frac{d\sigma}{d \cos \theta ^*} \propto \frac{3}{8} (1 + \cos^2 \theta ^*)
+ A_{FB}^{\ell} \cos \theta ^*}
\end{equation}
$A_{FB}^{\ell}$ can be determined with an unbinned maximum likelihood fit to
the $\cos \theta ^*$ distribution.  Unfortunately, as the original quark
direction in a proton-proton  collider is not known, $A_{FB}^{\ell}$  cannot be
used directly. However, it can be extracted from  the kinematics of the
dilepton system, as was shown in detail in~\cite{Dittmar:1997my}. The initial quark
direction is assumed to be  the boost direction of the $\ell \ell$ system with
respect to the beam axis. The probability to assign the correct quark direction
increases for larger rapidities of the  dilepton system. A purer, though
smaller signal sample, can thus be obtained by introducing a rapidity cut. For
the following study we will require $|Y_{\ell\ell}|>0.8$.

\textit{The $Z'$ rapidity distribution}:  To complete the $Z'$ analysis, one
can obtain some information  about the fraction of $Z'$'s produced from
$u\bar{u}$ and $d\bar{d}$ by analyzing the $Z'$ rapidity distribution. 
Assuming that the $W^{\pm}$ and $Z$ rapidity distribution has been measured  in
detail, following the ideas given in~\cite{Dittmar:1997md}, relative parton
distribution functions for $u$ and $d$ quarks as well as for the corresponding 
sea quarks and antiquarks are well known. Thus, the rapidity spectra can be
calculated separately for  $u\bar{u}$ and $d\bar{d}$ as well as for  sea quark
anti-quark annihilation and for the mass region of interest to analyze the $Z'$
rapidity distribution~\cite{delAguila:1993ym}.  Using these distributions a fit can be
performed to the $Z'$ rapidity distribution which allows to obtain the
corresponding fractions of $Z'$'s produced from $u\bar{u}$, $d\bar{d}$ as well
as for  sea quark anti-quark annihilation.  This will thus reveal how the $Z'$
couples to different quark flavors in a particular model.

\section{DISTINCTION BETWEEN MODELS AND PARAMETER DETERMINATION}
  
In the present analysis, {\tt PYTHIA} events of the type $pp \to \gamma,Z,Z'
\to ee,\mu\mu$ were simulated at a center of mass energy of 14~TeV, and for
different $Z'$ models. These events were analyzed using simple acceptance cuts
following  the design criteria of {\tt ATLAS} and {\tt CMS}.  The SM background
relative to the signal cross section is found to  be essentially negligible for
the considered $Z'$ models. We thus reconfirm the known $Z'$ boson LHC
discovery potential,  to reach masses up to about 5~TeV for a luminosity of 100
fb$^{-1}$ \cite{Djouadi:1992sx}. 

Let us discuss how well one can distinguish experimentally the different $Z'$
models  using the observables as defined before: $\sigma^{3\Gamma}_{\ell\ell}
\times \Gamma$, $A_{FB}^{\ell}$ as well as $R_{u\bar{u}}$ as obtained from the
rapidity distribution.  As a working hypothesis, a luminosity of 100~fb$^{-1}$
and a $Z'$ mass of 1.5~TeV  will be assumed in the following. 

A precise knowledge of the cross section and the total width allows to make a
first good distinction between the different models  as we will be discussed
later.  It is not obvious how accurate cross sections can  be measured and
interpreted at the LHC. Following the procedure outlined in~\cite{Dittmar:1997md},  an
accuracy of  $\pm$1\% could be assumed. It is however necessary to consider the
other observables. 

Very distinct forward backward charge asymmetries are expected as a function of
the dilepton mass and   for the different $Z'$ models, as shown in
Figure~\ref{modelscomp}a. One finds that additional and complementary
informations are also obtained  from $A_{FB}^{\ell}$ measured in the
interference region. The $Z'$ rapidity distribution is also analyzed. 
Figure~\ref{modelscomp}b shows the expected rapidity distribution  for the
$Z'_{\eta}$ model. A particular $Z'$ rapidity distribution is fitted using a
linear combination of the three pure quark-antiquark rapidity distributions.
The fit output gives the  $u\bar{u}$, $d\bar{d}$ and $sea$ quarks fraction in
the sample. In order to demonstrate the analysis power of this method  we also
show the $Z'_{\psi}$ rapidity distribution which has equal couplings to 
$u\bar{u}$ and $d\bar{d}$ quarks. 

\begin{figure}[htbp]
\vspace*{-2mm}
\begin{center}
\mbox{
\includegraphics[width=0.52\textwidth,height=8cm]{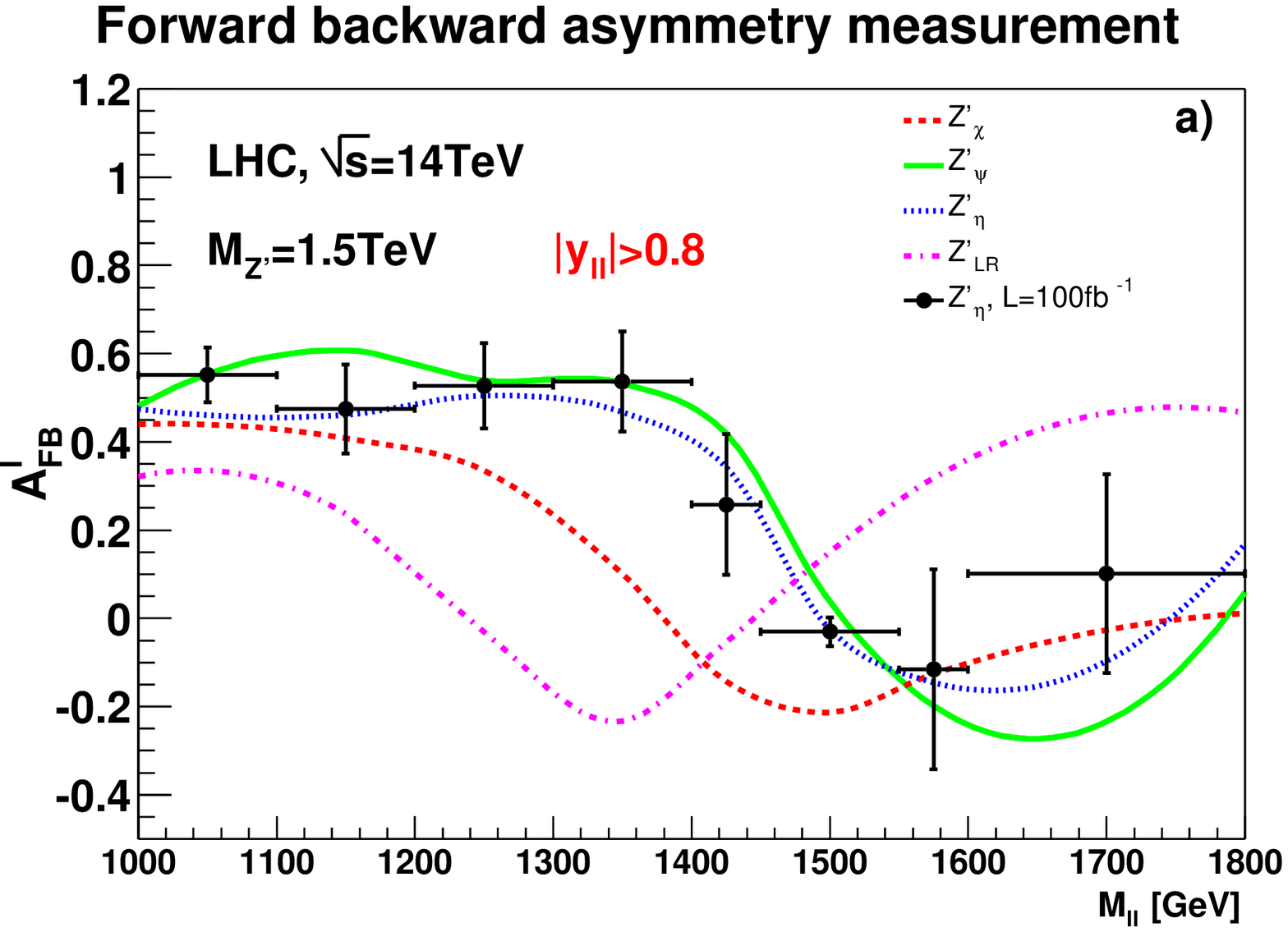}
\includegraphics[width=0.52\textwidth,height=8cm]{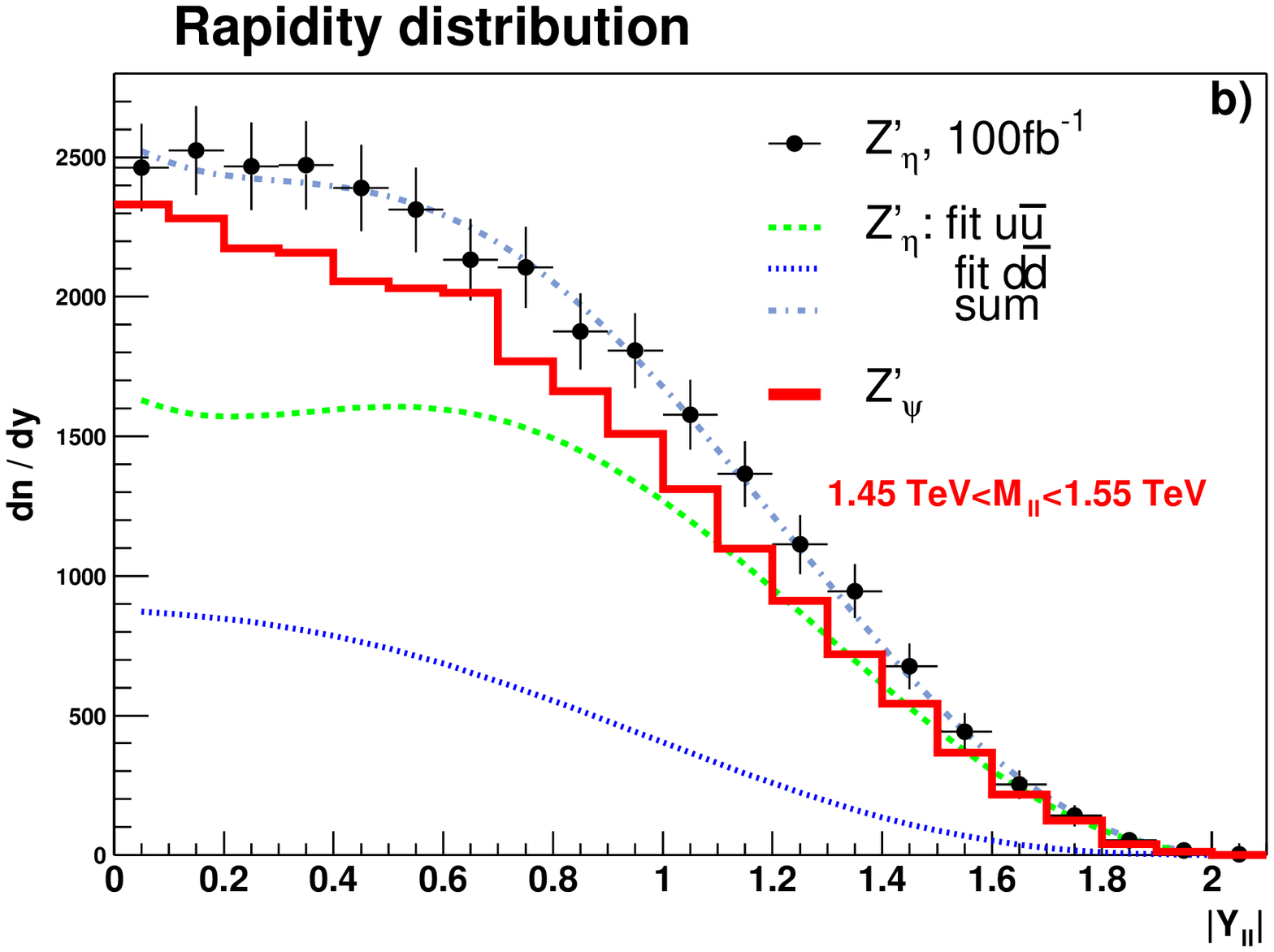}
}
\end{center}
\vspace*{-7mm}
\caption{\it $\mathrm{A^{\ell}_{FB}}$ (a) as a function of $M_{\ell \ell}$ for
four $Z'$ models.  The rapidity of the dilepton system is required to be larger
than 0.8. The observable rapidity distribution for two $Z'$ models is
shown in (b),  including the fit results which determine the  types of
$q\bar{q}$ fractions. A simulation of the statistical errors, including random
fluctuations of the  $Z'_{\eta}$ model and with errors  corresponding to a
luminosity of 100~fb$^{-1}$ has been included in both plots.}

\label{modelscomp}
\vspace*{-3mm}
\end{figure}

In a next step, assuming that a particular model has been selected, one would
like to know  how well the parameter(s), like $\cos\beta$ or $\alpha_{LR}$ can
be constrained.  Figure~\ref{varyparam} shows how the previously defined
observables vary as the model parameters are varied. In the case of the $E_6$
model for instance, one finds that $\cos\beta$ can not always be determined
unambiguously. Very similar results can be expected for different observables 
but using very different values for $\cos\beta$. Obviously, the combination of
the various measurements, helps to reduce some ambiguities.

\begin{figure}[htbp]
\begin{center}
\vspace*{-7mm}
\mbox{
\includegraphics[width=0.5\textwidth, height=5.5cm]{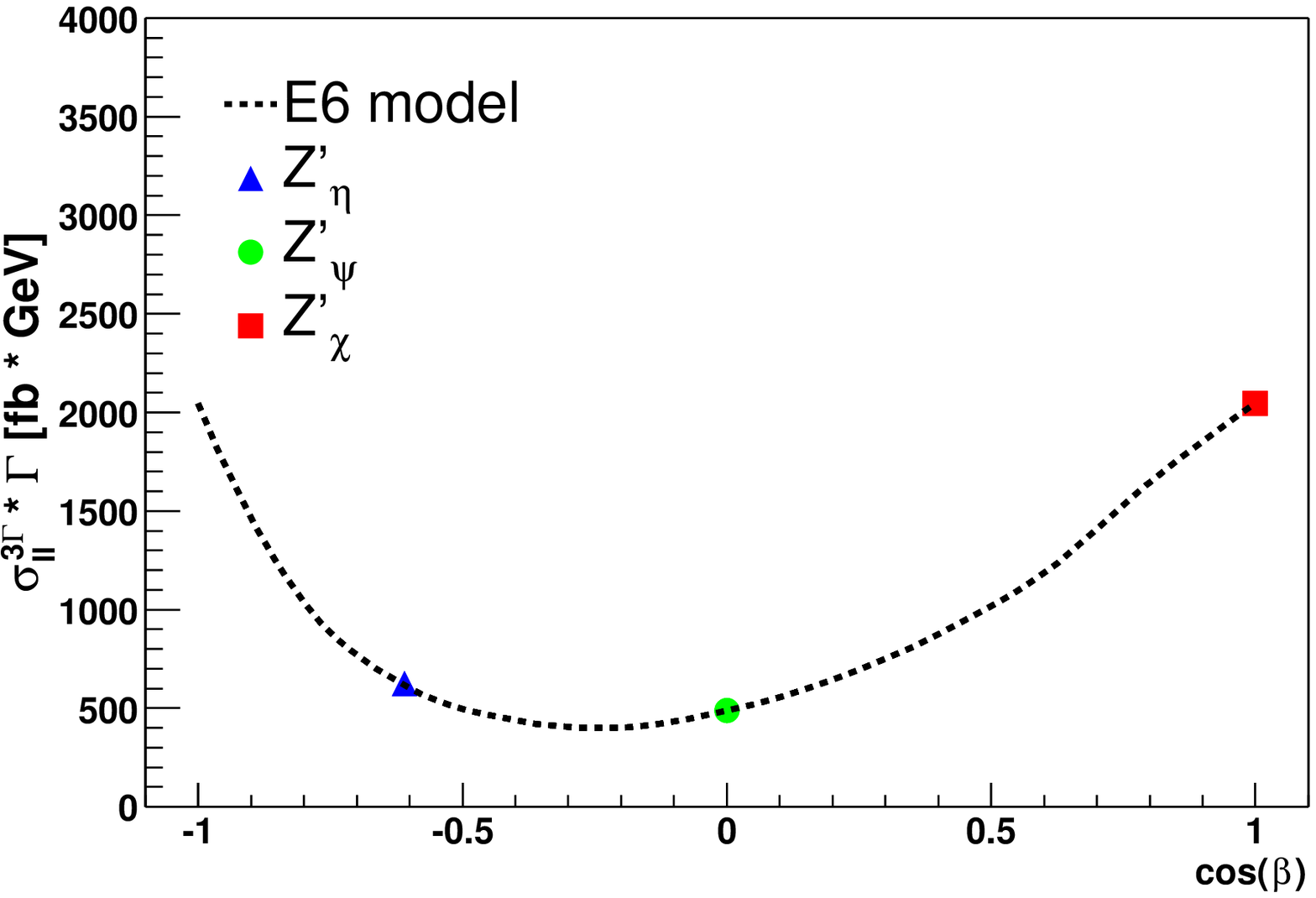}
\includegraphics[width=0.5\textwidth, height=5.5cm]{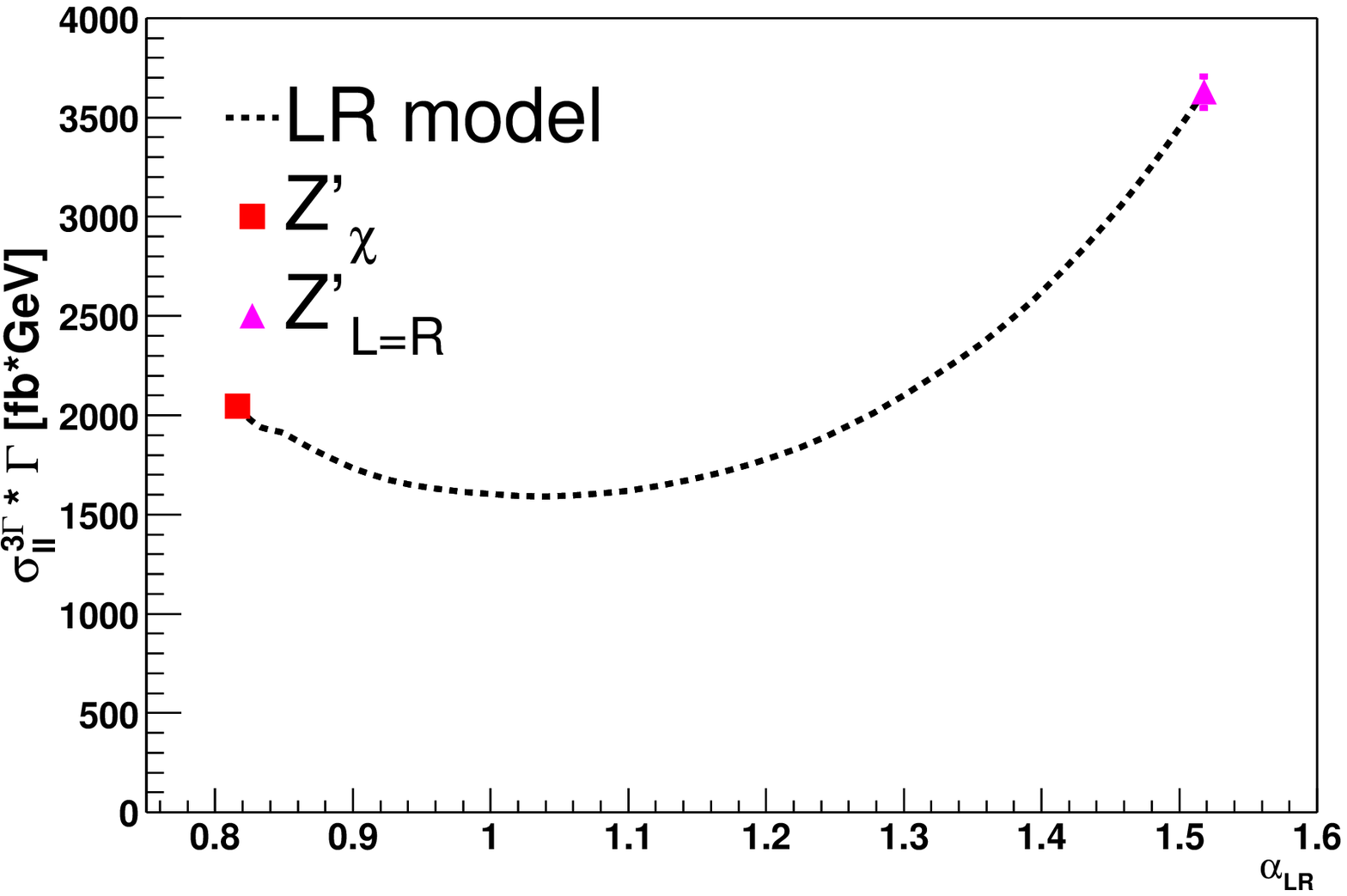}
}
\vspace*{-3mm}
\mbox{
\includegraphics[width=0.5\textwidth, height=5.5cm]{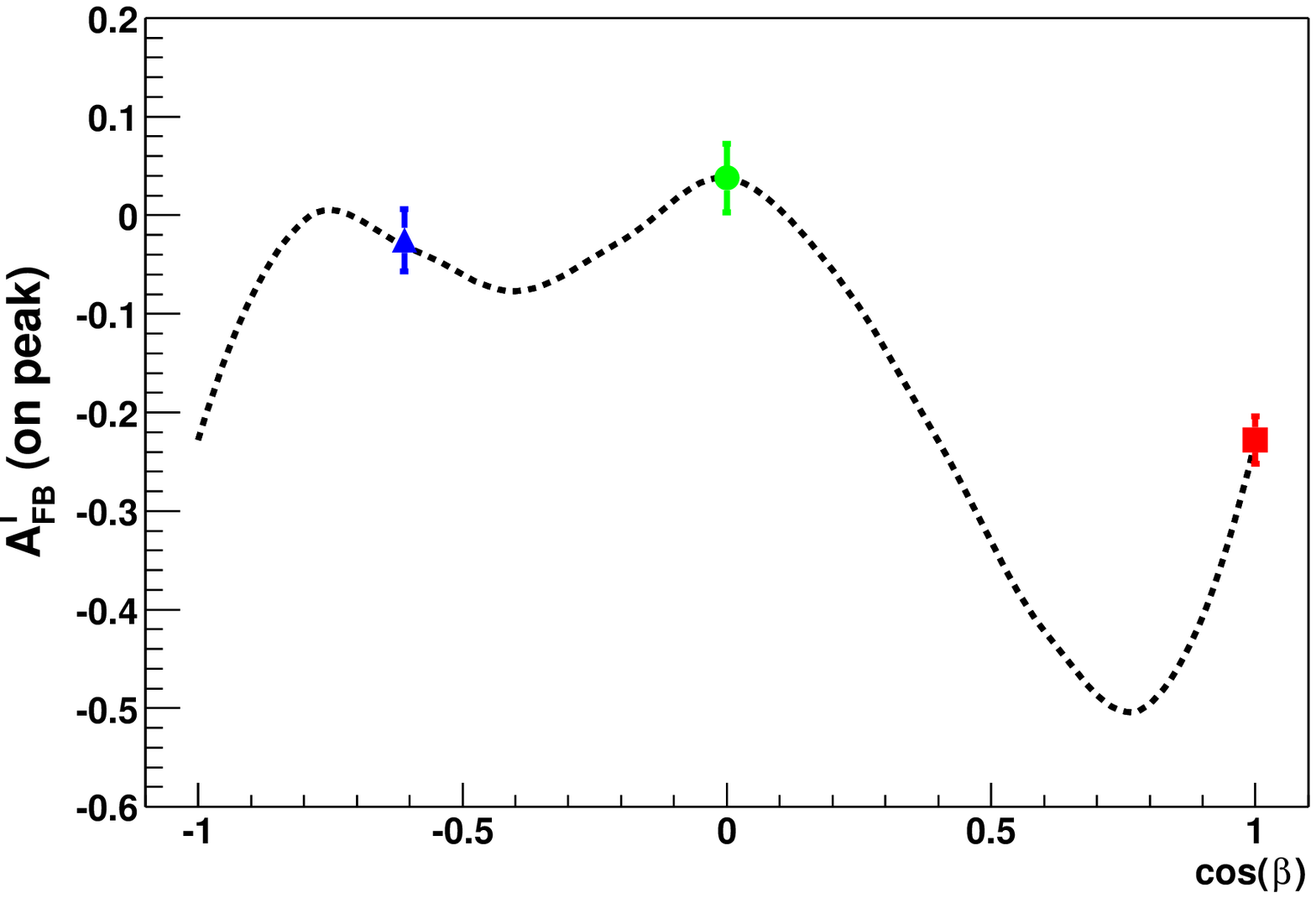}
\includegraphics[width=0.5\textwidth, height=5.5cm]{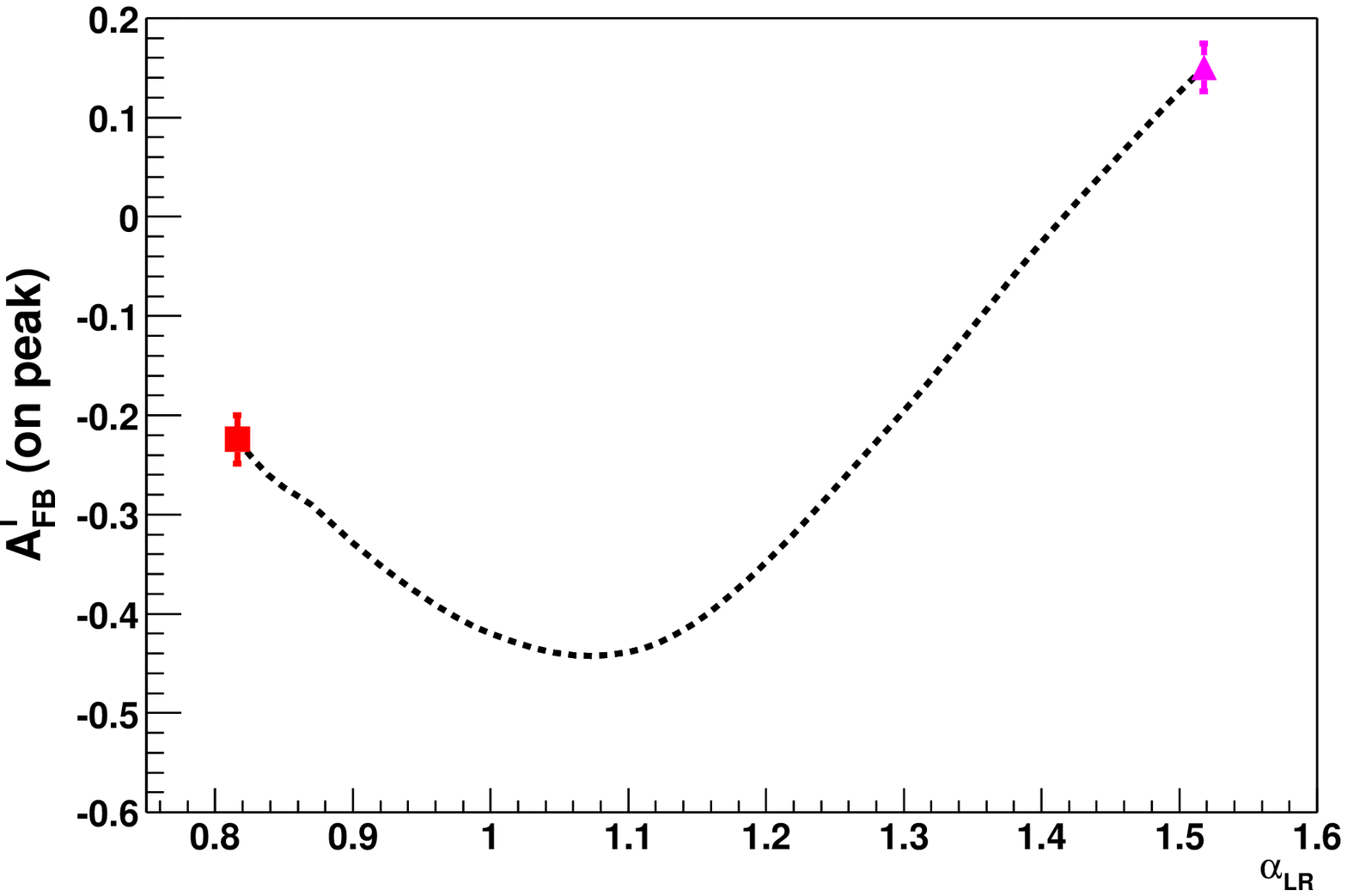}
}
\vspace*{-3mm}
\mbox{
\includegraphics[width=0.5\textwidth, height=5.5cm]{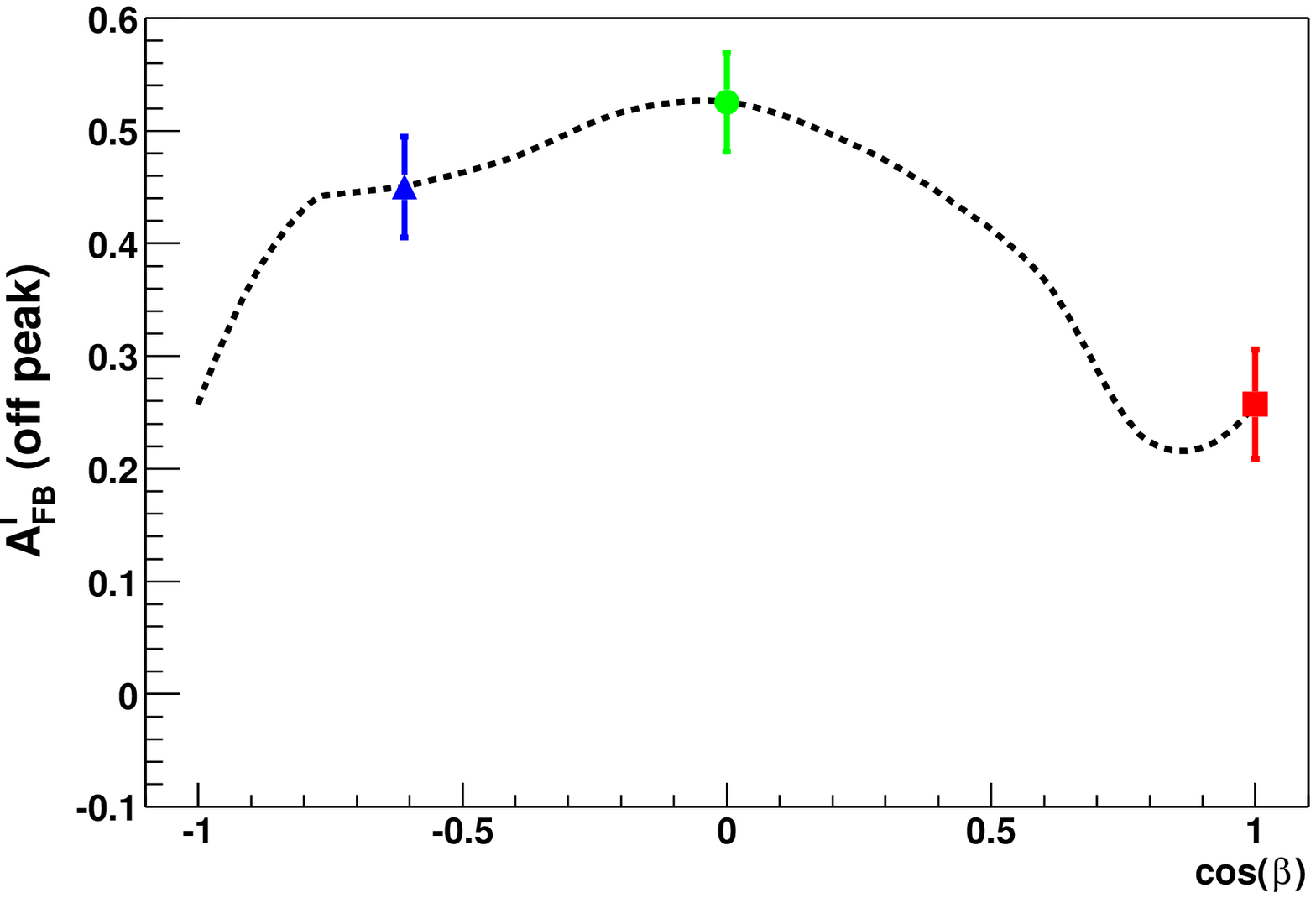}
\includegraphics[width=0.5\textwidth, height=5.5cm]{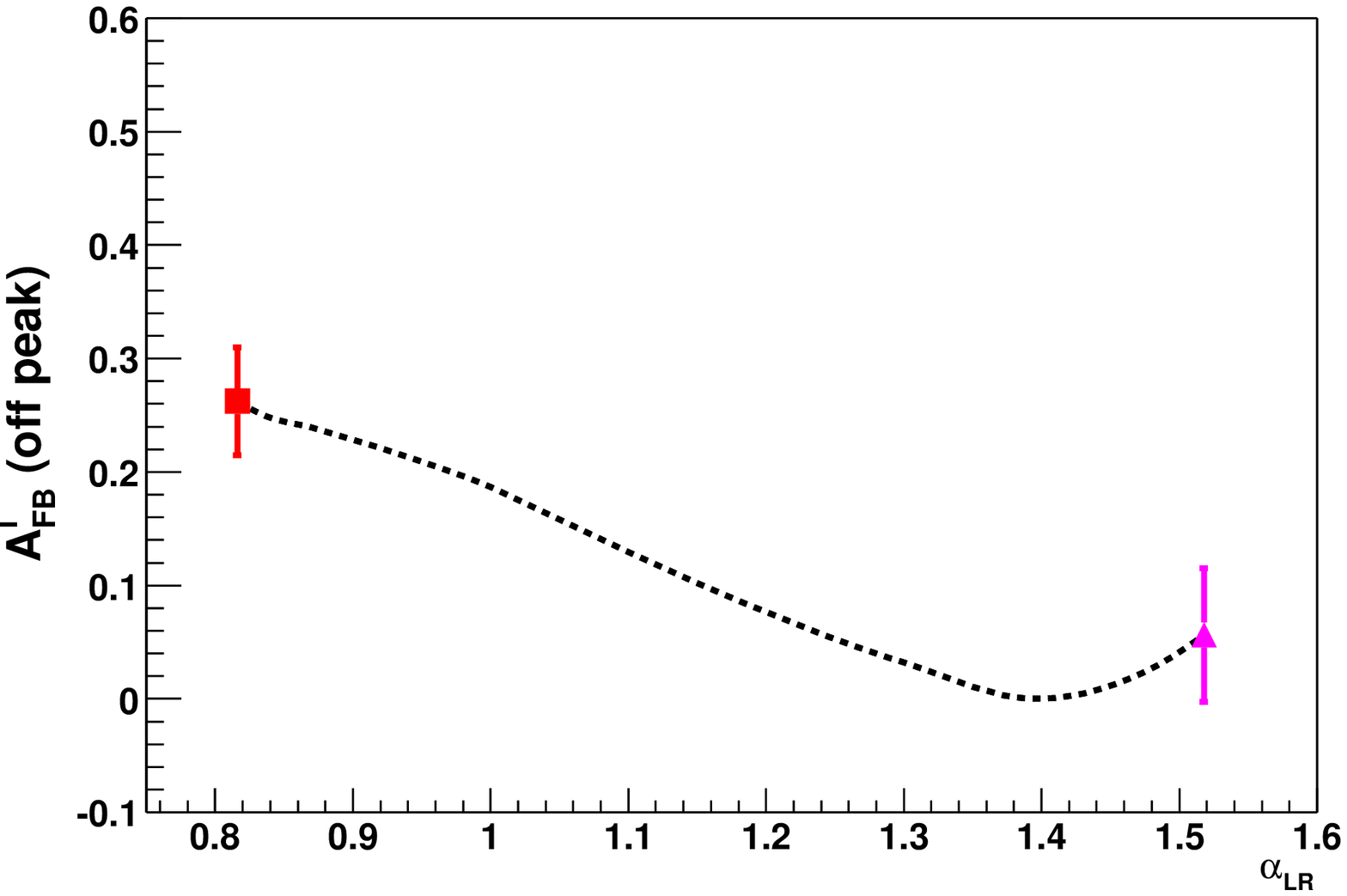}
}
\vspace*{-3mm}
\mbox{
\includegraphics[width=0.5\textwidth, height=5.5cm]{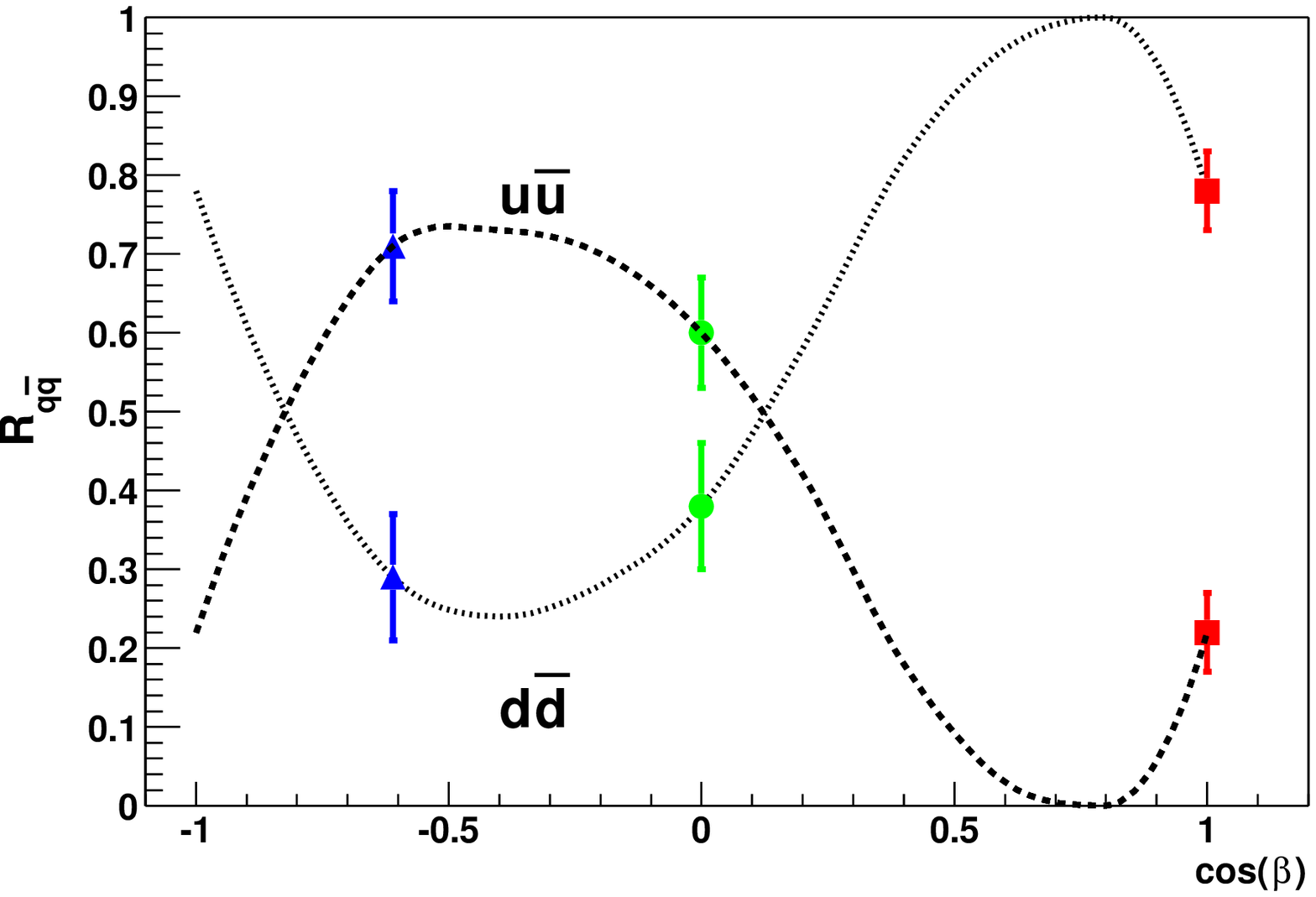}
\includegraphics[width=0.5\textwidth, height=5.5cm]{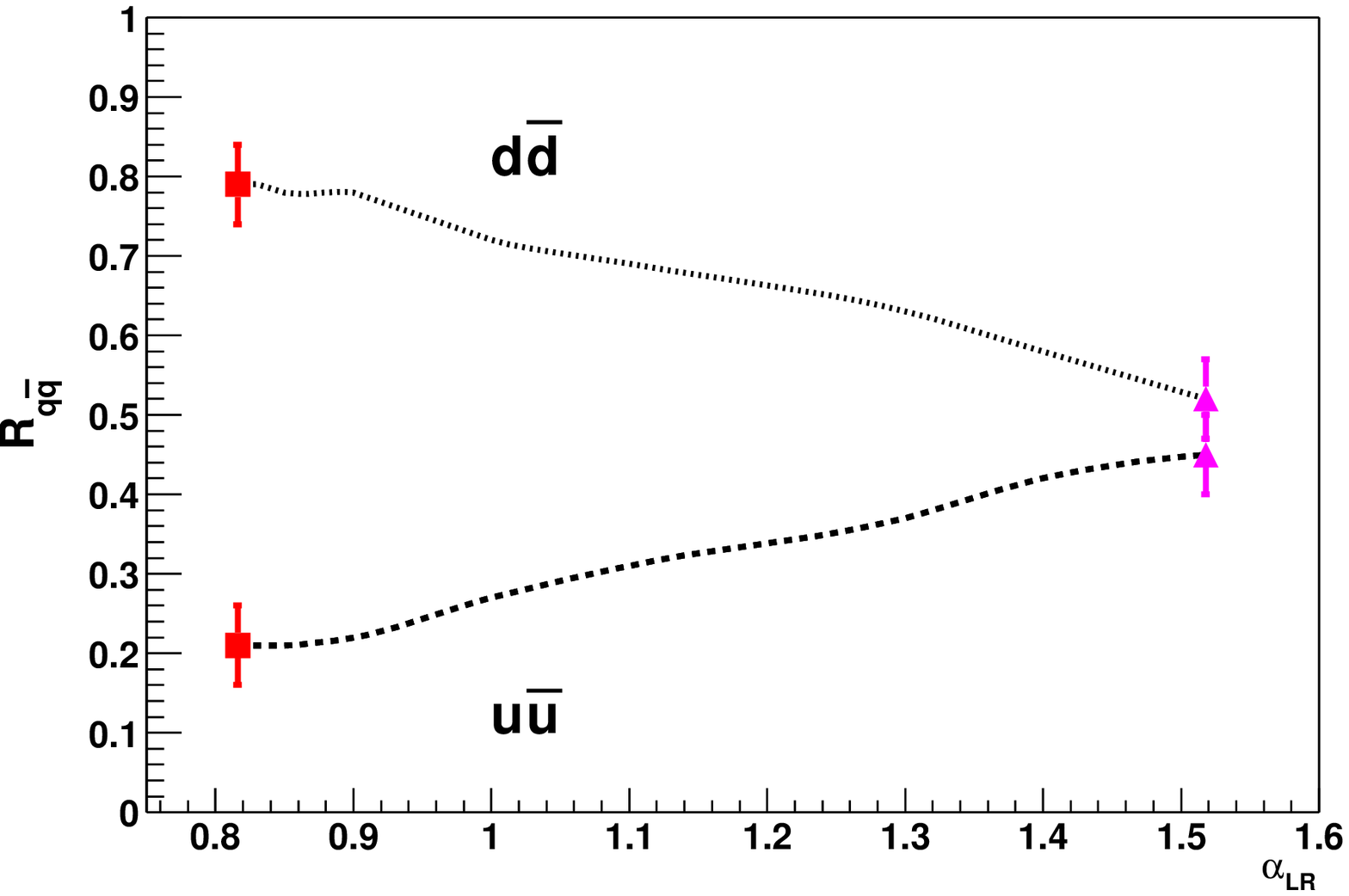}
}
\end{center}
\vspace*{-1mm}
\caption{\small \it Variation of $\sigma^{3\Gamma}_{\ell\ell} \cdot \Gamma$, $A^{\rm on-peak}_{FB}$,
$A^{\rm off-peak}_{FB}$ and the ratio $R_{q\bar{q}}$ as a function of the $E_6$
model parameter $\cos\beta$ (left) and the LR-model parameter $\alpha_{LR}$. The
points corresponding the  particular $Z'$ models are also shown. }
\label{varyparam}
\end{figure}

If the $Z'$ mass is increased, the number of events decreases drastically and
the differences between the models start to become covered within the statistical
fluctuations. For the assumed luminosity of 100~fb$^{-1}$, one could still distinguish a
$Z'_{\chi}$ from a $Z'_{LR}$ over a large parameter range and the 
$A_{FB}^{\ell}$ measurements provide some statistical significance up to $M_{Z'}=2-2.5$~TeV.
On the contrary, a $Z'_{\eta}$ could be differentiated from a $Z'_{\psi}$ only
up to a $Z'$ mass of at most 2~TeV, as in that case, the dependence of $A_{FB}^{\ell}$
on the $Z'$ mass is almost identical for the two models.

\section{CONCLUSIONS}

A realistic simulation of the study of the properties of  $Z'$ bosons in $E_6$
and LR models has been performed for the LHC. We have shown that, in addition 
to the $Z'$ production cross section times total decay width,  the measurement
of the forward-backward lepton charge asymmetry, both on the $Z'$ peak and in
the interference region, provide complementary information. We have also shown
that a fit of the rapidity distribution can provide a sensitivity to the $Z'$
couplings to up-type and down-type quarks. The combination of all these
observables would allow us to discriminate between $Z'$ bosons of different
models or classes of models for masses up to 2--2.5 TeV, if a luminosity of 100
fb$^{-1}$ is collected.

\setcounter{figure}{0}
\setcounter{table}{0}
\setcounter{section}{0}
\setcounter{equation}{0}
\clearpage

\providecommand{\href}[2]{#2}\begingroup\raggedright\endgroup
\end{document}